\pdfoutput=1
\documentclass[12pt]{article}
\usepackage{nameref}

\usepackage{amsmath,amsfonts,amsthm,amssymb}
\newcommand{\K}{\mathcal{K}}
\newcommand{\prop}{S}
\newcommand{\base}{b}
\newcommand{\ratio}{E}

\usepackage{newtxtext,newtxmath}

\usepackage{graphicx}

\usepackage[letterpaper,margin=1in]{geometry}

\renewenvironment{abstract}
	{\quotation}
	{\endquotation}

\date{}

\makeatletter
\renewcommand{\fnum@figure}{\textbf{Figure \thefigure}}
\renewcommand{\fnum@table}{\textbf{Table \thetable}}
\makeatother

\usepackage{scicite}

\usepackage{url}

\makeatletter
\newcommand{\continuedfigure}{%
	\addtocounter{figure}{-1}%
	\renewcommand{\fnum@figure}{\textbf{Figure \thefigure, continued}}%
}
\newcommand{\continuedtable}{%
	\addtocounter{table}{-1}%
	\renewcommand{\fnum@table}{\textbf{Table \thetable, continued}}%
}
\makeatother

\title{\bfseries Filling holes in science draws collective attention, but most higher-order holes remain unexplored}

\author{
	Jiajie~Luo$^{1}$\and
	James A.~Evans$^{1, 2, *}$\and
	\small$^{1}$Knowledge Lab, University of Chicago, Chicago, Illinois, United States of America.\and
	\small$^{2}$Santa Fe Institute, Santa Fe, New Mexico, United States of America.\and
	\small$^\ast$Corresponding author. Email: jevans@uchicago.edu\and
}

\begin{document}

\maketitle

\begin{abstract} \bfseries \boldmath

Most scientific discoveries justify their significance by filling holes between existing ideas and arguments. Representing knowledge as high-dimensional concept embeddings, we use persistent homology to detect scientific holes of increasing order, from lower-order gaps between disconnected ideas to higher-order cavities within unevenly connected ones; then we identify the research that fills them. Researchers who close holes draw outsized collective attention, staging novelty that the citing literature rewards, most strongly in discovery-oriented fields and least in design-oriented fields. Closing a hole is followed by persistent work on the same combinations of concepts at many times the rate of comparable holes left open. Yet as knowledge grows, higher-order holes explode while the fraction of research that fills them collapses, leaving most higher-order combinations unexplored. These results call for a richer science of holes and mark a frontier where contemporary AI might help fill the gaps human science opens.
\end{abstract}

\noindent

\noindent
Science advances by finding and filling ``holes'' in the fabric of our understanding about the world \cite{shi2015weaving}. A new result connects ideas that stood apart, resolves an anomaly, or bridges a question earlier work left open. In many cases, discovery closes a hole in the fabric of what is known \cite{kuhn1962, foster2015}. Most advances are recombinant, joining concepts that had not previously been brought together \cite{weitzman1998, youn2015}: papers built on atypical combinations of prior work are disproportionately likely to become highly cited \cite{uzzi2013, shi2023surprising}, and new discoveries lie latent in the unmade connections between disconnected literatures \cite{swanson1986,evans2010machine}.
Yet despite the centrality of gaps to how we describe scientific progress, systematic methods for locating them have only recently emerged \cite{kedrick2025, yadav2025}, with limited work that analyze gaps of different order or assessing the impact of works that close those gaps.
A rich science of discovery requires a sophisticated science of holes.

Representing knowledge geometrically brings these gaps into sharper view. When scientific concepts are embedded as points in a high-dimensional space constructed from co-use across the global corpus of scientific research \cite{mikolov2013}, the distance between two points reflects how rarely the concepts have been used together, directly and indirectly. Thus, empty regions of the space correspond to combinations of concepts that have not yet been realized. Such embeddings are not merely descriptive. Neural network models trained on the materials-science literature encode latent relationships that anticipate discoveries years before they are reported \cite{tshitoyan2019,sourati2023accelerating}.
A hole in the embedding is therefore a candidate for future work, and a work that ``fills'' it is one whose concepts span the void, knitting the surrounding region together.

Filling such a gap often involves an act of anticipation. Scientists extend the dense, well-trodden frontier of their fields by choosing problems adjacent to what is already known and avoiding distant leaps \cite{kauffman2000investigations,foster2015, rzhetsky2015}. Against this incrementally evolving backdrop, a work that fills a large void to connect concepts scientific fields have kept apart is a wager that connections nonobvious to those fields will come to matter to them, catalyzing a convergence of collective attention. Those who place such bets well, we will show, are rewarded: they anticipate and fill holes the community is poised to recognize.

Whether the gaps that go unfilled represent a genuine loss depends on how discovery accumulates. If the value of a higher-order combination is built up from its lower-order parts---if profitable pairs lead to profitable triples, and so on---then incremental science, filling the smallest holes first, will ladder its way toward higher combinations in time. But if some combinations are emergent, functional only when several elements are joined at once and offering no advantage in smaller subsets, then no gradient of partial success will point toward them, and a science that explores the landscape gap by gap will systematically pass them by. We will show that the prevalence of higher-order holes explodes as conceptual space grows, but they are rarely filled. We return to this contrast and its consequences in the Discussion.

To make holes precise, we turn to persistent homology, a tool from topological data analysis built to detect holes in data and measure their persistence or robustness \cite{carlsson2009, edelsbrunner2010, ghrist2008, zomorodian2005}. The method grows a ball of radius $r$ around each concept and connects concepts whose balls overlap. As $r$ increases, the concept cloud knits together through a nested sequence of complexes, and holes of successive order appear and then close (see Figure 1). A $0$-dimensional hole is a gap between two disconnected concepts, closed when a single discovery links them into a single component, thereby bridging a one-dimensional gap with an edge. A $1$-dimensional hole is a loop of connected concepts encircling an empty region, closed when a work spans three concepts at once, tiling a two-dimensional gap with the area of a triangle. A $2$-dimensional hole is a shell of concepts enclosing an empty cavity, closed when a work brings four concepts together, filling the three-dimensional void with the volume of a tetrahedron. Higher-dimensional holes exist, and trace the simultaneous discovery of profitable connections between five, six, seven, and more distant concepts.

Persistent homology records, for every hole, the radius at which it forms (the ``birth'' simplex) and the radius at which it is filled (the ``death'' simplex), the latter measuring the size of the conceptual leap required to close it.
The death radius therefore measures the size of the conceptual leap a hole demands. Among holes still open in a given year, a larger death value predicts a lower probability of being closed in each of the following five years, monotonically and at every order (Supplementary Text, Table \ref{tab:claude-schedule}). The death radius thus captures the reach of a field's incremental densification: how far a gap sits from the concepts that surround it.

\begin{figure}
    \centering
    \includegraphics[width = \linewidth]{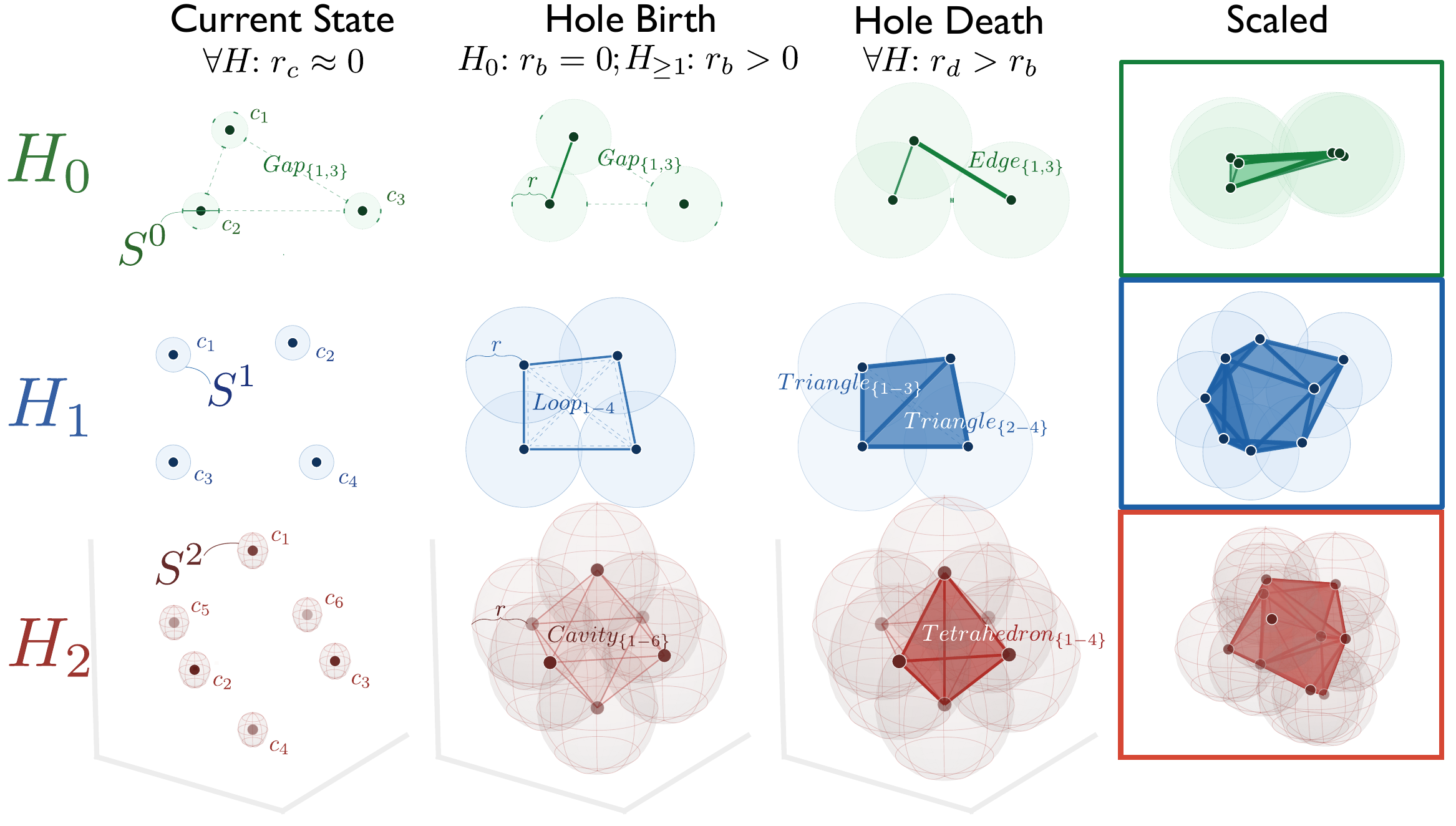}
    \caption{\textbf{Conceptual holes in embedding space.}
    \small{Illustrating zeroth-, first-, and second-order holes ($H_0$, $H_1$, $H_2$) in an embedding space of scientific concepts, where distances between concepts reflect their frequency of co-occurrence in prior research (learned via Word2Vec). Following persistent homology, holes are first ``born'' and later ``die'' as a growth parameter $r$ increases.
    We interpret these holes as opportunities for scientific discovery that researchers can ``see'' by simulating local growth in knowledge, approximated by $r$. $H_0$ holes are gaps between disconnected points: all $H_0$ holes are born at $r=0$ and die when the growing neighborhoods of two points intersect. $H_1$ holes are loops: a loop is born when four or more points form an empty cycle, and dies when that cycle becomes tiled by triangles as neighborhoods intersect. $H_2$ holes are cavities: a cavity is born when six or more points enclose an empty void, and dies when that void becomes filled by tetrahedra as neighborhoods intersect.}
    }
    \label{fig:conceptual_holes}
\end{figure}

We embed the concepts of 19 academic disciplines across the years of their development using a large bibliographic corpus, compute the persistent homology of each embedding in dimensions $0$, $1$, and $2$, and identify, for every hole, the works whose combination of concepts fills it.

\begin{figure}
    \centering
    \includegraphics[width = \linewidth]{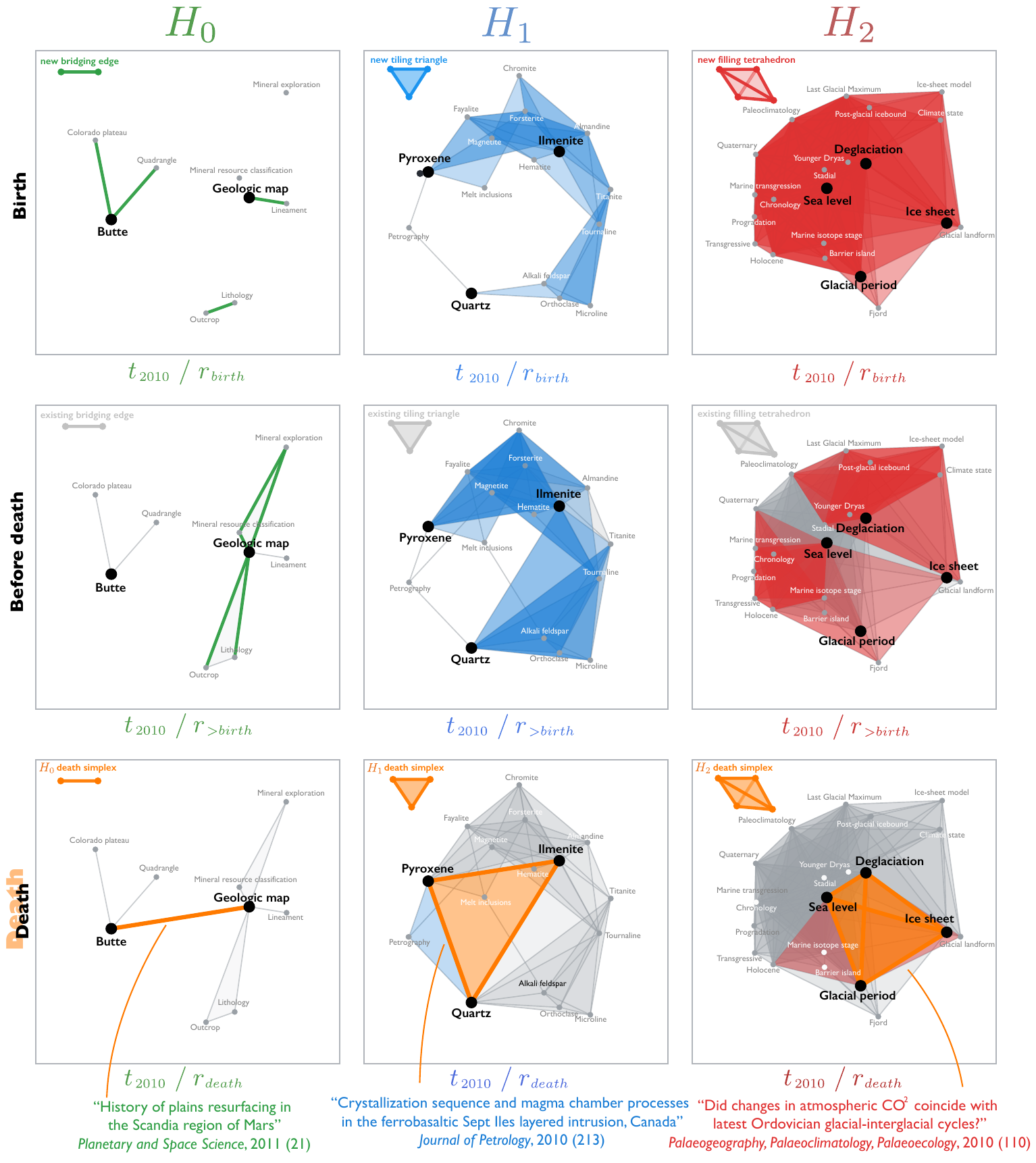}
    \caption{\textbf{Empirical holes in embedding space.}
    \small{We illustrate holes of dimensions 0 (gaps), 1 (loops), and 2 (voids) within an embedding of Geology concepts for the year 2010, for the associated Vietoris--Rips filtration.
    For each hole we show three steps of the filtration: at (or, for dimension 0, near) its birth, just before it dies, and at its death.
    At each step, we highlight simplices absent from the previously shown step in the hole's relevant color.
    Each hole is filled by a work published in 2010 whose concepts realize its death simplex, represented in orange. We give one example hole-filling work per dimension.}
    }
    \label{fig:empirical_holes}
\end{figure}

Two core findings follow. First, works that fill holes attract outsized attention. Across disciplines, hole-filling papers are both more cited and more surprising than their contemporaries, and the effect grows for higher-order holes.
This premium is not uniform across science. It grows with the order and size of the hole in discovery-oriented empirical and data-driven fields, and shrinks or reverses in design-oriented formal and theoretical fields, showing that disciplines differ systematically in whether they reward closing large and complex conceptual gaps. Filling an anticipated gap often proves a good investment for scientists. And once holes are filled, the concepts they inscribe remain close in scientific space.

Second, the supply of opportunities and their exploitation diverge sharply. As a field's concepts multiply, the number of holes grows combinatorially with their order, yet the fraction that any work fills collapses, so higher-order opportunities that proliferate in principle remain largely unrealized in practice.

\section*{Results}

\subsection*{Holes in embedding spaces}

\begin{figure}
    \centering
    \includegraphics[width = \linewidth]{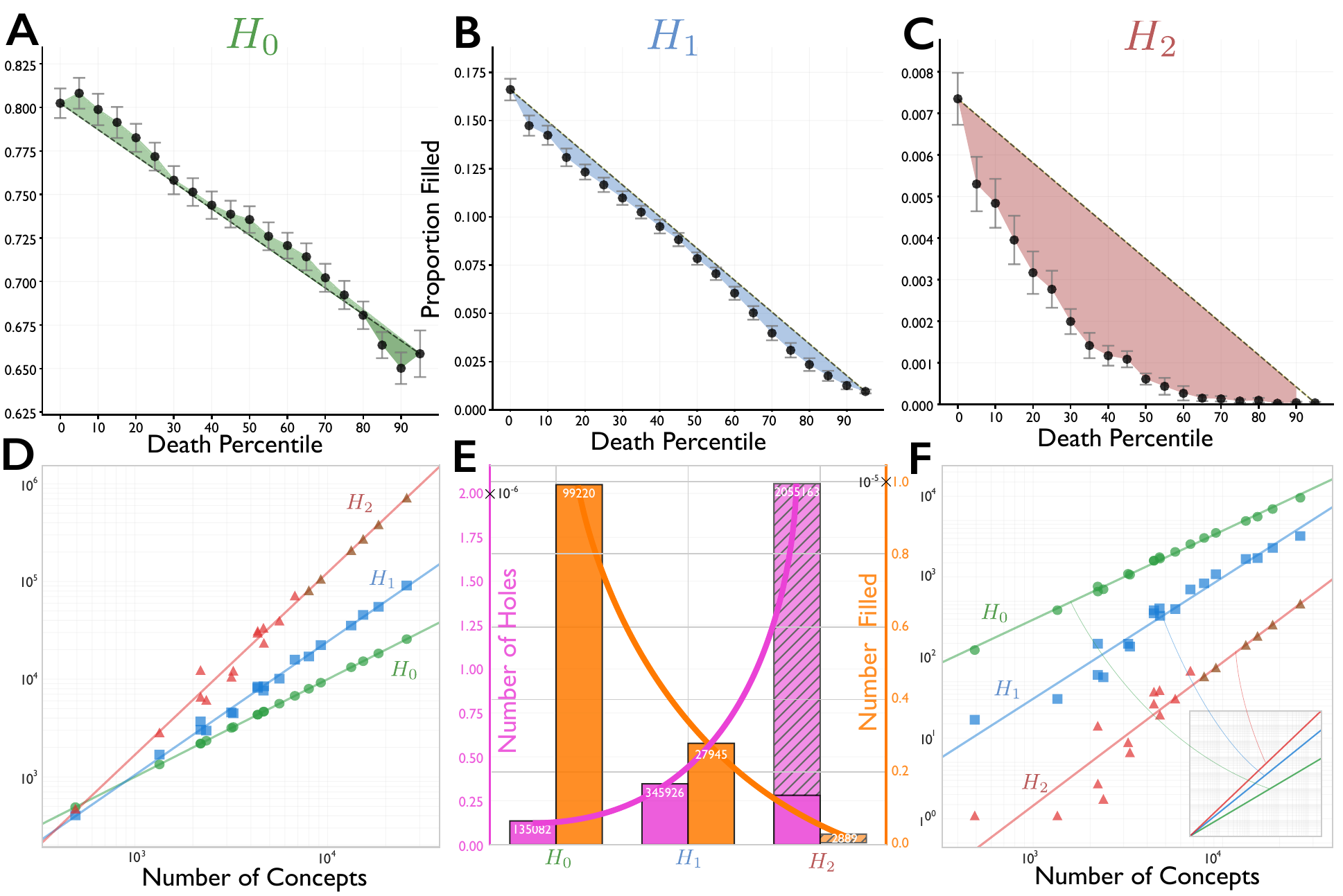}
    \caption{\textbf{Proportion of holes filled in.}
    \small{In Panels A, B, and C, we plot the proportion of holes filled against hole size (death value) for dimensions $0$, $1$, and $2$, respectively. The proportion of holes filled decreases with hole size. Relative to its scale, the curve is slightly concave in dimension $0$, slightly convex in dimension $1$, and convex in dimension $2$.
    For dimensions $0$, $1$, and $2$, Panels D and F plot, respectively, the mean number of holes and the mean number of holes filled by works against the mean number of concepts per discipline.
    The exponents of these power laws (i.e., the slopes in these log-log plots) are similar between the number of holes D and the number filled F.
    The two relationships differ chiefly in their prefactors rather than their exponents:
    the ratio $C_F/C_D$ falls from $0.29$ at dimension $0$ to $0.0032$ at dimension $2$
    (Table \ref{tab:claude-prefactors}). Note that the inset, which sets the $y$-intercept to
    zero, displays the exponents only and by construction removes the prefactor.
    To highlight this similarity, the inset in Panel F shows the same slopes as Panel D but with $y$-intercept set to zero for dimensions $0$, $1$, and $2$.
    Panel E shows the totals across all disciplines and years. For computational reasons, we did not compute dimension-$2$ persistent homology for biology, chemistry, computer science, engineering, medicine, and physics. For these disciplines, we infer the number of holes and holes filled from the power-law relationship observed in the other disciplines.}
    }
    \label{fig:hole_fill}
\end{figure}

We analyze the number of holes for the concept embeddings of each discipline, as well as the number of holes filled (i.e., those with at least one corresponding scientific work).
In Figure \ref{fig:hole_fill}$E$, we present, for dimensions $k = 0, 1, 2$, the total number of holes for each discipline alongside the number of holes filled.
For nearly all disciplines, as the dimension increases, the number of holes grows exponentially, while the number of filled holes decays exponentially.
Furthermore, fixing any dimension $k = 0, 1, 2$, comparing the (mean) number of holes in dimension $k$ with the (mean) number of concepts in a discipline reveals an exponential relationship.

The growth in the number of holes with increasing dimension $k$ stems from the combinatorial nature of the Vietoris--Rips complex.
A $k$-dimensional hole is formed by $k$-simplices.
Because there are $\binom{n}{k+1}$ possible $k$-simplices, the number of $k$-dimensional holes increases roughly exponentially for small $k$.
This behavior is well documented for Vietoris--Rips filtrations (see, e.g., \cite{Giustietal2015}) and is supported by theoretical guarantees on Vietoris--Rips complexes (e.g., \cite{Kahle2011,Kahle2014,BauerandPausinger2018}).
These guarantees are asymptotic and thus most clearly observed for large point clouds, in our case of scientific concepts.
Notably, environmental science, the smallest discipline in our analysis with $496$
concepts, is the only field that does \emph{not} exhibit an exponential increase in the number of
holes as dimension increases.

The decline in number of filled holes with increasing dimension is also expected.
To fill a $k$-dimensional hole, a work must contain the concepts that form its death simplex.
Recall that a $0$-dimensional death simplex is a bridging edge, a $1$-dimensional death simplex is a tiling triangle, and a $2$-dimensional death simplex is a filling tetrahedron.
That is, a work that fills in a $0$-dimensional hole connects $2$ concepts, a work that fills a $1$-dimensional hole connects $3$ concepts, a work that fills a $2$-dimensional hole connects $4$ concepts, and so on.

Additionally, the death simplices in dimensions $1$ and $2$ have broadly comparable side
lengths: the ratio of shortest to longest edge has median $0.87$ at dimension $1$ and $0.77$ at
dimension $2$ over all holes, with $69\%$ and $43\%$ respectively falling between $0.8$ and $1.0$.
Among the holes that are actually filled the ratio is systematically \emph{lower}, with median
$0.80$ at dimension $1$ and $0.65$ at dimension $2$, with half of filled dimension-$2$ voids
between $0.6$ and $0.8$. In this way, filling favors less regular simplices rather than the
most equilateral ones.
This suggests that filling in a $1$-dimensional (or $2$-dimensional) hole requires connecting $3$ (or $4$) similarly distanced concepts. This is more common with $2$ concepts, both from the rarity of scientists with expertise in three or four fields required to conceive and write-up such an investigation, and from the rarity of reviewers required to evaluate and certify them for publication.

We further analyze the relationship between the number of holes, the number of filled holes, and the number of concepts in a discipline.
For a fixed dimension $k$, the number of holes follows a power law in the number of
concepts (see Figure \ref{fig:hole_fill}D), with the exponent increasing with dimension: $1.00$,
$1.37$ and $1.87$ for dimensions $0$, $1$ and $2$, respectively. The number of filled holes follows a power law
as well (Figure \ref{fig:hole_fill}F), with exponents $1.10$, $1.51$ and $1.78$. We describe both
as power laws throughout.
More concepts yield more holes, and in turn more filled holes.

Reporting only the exponents, however, understates how few holes are filled, because the two
relationships differ far more in their \emph{prefactors} than their exponents. Writing the
number of holes as $D = C_D\,n^{a_D}$ and the number filled as $F = C_F\,n^{a_F}$ in the number of
concepts $n$, the filled fraction is
\[
  \frac{F}{D} \;=\; \frac{C_F}{C_D}\,n^{\,a_F-a_D},
\]
so with $a_F\approx a_D$ the filled fraction is nearly constant in concept count and is set by the
prefactor ratio. That ratio collapses across dimension --- $0.29$, $0.023$, $0.0032$ for
dimensions $0$, $1$ and $2$ (Table \ref{tab:claude-prefactors}). The ratio is the filled fraction
extrapolated to a single concept rather than the fraction itself, but its collapse tracks the
directly counted fill rates of $73.5\%$, $8.1\%$ and $0.2\%$. It is this collapse, and not any difference
in growth rate, that makes the overwhelming majority of higher-order combinations go unrealised.
Indeed the exponent difference $a_F-a_D$ is \emph{positive} at dimensions $0$ and $1$ ($+0.10$ and
$+0.14$), so the filled count there grows slightly faster than the total; the dimension-2 deficit
of $-0.09$ stands against a standard error of $0.31$ on $a_F$ and is not distinguishable from
zero.

Moreover, in the supplementary information we show that the rate at which holes fill in is not determined by the topology of the embedding space, but rather by how human research papers combine concepts.
Specifically, we produce 9 null models to generate simulated papers (represented as a set of concepts) and find that, in almost all cases, the fill-rate of holes by simulated papers is a tiny fraction of that which we observe with actual papers.
The exceptions are null models in which simulated papers combine concepts close by one another in embedding space. For these, we observe comparable fill rates for simulated and empirical papers (see Figures \ref{fig:filled_by_similar_concepts_papers_T=0.1} and \ref{fig:filled_by_similar_concepts_papers_T=0.05}), although constraining paper construction too strongly for similar concepts also eventually lowers the fill rate (e.g., see Figure \ref{fig:filled_by_similar_concepts_papers_T=0.01} in the SI). These results suggest that human scientists focus their research around disciplinary commitments, but are simultaneously incentivized to produce novel work by anticipating and filling modest holes within and near those regions \cite{foster2015}.

We also analyze the relationship between the size (i.e., death value) of a hole and whether or not it is filled. In Panels A, B, and C of Figure \ref{fig:hole_fill} --- corresponding to dimensions $0$, $1$, and $2$, respectively --- we examine the proportion of holes filled as a function of their relative size.
For each dimension $k=0,1,2$, we observe that, as we consider larger holes, a smaller proportion of holes are filled.
In other words, larger holes are less likely to have papers that combine the associated concepts.
We consistently observe this trend across dimensions, as well as at the level of individual disciplines.
This observation aligns with the intuition that smaller knowledge gaps are more easily and frequently bridged.
Moreover, because the death simplices in dimensions $1$ and $2$ overwhelmingly have similar-sided lengths, finding works that combine comparably distant concepts becomes less likely.
Relative to its scale, the corresponding curve is slightly concave in dimension $0$, slightly convex in dimension $1$, and convex in dimension $2$.
We also find that the proportion of holes filled in dimension $0$ remains largely between $0.65$ and $0.8$ across death percentiles.
In contrast, for dimensions $1$ and $2$, the proportion of filled holes decreases to nearly $0$ as death percentile increases.
Thus, while even large dimension-$0$ holes are filled at a substantial rate, large dimension-$1$ are rarely filled, and dimension-$2$ holes are very rarely filled.
These trends are consistent across disciplines.

\subsection*{Works that bridge holes}

\begin{figure}
    \centering
    \includegraphics[width = \linewidth]{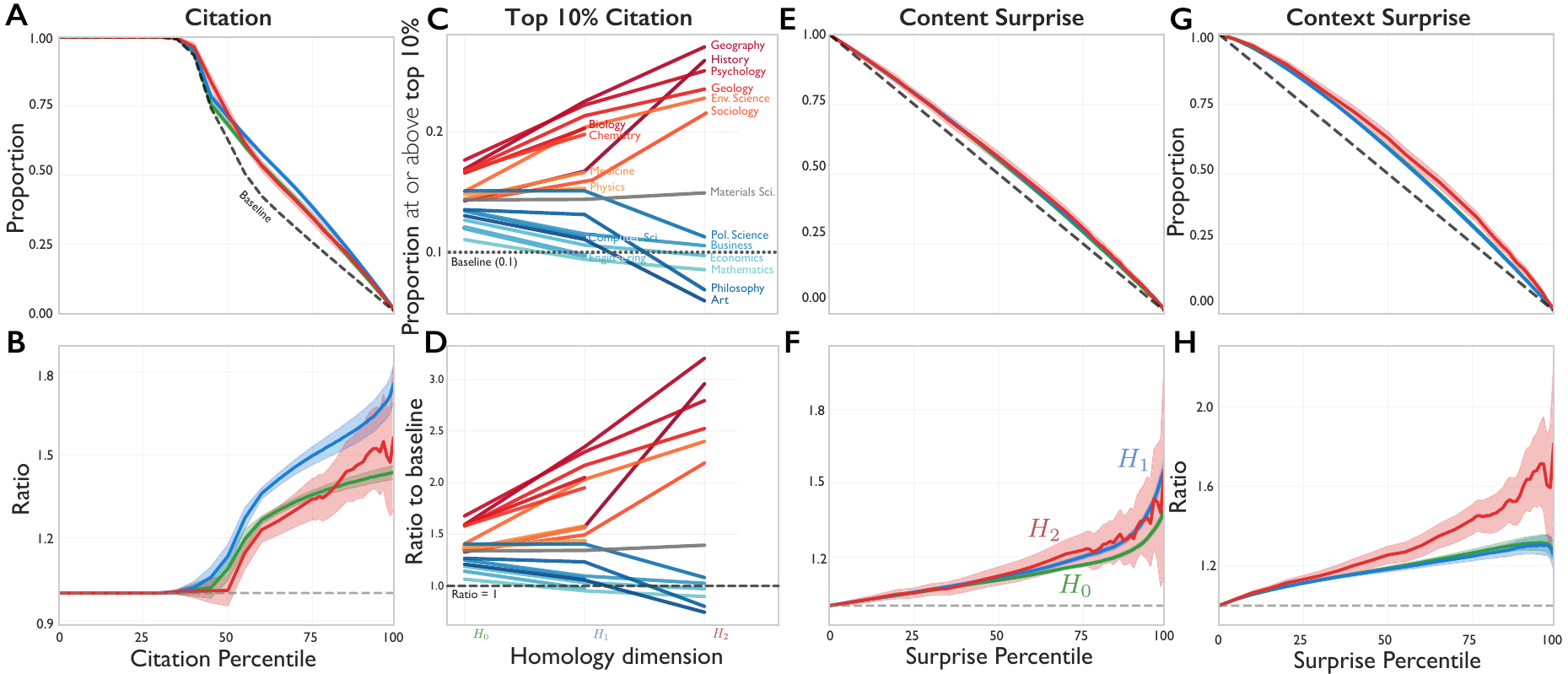}
    \caption{\textbf{Properties of work that fill holes.}
    \small{In Panel $A$, we plot, for dimensions $0$, $1$, and $2$, the proportion of hole-filling works at or above a given citation percentile threshold. For a dimension $k$ and threshold $N$, this is the proportion of works filling $k$-dimensional holes with citation count $\geq N$th percentile. Panel $B$ takes the ratio of this proportion to the baseline for each percentile. Panels $C$ and $D$ show the proportion of papers that fill 0th, 1st, and 2nd order holes within the top 10\% of citations, and the ratio of that proportion to the baseline, respectively, for each field. $C$ and $D$ demonstrate that while all fields manifest an increase in hit papers for bridging zeroeth-order gaps, and almost all fields for tiling first-order holes, only discovery-oriented fields systematically grow in high citation likelihood as hole dimensionality increases, while design-oriented fields symmetrically decrease in hit citation likelihood (see Figure
    \ref{fig:citation_hit_ratio_by_field}
    for the 25\%, 5\% and 1\% citation thresholds that reveal similar patterns. Panels $E$ through $H$ show the same as $A$ and $B$, but for the content- and context-surprise indices, respectively. Panels $E$ and $G$ plot the proportion of hole-filling works at or above threshold $N$ within the content and context surprise index distributions, which trace surprising combinations of curated concepts and surprising combinations of journal and workshop contexts from which the science draws, divided by the proportion of all works at that threshold. For larger $N$, the denominator is approximately $(100-N)/100$; for smaller $N$, the discrete nature of the citation counts causes deviations. Panels $F$ and $H$ show the ratio of that proportion to the baseline.}
    }
    \label{fig:hole_fill_papers}
\end{figure}

We now examine works that fill in holes, across dimensions $0$, $1$, and $2$.
For each dimension $k=0,1,2$, we compare works that fill holes with the full set of works, and, specifically, with works that do not fill holes within the same year and discipline.
We analyze two broad quantitative characteristics of these works. First, we examine the likelihood of high impact within the field, as assessed by percentiles of forward citation counts.

Second, we assess the surprisingness of the scientific work to the field. We assess this as the improbability of concept combination (``content surprise'') and the improbability of referenced venue combination, the institutional contexts from which concepts are drawn (``context surprise'') against highly predictive generative base models. These measures generalize the atypical-combination framework of \cite{uzzi2013} and are twice as predictive of high citations, field awards, and expert attributions of novelty \cite{shi2023surprising}.

For each metric, we compare hole-filling works with the broader corpus of works by taking the ratio of the corresponding index across different thresholds.
For example, we take the proportion of hole-filling works within the top $20\%$ citation count (for a given year and discipline) and divide it by the proportion of general works within the top $20\%$ citation count (roughly equal to $0.2$).

We assess the performance and reception of hole-filling works by examining their intensity of citation within the scientific community.
In Panels $A$ and $D$ of Figure \ref{fig:hole_fill_papers}, we compare the citation counts of hole-filling works with general works.
We observe, on average, that for each dimension $k=0,1,2$, works that fill in $k$-dimensional holes yield higher citation counts than general works.
This citation premium varies across disciplines by dimension, however, and the variation is structured rather than idiosyncratic. In dimension $0$, hole-filling works out-cite their peers in every discipline. In dimension $1$, exceptions are mathematics and engineering, and in dimension $2$, art, philosophy, mathematics, business and economics. Sorting disciplines by whether their citation premium rises or falls with hole order divides them along the boundary between discovery- and design-oriented fields, with the division widening by order (Table~\ref{tab:pd_main}). Averaged within group, the top-quintile citation enrichment of hole-filling works in the discovery-oriented natural and empirical sciences climbs from roughly $1.5$ in dimension $0$ to $1.9$ in dimension $1$ and $2.6$ in dimension $2$; in the formal and design fields it starts lower and declines, from roughly $1.2$ to $1.1$ to $0.9$, crossing below the no-enrichment line by dimension $2$.
We note that these group averages are taken over the disciplines for which each dimension is
available, so the dimension-$2$ averages rest on a smaller and slightly differently composed set than the
dimension-$0$ and dimension-$1$ averages: 4 of the 8 discovery-oriented fields and 9 of the 11 design-oriented
fields, since dimension-$2$ homology was not able to be computed for biology, chemistry, medicine, physics,
computer science or engineering. Per-discipline values, with unavailable dimensions marked, are
given in Table~\ref{tab:pd_main}. Nevertheless, all those measured were consistent with the difference in trajectory between dimension 0 and 1. Measured as the amplification from dimension $0$ to dimension $2$, the split is nearly categorical: history ($2.26\times$), geography ($2.00\times$), psychology ($1.67\times$), sociology, geology and environmental science all amplify by roughly $1.6\times$ or more, materials science is neutral ($1.01\times$), and mathematics ($0.86\times$), economics, political science, business, philosophy and art ($0.58\times$) all attenuate. Among the large fields for which we compute only dimensions $0$ and $1$, biology, chemistry, medicine and physics rise while computer science and engineering fall. Psychology and mathematics mark the extremes: the boost among psychology works filling holes of dimensions $0$, $1$, and $2$ reaches roughly $1.7$, $2.8$, and $3.2$ times, whereas in mathematics only about $22\%$ of works filling $0$-dimensional holes, and fewer than $20\%$ of those filling $1$- and $2$-dimensional holes, land in the top $20\%$ by citation. Where hole-filling works do outperform, their relative advantage typically grows at higher citation thresholds.

The same division governs the \emph{size} of the holes that confer citation attention by the community. Grouping the 19
disciplines into 8 discovery-oriented natural and empirical fields and 11 design-oriented formal and theoretical
fields --- a grouping we report because the fitted slopes separate along it, not one fixed in
advance as a hypothesis --- we regress the top-decile citation enrichment of hole-filling works on the death percentile of the hole they fill (Supplementary Text; Tables~\ref{tab:grouped-slopes-main-h0}--\ref{tab:grouped-coef-citation-h1}). In dimension $0$, the mean slope is $+0.50$ per 100 death percentiles in discovery-oriented fields against $-0.56$ in design-oriented fields: every discovery-oriented field has a positive slope and 10 of 11 design-oriented fields a negative one, with sociology at $0.00$, the sole field on the boundary. Dimension $1$ repeats the split more weakly ($+0.52$ against $-0.25$). A two-dimensional model over death percentile and citation percentile agrees, locating the citation optimum at large gaps and extreme percentiles in discovery-oriented fields and at small gaps and extreme percentiles in formal ones. In discovery-oriented fields, then, the works that close the largest anticipated gaps are ones the citing literature rewards most; in design-oriented fields, the citation advantage is concentrated in the smallest gaps and decays as gaps grow.

Once researchers fill holes in science, the concepts inscribing those holes remain much closer in future embeddings than those for unfilled holes, which rapidly drift apart (see Table \ref{tab:claude-drift} and Figures \ref{fig:claude-drift} and \ref{fig:claude-size}). This phenomenon intensifies with hole-filling order, from $0$ to $2$. The result corroborates our citation analysis by suggesting that later work not only references hole-filling work but also directly and indirectly replicates its combinations, resulting in conserved conceptual proximity within embeddings.

To analyze the novelty of hole-filling works, we analyze their context- and content-surprise indices.
In Panels $B$ and $E$ of Figure \ref{fig:hole_fill_papers}, we compare the content-surprise-index percentiles of hole-filling works with that of general works. We make the same comparison for context-surprise-index percentiles in Panels $C$ and $F$.
As with our analysis of citation counts, a notable trend for both surprise indices is that, on average, for each dimension $k=0,1,2$, hole-filling works are more surprising than the full population of published work.
This trend is consistent (with the singular exceptions of business and economics in dimension $2$ for content surprise, and business and environmental science, the smallest field, for context surprise) across disciplines in all dimensions.

This suggests that, on average, hole-filling works tend to be more novel than general works.
We also find that content surprise rises from Dimension 0 to 1 in every one of the natural sciences we evaluate---biology, chemistry, physics, medicine, engineering---as well as in mathematics and philosophy, so filling a bigger gap reads as more unexpected almost everywhere, even where it earns no additional citation.
Context surprise in the natural and engineering sciences, however, declines with dimension from an already low base (for example, chemistry, engineering, medicine and physics all fall from Dimension 0 to 1).
Nevertheless, for both surprise metrics, we observe that as we consider larger percentile thresholds, we tend to observe larger relative surprise in hole-filling works than for scientific works in general.

Neither surprise measure follows the discovery--design division that structures citation: content surprise rises with hole order on both sides of it, and the death-dependence of context surprise, where present, is led by engineering and the physical and life sciences, cutting across the division (Supplementary Text). Filling a large hole violates expectations everywhere; whether greater novelty is rewarded with more attention depends on whether a field is primarily discovery versus design-oriented.

\section*{Discussion}

Using persistent homology, we identify perceivable, anticipated holes in the concept space of scientific disciplines and ask which were filled by published work, and what became of the works that filled them. Two regularities emerged. Works that fill holes are, on average, more cited, repeated, and surprising than their contemporaries. Moreover, as concepts accumulate, holes proliferate combinatorially with their order while the fraction filled collapses.

The first finding encourages filling anticipated gaps. Across nearly every discipline, a work that closes a hole is more likely than its peers to rank among the most surprising of its year. It is also usually among the most discussed, referenced, and redone in future work.

The citation advantage is widespread but not universal, and its exceptions reveal a sharp distinction in how discovery vs. design-oriented fields search the space of scientific possibility. Disciplines divide along the boundary between discovery-oriented empirical and data-driven fields versus design-oriented formal and theoretical. In the discovery-oriented sciences, the citation premium for filling a hole grows with both the order of the hole and its size: works that close the largest, highest-order gaps are the ones most over-represented among the most cited. In the design-oriented fields---mathematics, economics, business, political science, philosophy, art, and, in dimension $1$, computer science and engineering---the premium is concentrated in the smallest gaps, decays as gaps grow, and vanishes or reverses for higher-order holes. Mathematics is the limiting case, where filling higher-order holes confers no citation premium at all. This aligns with the recent finding that mathematicians intellectually ``age'' twice as fast as their scientific contemporaries. This is presumably at least in part unlike their peers in discovery-oriented disciplines; their favorite citations, published on average two years before their own first publication, are rarely disproven \cite{cui2026aging}.

One reading of this divide is that discovery- and design-oriented fields value hole-filling for different reasons. In an discovery-oriented empirical field, a work that spans several distant concepts consolidates evidence from literatures that had not been read together, and the citing literature rewards the consolidation; the larger the gap and the more concepts joined, the more communities have reason to cite it. In a design-oriented formal or theoretical field, value resides in the depth of a proof, policy, model or artifact rather than in the breadth of what it connects. In this way, a work spanning distant concepts is more likely to be read as a survey, an application, or an arbitrary juxtaposition than as an advance, and higher-order combinations are treated as increasingly routine. Consistent with this, the divide is specific to attention: content surprise rises with hole order on both sides of the disciplinary divide. Filling a large hole is recognized as novel almost everywhere; whether that novelty is rewarded depends on the field. That novelty is recognized more reliably than it is rewarded with increased attention echoes the documented bias against novelty in science \cite{wang2017}, and especially in mathematics, where the most original work is appreciated only slowly \cite{uzzi2013} and aging scientists cleave most closely to their earliest influence \cite{cui2026aging}. Nevertheless, on balance, the bet on identifying and filling anticipated gaps, holes, and voids the surrounding literature is already converging toward tends to produce work the community comes to recognize and value---and does so most reliably where science is most discovery-oriented.

Closing a hole also leaves a trace in the geometry of the embedding, suggesting not only citation but replication. In the five
years after a hole is closed, the concepts it joins move apart much more slowly than those of
comparable holes left open --- at about a fifth the rate for gaps and two fifths for loops (Supplementary Text, Table \ref{tab:claude-drift} and
Figure \ref{fig:claude-drift}). The effect is relative, as both sets of concepts end somewhat further apart than they began. This is not wholly produced by the closing: the concepts of holes that are
closed are also the steadier ones in the years \emph{before} closure
(Figure \ref{fig:claude-size}), suggesting that the hole is ripe and ready to be filled.

Our second finding is more troubling for science. The number of higher-order holes
explodes as a field grows: holes scale as a power law in a discipline's concept count, with the
exponent increasing with order. The number filled obeys a power law with \emph{similar}
exponent but far smaller prefactor, and it is that gap in a ratio of $0.29$ at
dimension $0$ falling to $0.0032$ at dimension $2$, which leaves the overwhelming majority of
higher-order combinations unattempted. The shortfall is a matter of level rather than of growth
rate, and it deepens across orders rather than across field size. Whether this represents a real loss depends on the structure of discovery: are higher-order combinations cumulative on lower-order ones?

If and where they are---if the payoff to joining several concepts is assembled from the payoffs of their pairs and triples---then the relevant unfilled higher-order holes will largely fill with time. Incremental science, which fills the smallest and nearest gaps first, climbs a smooth landscape and will reach the higher combinations as a byproduct of filling lower ones. But if some combinations are emergent, valuable only when all of their elements are present together with no profitable sub-combination to point the way, then the landscape is rugged, and there is no gradient of partial success to climb \cite{kauffman1993}. Such combinations sit in holes that a field's natural densification may never close, because no step toward them is rewarded. Precisely the most novel higher-order opportunities, those least anticipated by current work, are the ones to which incremental search is structurally blind.

Emergent, non-additive combinations are not hypothetical. In genetics, higher-order epistasis describes phenotypes that appear only when several genes act in concert, and which cannot be predicted or detected from the effect of single genes or pairs \cite{phillips2008, domingo2019}. One can imagine four otherwise unrelated genes whose joint activation drives metastasis while no pair or triple among them does anything at all. Evolution itself struggles to cross such landscapes. Only a small minority of mutational trajectories toward a fitter protein pass through uphill intermediate steps \cite{weinreich2006} and incremental science labors under the same constraint. The same logic holds in chemistry, where individually inert reagents can combine to yield a useful product, and in pharmacology, where the effect of a drug combination emerges only at third- and higher-order interactions and cannot be extrapolated from pairwise tests \cite{tekin2018}. Connections of this kind lie dormant between unrelated literatures, in the unmade links that define undiscovered public knowledge \cite{swanson1986, evans2010machine}.

For human science, non-additive higher-order holes are hard to fill for three compounding reasons. First, they are hard to attempt: assembling an experiment that manipulates four or more distant concepts no one has combined is costly and without precedent, with no protocol to borrow and no pilot result to justify it. Second, they are hard to recognize: the value of such a combination can be judged only by an evaluator fluent in all of its disparate parts, and the very breadth that makes the combination novel makes such a reviewer rare. The work risks being read as arbitrary rather than profound by myopic reviewers. Our result that surprise is recognized more consistently than it is rewarded with attention and citations is the mild form of this problem; the severe form is a combination so unfamiliar that no one recognizes it at all. Third, and most fundamental, is the problem of prior probability: when lower-order subsets do not profitably combine, nothing raises the odds that anyone should profitably attempt the full combination in the first place. The breadcrumbs that lead incremental, attention-following science from one result to the next do not exist for an emergent combination, so the most valuable higher-order holes are also least likely to be probed.

These obstacles should not obscure our central message: anticipated holes are, in general, profitable for scientists to fill, and the field rewards those who fill them successfully. Our results invite a quantitative science of holes, one that models the probability that a hole of a given order, size, and surrounding density will be filled, and the accumulation dynamics by which filling lower-order holes raises the probability of filling higher-order ones. The scaling laws we report, between number of holes and number of concepts, and between holes and works that fill them, are first elements of such a science. They make it possible to ask, for any field, how far its realized discoveries lag the combinatorial opportunity its concepts afford, and which of its open holes are reachable by incremental accumulation and which are not.

Finally, the same features that make emergent higher-order holes hard for human science, including the breadth of expertise they demand, the combinatorial cost of attempting them, and the absence of any incremental gradient to follow, mark them as an opportunity for artificial intelligence. Systems that can hold many distant concepts in view simultaneously, that are not bound to a single discipline's frontier, and that need not follow the trail of accumulating attention could search the high-dimensional gaps that human incentives and cognition lead us to skip, especially when they are built to avoid the inferences people would already make, and so to target what we would otherwise miss \cite{sourati2023accelerating}. The unfilled higher-order frontier, on this view, is less a failure of science than a standing invitation: a natural division of labor in which machines probe the emergent combinations and human judgment selects and interprets what they return \cite{evans2025after}.

\newpage

\section*{Materials and Methods}

\subsection*{Data}
In this paper, we use the OpenAlex dataset \cite{Priem_OpenAlex_A_fullyopen}, a large-scale bibliometric database containing over 250 million academic works (we use a snapshot from December 2024). OpenAlex metadata includes concept labels assigned to academic works. These concepts are organized hierarchically into levels 0 through 5: level 0 concepts correspond to broad disciplines (e.g., mathematics, physics), level 1 concepts represent subdisciplines (e.g., pure mathematics), and levels 2 through 5 denote increasingly specific topics. Each concept at a given level has at least one ancestor in the previous level; that is, every level $k$ concept has a level $k-1$ ancestor.
We consider the following 19 disciplines, which are precisely the level-0 concepts: art, biology, business, chemistry, computer science, economics, engineering, environmental science, geography, geology, history, materials science, mathematics, medicine, philosophy, physics, political science, psychology, and sociology.

For each discipline $D$ (i.e., each level 0 concept), we consider its set of descendant concepts in levels 2 through 5, which we refer to as the concepts of $D$. We then compile all works associated with these concepts, which we refer to as the works of $D$. Each work $w$ in discipline $D$ is associated with the concepts in $D$ at levels 2 through 5.
Works associated with concepts from multiple (level 0) disciplines are included in each relevant discipline's set.

\subsection*{Embeddings of Concept Spaces}

For each discipline-year pair $(D, y)$, we consider all works in discipline $D$ published in year $y$ or earlier. We treat each work as a bag of concepts from $D$ (specifically, the level 2--5 concepts associated with that work) and use these bags to train a Word2Vec model \cite{mikolov2013} with vector size 100.
We consider a minimum concept frequency of $10$, which excludes concepts too rare to embed reliably. (This floor retains at least $97.7\%$ of concepts in every discipline--year, with no systematic variation across disciplines; see SI.)
This yields an embedding of the concepts of $D$ as of year $y$. For each discipline, we constructed embeddings for years 2005--2019.

\subsection*{Persistent Homology}

In this section, we provide the background for persistent homology, the primary tool in topological data analysis, which we use to analyze the topological structure of our concept embeddings.
Persistent homology relies on homology theory, an area of algebraic topology that concerns holes in topological spaces.
Homology is an algebraic tool that detects holes in a topological space.

We first discuss how one computes the homology of a simplicial complex.
A simplicial complex is a combinatorial description of a topological space, consisting of simplicies (i.e., vertices, edges, triangles, and higher-dimensional analogs) with certain requirements on simplex boundaries and pairwise intersection.
For a simplicial complex $K$, homology identifies features of different dimensions: $0$-dimensional holes correspond to connected components, $1$-dimensional holes correspond to cycles (loops not filled by area-tiling triangles), and $2$-dimensional holes correspond to voids (cavities enclosed by triangles and not filled by space-filling tetrahedra).

Formally, for a fixed dimension $k$, let $C_k(K)$ be the vector space of $k$-chains, where a $k$-chain is a formal linear combination of $k$-simplices in $K$. Each $k$-simplex is assigned an orientation, and the boundary map $\partial_k: C_k(K) \to C_{k-1}(K)$ sends each oriented $k$-simplex to the signed sum of its $(k-1)$-dimensional faces, with signs determined by the orientation.
For example, the boundary of an oriented edge $[v_0, v_1]$ is $[v_1] - [v_0]$, and the boundary of an oriented triangle $[v_0, v_1, v_2]$ is $[v_1, v_2] - [v_0, v_2] + [v_0, v_1]$.

A $k$-chain $c$ is called a $k$-cycle if it has no boundary; that is, $\partial_k(c) = 0$.
A $k$-chain $c$ is called a $k$-boundary if it is the boundary of some $(k+1)$-chain; that is, there exists $d \in C_{k+1}(K)$ such that $\partial_{k+1}(d) = c$.
One can readily verify that, for any dimension $k$, we have $\partial_{k}\circ\partial_{k+1}=0$, which implies that every boundary is a cycle.
Nevertheless, not every cycle is a boundary (i.e., is filled); cycles that are not boundaries enclose holes.
The $k$th homology group is defined as the quotient space $H_k(K) = \ker (\partial_k) / \operatorname{im} (\partial_{k+1})$.
The $k$th Betti number $\beta_k = \dim(H_k(K))$ counts the number of independent $k$-dimensional holes in $K$.
Intuitively, $H_k(K)$ captures cycles that are not boundaries.
We refer to the elements of $H_k(K)$ as the \emph{homology classes} of $K$ in dimension $k$; the $k$-dimensional homology classes of $K$ represent $k$-dimensional holes.

To compute persistent homology on a point cloud $X$, we first construct a ``filtration'', which is a collection of simplicial complexes that represent $X$ at different scales.
Formally, a \emph{filtration} is a nested sequence $\K_{\alpha_0} \subseteq \K_{\alpha_1} \subseteq \cdots \subseteq \K_{\alpha_n}$ of simplicial complexes, such that $\alpha_0 < \alpha_1 < \cdots < \alpha_n$; the simplicial complex $K_{\alpha_i}$ represents $X$ at scale $\alpha_i$.

Consider a finite point cloud $X = \{x_1, \ldots, x_r\}$ in a metric space $(M, d)$. One of the most common examples is the \emph{Vietoris--Rips (VR) filtration}.
The VR complex $\mathrm{VR}(X, M, d)$ at parameter $r$ includes a simplex with vertices $[x_{i_0}, \ldots, x_{i_k}]$ if $d(x_{i_j}, x_{i_{\ell}}) \leq 2r$ for all $j, \ell$.
As $r$ increases, we obtain the nested sequence that defines the \emph{VR filtration}.

Applying homology to a filtration yields a \emph{persistence module}, which consists of a sequence of vector spaces $H_k(K_{\alpha_0}) \to H_k(K_{\alpha_1}) \to \cdots \to H_k(K_{\alpha_n})$, where the maps are induced by the inclusions of simplicial complexes.
As the filtration parameter increases, homology classes (which represent holes) are ``born'' and subsequently ``die'' when filled.
A homology class is \emph{born} at $\alpha_i$ if it first appears in $H_k(K_{\alpha_i})$.
That is, the corresponding hole forms.
A homology class \emph{dies} at $\alpha_j$ ($j \geq i$) if its image in $H_k(K_{\alpha_j})$ becomes trivial.
That is, the corresponding hole fills in.
The \emph{birth value} is $\alpha_i$, and the \emph{death value} is $\alpha_j$.
If a class never dies, we say its death value is infinity.
The \emph{persistent homology} of the filtration is the collection of all such birth--death pairs, often visualized as a persistence diagram or barcode.
We also refer to the simplex whose addition causes a homology class to be born as its \emph{birth simplex}, and the simplex whose addition causes a class to die as its \emph{death simplex}.
When several simplices enter the filtration at the same value of $r$ --- which occurs frequently for a Vietoris--Rips filtration --- the birth (respectively, death) simplex of a homology class is not determined by the birth (respectively, death) value alone.
We therefore fix a total order on simplices refining the filtration --- first by filtration value, then lexicographically by sorted vertex labels, with lower-dimensional simplices preceding higher-dimensional ones on the same vertex set.
This produces a filtration in which exactly one simplex is added at each step, so that every birth and every death is attributable to a unique simplex.
Note that the birth and death \emph{values} are independent of this choice; only the birth and death simplices depends on the tie-breaking rule.
To compute persistent homology, we use the Open Applied Topology software \cite{open_applied_topology}, which computes homology with coefficients in $\mathbb{Q}$.

For each embedding (of a discipline $D$ at year $y$), we compute the persistent homology of the associated Vietoris--Rips complex, in dimension $0$ and $1$.
We also compute the persistent homology in dimension $2$ for most of our disciplines.
Due to computational limitations, however, we do not compute persistent homology in dimension 2 for the following (large) disciplines: biology, chemistry, computer science, engineering, medicine, and physics.

We represent each homology class (hole) by its death simplex and the vertices of that simplex.
(Recall that the death simplex is the simplex that fills the hole.)
We interpret the concepts corresponding to these vertices as the concepts that constitute the hole.
We say that a hole has an associated work $w$ in year $y$ if the set of concepts that form the death simplex is a subset of the concepts associated with $w$.
In other words, the work $w$ combines the concepts that form the death simplex of a given homology class.

One can alternatively measure the filling in of holes using mixup barcodes \cite{wagner2024}, which quantify how much the inclusion of a second point cloud shortens the persistence of features in the first \cite{yadav2025}.
We do not take this approach, because we seek to attribute each filling to the specific works whose concepts realize it, a level of resolution that aggregate interaction statistics do not provide.

\subsection*{Quantities Considered}

\subsubsection*{Holes Filled}

To determine what proportion of $k$-dimensional holes are filled, we proceed as follows. For a fixed year and discipline, we compute the size (death value) percentiles of all holes at thresholds $0,5,10,\ldots,100$. For each discipline $D$, year $y$, and percentile bin $[m, m+5)$ (with $m \in \{0,5,10,\ldots,95\}$), we consider holes whose sizes fall within that bin and calculate the proportion that have associated works. We then take a weighted average of these proportions across all years and disciplines.

\subsubsection*{Assessing Hole-Filling Papers}
We now discuss how we assess works that fill holes.
The metrics we consider in determining the success of works are citation count across all citation thresholds and the content- and context-surprise indices.

Citation count measures the scientific impact of a work, while the content- and
context-surprise indices \cite{uzzi2013,shi2023surprising} capture the novelty of its intellectual
combinations.
Content surprise considers the concepts (or keywords) of cited works, while context surprise considers the sources (e.g., journals or conferences) from which these ideas appear.
Both measure whether the observed combinations of cited items occur more or less frequently than expected under a randomized baseline, with less frequent combinations indicating greater novelty.

Given a discipline and year, we compute metric percentiles for all papers in that discipline and year, using thresholds $p \in \{5,10,\ldots,95,96,97,98,99\}$. For each threshold $p$ and each hole dimension $d \in \{0,1,2\}$, we calculate the proportion of hole-filling papers (i.e., papers associated with $d$-dimensional holes) whose citation count is at or above the $p$th percentile of all papers. We then compute the ratio of this proportion to the proportion of all papers in that discipline and year that are at or above the same $p$th percentile. This ratio indicates the relative over- or under-representation of hole-filling papers among highly cited works.

We perform a similar analysis for the content- and context-surprise indices.
Unlike citation counts, however, the content- and context-surprise indices are already provided as percentile ranks by \cite{team_embeddedness_2026}, so we use these values directly without additional percentile computation.

\clearpage

\bibliography{references}
\bibliographystyle{sciencemag}

\setcounter{figure}{0}
\setcounter{table}{0}
\renewcommand{\thefigure}{S\arabic{figure}}
\renewcommand{\thetable}{S\arabic{table}}

\newpage

\subsection*{Supplementary Text}

\subsubsection*{Concept Retention Under the Frequency Threshold}
When we trained our concept embeddings, we only considered concepts used at least $10$ times.
Because the frequency floor could, in principle, prune concepts unevenly across disciplines, we computed the fraction of each discipline's concepts retained at each embedding year.
Retention increases monotonically with year as the corpus accumulates, so the 2005 values bound the analysis window.
In Table \ref{tab:concept-retention}, we show the number and proportion of concepts kept in each discipline for years 2005, 2010, and 2019.
Within the analysis window (2005--2019), retention exceeds $97.7\%$ in every discipline--year, and $99.3\%$ from 2010 onward.
Retention is also unrelated to the discovery/design division.
The formal and design fields whose citation premium attenuates at higher hole order retain as large a share of concepts as the discovery-oriented fields, so differential pruning does not account for the contrast.

\begin{table}[t]
\centering
\caption{\textbf{Concept vocabulary retained by the Word2Vec frequency cutoff.} For each discipline and embedding year, the number of distinct concepts kept by the $\texttt{min\_count} = 10$ cutoff, the number present in the training corpus, and their ratio.
The corpus for year $Y$ is cumulative (all papers with publication year $\le Y$ carrying more than one concept), so a concept is kept when it occurs at least ten times in that corpus.}
\label{tab:concept-retention}
\footnotesize
\setlength{\tabcolsep}{3pt}
\begin{tabular}{lccccccccc}
\hline
 & \multicolumn{3}{c}{2005} & \multicolumn{3}{c}{2010} & \multicolumn{3}{c}{2019} \\
\cline{2-4}\cline{5-7}\cline{8-10}
Discipline & Kept & Total & Prop. & Kept & Total & Prop. & Kept & Total & Prop. \\
\hline
Environmental Science & 484 & 495 & 0.978 & 492 & 495 & 0.994 & 494 & 496 & 0.996 \\
Art & 1,335 & 1,347 & 0.991 & 1,341 & 1,347 & 0.996 & 1,346 & 1,347 & 0.999 \\
Sociology & 2,180 & 2,199 & 0.991 & 2,195 & 2,200 & 0.998 & 2,199 & 2,200 & 1.000 \\
History & 2,337 & 2,354 & 0.993 & 2,347 & 2,354 & 0.997 & 2,353 & 2,354 & 1.000 \\
Political Science & 4,643 & 4,674 & 0.993 & 4,673 & 4,677 & 0.999 & 4,676 & 4,677 & 1.000 \\
Computer Science & 9,164 & 9,221 & 0.994 & 9,217 & 9,224 & 0.999 & 9,224 & 9,224 & 1.000 \\
Philosophy & 3,160 & 3,174 & 0.996 & 3,168 & 3,174 & 0.998 & 3,174 & 3,174 & 1.000 \\
Economics & 4,324 & 4,341 & 0.996 & 4,338 & 4,341 & 0.999 & 4,341 & 4,341 & 1.000 \\
Chemistry & 15,234 & 15,289 & 0.996 & 15,280 & 15,289 & 0.999 & 15,289 & 15,289 & 1.000 \\
Psychology & 5,601 & 5,621 & 0.996 & 5,620 & 5,622 & 1.000 & 5,621 & 5,622 & 1.000 \\
Medicine & 18,302 & 18,365 & 0.997 & 18,357 & 18,365 & 1.000 & 18,365 & 18,365 & 1.000 \\
Geography & 3,231 & 3,242 & 0.997 & 3,240 & 3,242 & 0.999 & 3,241 & 3,242 & 1.000 \\
Engineering & 7,961 & 7,987 & 0.997 & 7,986 & 7,990 & 0.999 & 7,990 & 7,990 & 1.000 \\
Biology & 25,637 & 25,714 & 0.997 & 25,704 & 25,716 & 1.000 & 25,716 & 25,716 & 1.000 \\
Mathematics & 6,744 & 6,763 & 0.997 & 6,759 & 6,763 & 0.999 & 6,763 & 6,763 & 1.000 \\
Materials Science & 4,323 & 4,335 & 0.997 & 4,335 & 4,336 & 1.000 & 4,336 & 4,336 & 1.000 \\
Business & 2,187 & 2,193 & 0.997 & 2,192 & 2,193 & 1.000 & 2,192 & 2,193 & 1.000 \\
Physics & 13,221 & 13,251 & 0.998 & 13,249 & 13,252 & 1.000 & 13,252 & 13,252 & 1.000 \\
Geology & 4,637 & 4,646 & 0.998 & 4,644 & 4,646 & 1.000 & 4,646 & 4,646 & 1.000 \\
\hline
\end{tabular}
\end{table}

\subsubsection*{Holes Filled by Discipline}

\begin{figure}
	\centering
	\includegraphics[width=0.8\textwidth]{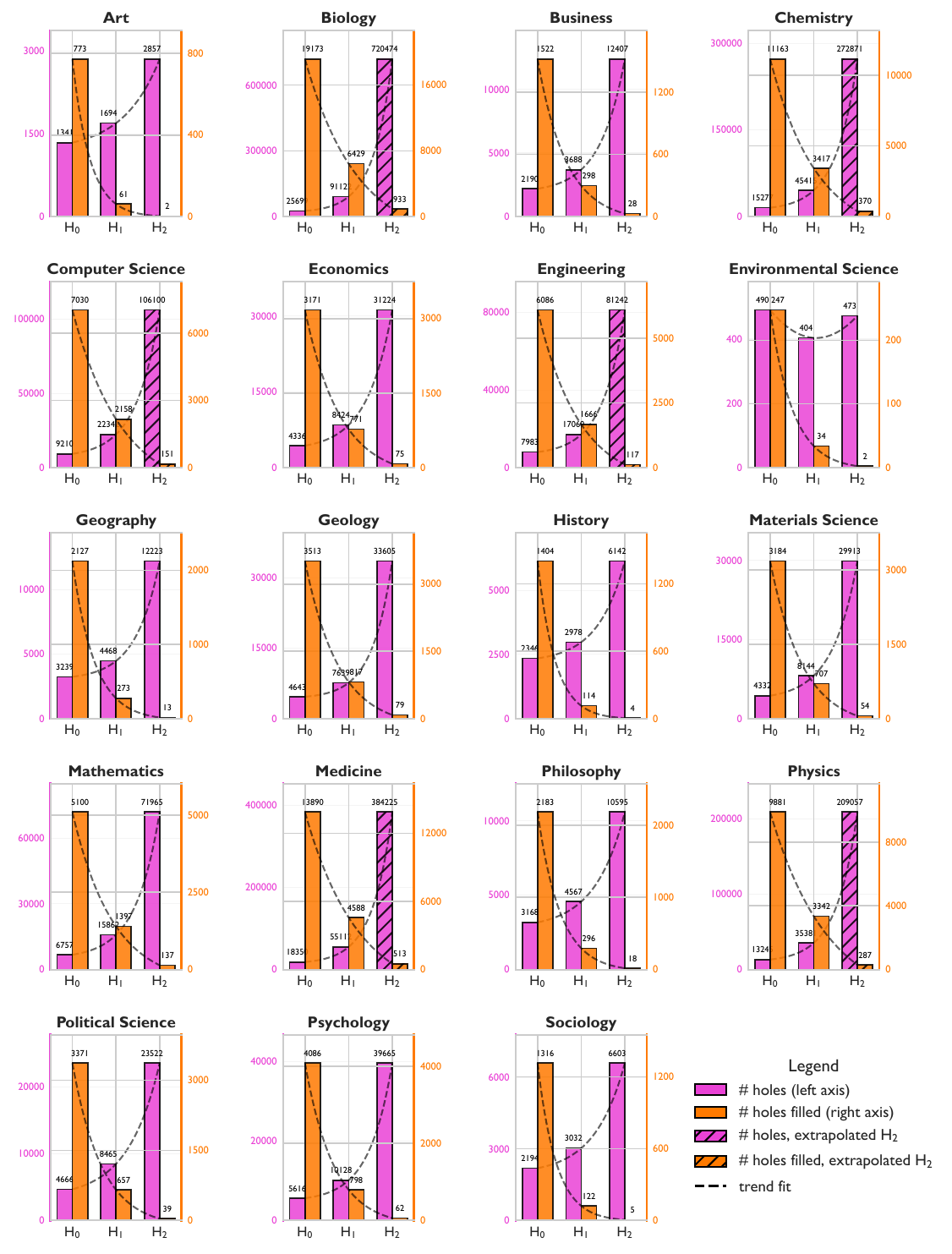}

	\caption{\textbf{Comparison of the number of holes and the number of holes filled, by discipline.}
	Same as Panel (E) of Figure~\ref{fig:hole_fill}, but for each discipline. Recall that for biology, chemistry, computer science, engineering, medicine, and physics, we did not compute persistent homology in dimension 2 due to computational limitations.
    For these disciplines, we used the power laws in the number of holes and number of holes filled to extrapolate these quantities (see Panels $D$ and $F$ of Figure \ref{fig:hole_fill}).}
	\label{fig:double_bar_plot_by_discipline}
\end{figure}

In Figure \ref{fig:double_bar_plot_by_discipline}, we show the average number of holes and holes filled for each discipline.
Recall that, due to computational limitations, we do not compute dimension-2 persistent homology for the largest fields (i.e., biology, chemistry, computer science, engineering, medicine, and physics.)
For these disciplines, we extrapolate the number of holes and the number of holes filled using the power law described Figure \ref{fig:hole_fill}.
For every discipline other than environmental science, the number of holes increases exponentially with dimension.
For every discipline, the number of holes decreases exponentially with dimension.
We attribute the discrepancy in the number of holes for environmental science to its relatively small number of concepts, of fewer than 500 concepts.
The theoretical result for the Vietoris--Rips complex, that the number of holes grows exponentially with dimension, holds only in the asymptotic regime of a large number of points \cite{BauerandPausinger2018,Kahle2011,Kahle2014}.

\begin{figure}
	\centering
	\includegraphics[width=0.8\textwidth]{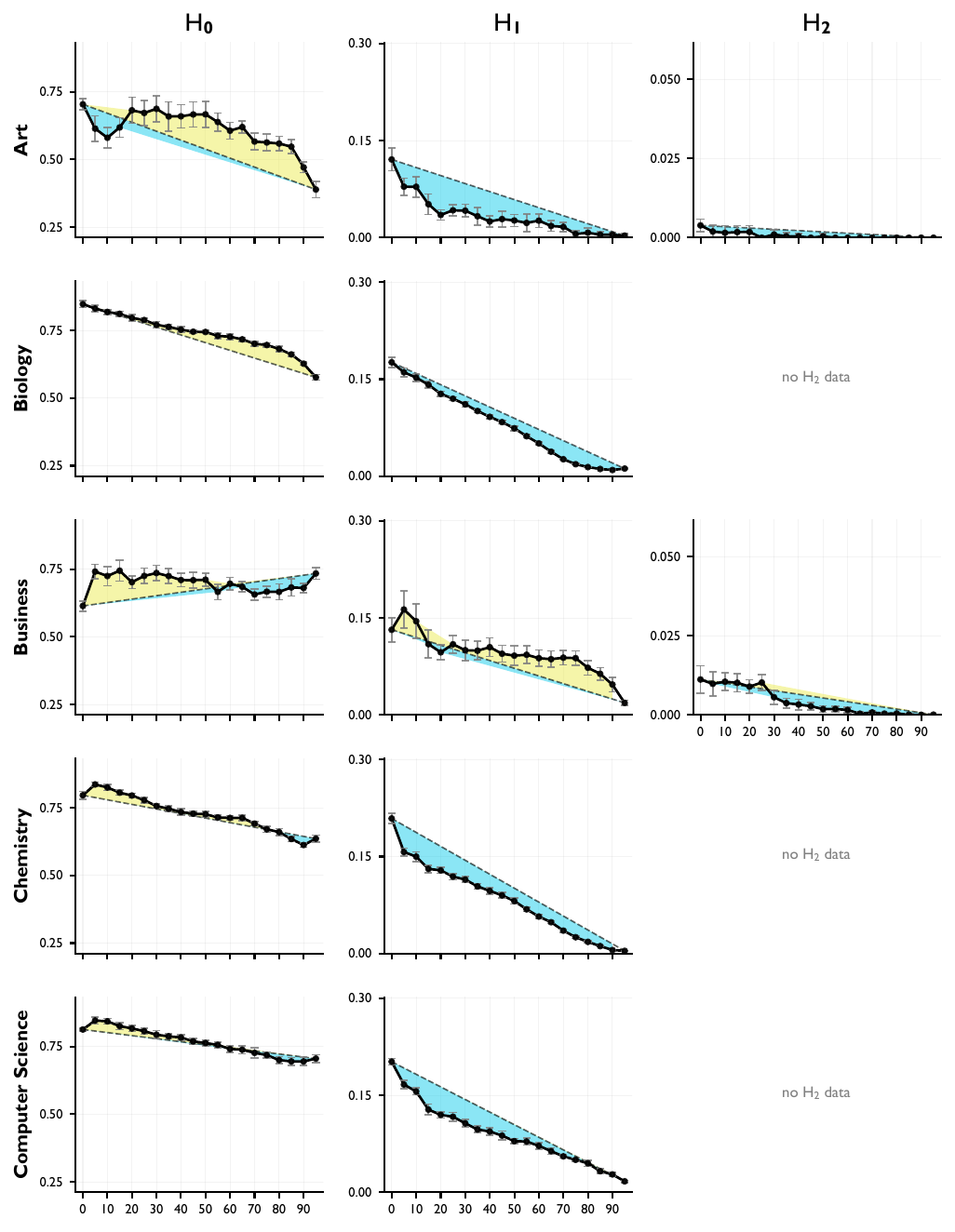}

	\caption{\textbf{Proportions of holes filled in relation to death values, by discipline.}
	Same as Panels $A$, $B$, and $C$ of Figure~\ref{fig:hole_fill}, but for each discipline. The full set of disciplines does not fit on a single page and continues on the pages that follow.}
	\label{fig:holes_filled_by_death_per_discipline_pg1}
\end{figure}

In Figure \ref{fig:holes_filled_by_death_per_discipline_pg1}, we show relationships between the proportions of holes filled, with respect to the size (i.e., death value) of the holes, for each discipline.
Read purely as curve shapes, the following descriptive features vary systematically across disciplines
and dimensions.

\paragraph{Trend}
In dimension 0, 15 of 19 disciplines
\emph{decrease} (small gaps fill in faster than large); the 4 that \emph{increase} are environmental science, business, economics and geology. In dimension 1, the
decrease is universal (19 of 19).
In dimension 2, the fill rate decreases for all the disciplines that we consider dimension-2 persistent homology.

\paragraph{Curvature} To capture the pattern of filled holes we classify each curve by its
position relative to the chord joining its endpoints: \emph{concave} $=$ curve above the chord,
indicating fillability sustained as hole size grows, and \emph{convex} $=$ curve below it,
Computing the signed area between curve and chord with the same
weighting used for the figures gives $+0.41$ at dimension $0$ (curve above the chord, concave),
$-0.70$ at dimension $1$ and $-0.19$ at dimension $2$ (both below, convex), matching the Results
text and Panels A--C of Figure \ref{fig:hole_fill}. The per-discipline classification below
reproduces exactly under the same computation: 19 of 19 at dimension $0$.
 In Dimension, 0 there are more concave than convex cases, suggesting a back-loaded regime where holes collapse in their relative likelihood of being filled at high points of the death percentile.
Concave: art, biology, business, chemistry, computer science, economics, environmental science, history, medicine, philosophy, physics.
Convex: engineering, geography, geology, materials science, mathematics, political science, psychology, sociology.

In Dimension 1, most curves fall into the front-loaded regime, with a minority whose
curves bow upward, sustaining or peaking fillability at moderate death values.
Concave: business, economics, engineering, environmental science, geography, geology, physics.
Convex: art, biology, chemistry, computer science, history, materials science, mathematics, medicine, philosophy, political science, psychology, sociology.
In Dimension 2, every curve is very convex except environmental science, the sole concave case, consistent with its interior peak (and small concept pool).
Concave: environmental science.
Convex: art, business, economics, geography, geology, history, materials science, mathematics, philosophy, political science, psychology, sociology.

In Figure \ref{fig:death_simplex_ratio_plot}, we show the shape of death simplices for holes in dimensions $1$ and $2$. Each point represents a hole; the horizontal axis denotes its death value, and the vertical axis denotes the min--max distance ratio, which is the ratio of the shortest to the longest edge length of the death simplex.
(Recall that a dimension-$0$ death simplex consists of only two points, so its min--max ratio is always $1$; we therefore omit dimension $0$.)
We produce separate plots for all holes and for filled holes.

In dimension $1$, the min--max ratios for both all holes and filled holes are predominantly between $0.8$ and $1$.
In dimension $2$, the ratios for all holes are also predominantly between $0.8$ and $1$, whereas the ratios for filled holes are predominantly between $0.6$ and $0.8$.
We also observe that, for both dimensions, filled holes have, on average, a smaller min--max ratio than holes in general.
This indicates that filling a hole requires combining concepts that are at comparable distances from one another --- three concepts in dimension $1$, four in dimension $2$ --- but that holes with slightly more uneven geometry are more likely to be filled than those with roughly equal edge lengths. This is likely because holes filled with uneven vertices approximate the projection to a lower-dimensional hole, which is likely easier to identify and verify. For example, if a paper newly combines three concepts $a$, $b$, and $c$, and $a$ and $b$ are close to one another in embedding space, while $c$ is distant (indicating that others have previously combined things nearby $a$ and $b$, but not $c$), then combining the three with a 2D triangle approximates combining the complex of $a-b$ with $c$ as a 1D edge.

\paragraph{Citation Impact} In Figure \ref{fig:citation_hit_ratio_by_field}, we show the proportion of papers that fill 0th, 1st, and 2nd order holes within the top $N$\% of citations, and the ratio of that proportion to the baseline, respectively, for each field, where N equals 25\%, 5\% and 1\% (see Figure~\ref{fig:hole_fill_papers}, panels $C$ and $D$, for 10\%). For the 25\% threshold, these figures demonstrate that while all fields manifest an increase in hit papers for bridging zeroeth-order gaps and tiling first-order holes, only discovery-oriented empirical fields systematically grow in high citation likelihood as hole dimensionality increases. In contrast, formal and design fields symmetrically decrease in hit citation likelihood. While the same increasing/decreasing pattern holds for the 10\%, 5\% and 1\% thresholds, the likelihood that first and second-order holes exhibit higher likelihoods of hit papers than baseline decreases with more restrictive citation thresholds. Nevertheless, all fields besides engineering and art show a higher likelihood of $H_1$ holes filled achieving higher-than-average 1\% citations than would be expected at baseline.

\subsubsection*{Holes Filled by Simulated Papers}

Recall that a topological hole is counted as \emph{filled} when at least one paper's concept set contains every concept lying on the hole's death simplex.
That is, when a real paper spans the conceptual gap the hole represents.
To understand the mechanisms behind how holes are filled, we use a battery of null models to generate synthetic papers and compare the proportion of holes filled by synthetic papers against real papers.
Each null model generates papers that preserve some features of real papers and randomize others; contrasting fill rates across these null models helps us isolate which structural ingredient is responsible for a hole being filled.

Each model preserves the empirical distribution of the number of concepts per paper, so that differences cannot be attributed to papers simply being larger or smaller.
Throughout, the persistent-homology structures of the concept embeddings are held fixed; only the papers that fill their holes vary.

\paragraph{Notation.}
Fixing a discipline and year, let the discipline--year concept vocabulary be $\mathcal{C}=\{1,\dots,N\}$, and let $\{k_p\}$ be the observed lengths (concept counts) of the real papers, which every generative null preserves.
Write $f_j$ for the total number of
(paper,\,concept) incidences involving concept $j$ across that discipline--year pair, and $p_j = f_j / \sum_{l\in\mathcal{C}} f_l$ for the corresponding empirical concept frequency.
Recall that for each discipline--year pair, we have a word2vec concept embedding.
For concepts $i$ and $j$, we use the cosine distance $d_{ij} = 1 - \dfrac{\mathbf{v}_i \cdot \mathbf{v}_j}{\lVert \mathbf{v}_i\rVert\,\lVert \mathbf{v}_j\rVert}$ and the induced similarity $s_{ij} = \dfrac{1}{1 + d_{ij}}$.

\paragraph{Uniform random concepts.}
Each simulated paper draws its concepts as a uniform random sample, without replacement, from the full concept vocabulary of that discipline--year.
This model preserves the concept-count frequencies, but retains neither concept popularity nor any relationship between concepts.
This is the most permissive null model, in that a hole is filled only if an arbitrary, unstructured combination of concepts happens to coincide with the concepts in its death simplex.
We show the fill rate for the random-concepts null model in Figure \ref{fig:filled_by_random_concepts_papers}.

\paragraph{Frequency-preserving resample (``permuted'').}
We form the full multiset of $\sum_p k_p$ concept slots implied by the real paper lengths and assign each slot an independent draw from $\mathrm{Multinomial}(p)$.
That is, concept $j$ is drawn with probability $p_j$.
Slots are then grouped back into papers so that paper $p$ receives exactly $k_p$ concepts.
This method resolves within-paper collisions (a concept drawn more than once for the same paper) by performing frequency-preserving swaps with other papers, so that each simulated paper is a proper set of distinct concepts. This construction preserves both each paper's length $k_p$ and each concept's marginal frequency $f_j$ (in expectation), while ignoring co-occurrence structure (i.e., which concepts tend to appear \emph{together}).
Comparing against the uniform null helps identify how much of the fill rate is attributable to concept popularity alone.
We show the fill rate for the permuted-concepts null model in Figure \ref{fig:filled_by_permuted_concepts_papers}.

\paragraph{Concept-relabeling control (``shuffled'')}
This null model leaves both the real papers and the real homology entirely untouched and instead randomly permutes the correspondence between holes and the concepts on their death simplices.
This null model leaves the real papers untouched but randomly permutes the labels of each concept.
Concretely, we shuffle the concept labels attached to each embedding's death simplices, so that each hole is re-associated with an arbitrary set of concepts of the same size before we check whether any real paper contains that set.
This asks whether the fill rate of holes is specific to the actual concepts bounding it, or is merely what one would expect for an arbitrary concept combination of the same size.
Because the real papers are retained verbatim, it isolates the identity of a hole's bounding concepts as the variable of interest, complementing the generative nulls that instead resample the papers.
We show the fill rate for the shuffled-concepts null model in Figure \ref{fig:filled_by_shuffled_concepts_papers}.

\paragraph{Similarity-weighted growth (``similar,'' with temperature).}
This model generates papers one concept at a time, using the embedding space to determine distance and similarity between concepts, with a temperature
parameter $T>0$ controlling how strongly the growth favors conceptual coherence.
A paper of length $k$ is built as follows.
The first concept $c_1$ is drawn from the temperature-tilted frequency distribution
\begin{equation}
  \Pr(c_1 = j) \;=\; \frac{f_j^{\,1/T}}{\sum_{l\in\mathcal{C}} f_l^{\,1/T}} .
\end{equation}
For each subsequent step $t = 2,\dots,k$, let $S_{t-1} = \{c_1,\dots,c_{t-1}\}$ be the concepts chosen so far.
Every remaining concept $a \notin S_{t-1}$ is assigned an affinity equal to its maximum similarity to the current set, $\sigma_a = \max_{c\in S_{t-1}} s_{ac}$, and the next concept is drawn with probability
\begin{equation}
  \Pr(c_t = a \mid S_{t-1})
  \;=\; \frac{\sigma_a^{\,1/T}}{\sum_{a'\notin S_{t-1}} \sigma_{a'}^{\,1/T}}\,,
  \qquad a \notin S_{t-1}
\end{equation}
(reducing to a uniform draw over the remaining concepts in the degenerate case $\sigma_a = 0$ for all $a$). The exponent $1/T$ tilts both distributions.
As $T \to 0$ the process becomes greedy, always adding the concepts most similar to those already present and producing tightly coherent, homophilous
papers; $T = 1$ samples in direct proportion to embedding similarity; and larger $T$ flattens the weights toward the uniform-random null.
We report six temperatures, $T \in \{0.01,\, 0.05,\, 0.1,\, 0.5,\, 1,\, 2\}$, tracing the
transition from strongly coherent papers ($T = 0.01$) to increasingly diffuse sampling ($T = 2$).
This null model preserves paper length, concept popularity, and the local geometry of the concept embedding.
We show the fill rate for the similar-concepts null model in Figures \ref{fig:filled_by_similar_concepts_papers_T=0.05}--\ref{fig:filled_by_similar_concepts_papers_T=2}.

\paragraph{Summary}
For all null models other than the similar-concept null model, we observe that the fill rate of papers is substantially lower than for real papers.
In dimension $0$, the fill rate is a fraction of that of the real papers, and in dimensions $1$ and $2$, the fill rate is virtually $0$ for all disciplines.
However, for the similar-concept null model, as we lower the temperature $T$ from $2$ to $0.05$, the fill rate starts to resemble that of real papers.
In particular, the fill rate across all dimensions increases across all fields.
When we consider $T=0.01$, however, the fill rate across all dimensions drops.
This shows that the fill rate of holes is sensitive to the similarity of concepts in papers.
Only when concepts are arranged with a structure sufficiently similar to real data do the fill rates approach those observed empirically.
At very low temperatures ($T=0.01$), the model becomes too constrained, which consequently limits the ability to fill holes.

\subsubsection*{Papers That Fill Holes}

In Figure \ref{fig:fill_papers_art}, we show the performance of hole-filling papers in each discipline with respect to citation count, content surprise, and context surprise.
For each metric (citation, content surprise, and context surprise), we determine each discipline's enrichment ratio (i.e., the proportion of hole-filling papers above a given metric percentile, compared with the baseline) trajectory across the dimensions available to it.
Six disciplines---biology, chemistry, computer science, engineering, medicine, physics---have no dimension 2 data (from size), so we can only show their dimension 0 and dimension 1 information.
We provide per-discipline enrichment values of each metric, for each discipline and in each dimension, in Table~\ref{tab:pd_main}.

\paragraph{Citation Count}
The disciplines that gain citations from filling higher-dimensional gaps are not a random subset. They
divide along the boundary between discovery-oriented empirical and data-driven fields versus design-oriented formal and theoretical
fields, and the division widens with dimension. Averaging the enrichment within each group, the discovery-oriented sciences run $1.47 \to 1.89 \to 2.62$ across Dimensions 0, 1 and 2, while the design-oriented
fields run $1.20 \to 1.08 \to 0.89$---starting lower, and crossing below the no-enrichment line by
Dimension 2.

Measured as the amplification from Dimension 0 to Dimension 2, the split is close to categorical. Six
fields amplify by at least $1.58\times$---history ($2.26\times$), geography ($2.00\times$), psychology
($1.67\times$), sociology, geology and environmental science---materials science is neutral
($1.01\times$), and the remaining six all attenuate: mathematics ($0.86\times$), economics, political
science, business, philosophy and art ($0.58\times$). In a discovery-oriented field, closing a void appears to
consolidate a body of evidence in a way the citing literature rewards. In a design-oriented field, the
higher-order gap is closed by work the literature treats as increasingly routine. Mathematics is the
limiting case, where the enrichment sits at approximately $1$ at Dimension 0 and drifting below it thereafter.

\paragraph{Content Surprise}
Content surprise does not respect this boundary. It rises from Dimension 0 to 1 in every one of the
natural sciences with data---biology, chemistry, physics, medicine, engineering---as well as in
mathematics and philosophy, so filling a bigger gap necessarily involves pulling together surprising combinations of contents almost everywhere, even
where it earns no additional citation. The exceptions are economics and history, whose
higher-dimensional fillers become progressively less surprising.

\paragraph{Context Surprise}
Context surprise inverts the pattern.
Under the journal-based definition, the natural and engineering sciences decline with dimension from an already low base---chemistry, engineering, medicine and physics all fall from Dimension 0 to 1---and the overall dimension trend is carried by just two fields, geology and philosophy.
The fields that the content variant marks as mildly novel are precisely those the context variant marks as conventional, and vice versa.

\begin{table}
    \centering
    \caption{\textbf{Citation and surprise enrichment of hole-filling papers, by discipline.}
        Each entry is the mean enrichment ratio over metric percentiles $p=80-99$. Rows are sorted by mean citation enrichment and ``---'' marks a dimension not available for that discipline.}
    \label{tab:pd_main}

    \footnotesize
    \setlength{\tabcolsep}{4pt}
    \begin{tabular}{lccccccccc}
        \\
        \hline
        & \multicolumn{3}{c}{Citation} & \multicolumn{3}{c}{Surprise (content)}
        & \multicolumn{3}{c}{Surprise (context)}\\
        \cline{2-4}\cline{5-7}\cline{8-10}
        Discipline & $H_0$ & $H_1$ & $H_2$ & $H_0$ & $H_1$ & $H_2$ & $H_0$ & $H_1$ & $H_2$\\
        \hline
        Geography & 1.57 & 2.32 & 3.13 & 1.30 & 1.33 & 1.92 & 1.51 & 1.47 & 1.14\\
        Psychology & 1.67 & 2.33 & 2.79 & 1.28 & 1.38 & 1.49 & 1.58 & 1.71 & 1.57\\
        Geology & 1.55 & 2.14 & 2.45 & 1.24 & 1.19 & 1.19 & 1.58 & 1.82 & 2.37\\
        History & 1.32 & 1.59 & 2.98 & 1.43 & 1.36 & 1.17 & 1.93 & 2.65 & 2.39\\
        Environmental Science & 1.42 & 2.06 & 2.23 & 1.19 & 1.60 & 2.54 & 1.10 & 1.67 & 0.85\\
        Biology & 1.58 & 2.04 & --- & 1.28 & 1.38 & --- & 1.34 & 1.35 & ---\\
        Chemistry & 1.58 & 1.96 & --- & 1.23 & 1.32 & --- & 1.09 & 1.08 & ---\\
        Sociology & 1.33 & 1.50 & 2.10 & 1.34 & 1.25 & 2.30 & 1.97 & 1.85 & 2.30\\
        Medicine & 1.34 & 1.57 & --- & 1.20 & 1.30 & --- & 1.19 & 1.17 & ---\\
        Physics & 1.37 & 1.44 & --- & 1.21 & 1.25 & --- & 1.03 & 1.02 & ---\\
        Materials Science & 1.33 & 1.34 & 1.35 & 1.26 & 1.38 & 1.25 & 1.10 & 1.07 & 1.11\\
        Political Science & 1.42 & 1.42 & 1.11 & 1.38 & 1.53 & 1.29 & 2.16 & 2.44 & 1.38\\
        Computer Science & 1.27 & 1.11 & --- & 1.28 & 1.31 & --- & 1.93 & 2.05 & ---\\
        Philosophy & 1.25 & 1.26 & 0.86 & 1.22 & 1.37 & 1.51 & 1.92 & 2.38 & 4.32\\
        Business & 1.27 & 1.11 & 0.98 & 1.16 & 1.17 & 1.04 & 1.92 & 1.66 & 0.94\\
        Economics & 1.21 & 1.06 & 0.99 & 1.21 & 1.10 & 1.03 & 2.02 & 1.79 & 1.43\\
        Engineering & 1.14 & 0.98 & --- & 1.32 & 1.40 & --- & 1.31 & 1.07 & ---\\
        Mathematics & 1.07 & 0.97 & 0.92 & 1.11 & 1.17 & 1.29 & 1.42 & 1.36 & 1.26\\
        Art & 1.20 & 1.04 & 0.70 & 1.37 & 1.15 & 2.18 & 1.56 & 1.82 & 0.00\\
        \hline
    \end{tabular}
\end{table}

\subsubsection*{Prescience of hole-filling papers}

To analyze the impact of hole-filling works, we analyze their prescience index.
Prescience captures whether a work anticipated future research directions \cite{Lockhart2026China}.
Like surprise, it has content and context variants, but rather than measuring how unexpected a work's combinations were at the time of publication, prescience measures whether those initially improbable combinations became typical or probable under the model in subsequent research.
A highly prescient work seems unusual when published but later proves to be ahead of its time, indicating that it helped redirect rather than simply follow the scientific frontier.
We provide per-discipline enrichment values of prescience, for each discipline and in each dimension, in Table~\ref{tab:pd_pres}.
We show the overall prescience of hole-filling works in Figure \ref{fig:fill_papers_prescience}, and for each discipline in Figure \ref{fig:fill_papers_art_prescience}.

Similar to our analysis of citation count and surprise indices, we observe a notable trend: on average, for each dimension $k=0,1,2$, hole-filling works have higher prescience than general works.
This trend is consistent among most disciplines and dimensions (with the exception of environmental science, history, and philosophy in dimension $2$, and art in dimensions $1$ and $2$ for content prescience; and business, philosophy, and political science in dimension $2$, and art in dimensions $1$ and $2$ for context prescience).
This suggests that, on average, hole-filling works tend to be more prescient than general works.
When comparing works that fill $0$-dimensional holes and works that fill $1$-dimensional holes, we note that --- with the exception of art --- the relative prescience among hole-filling works in dimension $1$ has similar or higher content prescience than hole-filling papers in dimension $0$.
However, when comparing works that fill $2$-dimensional holes and works that fill $1$-dimensional holes, there is a mix of disciplines that rise and fall in prescience
Of particular note, the content and context prescience for $2$-dimensional hole-filling works compared to 1-dimensional hole-filling works drops for environmental science, history, philosophy, and art; for context prescience, we see this drop also occurs for business, economics, and political science.
Similar to our other metrics, we often observe that, as we consider larger prescience-index-percentile thresholds, we frequently observe larger relative prescience in hole-filling works than for general works.

\begin{table}
    \centering
    \caption{\textbf{Prescience enrichment of hole-filling papers, by discipline.}
        Layout and conventions as in Table~\ref{tab:pd_main}; rows keep the same citation-based
        ordering so the two tables can be read against each other.}
    \label{tab:pd_pres}

    \footnotesize
    \setlength{\tabcolsep}{5pt}
    \begin{tabular}{lcccccc}
        \\
        \hline
        & \multicolumn{3}{c}{Prescience (content)} & \multicolumn{3}{c}{Prescience (context)}\\
        \cline{2-4}\cline{5-7}
        Discipline & $H_0$ & $H_1$ & $H_2$ & $H_0$ & $H_1$ & $H_2$\\
        \hline
        Geography & 1.19 & 1.30 & 1.40 & 1.11 & 1.07 & 1.11\\
        Psychology & 1.19 & 1.27 & 1.22 & 1.43 & 1.57 & 1.50\\
        Geology & 1.14 & 1.21 & 1.18 & 1.17 & 1.21 & 1.39\\
        History & 1.22 & 1.38 & 0.28 & 1.23 & 1.38 & 1.13\\
        Environmental Science & 1.06 & 1.22 & 0.07 & 1.24 & 1.39 & 1.18\\
        Biology & 1.16 & 1.23 & --- & 1.38 & 1.38 & ---\\
        Chemistry & 1.14 & 1.18 & --- & 1.26 & 1.24 & ---\\
        Sociology & 1.24 & 1.36 & 3.31 & 1.49 & 1.36 & 1.29\\
        Medicine & 1.11 & 1.21 & --- & 1.31 & 1.29 & ---\\
        Physics & 1.15 & 1.23 & --- & 1.11 & 1.10 & ---\\
        Materials Science & 1.14 & 1.18 & 1.30 & 1.19 & 1.12 & 1.09\\
        Political Science & 1.44 & 1.57 & 1.81 & 1.50 & 1.82 & 0.62\\
        Computer Science & 1.33 & 1.34 & --- & 1.50 & 1.49 & ---\\
        Philosophy & 1.19 & 1.30 & 1.21 & 1.26 & 1.28 & 0.37\\
        Business & 1.26 & 1.32 & 1.72 & 1.38 & 1.37 & 1.05\\
        Economics & 1.30 & 1.27 & 1.22 & 1.39 & 1.17 & 1.03\\
        Engineering & 1.30 & 1.40 & --- & 1.41 & 1.33 & ---\\
        Mathematics & 1.10 & 1.12 & 1.09 & 1.24 & 1.16 & 1.04\\
        Art & 1.22 & 0.72 & 1.14 & 1.12 & 0.94 & 0.00\\
        \hline
    \end{tabular}
\end{table}

\subsubsection*{Methods for supporting analyses}

The analyses below share a common setting. Holes are the persistent-homology classes already
described, restricted to the analysis window $2005$--$2019$. The class of infinite death value,
the single $H_0$ class that never dies, whose death simplex is empty and which would therefore be
matched by every work, is excluded throughout, as is any class whose death simplex does not
resolve to exactly $k+2$ distinct concepts. A work \emph{closes} a hole when its concept set
contains the entire death simplex, the same superset rule used elsewhere in this paper. Where a
standard error is quoted, it treats each discipline--year as a single observation and measures the
dispersion of rates across those cells, matching the estimator used for the enrichment curves in
the main text. A binomial error on the pooled proportion would understate the uncertainty roughly
sixfold at dimension $0$, because rates differ substantially between discipline--years. Every rate
reported below is \emph{weighted}: a plotted or tabulated proportion is the pooled count over all
discipline--years, which is the count-weighted mean of the per-cell rates, and the standard error
weights each cell by the same counts. This follows the main text, and the choice is consequential
rather than cosmetic. Weighting the top-$1\%$ eligible-baseline enrichment gives $1.261$,
$1.411$ and $1.185$ across dimensions $0$, $1$ and $2$, where weighting each discipline--year
equally instead gives $1.270$, $1.388$ and $1.425$, turning the dimension-2 figure into the
largest of the three.

\paragraph{Later closure of open holes (Table \ref{tab:claude-schedule}).} For a hole open in year
$y$ we count the works published in each of $y+1,\dots,y+5$ whose concepts contain its death
simplex. Horizons of one and two years are recorded directly in the homology output; horizons
three to five were computed for this analysis on a stratified subsample of $2{,}500$ holes per
(year, dimension, closed-status) cell, which suffices because every quantity reported is a rate
within a (year, dimension, death-decile) cell. Holes are ranked into death deciles within each
discipline--year--dimension fiber, and the table gives the rank correlation between decile and
closure probability at each horizon.

\paragraph{Recurrence.} The same counts support the recurrence claim in the abstract and
discussion: the probability that a hole closed in year $y$ has its death simplex realized again in
$y+1$, set against the corresponding probability for holes left open in $y$. Two caveats belong
with these figures. The comparison is made within death decile and within discipline--year but is
\emph{not} matched on concept frequency, and part of the association is mechanical, since a
combination realized once involves concepts that co-occur and such concepts are combined again.
And the ratio between the two probabilities depends strongly on order, because the baseline for
open holes differs by three orders of magnitude between dimensions; the absolute rates are the
more informative quantity, and the elevated rate is stable across all five horizons.

\paragraph{Eligible baseline for citation enrichment (Table \ref{tab:claude-eligible},
Figure \ref{fig:claude-eligible}).} Closing a
hole of order $k$ requires containing a $(k+2)$-vertex simplex, so a work carrying fewer than
$k+2$ of the discipline--year's concepts cannot close one. The eligible baseline restricts the
comparison population to works carrying at least $k+2$ such concepts, counting only concepts
present in that discipline--year's concept list, since concepts outside it cannot belong to any
simplex. Enrichment is otherwise computed exactly as in the main text: the proportion of closing
works at or above a citation percentile, divided by the same proportion in the comparison
population, taken over all works rather than a single document type.

\paragraph{Filling horizon (Figure \ref{fig:claude-horizon}).} A year-$y$ embedding is trained on
the cumulative corpus through $y$, so a year-$y$ work that closes a year-$y$ hole is itself part of
the corpus that fixed the geometry of that hole. To check that the citation premium does not depend
on this, we repeat the measurement requiring the closing work to be published in $y+1$ or $y+2$,
which places it outside the training corpus by construction. Two details matter. A horizon-$h$
closing work is compared against the $y+h$ population, since comparing it against the $y$ cohort
would penalise it for being a year younger; and where eligibility enters it is counted against
\emph{year $y$'s} concept vocabulary, which is the vocabulary the simplex lives in. The single-year
filler sets are read from the stored homology output, so no embedding is retrained. The comparison
population here is \emph{all} works, matching the enrichment figures in the main text rather than
the eligible baseline of Figure \ref{fig:claude-eligible}, so that the horizons can be read
directly against them; panel (A) accordingly reproduces the main text's same-year values of
$1.437$, $1.758$ and $1.564$.

\paragraph{Scaling of holes and closed holes (Table \ref{tab:claude-prefactors}).} For each
discipline we take the mean number of classes and the mean number of closed classes over the
window, and regress $\log_{10}$ of each on $\log_{10}$ of the discipline's concept count by
ordinary least squares, giving the exponent $a$ and prefactor $C$ of $y = C\,n^{a}$. Dimension 2
uses the thirteen disciplines for which it was computed. Concept count is obtained as the number
of finite dimension-0 classes plus one; in a Vietoris--Rips filtration every point is a
dimension-0 class born at radius zero and the essential class is excluded, so this is an identity
up to the $+1$ offset, and the dimension-0 row of the table is definitional rather than measured.

\paragraph{Later motion of the joined concepts (Table \ref{tab:claude-drift},
Figures \ref{fig:claude-drift} and \ref{fig:claude-size}).} Each year's embedding is trained
separately, so the distances among the $k+2$ concepts of a death simplex change from year to year.
We summarize a simplex in a given year by its mean pairwise edge length and follow that quantity
over the years around the year in which the hole is recorded. Holes closed in year $y$ are
compared with holes still open in $y$ from the same discipline, year, order and death decile,
matched in addition on the mean edge before filling, on the frequency of the concepts involved,
and on the trend before filling. Only holes closed by exactly one work are used: one document is a
negligible part of a cumulative corpus of millions, so the motion measured is not the closing
work's own contribution to the training data. The concepts of both groups move apart over time,
although randomly chosen concepts do not (see below), so the quantity reported is the difference
between the two groups' changes --- a statement about relative, not absolute, motion. The trend
before filling is matched on the change from $y-3$ to $y-1$, which makes the $y-1$ column close to
zero by construction; the $y-2$ column is the informative placebo. At order 0 it is an eighth of
the five-year estimate, so that estimate reads as a break at filling. At order 1 it is two fifths
of it ($-1.24$ against $-2.94$), so part of the order-1 difference was already present before the
hole was closed.

Figure \ref{fig:claude-drift} shows one closed hole beside one matched open hole, and two things
limit what it can carry. The pair was selected \emph{because} it displays the average pattern
clearly, from $11{,}093$ matched order-1 pairs across five disciplines; and an individual hole's
mean edge moves by around $0.03$ from one year to the next, against an average effect of $0.005$,
so no single pair can speak to the size of the effect. Both triples were chosen to be recognisable
subject matter --- theatre on one side, cultural history on the other --- so that the contrast
cannot be read as coherent concepts set against incoherent ones, and the two were the same
distance apart at $y-3$. The nearby concepts in that figure are held in one position across the
panels. That is legitimate because the surrounding space does not, on average, move: over $4{,}699$
closed order-1 holes, the mean distance among randomly chosen concepts changes by $-0.0009$ over
five years and among near neighbours by $+0.0002$. What the data support is therefore not that the
surroundings spread while the hole holds, but that a closed simplex contracts slightly
\emph{relative} to its neighbourhood, by $0.0028$. In the example itself the backdrop's mean
pairwise distance changes by at most $0.03$ across the panels, against $0.12$ for the open hole.

Figure \ref{fig:claude-size} drops the matching and measures each hole against the year in which
it is recorded rather than a year before filling. For every hole of 2005--2019, closed in $y$ or
left open through $y+2$, it plots the mean edge in each year from $y-3$ to $y+5$ divided by its
value in $y$. Both groups are at their smallest in $y$ itself. In every other year an open hole has
moved further from its year-$y$ size than a closed one, by a factor of about two at order 0,
three to five at order 1, and three to nineteen at order 2, and that factor is larger before
closure than after it at every order. What the figure shows, then, is that the holes which get
closed are those whose concepts keep a steadier configuration, before closure as well as after. It
does not isolate an effect of closing; that is what the matched comparison of
Table \ref{tab:claude-drift} is for.

\paragraph{Order-filtered filler sets (Table \ref{tab:claude-overlap},
Figure \ref{fig:claude-order}).} Closing a dimension-$k$ hole means containing its $(k+2)$-vertex
death simplex, and a work carrying those concepts necessarily carries every sub-simplex they span.
One work can therefore close a void, a loop and a gap at once, and the three orders are not
separate populations of works. Table \ref{tab:claude-overlap} gives the overlap. It is strongly
one-directional: $84.8\%$ of the works that close a loop also close a gap, and $92.6\%$ of those
that close a void also close a gap, with $48.4\%$ also closing a loop --- while the works that
close a gap are $95.0\%$ gap-only, simply because there are twenty times more of them. Only $914$
works in the window close a void and nothing of lower order.

To separate the orders we reclassify each hole-closing work in two ways, shown alongside the
published comparison as the three panels of Figure \ref{fig:claude-order}. Panel (A) is the main
text's own measurement (Figure \ref{fig:hole_fill_papers}D): a work enters the dimension-$n$ curve
if it closes a hole of order $n$, whatever else it closes. Panels (B) and (C) classify the work
instead by the \emph{highest} and by the \emph{lowest} order hole it closes. Both require an
order-$n$ closure and both partition the filler population, since every work has exactly one
highest and one lowest order.

Classifying by the highest order barely changes anything, staying within $0.018$ of (A) at every
percentile. That is worth stating rather than hiding: it means the dimension-1 premium is not
carried by the rare works that also close voids. Classifying by the lowest order is what makes the
orders disjoint --- at dimension 1 it keeps only the $295{,}302$ works that close a loop but no
gap, where the unrestricted dimension-0 and dimension-1 sets share $1{,}645{,}748$ works. On those
disjoint sets the dimension-1 premium still exceeds the dimension-0 premium, $1.606\pm0.063$
against $1.437\pm0.025$ in the top $1\%$ against all works, so the step from gaps to loops is not
an artefact of the two filler sets overlapping. Dimension 2 cannot be placed: its disjoint set of
$914$ works gives $1.524\pm0.885$. Comparing a ``loop but no gap'' set against all works does mix
the order-1 effect with the property of closing nothing lower. Restricting the comparison
population to works that also close nothing below order $n$ gives $1.679\pm0.067$ at dimension 1
and does not change the ordering.

\begin{table}[htbp]
\centering
\caption{\textbf{Among holes still open in year $y$, a larger death value predicts a lower
probability of being closed, at every horizon.} Rank correlation between death decile and the
probability of closure in each subsequent year, over holes unfilled at $y$, 2005--2019. A reading
of the filtration radius as a schedule of discovery predicts the opposite sign, and predicts the
association weakening as the horizon lengthens; neither occurs. At dimension 1 the annual closure
probability falls from $0.062$ in the smallest death decile to $0.004$ in the largest.}
\label{tab:claude-schedule}
\begin{tabular}{lrrrrr}
\hline
Order & $y{+}1$ & $y{+}2$ & $y{+}3$ & $y{+}4$ & $y{+}5$ \\
\hline
$H_0$ & $-0.70$ & $-0.73$ & $-0.81$ & $-0.78$ & $-0.81$ \\
$H_1$ & $-1.00$ & $-1.00$ & $-1.00$ & $-1.00$ & $-1.00$ \\
$H_2$ & $-0.99$ & $-0.96$ & $-0.99$ & $-1.00$ & $-0.99$ \\
\hline
\end{tabular}
\end{table}

\begin{table}[htbp]
\centering
\caption{\textbf{Restricting the comparison population to works that could have filled a hole of
the given order removes two fifths to a half of the citation premium at dimensions 0 and 1, and two
thirds at dimension 2.}
Citation enrichment of hole-filling works in the top $1\%$, against all works and against all
works carrying at least $k+2$ of the discipline-year's concepts. Intervals are $95\%$, from a
standard error treating each discipline--year as one observation. Under the eligible baseline the
dimension-$2$ premium is distinguishable neither from one nor from the dimension-$0$ premium.}
\label{tab:claude-eligible}
\begin{tabular}{lrr}
\hline
Order & baseline: all works & baseline: $\geq k+2$ concepts \\
\hline
$H_0$ & $1.437$ & $1.261 \pm 0.025$ \\
$H_1$ & $1.758$ & $1.411 \pm 0.056$ \\
$H_2$ & $1.564$ & $1.185 \pm 0.204$ \\
\hline
\end{tabular}
\end{table}

\begin{table}[htbp]
\centering
\caption{\textbf{Holes and holes filled scale with concept count at similar exponents but very
different prefactors.} Ordinary least squares in $\log_{10}$ across disciplines, on yearly means
for 2005--2019; dimension 2 uses the thirteen disciplines for which it was computed. The
prefactor ratio $C_F/C_D$ falls roughly ninety-fold across dimension, which is what makes the
filled fraction small; the exponent difference is \emph{positive} at dimensions 0 and 1. Note
that concept count is obtained as the number of finite dimension-0 classes plus one, which in a
Vietoris--Rips filtration is an identity rather than a measurement, so the dimension-0 row is
definitional rather than measured. Its exponent is one up to the $+1$ offset, which is why the fit
returns $1.0004$ with $R^2 = 0.9999999$ rather than exactly $1$.}
\label{tab:claude-prefactors}
\begin{tabular}{lrrrrrrr}
\hline
Dim & $C_D$ & $C_F$ & $C_F/C_D$ & $a_D$ & $a_F$ & $a_F-a_D$ & $R^2$ (filled) \\
\hline
0 & $0.996$ & $0.291$ & $0.292$ & $1.000$ & $1.103$ & $+0.102$ & $0.997$ \\
1 & $0.0835$ & $1.89\times10^{-3}$ & $0.0226$ & $1.366$ & $1.505$ & $+0.139$ & $0.954$ \\
2 & $4.22\times10^{-3}$ & $1.36\times10^{-5}$ & $3.22\times10^{-3}$ & $1.867$ & $1.777$ & $-0.090$ & $0.751$ \\
\hline
\end{tabular}
\end{table}

\begin{table}[htbp]
\centering
\caption{\textbf{The filler sets of different orders overlap heavily, and in one direction: works
that close a higher-order hole almost always close a lower-order one as well.} Distinct works
closing a hole in the same year as the hole is recorded, pooled over the nineteen disciplines,
2005--2019, essential classes excluded. Rows are the works closing a hole of each order; columns
give the percentage of those works that also close a hole of another order, and the percentage
closing nothing else. A work containing a $(k+2)$-vertex death simplex also contains every
sub-simplex it spans, which is what produces the asymmetry; the reverse percentages are small only
because gap-closers outnumber loop-closers about twenty to one. Only $914$ works close a void and
nothing of lower order, which is why the dimension-2 row of
Figure \ref{fig:claude-order}(C) is uninformative.}
\label{tab:claude-overlap}
\begin{tabular}{lrrrrr}
\hline
A work closing & works & also $H_0$ & also $H_1$ & also $H_2$ & this order only \\
\hline
$H_0$ (gap)  & $33{,}370{,}963$ & ---      & $4.9\%$  & $0.04\%$ & $95.0\%$ \\
$H_1$ (loop) & $1{,}941{,}050$  & $84.8\%$ & ---      & $0.4\%$  & $15.2\%$ \\
$H_2$ (void) & $15{,}583$       & $92.6\%$ & $48.4\%$ & ---      & $5.9\%$ \\
\hline
\end{tabular}
\end{table}

\begin{table}[htbp]
\centering
\caption{\textbf{The concepts joined by a closed hole drift apart more slowly than those of
comparable holes left open, at orders 0 and 1.} Difference between the two groups' change in mean
pairwise edge length, in units of $10^{-3}$, for holes closed by exactly one work; like-for-like
groups matched on discipline, year, order, death decile, pre-filling level, concept frequency and
pre-existing trend. Columns $y{-}2$ and $y{-}1$ are measured before filling; because the trend is
matched on the change from $y{-}3$ to $y{-}1$, the $y{-}1$ column is near zero by construction and
$y{-}2$ is the informative placebo. A negative
entry means the closed holes' concepts separated \emph{less} than the open holes' over the same
interval, not that they moved closer: in absolute terms both groups separate, the closed holes by
$+1.35\times10^{-3}$ at order 0 and $+2.23\times10^{-3}$ at order 1 over five years against
$+6.47$ and $+5.17$ for the open holes. Order 2 shows no difference at any horizon, the two groups
separating by $+4.44\times10^{-3}$ apiece. The filling year is positive at every order, so
contamination of that year's embedding by that year's filling work cannot produce the later
pattern. Counts of closed holes are $198{,}930$, $185{,}209$ and $5{,}215$. The comparison is
observational: the pre-existing trend had to be matched away rather than being absent, and at
order 1 the $y{-}2$ placebo is two fifths the size of the five-year estimate.}
\label{tab:claude-drift}
\small
\setlength{\tabcolsep}{4pt}
\begin{tabular}{lrrrrrr}
\hline
Order & $y{-}2$ & $y{-}1$ & $y{+}0$ & $y{+}1$ & $y{+}3$ & $y{+}5$ \\
\hline
$H_0$ & $-0.62\pm0.08$ & $-0.18\pm0.04$ & $+1.92\pm0.09$ & $-2.38\pm0.10$ & $-4.08\pm0.11$ & $-5.12\pm0.12$ \\
$H_1$ & $-1.24\pm0.05$ & $-0.31\pm0.02$ & $+1.96\pm0.05$ & $-1.80\pm0.06$ & $-2.38\pm0.07$ & $-2.94\pm0.07$ \\
$H_2$ & $-1.08\pm0.21$ & $-0.02\pm0.13$ & $+6.42\pm0.25$ & $+0.35\pm0.28$ & $+0.18\pm0.34$ & $-0.00\pm0.40$ \\
\hline
\end{tabular}
\end{table}

\subsection*{Death-value analysis of hole-filling papers}

We now discuss the relationship between death values of holes and the performance
of works that fill them.

We characterize the performance of a hole-filling work by the  three paper-level metrics---its citation count, its content surprise, and its context surprise---and ask, for each metric, whether works that close larger holes perform differently from those that close smaller ones. These metrics are distinct from citation-based metrics in that they are calculated from the moment of publication and can be calculated based on the data and metadata of the research products alone.
(We also separately consider the content prescience and context prescience.)
Both quantities are naturally expressed as percentiles: the size of a hole through the percentile of its death value, and the performance of a work through the percentile of its metric. We therefore study their relationship through a two-dimensional surface whose axes are the \emph{death percentile} of the filled hole and a \emph{metric-percentile threshold}.
From the raw counts we build first a \emph{proportion surface} and then, after normalizing against a baseline, a \emph{ratio (enrichment) surface}; these are constructed separately for each discipline, each homology dimension ($H_0$, $H_1$, and, if available, $H_2$), and each metric.

The unit of analysis is the hole-filling work. Every filled hole is associated with works that fill it; a work that fills several holes (in a given discipline--year and dimension) is counted once and is assigned the largest death percentile of the holes it fills.
Death is a property of the hole: its death percentile $\delta \in [0,100]$ is the rank of the hole's death value within the death distribution of that discipline and dimension, so a larger $\delta$ denotes a larger hole among comparable holes.
We partition the death axis into bins $B_1,\dots,B_K$ and place each work in the bin containing its death percentile.
Each work falls in exactly one bin.
(Unless noted, the bin edges are the percentiles at which the data are compiled, $0,1,2,3,4,5,10,15,\dots,95,100$; because data is sparse in $H_2$, we coarsen consider fixed-width decile bins.)

Within a discipline, dimension and metric, let each filling work $w$ carry a metric percentile $\rho_w \in [0,100]$---the rank of its metric value within the appropriate comparison cohort, so that ``$\rho_i \ge p$'' means the work lies in the top $(100-p)\%$ on that metric.
For citation, $\rho$ is the percentile of the work's citation count among papers of the same discipline and publication year.
For surprise and prescience, $\rho$ is computed similarly. (Note that the surprise and prescience percentiles are computed in \cite{team_embeddedness_2026}.)
For a death bin $B_d$ (with representative death percentile $d$) and a threshold $p$, the proportion surface is the fraction of the works in that bin whose metric percentile is at or above $p$,
\begin{equation}\label{eq:prop}
  \prop(d,p)
  \;=\;\frac{\#\{\,w : \delta_w \in B_d \ \text{and}\ \rho_w \ge p\,\}}
            {\#\{\,w : \delta_w \in B_d\,\}}
  \;=\;\widehat{\Pr}\!\left(\rho \ge p \,\middle|\, \delta \in B_d\right),
\end{equation}
evaluated over the top quartile of the metric, $p = 75, 76, \dots, 99$.
This yields one proportion value per (death bin, threshold) cell: reading up a column shows how the works of a fixed hole-size range distribute across the metric's upper tail, and reading across a row shows how a fixed tail fraction varies with hole size.

To judge whether a proportion is high or low, we compare it against a baseline, which measures the proportion a \emph{typical} work of the field would attain.
Because $p$ is a percentile threshold, a work drawn at random from a cohort whose metric is uniform by rank exceeds $p$ with probability $(100-p)/100$; this is the baseline for surprise and prescience, for which we use papers' pre-computed percentile ranks. For citation count, we use the discipline's own empirical citation distribution, taking $\base(p)$ to be the fraction of the discipline's papers at or above the $p$-th citation percentile (weighted by yearly paper counts); for larger values of $p$, this is close to $(100-p)/100$.
In every case, the baseline depends only on the threshold $p$ and not on the death bin, so the baseline surface is a plane replicated along the death axis.

The ratio, or enrichment, surface is the quotient of the two,
\begin{equation}\label{eq:ratio}
  \ratio(d,p) \;=\; \frac{\prop(d,p)}{\base(p)} .
\end{equation}
Here, $\ratio(d,p) = 1$ means that works filling holes of death bin $B_d$ reach the top $(100-p)\%$ of the metric exactly as often as chance; $\ratio > 1$ signals over-representation (an enrichment ``bonus'' for that hole-size / performance combination), and $\ratio < 1$ signals under-representation.

We show the correspondence between death percentile and death value in Figure \ref{fig:death_percentile_to_value}.
For citation, content surprise, and context surprise, we show the cumulative proportion plots for all disciplines in Figure \ref{fig:3d_proportion_plots}, the cumulative ratio plots for all disciplines in Figure \ref{fig:2D_ratio_plots_all_disciplines}, and the individual ratio plots for each discipline in Figure \ref{fig:2D_ratio_plots_all_disciplines_art}.
For content prescience and context prescience, we show the cumulative proportion plots for all disciplines in Figure \ref{fig:3d_proportion_plots_prescience}, the cumulative ratio plots for all disciplines in Figure \ref{fig:2D_ratio_plots_all_disciplines_prescience}, and the individual ratio plots for each discipline in Figure \ref{fig:2D_ratio_plots_all_disciplines_art_prescience}.

Surfaces are pooled over publication years 2005--2019: a per-discipline surface pools that discipline's years, and the all-disciplines surface pools every discipline and year, in both cases weighting each contributing group by its number of works.
We render them as filled contours, using a sequential color map for proportion surfaces and a diverging map centered at $\ratio = 1$ for ratio surfaces, with a contour drawn at $\ratio = 1$ to mark the boundary between over- and under-representation.

\subsubsection*{Analysis}
For each discipline, dimension and metric, to summarize a whole surface as a function of the death value, we average over the top decile of
a metric (i.e., percentiles $p \in \{90,\dots,99\}$), and define the band enrichment of a bin as the mean proportion divided by the mean corresponding baseline.
Every cell in the band is incorporated into this average; we use a bin whenever there is a hole-filling paper.
The death axis is cut
into 24 percentile bins of unequal width, one percentile wide from death 0 to death 5 and five
percentiles wide from death 5 to death 100. The \emph{death slope} is then the ordinary least-squares slope of band enrichment against the bin midpoint, fitted over whichever of those bins the discipline populates, and reported per 100 death percentiles.
Since $E = 1$ denotes chance, a slope of $+0.5$ means that a discipline's enrichment rises by half a unit
from the smallest gaps to the largest---for instance from $1.4$, filling papers being some 40\% more
likely than chance to fall in the top decile, to $1.9$. A slope is positive when the enrichment grows
with the death value and negative when the advantage is concentrated in the smallest gaps.

In addition to the slope analysis described above, we fit a reliability-weighted log-linear model
\begin{equation}\label{eq:model}
    \log E \;=\; b_0 \;+\; b_d\,D \;+\; b_p\,P \;+\; b_{dp}\,(D\!\cdot\!P)\,,
\end{equation}
with $D=d/100$ and $P=(p-75)/24$ both in $[0,1]$, weights equal to the expected top-tail hit count $\text{count}(d)\,(100-p)/100$, and cells kept if and only if that expected count is at least $E_{\min}$ ($10$ for dimensions 0--1, $5$ for dimension 2) and $E>0$.
Thus $b_d$ is the death main effect (bonus grows/shrinks with hole size), $b_p$ is the \emph{metric-tail concentration} (does enrichment intensify deeper into the tail), and $b_{dp}$ is the \textbf{interaction}: does the tail concentration itself strengthen for larger holes?
We also record the 2D cell of maximum reliable enrichment (peak death percentile/metric percentile) and a separability residual $s$ (the fraction of variance in $\log E$ left after the best rank-1 approximation; $s\approx0$ means the surface factorizes as $f(\text{death})\,g(\text{metric})$).

\paragraph{Citation}

Citation carries by far the strongest death structure of any channel, and its structure is not a single tendency but a division between different kinds of disciplines.
In $H_0$, the citation enrichment pooled across all disciplines seems to depend little on the death value.
However, this is because the relationship between citation enrichment and death value depends on the discipline, where 9 of the 19 disciplines exhibit a positive relationship and 10 exhibit a negative relationship.

Separating the disciplines by character of field makes the pattern explicit.
Two distinct partitions of the same 19 disciplines are in play here and should not be conflated: the
sign of the fitted slope, and the classification of a field as discovery-oriented and
natural-scientific (8 disciplines) or as design-oriented and formal or theoretical (11 disciplines). The
second was not fixed in advance as a hypothesis; we report it because the ordering of the fitted
slopes aligns with it, exactly at both ends, as set out below. The same delineation also emerges independently, without being imposed, when disciplines
are sorted by how their top-$n\%$ citation enrichment changes with dimension.
Under the second partition, the mean slope is +0.50 (for discovery-oriented disciplines) against -0.56 (for design-oriented disciplines), a difference of +1.06.

The two partitions almost coincide.
Every one of the 8 discovery-oriented fields has a positive slope and 10 of the 11 formal fields a negative one; ordered by slope, the ten most negative disciplines are design-oriented without exception and the eight most positive discovery-oriented without exception.
The sole crossover is sociology, whose slope is very close to $0$, which falls exactly at the boundary between the two blocks.
The same division recurs in $H_1$, although more weakly: +0.52 (for discovery-oriented disciplines) against -0.25 (for design-oriented disciplines), a difference of +0.77, with 6 of 8 discovery-oriented fields positive and 10 of 11 design-oriented fields negative.

Concretely, the largest positive $H_0$ slopes are Environmental Science +0.80, Geography +0.73, Biology +0.57, Geology +0.55, and the largest negative are Art -2.92, Computer Science -0.72, Philosophy -0.60, Business -0.55.
In the first group, papers that close a large gap accumulate citations; in the second, the citation advantage is greatest for papers closing the smallest gaps and decays as gaps grow.
Per-discipline values are given in Table~\ref{tab:grouped-slopes-main-h0} for $H_0$ and
Table~\ref{tab:grouped-slopes-main-h1} for $H_1$. The division is reproduced
independently by the two-dimensional model: the interaction coefficient $b_{dp}$ averages
-0.02 in discovery-oriented fields against -0.15 in
design-oriented ones in $H_0$, and +0.11 against -0.22 in $H_1$.
Since $b_{dp}$ locates the most enriched corner of the surface, this places the citation optimum at large gaps and extreme percentiles in discovery-oriented fields and at small gaps and extreme percentiles in design-oriented ones.

\paragraph{Content Surprise}

The effect of death value on content surprise appears to be minimal.
When a discipline's coefficients depart from zero, they tend to be barely negative, and where they are positive, they tend to be barely positive.

In $H_0$ the death coefficient $b_d$ is negative in 7 of the 19 disciplines, positive in 4 disciplines, and within 0.05 of 0 in the remaining 8.
The three groups are not comparable in size.
The negative ones are Art -0.41, Philosophy -0.16, Sociology -0.13, Computer Science -0.11, Engineering -0.11, Physics -0.09 and Economics -0.05, so in Art the enrichment of filling papers at the largest death values stands 34\% below its value at the smallest.
The positive ones are Psychology +0.08, History +0.08, Political Science +0.09 and Environmental Science +0.09; each sits just above the band of radius $0.05$ from $0$.
The disciplines inside the band (of radius $0.05$ from $0$) are Biology, Business, Chemistry, Geography, Geology, Materials Science, Mathematics and Medicine. Coefficients for every discipline in both
dimensions are given in Table~\ref{tab:grouped-coef-surprise}.

The interaction coefficient $b_{dp}$, which records whether the concentration of the advantage in the extreme tail strengthens or weakens as the death value grows, is negative in 6 disciplines, positive in 5 and close to zero in 8 for $H_0$.
Its two largest negative values, Art -0.23 and Philosophy -0.23, belong to the same two disciplines as the two largest negative death
coefficients.
Where the advantage of a filling paper erodes with the death value, its concentration in the extreme tail erodes with it.

An important relationship between content surprise and citation count is that the disciplines that move in content surprise are the disciplines that move in citation count.
Ordered by death slope, the two metrics place the 19 disciplines in nearly the same sequence in $H_0$: Art, Philosophy and Business fall at the negative end of both and Environmental Science at the positive end of both. The correspondence is one of order, not sign.
Only 11 of the 19 disciplines share a sign between the two metrics, which is close to random chance, because most content-surprise coefficients lie very close to $0$.

$H_1$ follows similar trends, but more weakly.
The death coefficient is negative in
9 disciplines, most steeply Sociology -0.56, Philosophy -0.43 and History -0.33, positive in
2 (Psychology +0.13 and Environmental Science +0.35) and close to zero in
8; the median discipline runs from 1.21 over the lower
half of the death axis to 1.13 over the upper. The interaction coefficient of this
dimension is dominated by History +2.59 and Art +1.91, values far outside the range of
the other 17 disciplines.
papers to $H_1$.

\paragraph{Context Surprise}

Context surprise carries a more clearly patterned death dependence than content surprise.
Its coefficients are negative in a substantial majority of the disciplines in both dimensions; the
negative group is led by the experimental and the physical sciences, and the few positive disciplines
are largely the same across the two dimensions.

In $H_0$, the death coefficient is negative in 10 of the 19
disciplines, positive in 5 disciplines, and close to zero in 4 disciplines.
The negative ones are Engineering -0.37, Physics -0.18, Computer Science -0.15, Materials Science -0.14, Mathematics -0.12, Chemistry -0.10, Geography -0.09, Medicine -0.08, Sociology -0.06 and Biology -0.05. The positive ones are
Art +0.09, Political Science +0.09, Psychology +0.14, Geology +0.34 and Environmental Science +1.49, and the disciplines that show little dependence are Business, Economics, History and Philosophy.
Two features of that arrangement should be kept apart. The negative group is led by engineering and by the physical and life sciences, which is a description of the coefficients and not a classification that could have been stated in advance: it cuts across the discovery-versus-design division of citation count, placing Engineering, Physics, Computer Science and Mathematics, all design-oriented disciplines, alongside Chemistry, Medicine and Biology, all discovery-oriented ones.
Separately, the largest positive coefficient, Environmental Science +1.49, is also the largest in
absolute value in any novelty channel in $H_0$ or $H_1$. The same discipline carries the largest (negative)
interaction coefficient of this dimension, -1.41, and since $b_{dp} = 0$ is
exactly the separable surface, its surface is also the furthest of the 19 disciplines from separable.

The interaction coefficient is the most consistent quantity in this channel. In $H_0$ it is negative
in 14 of the 19 disciplines, close to zero in
one and positive in only 4
(Art +0.08, Sociology +0.13, Political Science +0.20 and Geology +0.25). The concentration of the advantage in the extreme tail therefore weakens as
the death value grows almost everywhere, whatever the death coefficient of the discipline does. Taking
the two together, the death dependence within the top decile is negative in
11 disciplines, positive in 4 and
negligible in 4. These coefficients act on the largest enrichment of
any channel. The median discipline stands at 1.53 over the lower half of the death axis
and 1.39 over the upper, so its advantage over chance falls from
53\% to 39\%. In the humanities and the social sciences, a
filling paper is roughly twice as likely as chance to appear in an unexpected journal (History
2.01, Business 1.90, Political Science 1.89 and Philosophy
1.82 over the lower half), against Physics 1.01 and Chemistry
1.15 at the other extreme.

$H_1$ carries the pattern further. The death coefficient is negative in
12 of the 19 disciplines, led by History -0.95, Engineering -0.68, Business -0.33 and Economics -0.28,
positive in 4 (Art +0.08, Geology +0.49, Philosophy +0.51 and Environmental Science +0.55) and close to zero in
3. Within the top decile, the dependence is negative in
13 disciplines and positive in 2, and
here the coefficients and the death slope agree in sign in all 19 disciplines.
Geology and Environmental Science are positive in both dimensions and Philosophy joins them in $H_1$;
History moves the other way, having no death dependence in $H_0$ and the largest negative coefficient
of the channel in $H_1$. The median discipline runs from 1.59 to 1.33
between the halves of the death axis.

\paragraph{Content Prescience}

Content prescience seems to have little clear dependence on death, and the little structure it has points one
way. In $H_0$, not one discipline shows the advantage of a filling paper growing appreciably with the
death value.

In $H_0$ the death coefficient is negative in 7 of the 19
disciplines, close to zero in 11, and positive in exactly one,
Political Science +0.05, which is itself at the edge of the band. The negative values are small:
Art -0.20, Environmental Science -0.18, Philosophy -0.14 and Sociology -0.12 are the largest of them, so the steepest decline in the content prescience loses
18\% of the advantage over chance across the whole death axis. The
interaction coefficient is negative in 9 disciplines and positive in
3, and combining the two, the death dependence within the top decile is
negative in 12 disciplines, negligible in
6 and positive in one. The direction is
consistent and the magnitude is modest. The median discipline holds 1.11 over the lower
half of the death axis and 1.09 over the upper, so a filling paper is about
11\% more likely than chance to reach the top content-prescience decile, whatever
the death value of the gap it closes. Per-discipline coefficients for both prescience metrics are
given in Table~\ref{tab:grouped-coef-prescience}; the corresponding death slopes and peaks are in
Table~\ref{tab:grouped-slopes-prescience} for $H_0$ and
Table~\ref{tab:grouped-slopes-prescience-h1} for $H_1$.

In $H_1$, the death coefficient is negative in 8 disciplines, positive in 6 and close to zero in 5; within the top decile it is negative in 9, positive in 7 and negligible in 3.
The slope declines in 12 disciplines, as does the coefficient combination, and the 3 disciplines that change direction between the weightings are Geography, Medicine and Sociology.

The extremes of this dimension belong to its two smallest samples. Art holds the largest negative
death coefficient, -0.57, and History the largest positive,
+0.38, and the same two disciplines hold the two largest interaction
coefficients, Art +1.24 and History +1.12, on
148 and 542
filling papers. Setting those two aside, the death coefficients of the remaining
17 disciplines all lie between -0.16 and
+0.25. The median discipline runs from 1.18 over the lower half of the death axis to
1.10 over the upper.

\paragraph{Context Prescience}

Context prescience arranges the disciplines much as context surprise does in $H_0$ and loses the
arrangement in $H_1$.

In $H_0$ the death coefficient is negative in 8 of the 19
disciplines, positive in 4 and close to zero in
7. The negative ones are led by Engineering -0.21, Art -0.19, Physics -0.17 and Computer Science -0.09 and the
positive by Psychology +0.13, Philosophy +0.09 and Business +0.09. The interaction coefficient is negative in
13 disciplines and positive in 3, and
combining the two, the death dependence within the top decile is negative in
13 disciplines, positive in 4 and
negligible in 2: the same predominance as context surprise, and the
same engineering and physical disciplines at the negative end. The median discipline runs from
1.30 over the lower half of the death axis to 1.20 over the upper.

In $H_1$, the death coefficient is negative in
8 disciplines and positive in 7, with
History -0.51 and Engineering -0.36 at one end and Environmental Science +0.81 and Philosophy +0.63 at the other, and
of the 4 disciplines that were positive in $H_0$, only Philosophy is
positive. Within the top decile, the dependence is negative in 11
disciplines and positive in 5, and the median discipline runs from
1.25 to 1.11. The majority is negative in both dimensions, and beyond that
the two dimensions do not agree about which disciplines belong to it.

\begin{table}[htbp]
\centering
\caption{\textbf{Death slopes and peak enrichment for the main metric channels in $H_0$.} For each discipline and channel, Slope is the change in top-decile enrichment per 100 death percentiles, estimated over every death bin that contains at least one filling paper, and Peak is the largest band enrichment attained in any death bin. Positive slopes indicate that the advantage of filling papers grows with the size of the gap filled; negative slopes indicate that it is concentrated in the smallest gaps.}
\label{tab:grouped-slopes-main-h0}
\begin{tabular}{lcccccc}
\hline
 & \multicolumn{2}{c}{Citation} & \multicolumn{2}{c}{Surprise (content)} & \multicolumn{2}{c}{Surprise (context)} \\
Discipline & Slope & Peak & Slope & Peak & Slope & Peak \\
\hline
Art & -2.92 & 6.26 & -0.71 & 2.12 & -0.35 & 2.57 \\
Biology & +0.57 & 1.89 & +0.04 & 1.28 & -0.42 & 1.64 \\
Business & -0.55 & 2.33 & -0.38 & 1.71 & -0.01 & 2.39 \\
Chemistry & +0.47 & 1.76 & -0.03 & 1.29 & -0.33 & 1.32 \\
Computer Science & -0.72 & 2.39 & +0.01 & 1.26 & +0.06 & 2.19 \\
Economics & -0.24 & 1.50 & +0.07 & 1.34 & +0.22 & 2.22 \\
Engineering & -0.36 & 1.71 & +0.03 & 1.42 & -0.87 & 2.13 \\
Environmental Science & +0.80 & 1.70 & +1.16 & 2.02 & -1.92 & 12.73 \\
Geography & +0.73 & 1.85 & +0.08 & 1.40 & -0.28 & 2.08 \\
Geology & +0.55 & 1.79 & +0.07 & 1.31 & +0.73 & 2.02 \\
History & -0.02 & 1.70 & +0.19 & 1.79 & -0.35 & 2.69 \\
Materials Science & +0.18 & 1.48 & +0.09 & 1.31 & -0.43 & 1.56 \\
Mathematics & -0.24 & 1.38 & -0.14 & 1.31 & -0.07 & 1.60 \\
Medicine & +0.14 & 1.51 & -0.08 & 1.27 & -0.39 & 1.63 \\
Philosophy & -0.60 & 3.31 & -0.42 & 1.60 & +0.10 & 2.22 \\
Physics & -0.32 & 1.83 & -0.05 & 1.19 & -0.19 & 1.14 \\
Political Science & -0.21 & 2.07 & +0.34 & 1.56 & +0.41 & 2.83 \\
Psychology & +0.55 & 1.99 & +0.06 & 1.33 & +0.10 & 1.75 \\
Sociology & +0.00 & 1.73 & -0.10 & 1.69 & +0.64 & 2.27 \\
\hline
\\
\end{tabular}
\end{table}

\begin{table}[htbp]
\centering
\caption{\textbf{Death slopes and peak enrichment for the main metric channels in $H_1$.} Columns are as in Table~\ref{tab:grouped-slopes-main-h0}. The division of the citation slopes by discipline character recurs here, though more weakly than in $H_0$.}
\label{tab:grouped-slopes-main-h1}
\begin{tabular}{lcccccc}
\hline
 & \multicolumn{2}{c}{Citation} & \multicolumn{2}{c}{Surprise (content)} & \multicolumn{2}{c}{Surprise (context)} \\
Discipline & Slope & Peak & Slope & Peak & Slope & Peak \\
\hline
Art & -1.28 & 3.77 & -1.73 & 12.73 & -0.27 & 4.49 \\
Biology & +0.12 & 2.32 & -0.11 & 1.46 & -0.68 & 1.65 \\
Business & -0.52 & 2.54 & -0.13 & 2.10 & -0.67 & 2.81 \\
Chemistry & +0.09 & 2.20 & -0.17 & 1.44 & -0.50 & 1.41 \\
Computer Science & -0.33 & 1.59 & +0.00 & 1.29 & -0.60 & 2.32 \\
Economics & -0.21 & 1.48 & -0.28 & 1.48 & -1.14 & 3.09 \\
Engineering & -0.02 & 1.75 & -0.03 & 1.43 & -0.89 & 1.94 \\
Environmental Science & +0.60 & 5.10 & -1.48 & 16.36 & -1.42 & 9.09 \\
Geography & +1.96 & 4.48 & -0.62 & 2.47 & -0.49 & 4.00 \\
Geology & +1.28 & 3.58 & -0.28 & 1.65 & +0.81 & 2.76 \\
History & -0.13 & 3.43 & -0.84 & 6.06 & -2.16 & 6.36 \\
Materials Science & -0.16 & 1.67 & +0.20 & 1.62 & -0.45 & 1.71 \\
Mathematics & -0.14 & 1.15 & +0.13 & 1.39 & -0.00 & 1.54 \\
Medicine & -0.06 & 1.94 & -0.09 & 1.55 & -0.08 & 1.26 \\
Philosophy & -0.15 & 1.90 & -0.46 & 3.64 & +2.63 & 8.54 \\
Physics & -0.01 & 1.62 & -0.09 & 1.41 & -0.40 & 1.30 \\
Political Science & -0.94 & 1.92 & -0.33 & 2.54 & -0.90 & 3.35 \\
Psychology & +0.35 & 3.22 & +0.10 & 2.03 & -0.22 & 2.81 \\
Sociology & +1.01 & 4.18 & -0.58 & 4.85 & +5.62 & 18.18 \\
\hline
 \\
\end{tabular}
\end{table}

\begin{table}[htbp]
\centering
\caption{\textbf{Death slopes and peak enrichment for the prescience channels in $H_0$.} Columns are as in Table~\ref{tab:grouped-slopes-main-h0}. Both prescience channels show smaller slope magnitudes than citation.}
\label{tab:grouped-slopes-prescience}
\begin{tabular}{lcccc}
\hline
 & \multicolumn{2}{c}{Prescience (content)} & \multicolumn{2}{c}{Prescience (context)} \\
Discipline & Slope & Peak & Slope & Peak \\
\hline
Art & -0.12 & 1.64 & -0.29 & 1.84 \\
Biology & -0.08 & 1.37 & -0.19 & 1.78 \\
Business & -0.17 & 1.72 & +0.02 & 1.65 \\
Chemistry & -0.06 & 1.29 & -0.32 & 1.66 \\
Computer Science & -0.05 & 1.65 & +0.04 & 1.93 \\
Economics & -0.02 & 1.49 & -0.12 & 1.76 \\
Engineering & +0.05 & 1.24 & -0.55 & 2.62 \\
Environmental Science & +0.56 & 2.22 & -1.06 & 10.91 \\
Geography & +0.04 & 1.19 & -0.03 & 1.25 \\
Geology & -0.03 & 1.19 & +0.16 & 1.25 \\
History & +0.05 & 1.47 & +0.07 & 1.59 \\
Materials Science & +0.07 & 1.15 & -0.31 & 1.52 \\
Mathematics & -0.12 & 1.14 & -0.01 & 1.36 \\
Medicine & -0.13 & 1.31 & -0.30 & 1.98 \\
Philosophy & -0.34 & 1.48 & -0.32 & 1.84 \\
Physics & +0.01 & 1.11 & -0.22 & 1.54 \\
Political Science & -0.15 & 1.63 & +0.16 & 2.06 \\
Psychology & -0.11 & 1.50 & +0.03 & 1.79 \\
Sociology & -0.07 & 1.44 & +0.35 & 1.71 \\
\hline
 \\
\end{tabular}
\end{table}

\begin{table}[htbp]
\centering
\caption{\textbf{Death slopes and peak enrichment for the prescience channels in $H_1$.} Columns are as in Table~\ref{tab:grouped-slopes-main-h0}. Two further entries rest on fewer than five death bins and are flagged accordingly, as the corresponding surprise entries are in Table~\ref{tab:grouped-slopes-main-h1}.}
\label{tab:grouped-slopes-prescience-h1}
\begin{tabular}{lcccc}
\hline
 & \multicolumn{2}{c}{Prescience (content)} & \multicolumn{2}{c}{Prescience (context)} \\
Discipline & Slope & Peak & Slope & Peak \\
\hline
Art & -2.27 & 12.73 & +0.59 & 2.63 \\
Biology & -0.20 & 1.33 & -0.18 & 1.54 \\
Business & +0.11 & 2.02 & -0.44 & 2.01 \\
Chemistry & -0.16 & 1.61 & -0.36 & 1.50 \\
Computer Science & -0.09 & 1.56 & +0.07 & 1.70 \\
Economics & -0.68 & 1.97 & -0.60 & 1.78 \\
Engineering & +0.13 & 2.15 & -0.33 & 2.03 \\
Environmental Science & -4.41 & 16.36 & -0.57 & 4.73 \\
Geography & -0.09 & 2.18 & -0.13 & 2.54 \\
Geology & -0.05 & 1.58 & +0.33 & 1.52 \\
History & -0.67 & 6.67 & -1.25 & 2.22 \\
Materials Science & -0.16 & 1.43 & -0.18 & 1.34 \\
Mathematics & +0.05 & 1.31 & -0.24 & 1.51 \\
Medicine & -0.03 & 1.62 & +0.14 & 1.74 \\
Philosophy & -0.66 & 2.19 & +0.99 & 4.73 \\
Physics & +0.10 & 1.39 & -0.33 & 1.49 \\
Political Science & -0.69 & 2.04 & -0.79 & 3.12 \\
Psychology & -0.09 & 2.10 & +0.32 & 3.25 \\
Sociology & -0.31 & 2.81 & +0.58 & 9.09 \\
\hline \\
\end{tabular}
\end{table}

\begin{table}[htbp]
\centering
\small
\caption{\textbf{Two-dimensional log-enrichment coefficients for the citation channel in $H_0$.} Coefficients of Eq.~\eqref{eq:model}: $b_0$ the intercept, $b_d$ the death main effect, $b_p$ the concentration of enrichment toward the extreme metric tail, and $b_{dp}$ the interaction, which vanishes exactly when the surface is separable; $R^2$ is the weighted fit, $s$ the separability residual, and the final column gives the death and metric percentile of the most enriched cell; the death entry is a bin midpoint and so is often a half-integer, which the earlier document truncated to an integer. The sign of $b_d$ and of $b_{dp}$ follows the discovery/design division discussed in the text.}
\label{tab:grouped-coef-citation}
\begin{tabular}{lccccccc}
\hline
Discipline & $b_0$ & $b_d$ & $b_p$ & $b_{dp}$ & $R^2$ & $s$ & peak death/citation \\
\hline
Art & +0.46 & -0.56 & +0.34 & -0.41 & 0.53 & 0.03 & 0.5/98 \\
Biology & +0.27 & +0.26 & +0.04 & +0.10 & 0.89 & 0.06 & 87.5/99 \\
Business & +0.28 & -0.18 & +0.13 & +0.02 & 0.43 & 0.13 & 1.5/98 \\
Chemistry & +0.30 & +0.20 & +0.08 & +0.01 & 0.63 & 0.05 & 57.5/99 \\
Computer Science & +0.32 & -0.10 & +0.16 & -0.36 & 0.66 & 0.07 & 7.5/99 \\
Economics & +0.23 & -0.18 & +0.18 & -0.12 & 0.70 & 0.37 & 72.5/99 \\
Engineering & +0.16 & -0.08 & +0.17 & -0.31 & 0.47 & 0.06 & 4.5/99 \\
Environmental Science & -0.49 & +0.84 & +0.77 & -0.71 & 0.55 & 0.38 & 3.5/99 \\
Geography & +0.19 & +0.23 & +0.08 & +0.18 & 0.83 & 0.09 & 87.5/99 \\
Geology & +0.31 & +0.17 & -0.04 & +0.15 & 0.73 & 0.06 & 72.5/96 \\
History & +0.22 & -0.03 & +0.01 & +0.17 & 0.07 & 0.34 & 92.5/99 \\
Materials Science & +0.20 & +0.10 & +0.07 & -0.05 & 0.21 & 0.09 & 57.5/97 \\
Mathematics & +0.15 & -0.03 & +0.04 & -0.24 & 0.48 & 0.16 & 2.5/91 \\
Medicine & +0.22 & +0.11 & +0.02 & +0.02 & 0.52 & 0.14 & 2.5/99 \\
Philosophy & +0.14 & +0.10 & +0.10 & -0.12 & 0.06 & 0.03 & 0.5/99 \\
Physics & +0.32 & -0.02 & +0.11 & -0.17 & 0.33 & 0.08 & 0.5/99 \\
Political Science & +0.32 & -0.11 & +0.17 & -0.07 & 0.17 & 0.16 & 87.5/99 \\
Psychology & +0.27 & +0.24 & +0.11 & +0.11 & 0.80 & 0.14 & 97.5/97 \\
Sociology & +0.31 & -0.09 & +0.11 & -0.07 & 0.07 & 0.03 & 4.5/99 \\
\hline
\end{tabular}
\end{table}

\begin{table}[htbp]
\centering
\small
\caption{\textbf{Two-dimensional log-enrichment coefficients for the citation channel in $H_1$.} Coefficients of Eq.~\eqref{eq:model}: $b_0$ the intercept, $b_d$ the death main effect, $b_p$ the concentration of enrichment toward the extreme metric tail, and $b_{dp}$ the interaction, which vanishes exactly when the surface is separable; $R^2$ is the weighted fit, $s$ the separability residual, and the final column gives the death and metric percentile of the most enriched cell; the death entry is a bin midpoint and so is often a half-integer, which the earlier document truncated to an integer. Columns are as in Table~\ref{tab:grouped-coef-citation}.}
\label{tab:grouped-coef-citation-h1}
\begin{tabular}{lccccccc}
\hline
Discipline & $b_0$ & $b_d$ & $b_p$ & $b_{dp}$ & $R^2$ & $s$ & peak death/citation \\
\hline
Art & +0.19 & -0.63 & +0.05 & -0.58 & 0.23 & 0.12 & 0.5/97 \\
Biology & +0.52 & +0.19 & +0.22 & -0.04 & 0.74 & 0.45 & 47.5/99 \\
Business & +0.17 & -0.24 & +0.11 & -0.13 & 0.27 & 0.20 & 0.5/99 \\
Chemistry & +0.47 & +0.19 & +0.24 & -0.12 & 0.55 & 0.40 & 52.5/99 \\
Computer Science & +0.26 & -0.13 & -0.12 & -0.25 & 0.69 & 0.22 & 0.5/99 \\
Economics & +0.11 & -0.18 & +0.05 & -0.09 & 0.34 & 0.28 & 0.5/99 \\
Engineering & +0.05 & -0.06 & +0.00 & -0.32 & 0.21 & 0.13 & 92.5/99 \\
Environmental Science & +0.15 & +0.55 & -0.08 & +0.65 & 0.43 & 0.20 & 7.5/99 \\
Geography & +0.46 & +0.42 & +0.25 & +0.45 & 0.63 & 0.22 & 87.5/99 \\
Geology & +0.47 & +0.44 & +0.19 & +0.07 & 0.55 & 0.13 & 92.5/99 \\
History & +0.33 & -0.70 & +0.33 & -0.48 & 0.21 & 0.20 & 67.5/99 \\
Materials Science & +0.22 & +0.09 & +0.14 & -0.22 & 0.02 & 0.17 & 47.5/99 \\
Mathematics & +0.10 & -0.10 & -0.13 & -0.15 & 0.28 & 0.23 & 17.5/99 \\
Medicine & +0.39 & -0.01 & +0.08 & +0.04 & 0.17 & 0.15 & 0.5/99 \\
Philosophy & +0.24 & -0.32 & +0.03 & +0.18 & 0.18 & 0.39 & 67.5/99 \\
Physics & +0.29 & +0.12 & +0.08 & -0.10 & 0.20 & 0.12 & 82.5/99 \\
Political Science & +0.31 & -0.26 & +0.29 & -0.55 & 0.40 & 0.13 & 17.5/99 \\
Psychology & +0.53 & +0.23 & +0.35 & +0.04 & 0.37 & 0.31 & 87.5/98 \\
Sociology & +0.38 & +0.03 & -0.05 & +0.08 & 0.01 & 0.48 & 97.5/98 \\
\hline
\end{tabular}
\end{table}

\begin{table}[htbp]
\centering
\footnotesize
\caption{\textbf{Two-dimensional log-enrichment coefficients for the two surprise channels.} Definitions as in Table~\ref{tab:grouped-coef-citation}; the tail concentration $b_p$ is omitted here because it is positive throughout. The content and context blocks are separate measurements and are read separately.}
\label{tab:grouped-coef-surprise}
\begin{tabular}{lcccccccc}
\hline
 & \multicolumn{2}{c}{Content, $H_0$} & \multicolumn{2}{c}{Content, $H_1$} & \multicolumn{2}{c}{Context, $H_0$} & \multicolumn{2}{c}{Context, $H_1$} \\
Discipline & $b_d$ & $b_{dp}$ & $b_d$ & $b_{dp}$ & $b_d$ & $b_{dp}$ & $b_d$ & $b_{dp}$ \\
\hline
Art & -0.41 & -0.23 & -0.01 & +1.91 & +0.09 & +0.08 & +0.08 & -0.12 \\
Biology & +0.04 & -0.03 & -0.06 & -0.03 & -0.05 & -0.29 & -0.24 & -0.48 \\
Business & -0.04 & -0.12 & +0.01 & +0.11 & +0.02 & -0.14 & -0.33 & +0.05 \\
Chemistry & -0.01 & -0.06 & +0.02 & -0.19 & -0.10 & -0.24 & -0.06 & -0.38 \\
Computer Science & -0.11 & +0.12 & -0.10 & +0.17 & -0.15 & -0.07 & -0.21 & -0.36 \\
Economics & -0.05 & -0.01 & -0.07 & -0.15 & -0.01 & -0.03 & -0.28 & -0.52 \\
Engineering & -0.11 & +0.17 & -0.18 & +0.17 & -0.37 & -0.42 & -0.68 & -0.58 \\
Environmental Science & +0.09 & +0.46 & +0.35 & +0.59 & +1.49 & -1.41 & +0.55 & -2.68 \\
Geography & -0.02 & +0.01 & -0.01 & -0.36 & -0.09 & -0.17 & +0.01 & -0.83 \\
Geology & -0.01 & -0.05 & -0.05 & -0.27 & +0.34 & +0.25 & +0.49 & +0.25 \\
History & +0.08 & +0.06 & -0.33 & +2.59 & +0.00 & -0.06 & -0.95 & +0.26 \\
Materials Science & +0.02 & -0.03 & -0.07 & +0.09 & -0.14 & -0.28 & -0.14 & -0.12 \\
Mathematics & -0.00 & -0.02 & +0.05 & +0.16 & -0.12 & -0.10 & -0.03 & -0.09 \\
Medicine & -0.03 & -0.02 & +0.05 & -0.10 & -0.08 & -0.19 & +0.02 & -0.09 \\
Philosophy & -0.16 & -0.23 & -0.43 & +0.24 & -0.00 & -0.25 & +0.51 & +0.18 \\
Physics & -0.09 & +0.03 & +0.05 & -0.11 & -0.18 & -0.14 & -0.21 & -0.27 \\
Political Science & +0.09 & +0.19 & +0.03 & -0.09 & +0.09 & +0.20 & -0.10 & +0.10 \\
Psychology & +0.08 & -0.06 & +0.13 & -0.17 & +0.14 & -0.20 & -0.16 & -0.33 \\
Sociology & -0.13 & +0.03 & -0.56 & +0.05 & -0.06 & +0.13 & -0.20 & +0.26 \\
\hline
\end{tabular}
\end{table}

\begin{table}[htbp]
\centering
\footnotesize
\caption{\textbf{Two-dimensional log-enrichment coefficients for the two prescience channels.} Definitions as in Table~\ref{tab:grouped-coef-citation}.}
\label{tab:grouped-coef-prescience}
\begin{tabular}{lcccccccc}
\hline
 & \multicolumn{2}{c}{Content, $H_0$} & \multicolumn{2}{c}{Content, $H_1$} & \multicolumn{2}{c}{Context, $H_0$} & \multicolumn{2}{c}{Context, $H_1$} \\
Discipline & $b_d$ & $b_{dp}$ & $b_d$ & $b_{dp}$ & $b_d$ & $b_{dp}$ & $b_d$ & $b_{dp}$ \\
\hline
Art & -0.20 & +0.14 & -0.57 & +1.24 & -0.19 & -0.02 & +0.20 & +2.39 \\
Biology & +0.00 & -0.04 & -0.04 & -0.22 & -0.03 & -0.08 & -0.01 & -0.20 \\
Business & -0.03 & -0.07 & +0.09 & -0.02 & +0.09 & -0.06 & -0.02 & -0.63 \\
Chemistry & -0.04 & -0.07 & -0.10 & -0.15 & -0.08 & -0.20 & -0.02 & -0.27 \\
Computer Science & -0.08 & -0.05 & -0.05 & +0.03 & -0.09 & -0.03 & +0.11 & -0.03 \\
Economics & -0.04 & -0.10 & -0.05 & -0.51 & +0.02 & -0.12 & -0.25 & -0.43 \\
Engineering & -0.05 & +0.07 & -0.05 & +0.04 & -0.21 & -0.20 & -0.36 & -0.41 \\
Environmental Science & -0.18 & -0.14 & +0.04 & -0.39 & +0.03 & -0.77 & +0.81 & -1.15 \\
Geography & -0.03 & +0.00 & -0.07 & +0.18 & -0.05 & -0.06 & +0.24 & -0.76 \\
Geology & -0.04 & -0.04 & -0.12 & -0.02 & +0.05 & +0.11 & +0.14 & +0.09 \\
History & +0.01 & +0.13 & +0.38 & +1.12 & +0.03 & +0.12 & -0.51 & +1.54 \\
Materials Science & +0.04 & -0.02 & -0.16 & -0.00 & -0.05 & -0.26 & -0.19 & +0.06 \\
Mathematics & -0.06 & -0.08 & +0.08 & -0.10 & -0.08 & -0.09 & +0.10 & -0.47 \\
Medicine & -0.02 & -0.02 & +0.10 & -0.03 & -0.05 & -0.09 & +0.03 & -0.01 \\
Philosophy & -0.14 & -0.27 & -0.09 & -0.06 & +0.09 & -0.24 & +0.63 & -0.71 \\
Physics & +0.01 & -0.03 & -0.01 & +0.19 & -0.17 & -0.06 & -0.16 & -0.09 \\
Political Science & +0.05 & -0.16 & +0.12 & -0.39 & +0.07 & +0.15 & -0.17 & -0.46 \\
Psychology & +0.03 & -0.15 & -0.10 & -0.32 & +0.13 & -0.09 & -0.09 & +0.11 \\
Sociology & -0.12 & -0.11 & +0.25 & +0.24 & -0.01 & +0.05 & -0.25 & +1.04 \\
\hline
\end{tabular}
\end{table}

\clearpage

\newpage

\newpage

\begin{figure}
	\centering
	\includegraphics[width=0.8\textwidth]{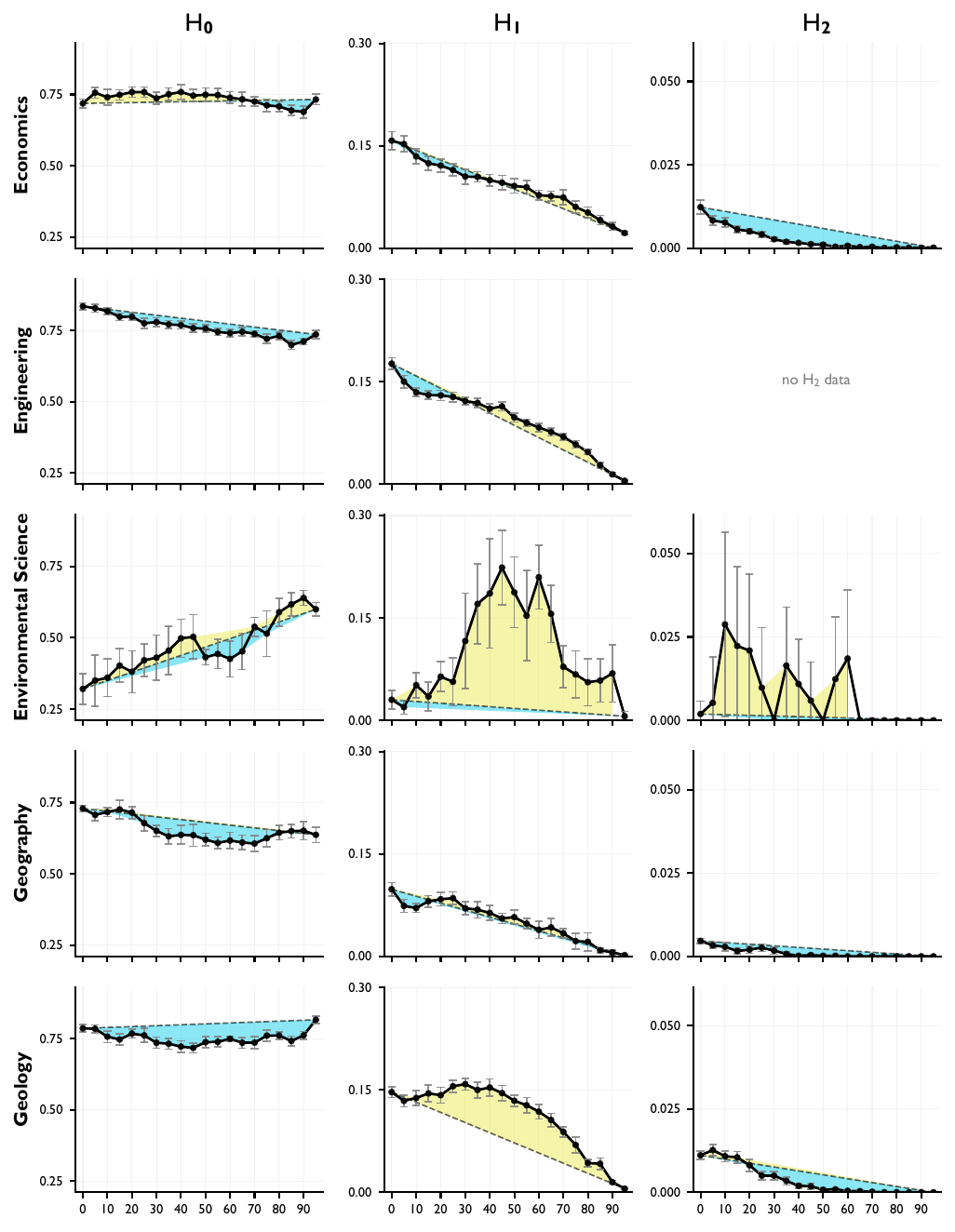}

	\continuedfigure
	\caption{\textbf{Proportions of holes filled in relation to death values, by discipline.}
	Page 2 of 4; further disciplines.}
	\label{fig:holes_filled_by_death_per_discipline_pg2}
\end{figure}

\begin{figure}
	\centering
	\includegraphics[width=0.8\textwidth]{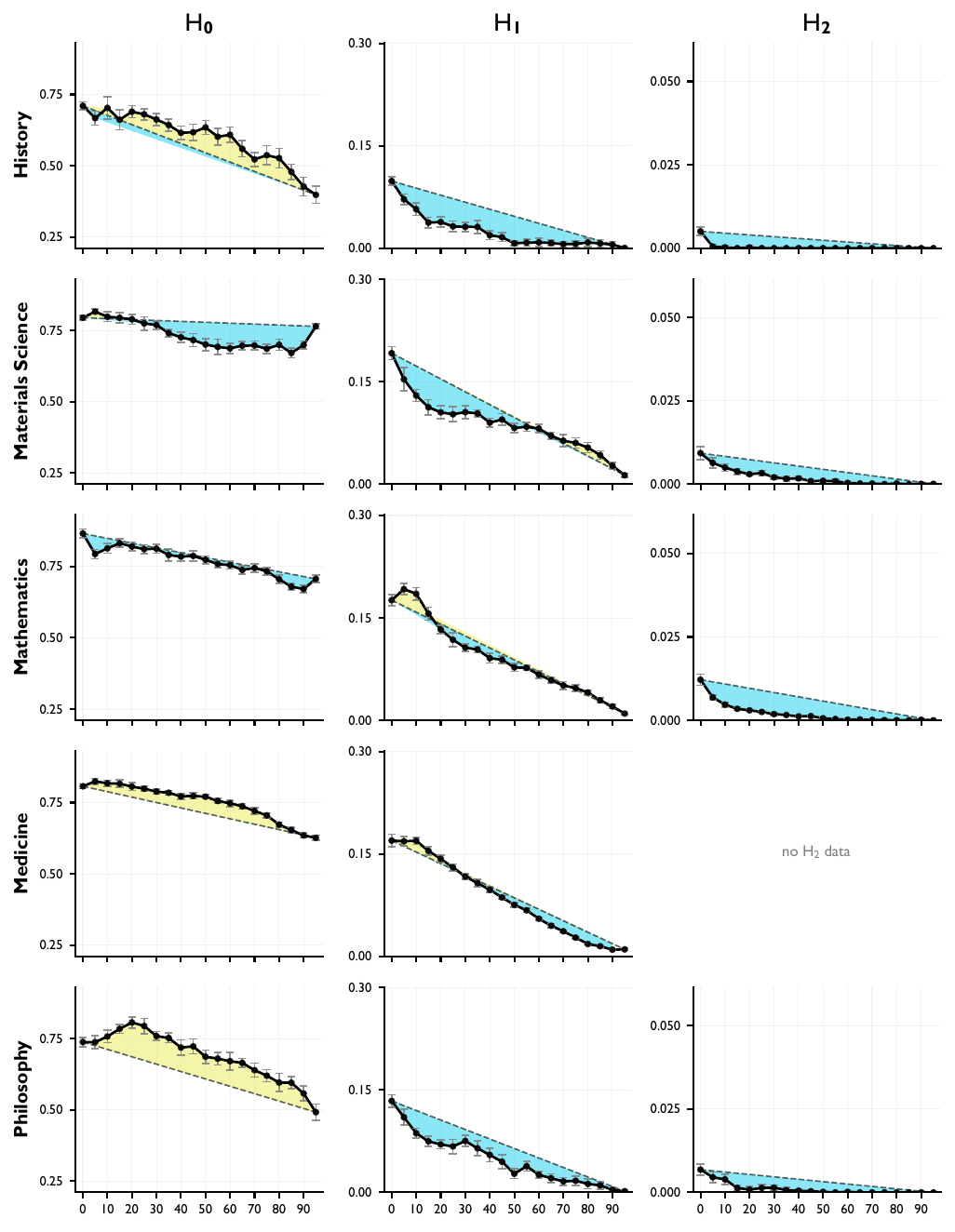}

	\continuedfigure
	\caption{\textbf{Proportions of holes filled in relation to death values, by discipline.}
	Page 3 of 4; further disciplines.}
	\label{fig:holes_filled_by_death_per_discipline_pg3}
\end{figure}

\begin{figure}
	\centering
	\includegraphics[width=0.8\textwidth]{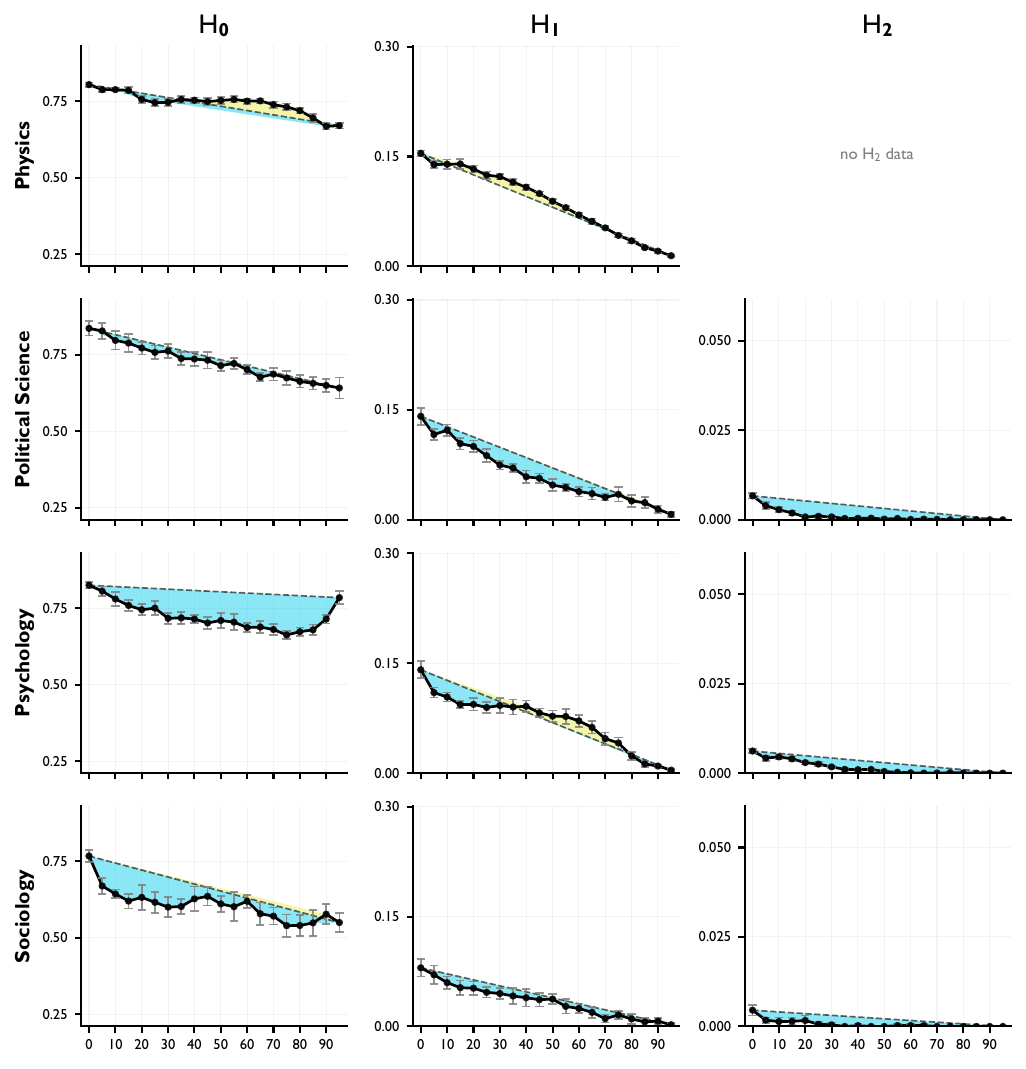}

	\continuedfigure
	\caption{\textbf{Proportions of holes filled in relation to death values, by discipline.}
	Page 4 of 4; further disciplines.}
	\label{fig:holes_filled_by_death_per_discipline_pg4}
\end{figure}

\newpage

\begin{figure}
	\centering
	\includegraphics[width=0.8\textwidth]{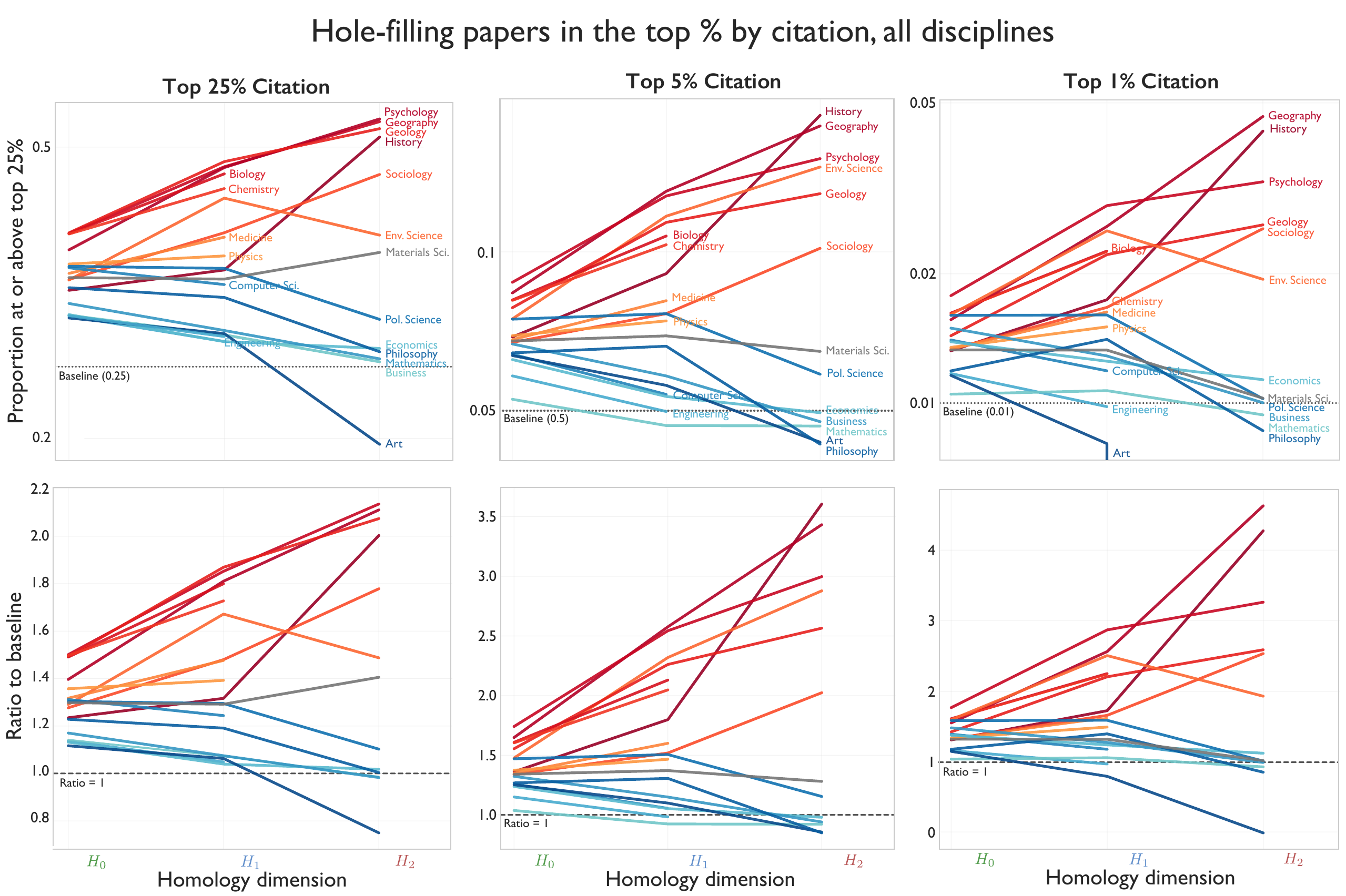}

	\caption{\textbf{Hit citations for papers that fill 0th, 1st, and 2nd order holes by field.} Proportion of papers that fill 0th, 1st, and 2nd order holes within the top $N$\% of citations, and the ratio of that proportion to the baseline, respectively, for each field, where N equals 25\%, 5\% and 1\% (see Figure~\ref{fig:hole_fill_papers}, panels $C$ and $D$, for 10\%). For the 25\% threshold, these figures demonstrate that while all fields manifest an increase in hit papers for bridging zeroeth-order gaps and tiling first-order holes, only discovery-oriented fields systematically grow in high citation likelihood as hole dimensionality increases, while design-oriented fields symmetrically decrease in hit citation likelihood.}
	\label{fig:citation_hit_ratio_by_field}
\end{figure}

\newpage

\begin{figure}
	\centering
	\includegraphics[width=0.8\textwidth]{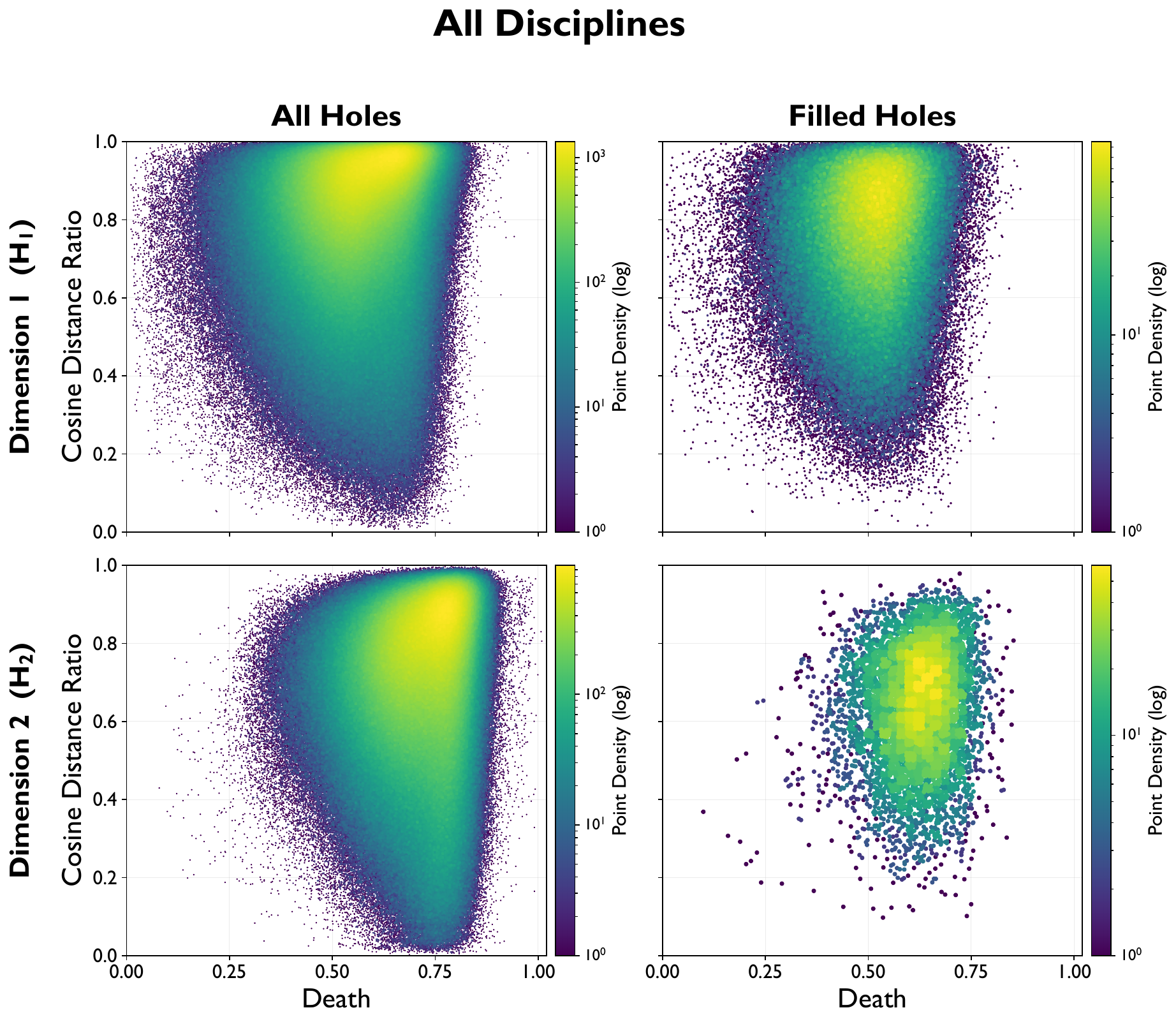}

	\caption{\textbf{Death value versus death-simplex edge-length ratio, by homology dimension and hole type.}
    Each panel shows the death simplex ratios of every hole as a density-colored scatter (viridis, log scale) of death value against its death-simplex ratio.
    The death-simplex ratio is the minimum-to-maximum ratio between the cosine distance among the concept embeddings forming the death simplex, where $1$ is an equilateral simplex and $\approx0$ is an elongated one.
    We show this for dimension $1$ and $2$ (recall that a death simplex in dimension $0$ only has one edge), and distinguish between all holes and holes that are filled in.}
	\label{fig:death_simplex_ratio_plot}
\end{figure}

\newpage

\begin{figure}
	\centering
	\includegraphics[width=0.8\textwidth]{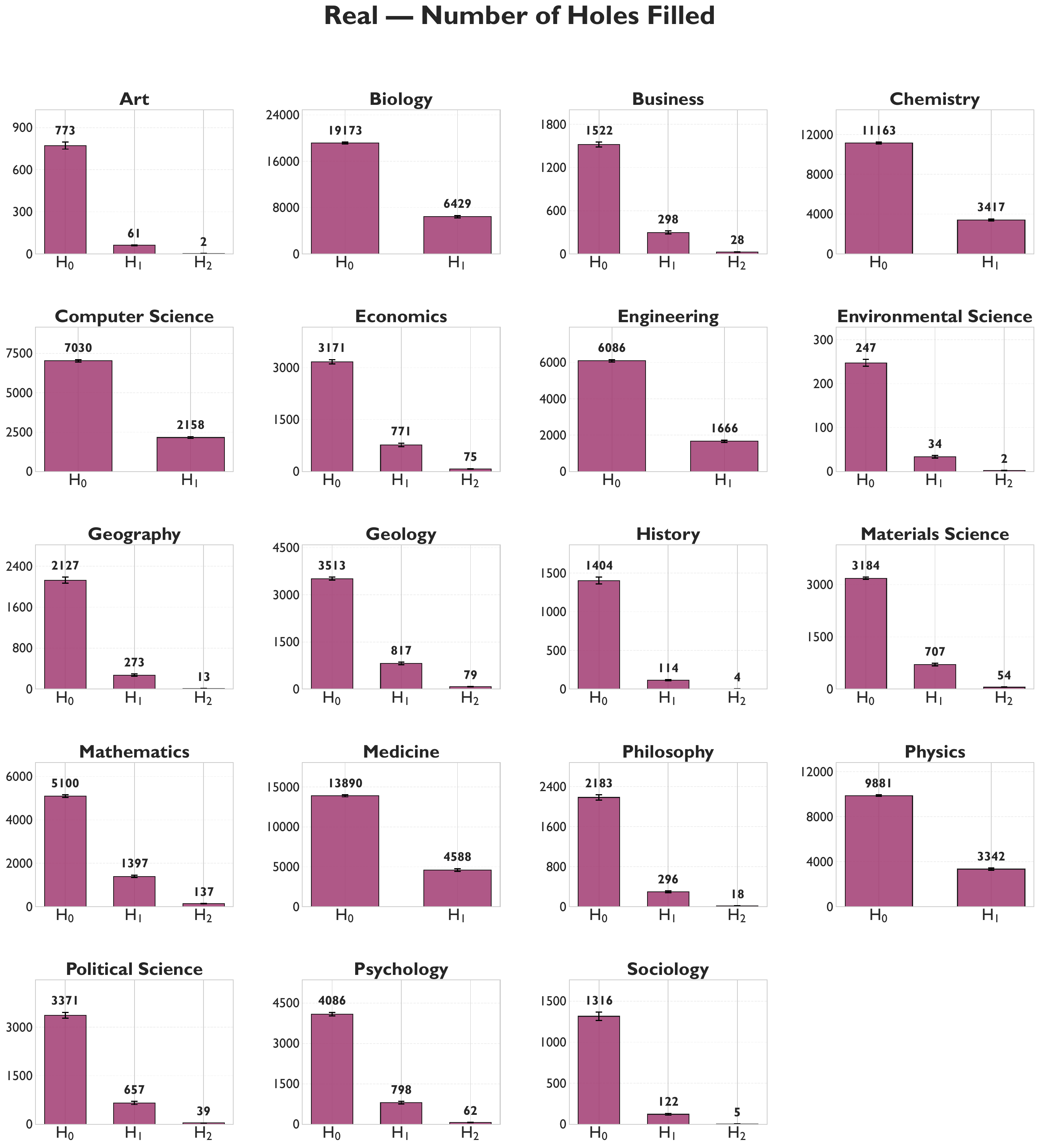}

	\caption{\textbf{Number of holes filled by papers.}
	We show bar plots of the number of holes filled by papers, for each discipline.}
	\label{fig:filled_by_real_papers}
\end{figure}

\begin{figure}
	\centering
	\includegraphics[width=0.8\textwidth]{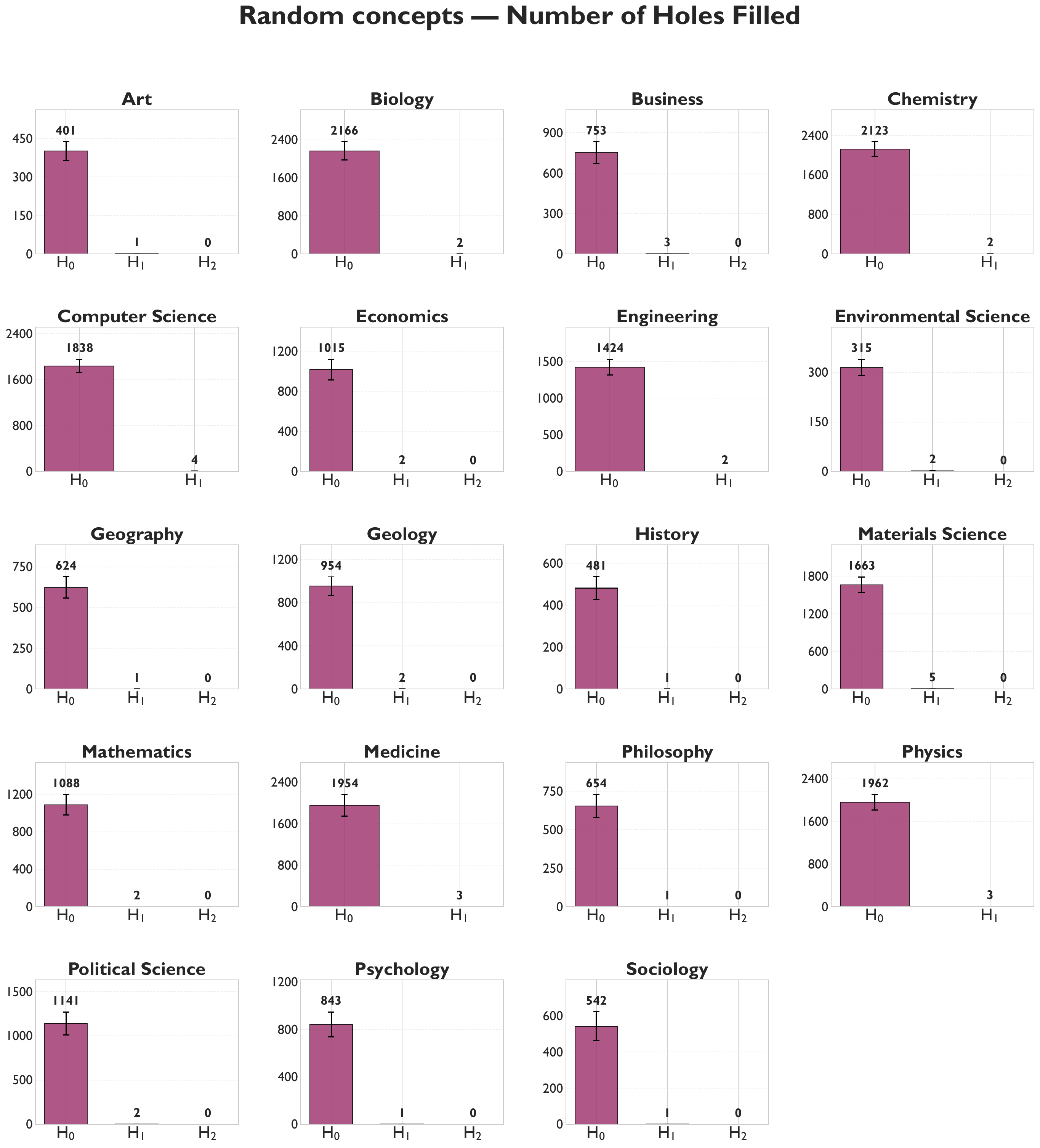}

	\caption{\textbf{Number of holes filled by simulated papers     with random sets of concepts, by discipline and homology dimension.}
	We show bar plots of the number of holes filled by simulated papers for each discipline. We generate simulated papers by replacing each real paper's concept set with the same number of concepts, drawn uniformly at random from the full set of concepts in that discipline. That is, a real paper with $N$ concepts becomes a ``synthetic paper'' of $N$ randomly chosen concepts.}
	\label{fig:filled_by_random_concepts_papers}
\end{figure}

\begin{figure}
	\centering
	\includegraphics[width=0.8\textwidth]{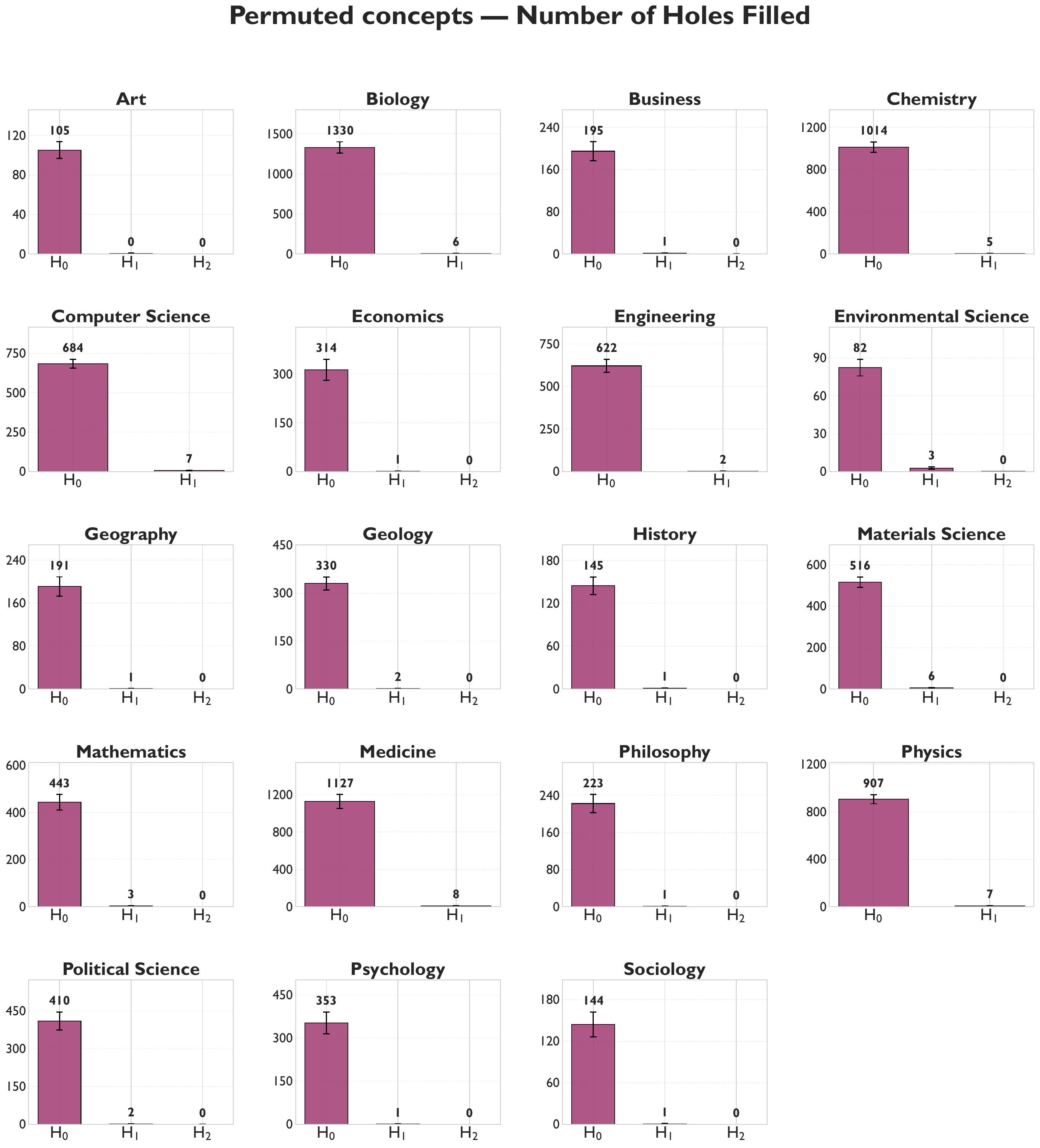}

	\caption{\textbf{Number of holes filled by simulated papers with frequency-matched random sets of concepts, by discipline and homology dimension.}
	We show bar plots of the number of holes filled by simulated papers, for each discipline. As with the simulated papers with random concepts, we generate simulated papers by replacing each real paper's concept set with the same number of concepts; here, however, each concept is drawn at random in proportion to how frequently it appears across papers in that discipline, without replacement so that concepts are not repeated within a paper. That is, a real paper with $N$ concepts becomes a ``synthetic paper'' of $N$ concepts sampled according to the empirical concept-frequency distribution. This preserves both each paper's number of concepts and the overall popularity of individual concepts, while randomizing which concepts co-occur.}
	\label{fig:filled_by_permuted_concepts_papers}
\end{figure}

\begin{figure}
	\centering
	\includegraphics[width=0.8\textwidth]{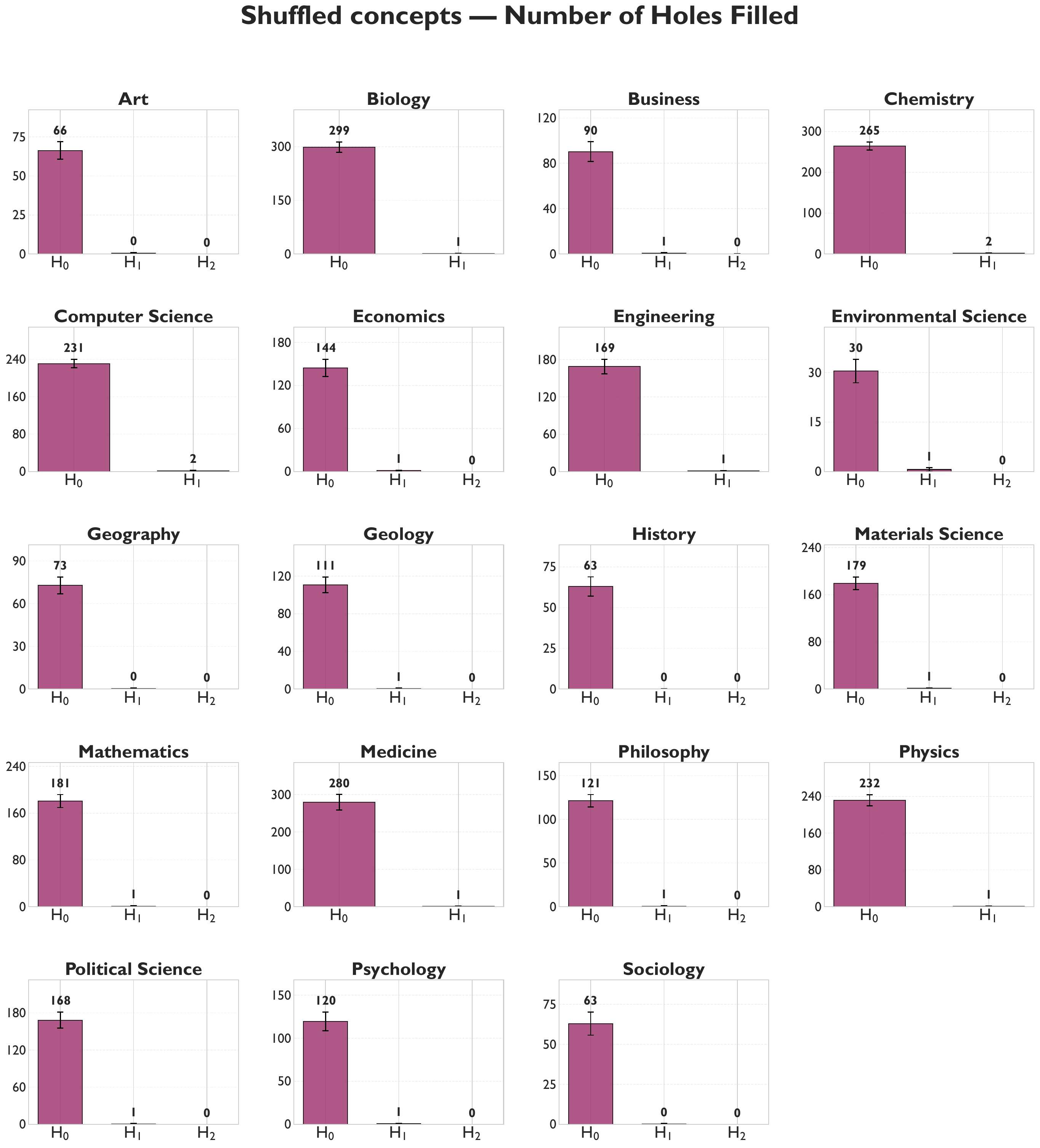}

	\caption{\textbf{Number of holes filled after randomly shuffling concept identities, by discipline and homology dimension.}
	We show bar plots of the number of holes filled, for each discipline. Rather than generating synthetic papers, we apply a permutation to the set of concepts in a given discipline, which, in effect, relabels each concept. We then count a hole as filled if any real paper contains that relabeled set of concepts. This preserves both the holes themselves and the real papers, while destroying the alignment between a hole and the concepts that gave rise to it.}
	\label{fig:filled_by_shuffled_concepts_papers}
\end{figure}

\newpage

\begin{figure}
	\centering
	\includegraphics[width=0.8\textwidth]{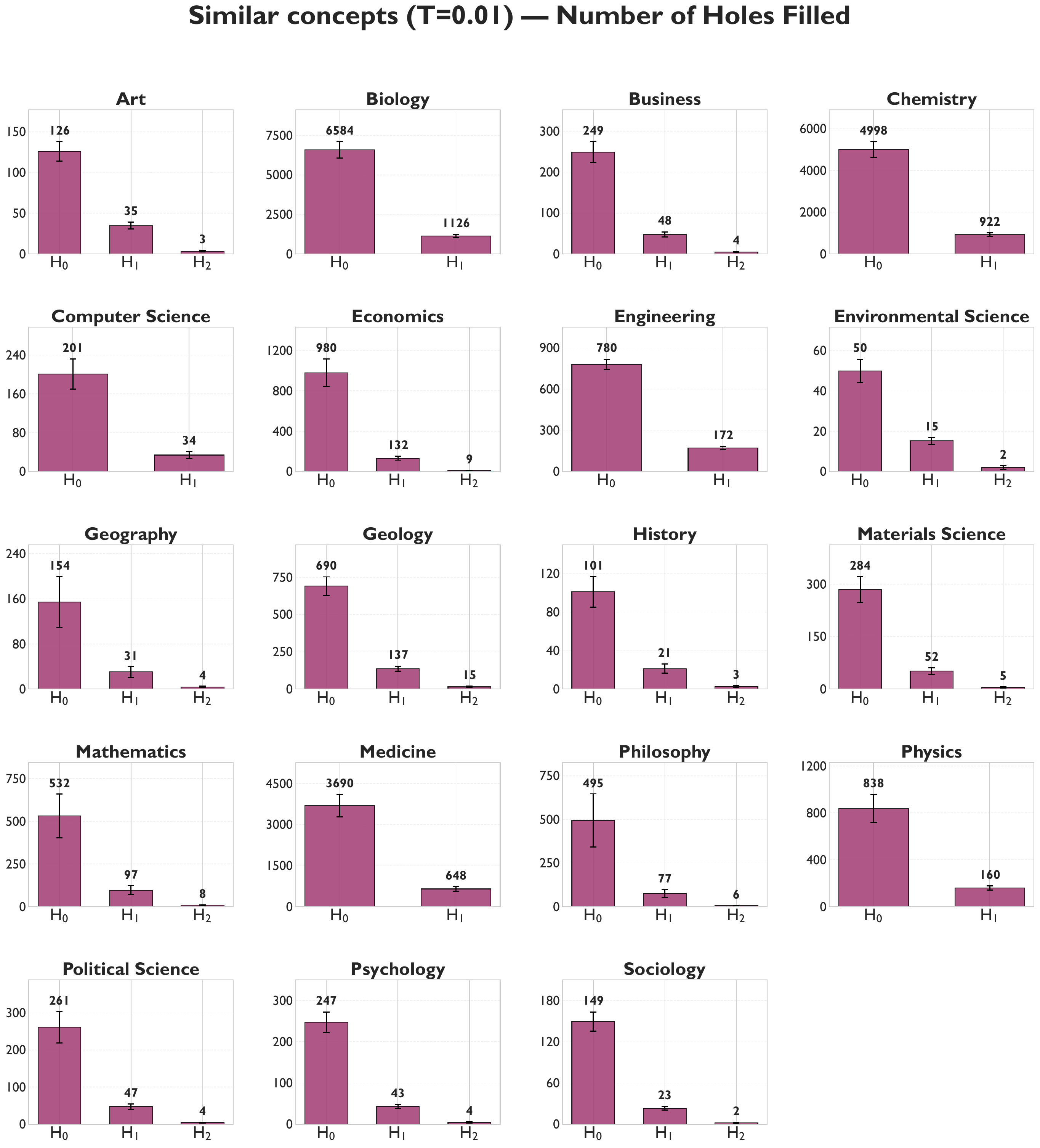}

	\caption{\textbf{Number of holes filled by simulated papers with semantically similar sets of concepts (temperature $T =  0.01$), by discipline and homology dimension.}
    We generate each simulated paper by replacing a real paper's concepts with the same number of concepts chosen to be semantically similar. Starting from a single concept sampled by frequency, we add concepts one at a time, preferring those whose embedding vector is close (high cosine similarity) to the concepts already chosen. The temperature controls the strength of this preference. As $T \to 0$, papers are built almost entirely from their most similar concepts (strong homophily), while large $T$ approaches uniformly random selection. Here, $T = 0.01$.
    }
	\label{fig:filled_by_similar_concepts_papers_T=0.01}
\end{figure}

\begin{figure}
	\centering
	\includegraphics[width=0.8\textwidth]{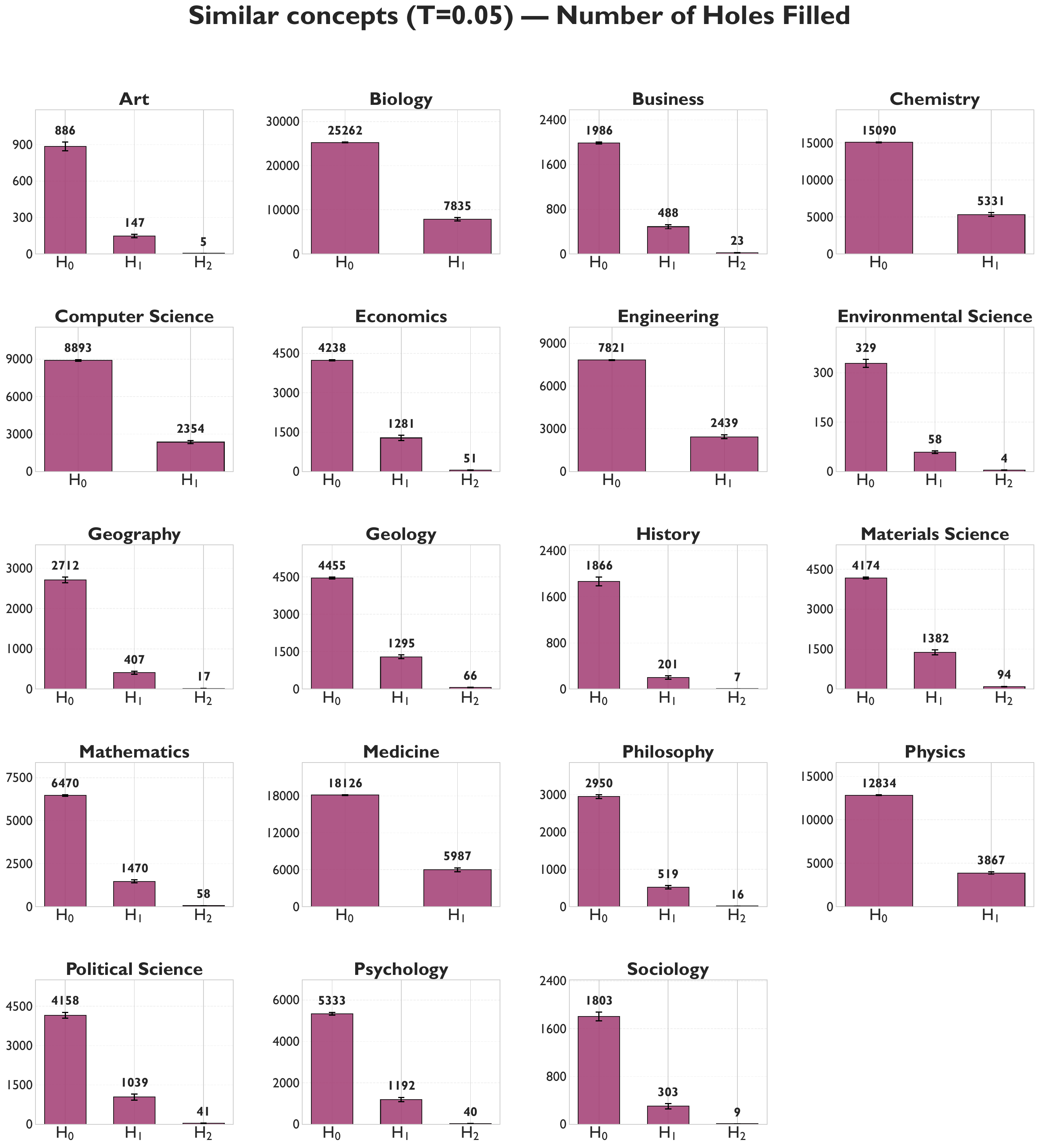}

	\caption{\textbf{Number of holes filled by simulated papers with semantically similar sets of concepts (temperature $T =  0.05$), by discipline and homology dimension.}
    Same as Figure~\ref{fig:filled_by_similar_concepts_papers_T=0.01}, but for $T = 0.05$.
    }
	\label{fig:filled_by_similar_concepts_papers_T=0.05}
\end{figure}

\begin{figure}
	\centering
	\includegraphics[width=0.8\textwidth]{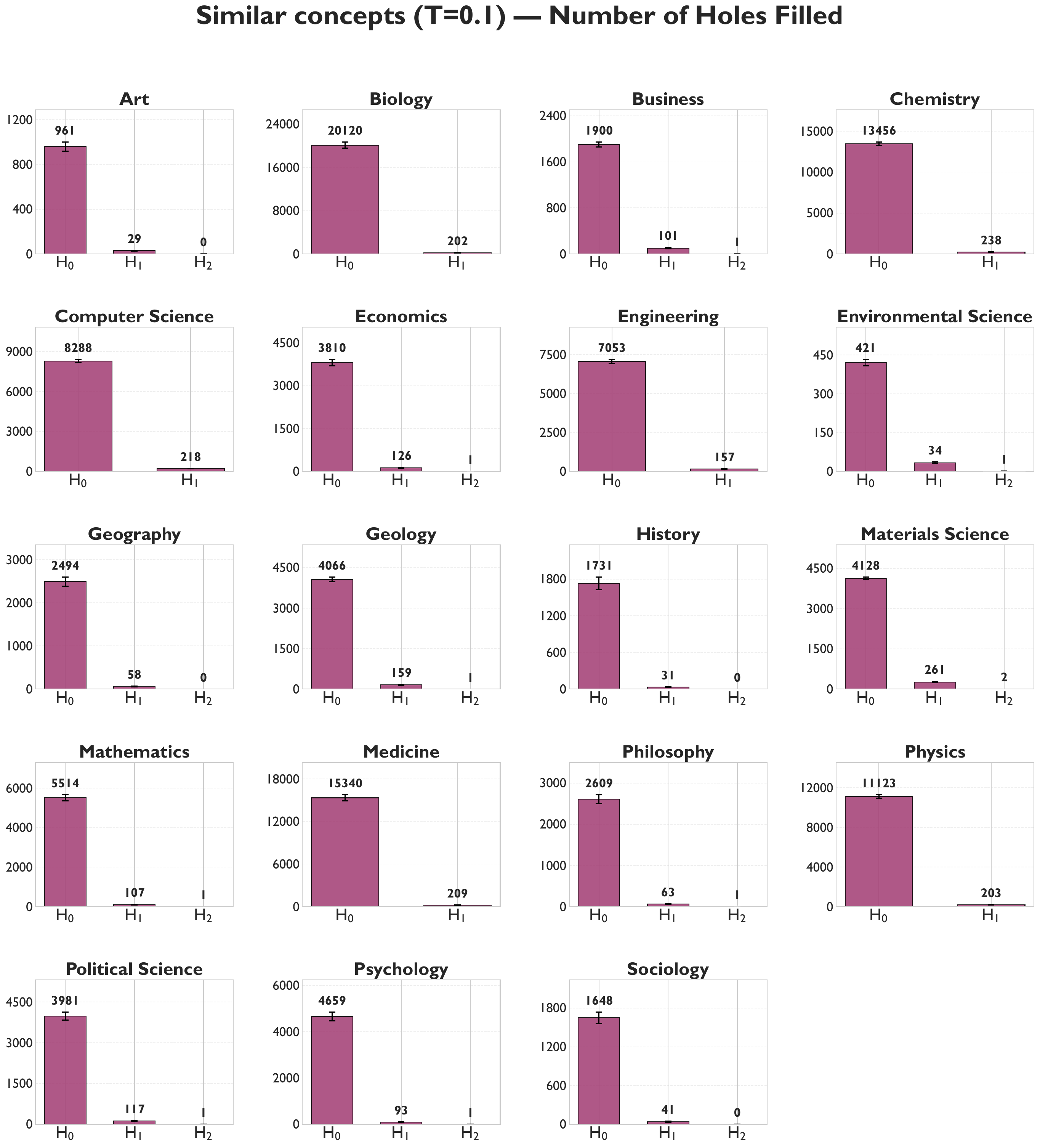}

	\caption{\textbf{Number of holes filled by simulated papers with semantically similar sets of concepts (temperature $T =  0.1$), by discipline and homology dimension.}
    Same as Figure~\ref{fig:filled_by_similar_concepts_papers_T=0.01}, but for $T = 0.1$.
    }
	\label{fig:filled_by_similar_concepts_papers_T=0.1}
\end{figure}

\begin{figure}
	\centering
	\includegraphics[width=0.8\textwidth]{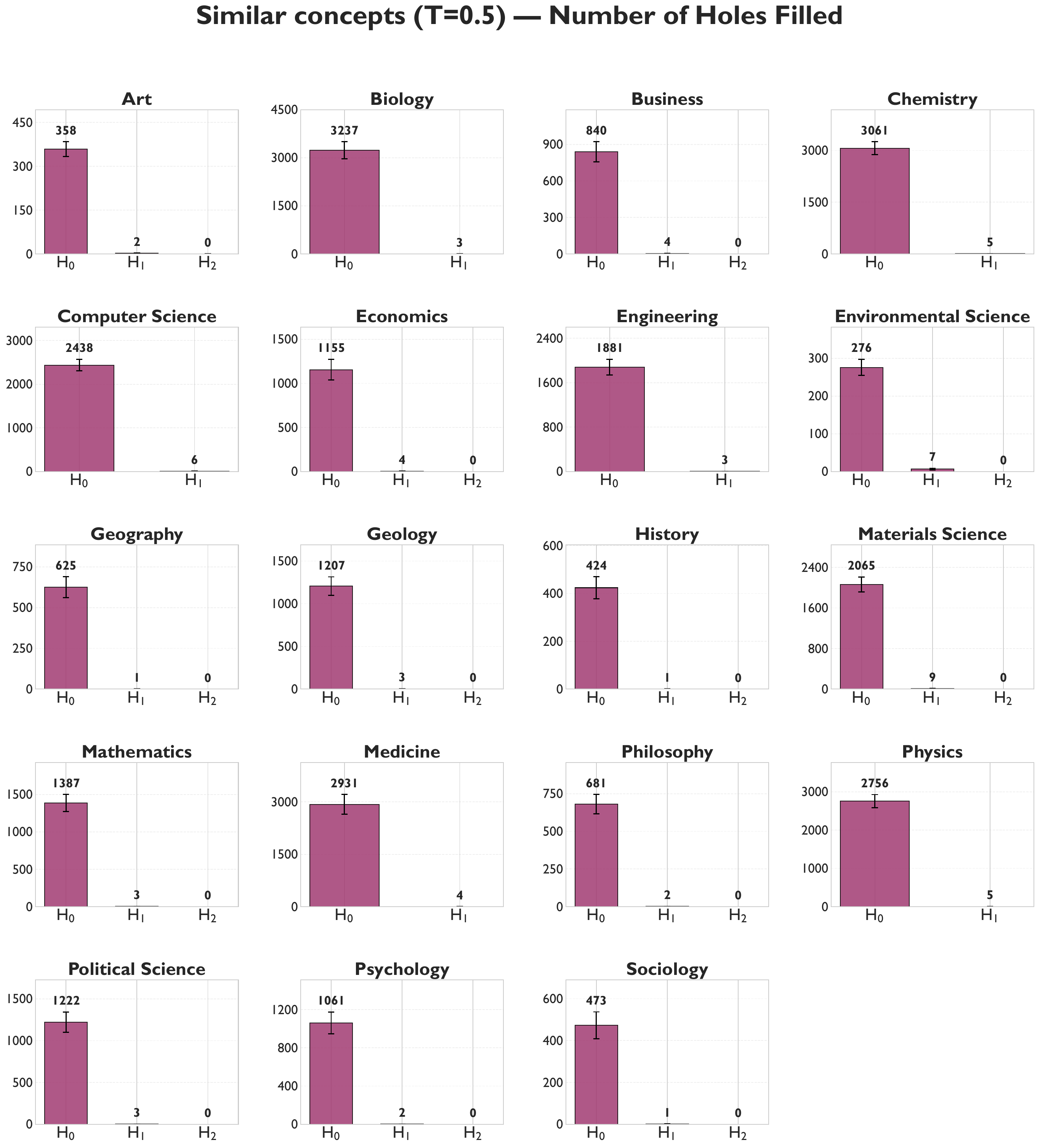}

	\caption{\textbf{Number of holes filled by simulated papers with semantically similar sets of concepts (temperature $T =  0.5$), by discipline and homology dimension.}
    Same as Figure~\ref{fig:filled_by_similar_concepts_papers_T=0.01}, but for $T = 0.5$.
    }
	\label{fig:filled_by_similar_concepts_papers_T=0.5}
\end{figure}

\begin{figure}
	\centering
	\includegraphics[width=0.8\textwidth]{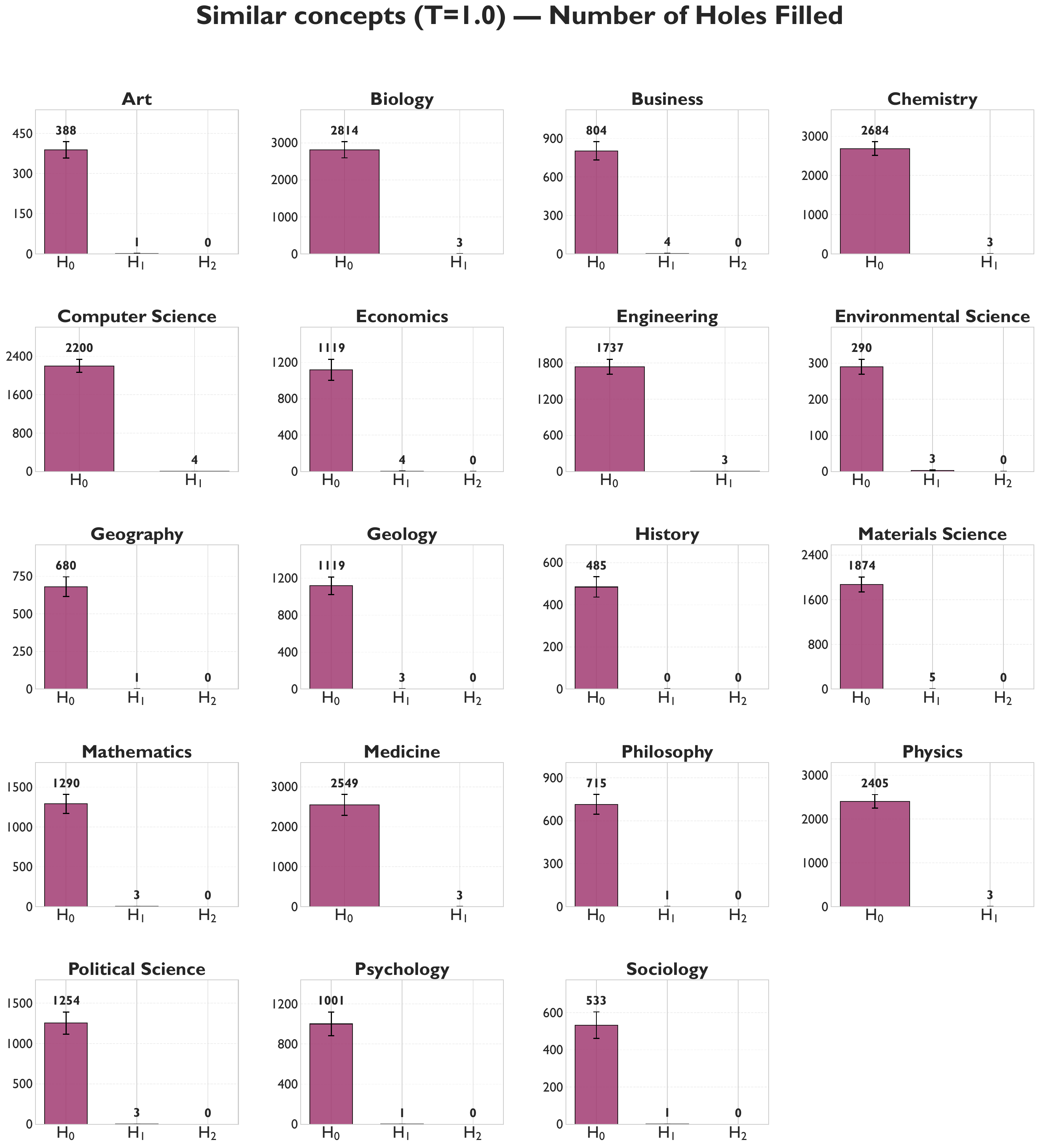}

	\caption{\textbf{Number of holes filled by simulated papers with semantically similar sets of concepts (temperature $T =  1$), by discipline and homology dimension.}
    Same as Figure~\ref{fig:filled_by_similar_concepts_papers_T=0.01}, but for $T = 1$.
    }
	\label{fig:filled_by_similar_concepts_papers_T=1}
\end{figure}

\begin{figure}
	\centering
	\includegraphics[width=0.8\textwidth]{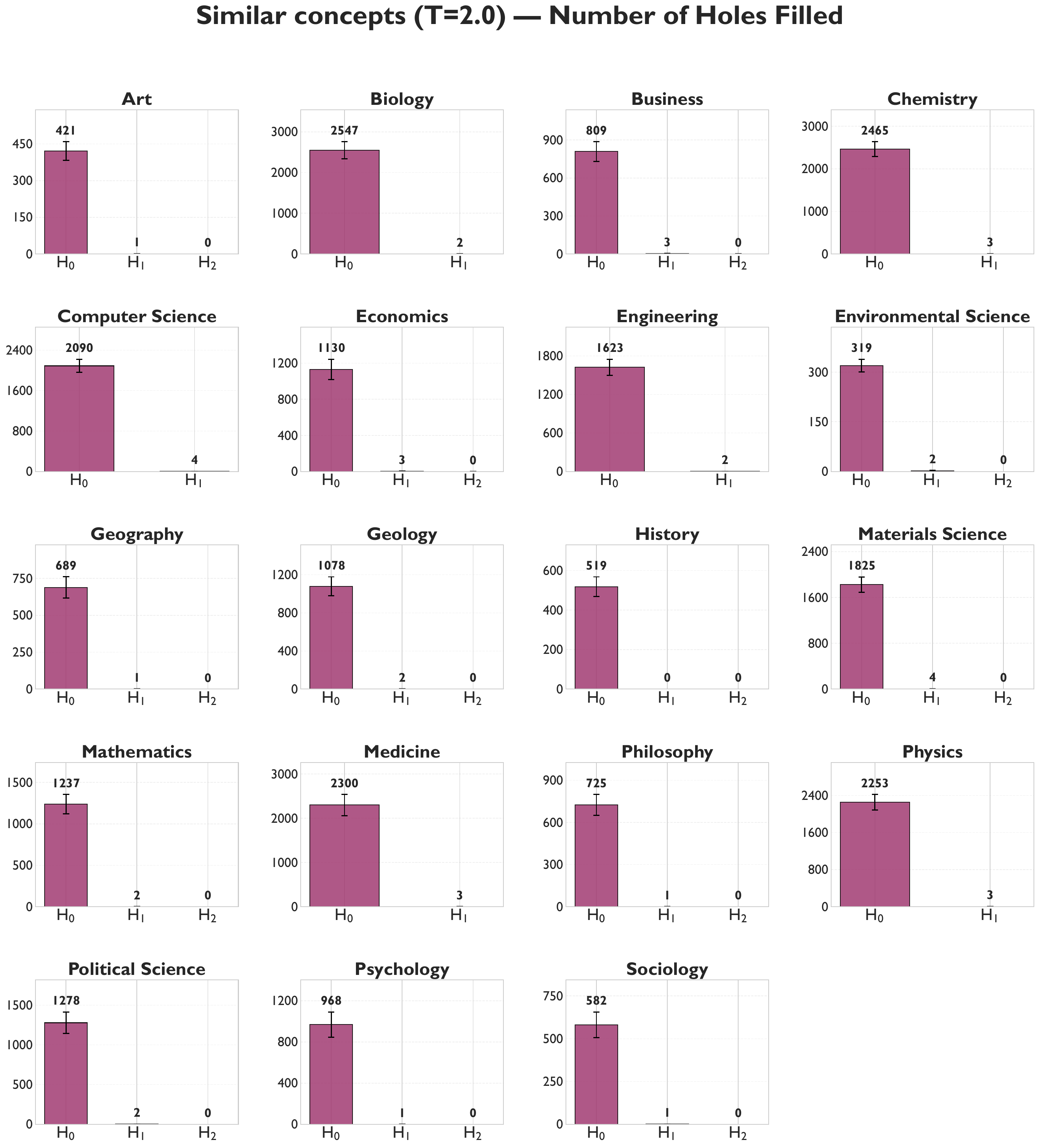}

	\caption{\textbf{Number of holes filled by simulated papers with semantically similar sets of concepts (temperature $T =  2$), by discipline and homology dimension.}
    Same as Figure~\ref{fig:filled_by_similar_concepts_papers_T=0.01}, but for $T = 2$.
    }
	\label{fig:filled_by_similar_concepts_papers_T=2}
\end{figure}

\newpage

\begin{figure}
	\centering
	\includegraphics[width=0.8\textwidth]{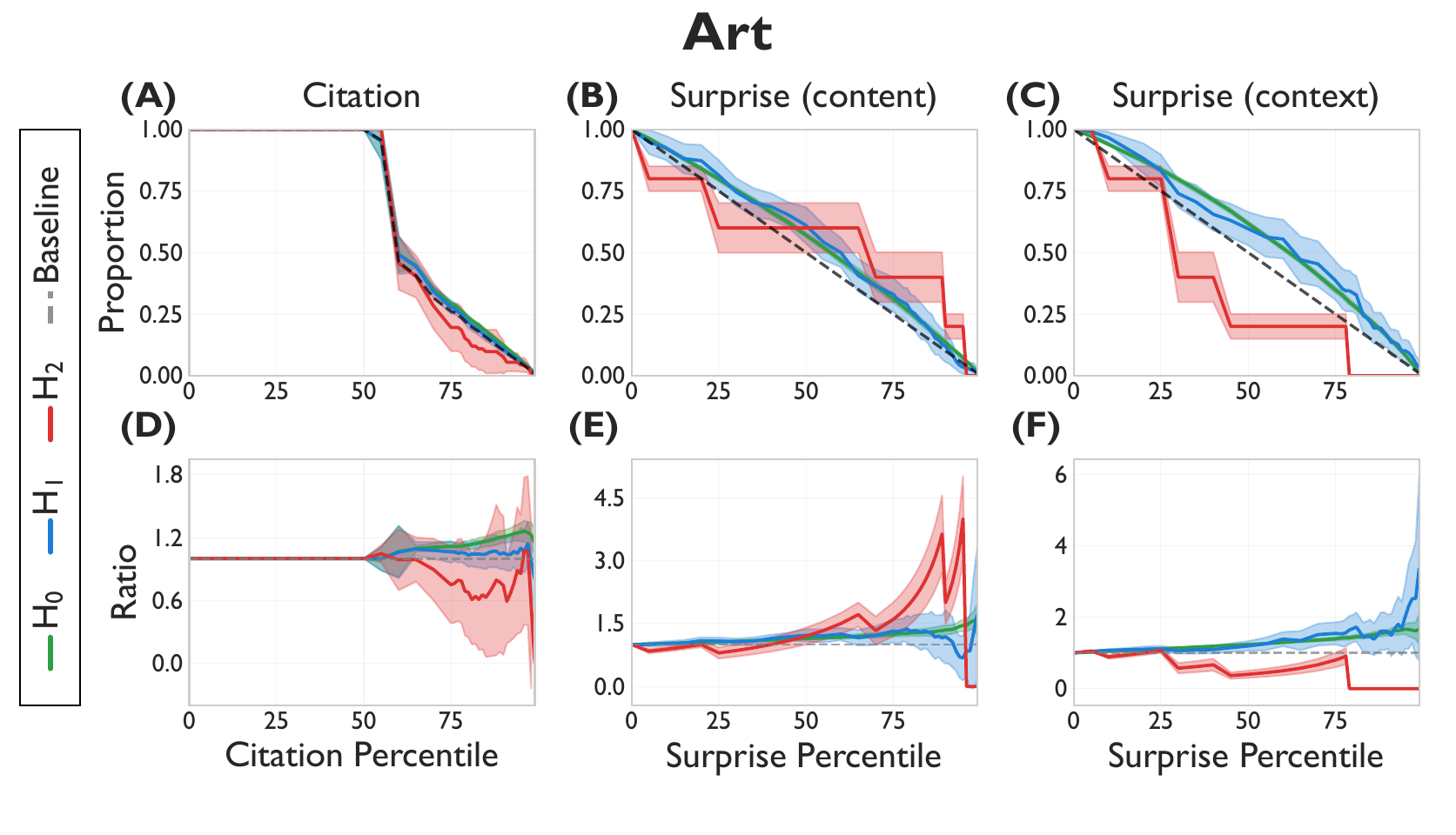}

	\caption{\textbf{Properties of works that fill in holes, by discipline.}
    Same as Figure~\ref{fig:hole_fill_papers}, but for individual disciplines, one discipline per page. This page: art.
    }
	\label{fig:fill_papers_art}
\end{figure}

\begin{figure}
	\centering
	\includegraphics[width=0.8\textwidth]{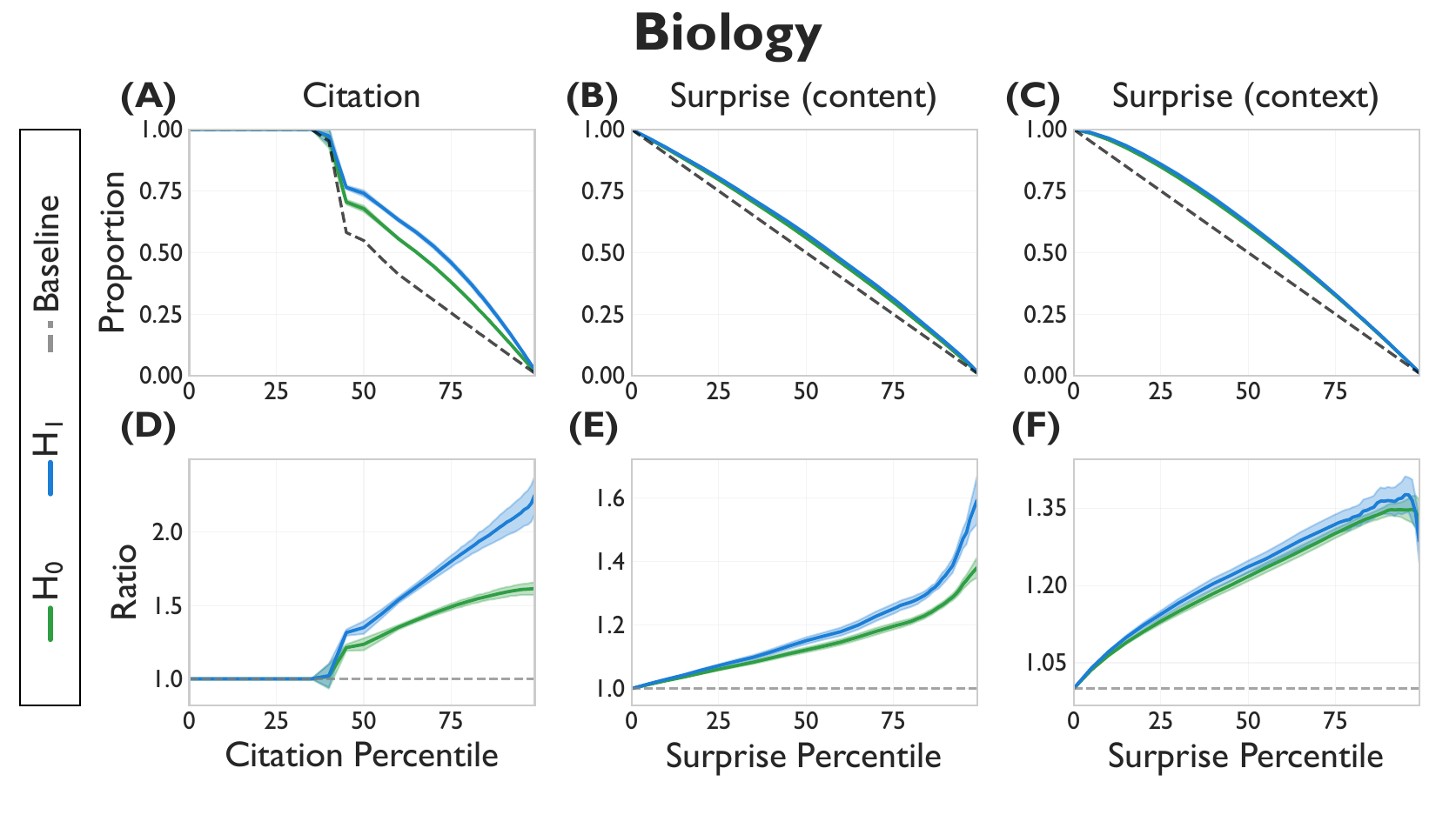}

	\continuedfigure
	\caption{\textbf{Properties of works that fill in holes, by discipline.}
    This page: biology.
    }
	\label{fig:fill_papers_biology}
\end{figure}

\begin{figure}
	\centering
	\includegraphics[width=0.8\textwidth]{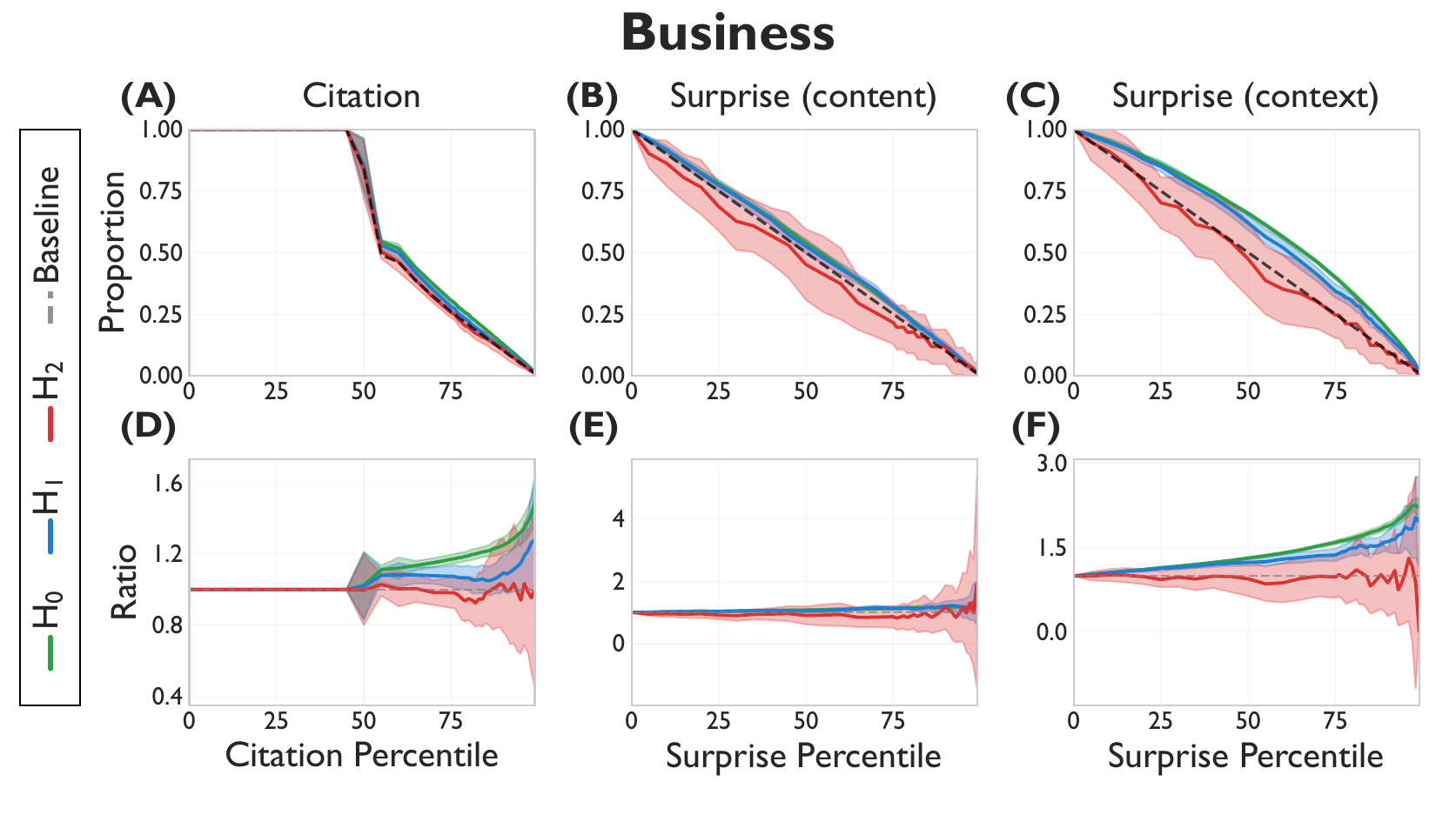}

	\continuedfigure
	\caption{\textbf{Properties of works that fill in holes, by discipline.}
    This page: business.
    }
	\label{fig:fill_papers_business}
\end{figure}

\begin{figure}
	\centering
	\includegraphics[width=0.8\textwidth]{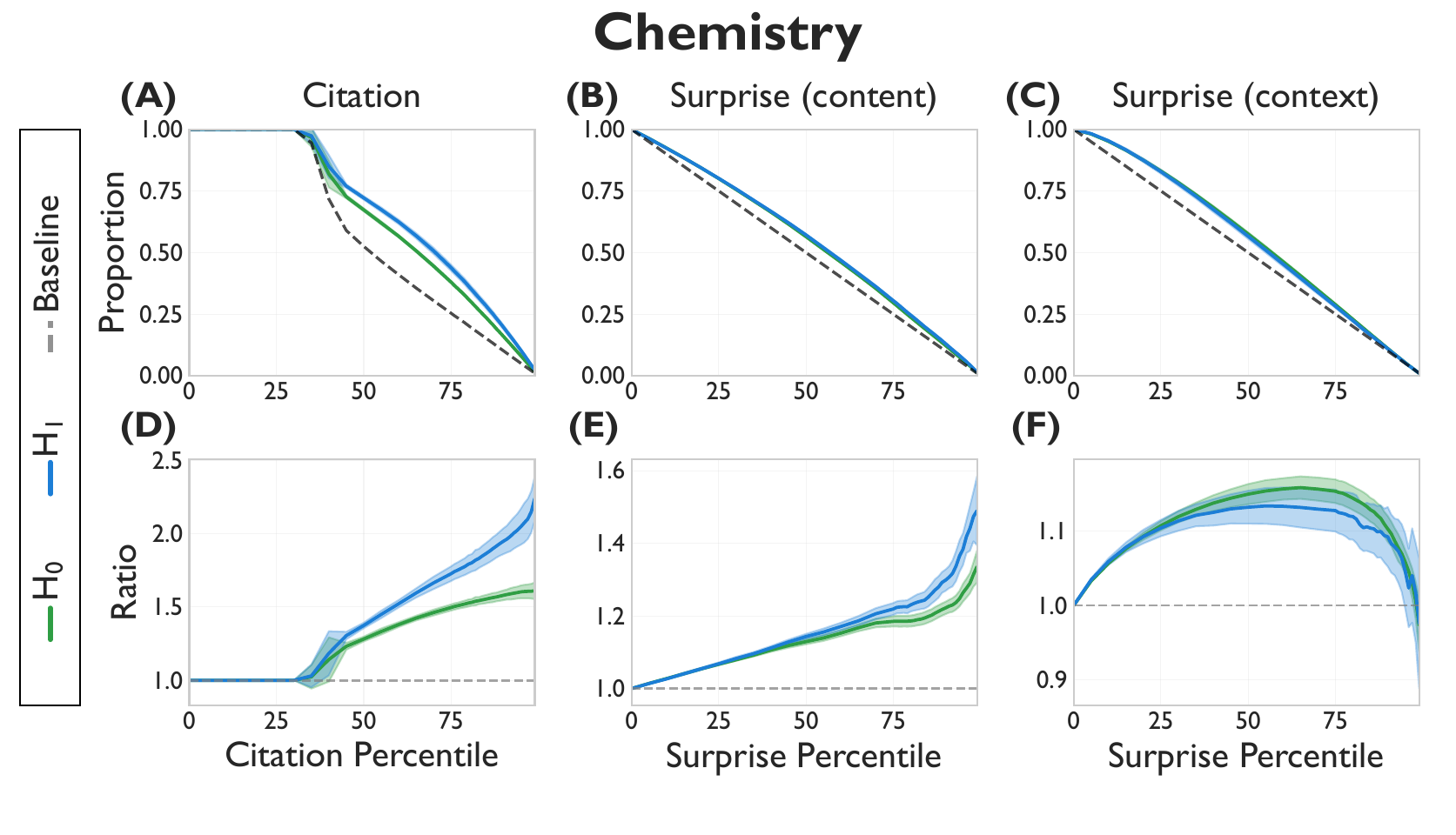}

	\continuedfigure
	\caption{\textbf{Properties of works that fill in holes, by discipline.}
    This page: chemistry.
    }
	\label{fig:fill_papers_chemistry}
\end{figure}

\begin{figure}
	\centering
	\includegraphics[width=0.8\textwidth]{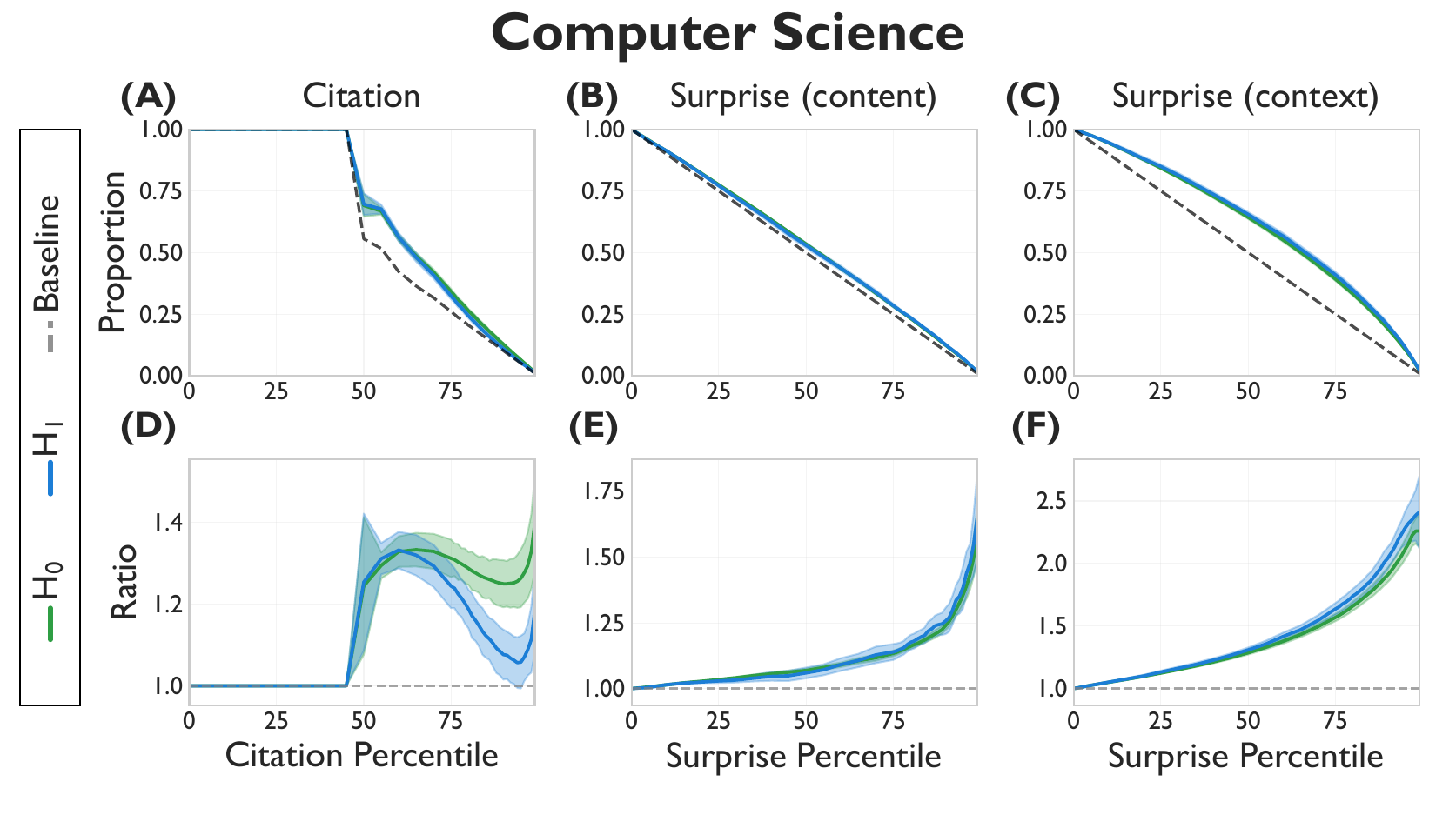}

	\continuedfigure
	\caption{\textbf{Properties of works that fill in holes, by discipline.}
    This page: computer science.
    }
	\label{fig:fill_papers_computer_science}
\end{figure}

\begin{figure}
	\centering
	\includegraphics[width=0.8\textwidth]{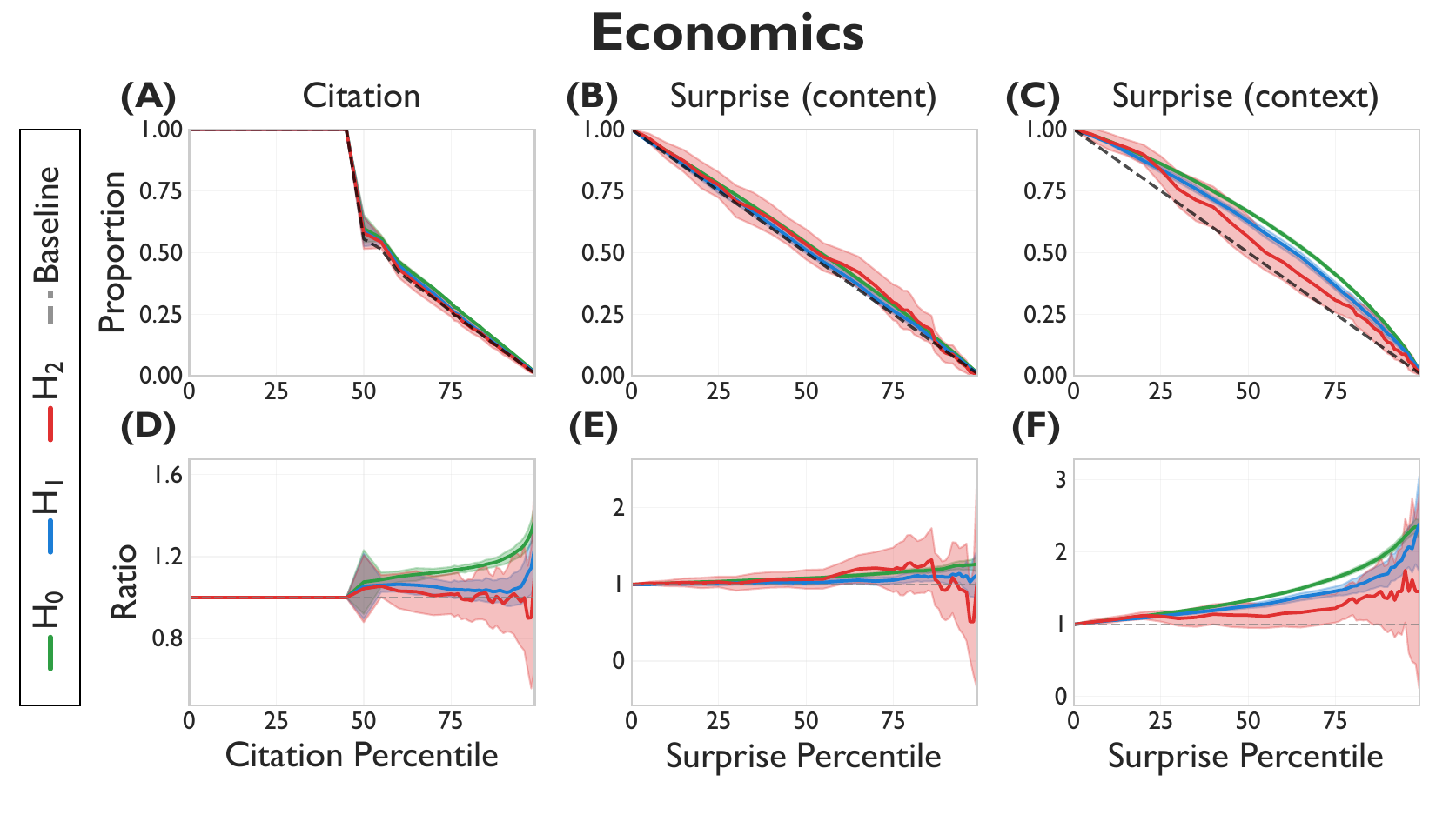}

	\continuedfigure
	\caption{\textbf{Properties of works that fill in holes, by discipline.}
    This page: economics.
    }
	\label{fig:fill_papers_economics}
\end{figure}

\begin{figure}
	\centering
	\includegraphics[width=0.8\textwidth]{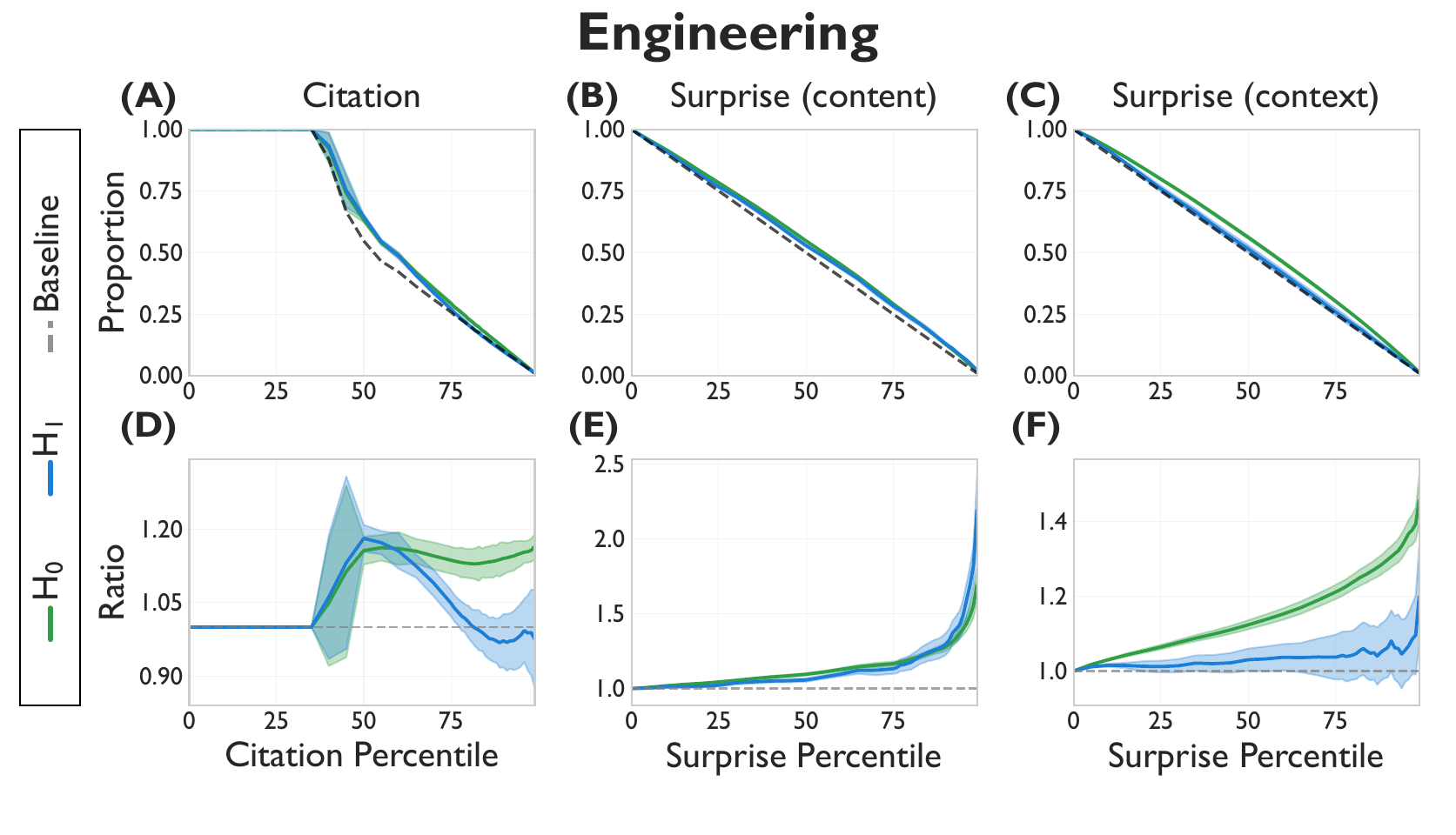}

	\continuedfigure
	\caption{\textbf{Properties of works that fill in holes, by discipline.}
    This page: engineering.
    }
	\label{fig:fill_papers_engineering}
\end{figure}

\begin{figure}
	\centering
	\includegraphics[width=0.8\textwidth]{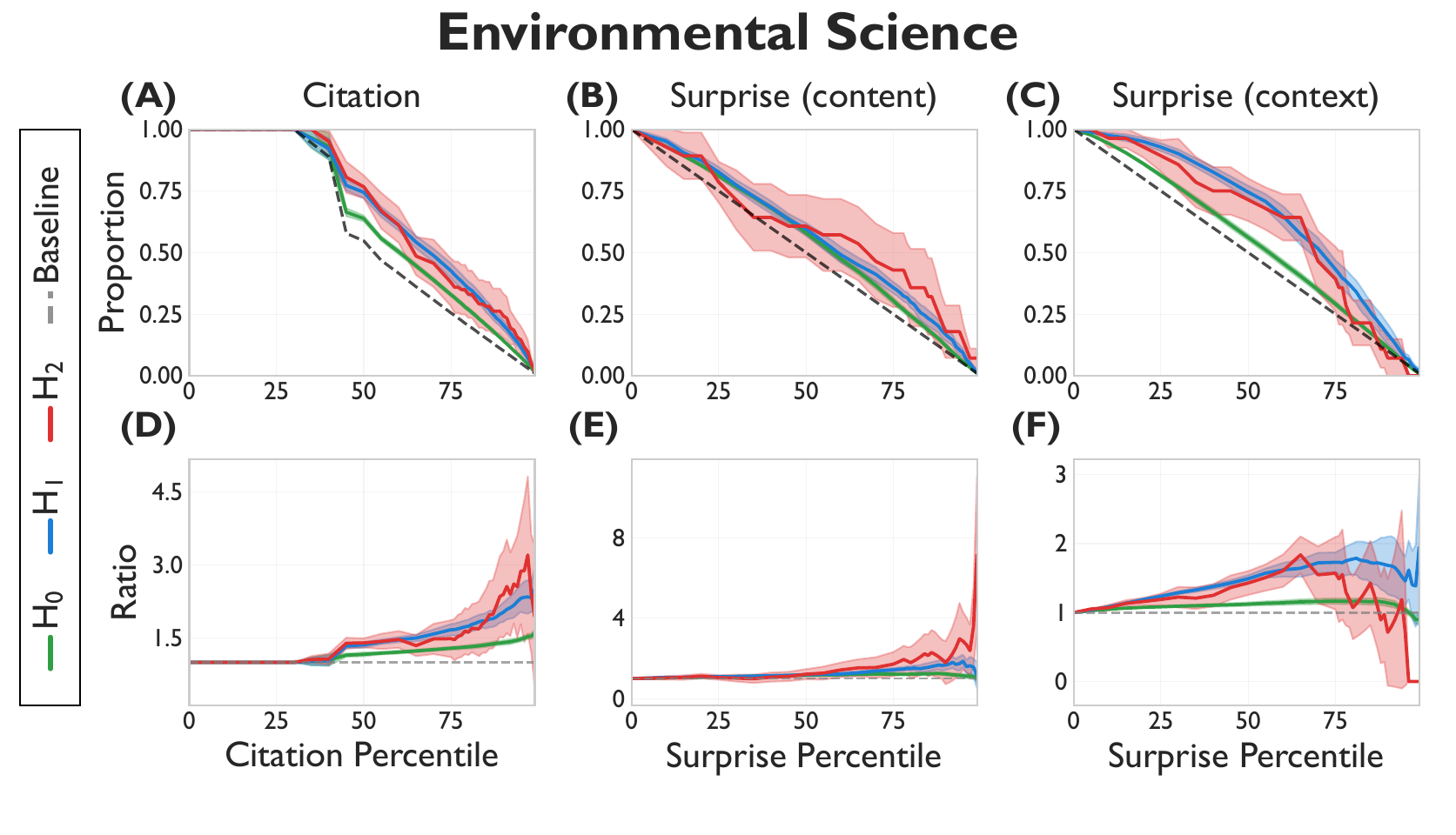}

	\continuedfigure
	\caption{\textbf{Properties of works that fill in holes, by discipline.}
    This page: environmental science.
    }
	\label{fig:fill_papers_environmental_science}
\end{figure}

\begin{figure}
	\centering
	\includegraphics[width=0.8\textwidth]{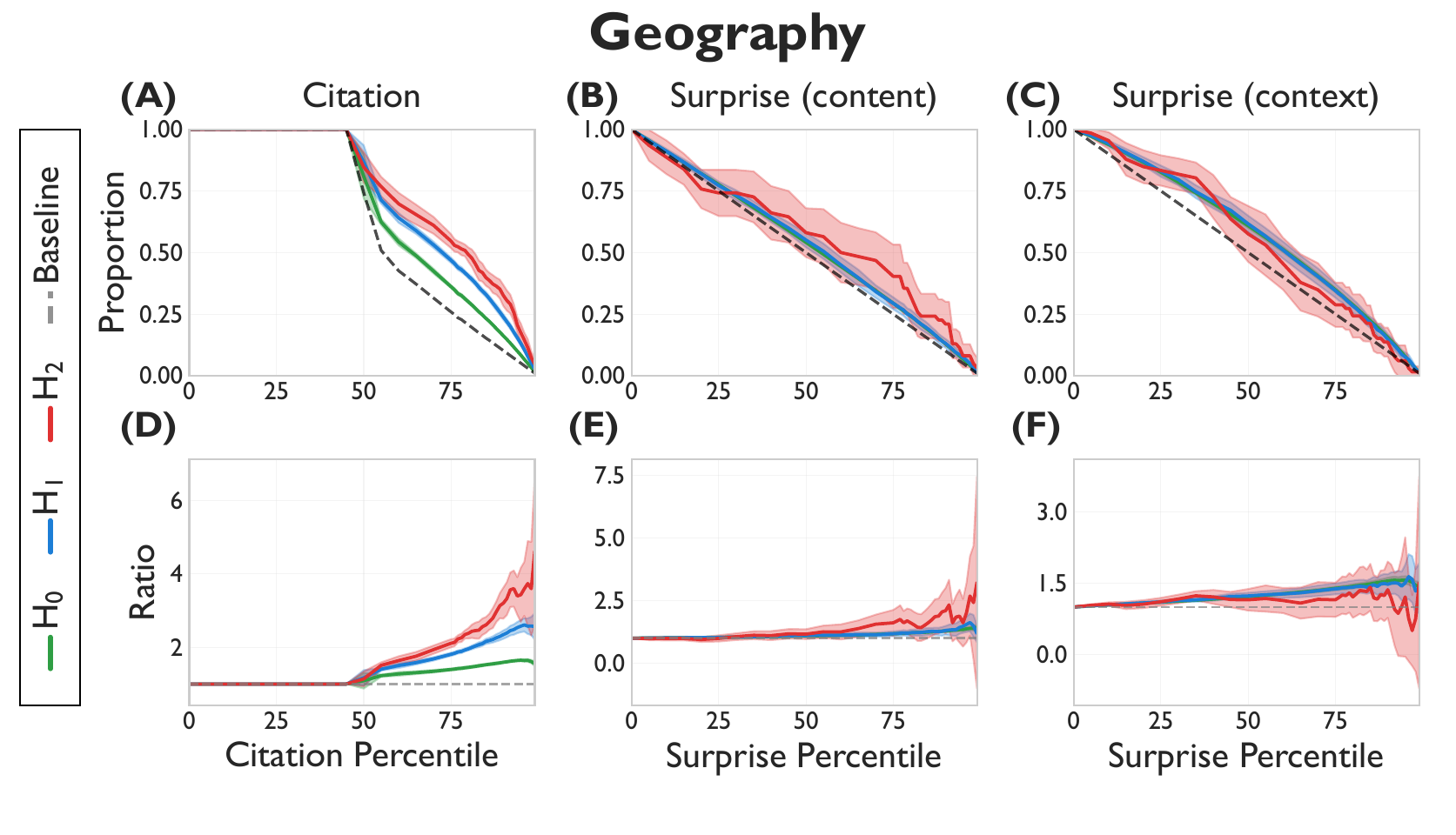}

	\continuedfigure
	\caption{\textbf{Properties of works that fill in holes, by discipline.}
    This page: geography.
    }
	\label{fig:fill_papers_geography}
\end{figure}

\begin{figure}
	\centering
	\includegraphics[width=0.8\textwidth]{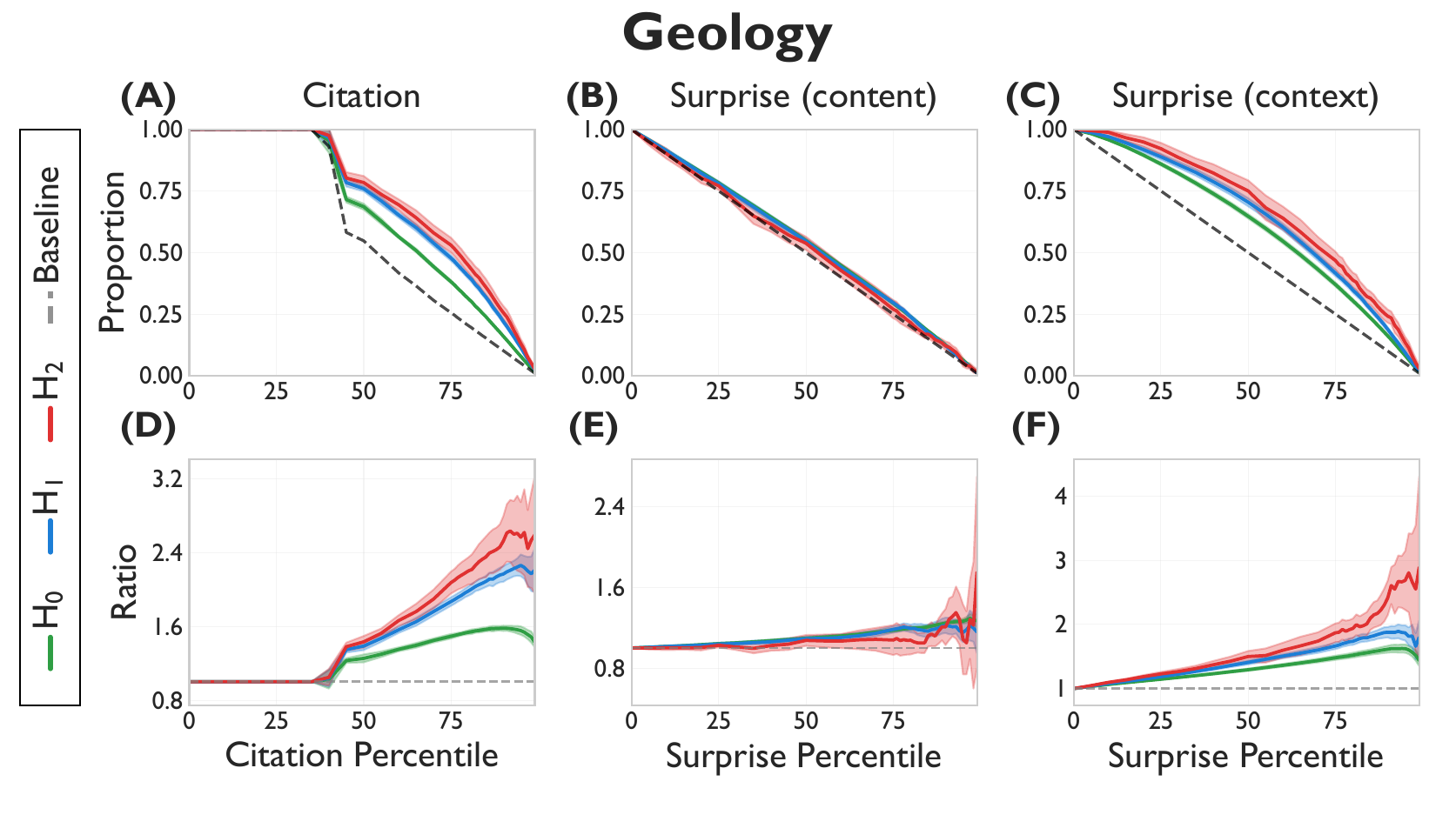}

	\continuedfigure
	\caption{\textbf{Properties of works that fill in holes, by discipline.}
    This page: geology.
    }
	\label{fig:fill_papers_geology}
\end{figure}

\begin{figure}
	\centering
	\includegraphics[width=0.8\textwidth]{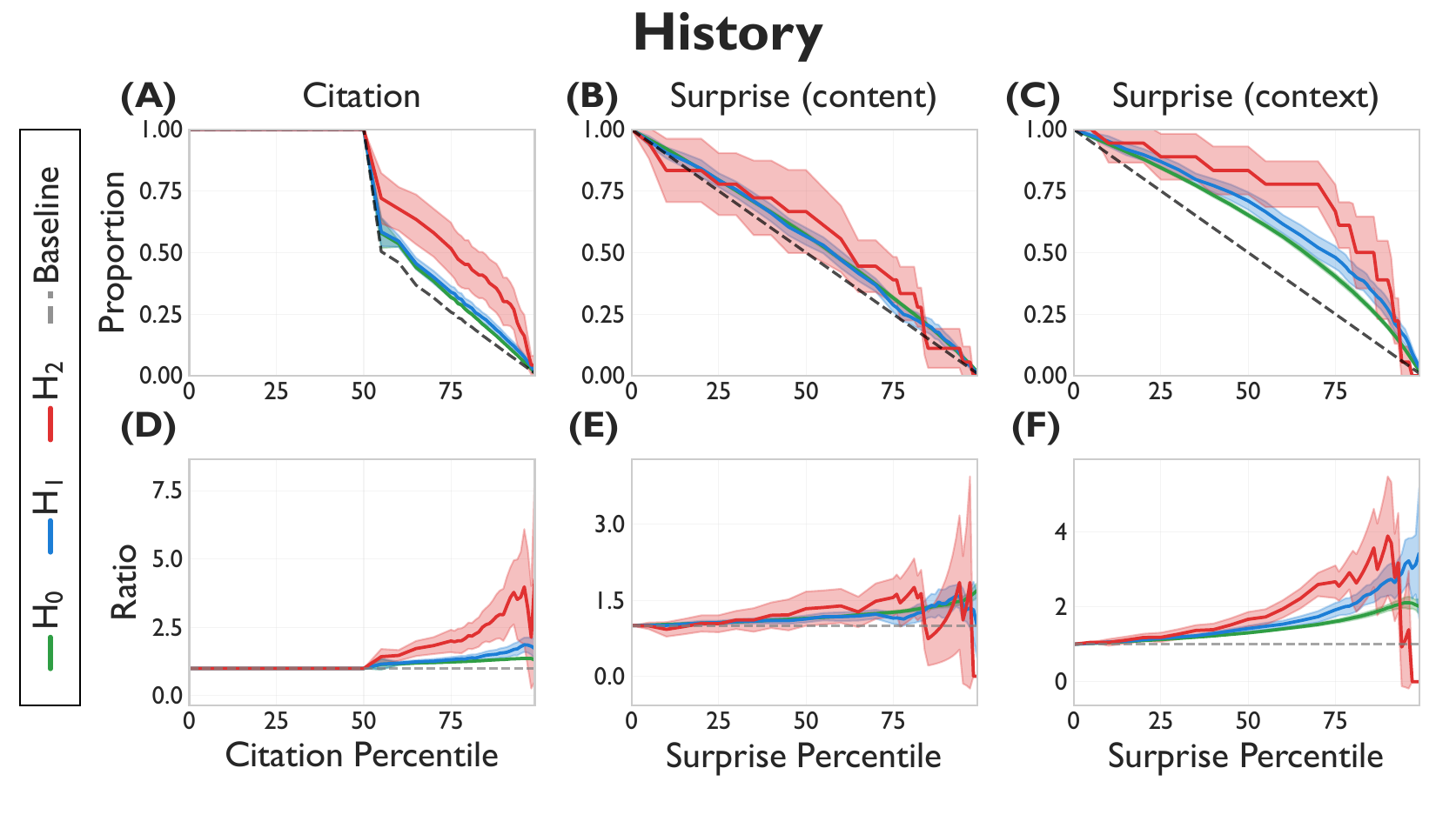}

	\continuedfigure
	\caption{\textbf{Properties of works that fill in holes, by discipline.}
    This page: history.
    }
	\label{fig:fill_papers_history}
\end{figure}

\begin{figure}
	\centering
	\includegraphics[width=0.8\textwidth]{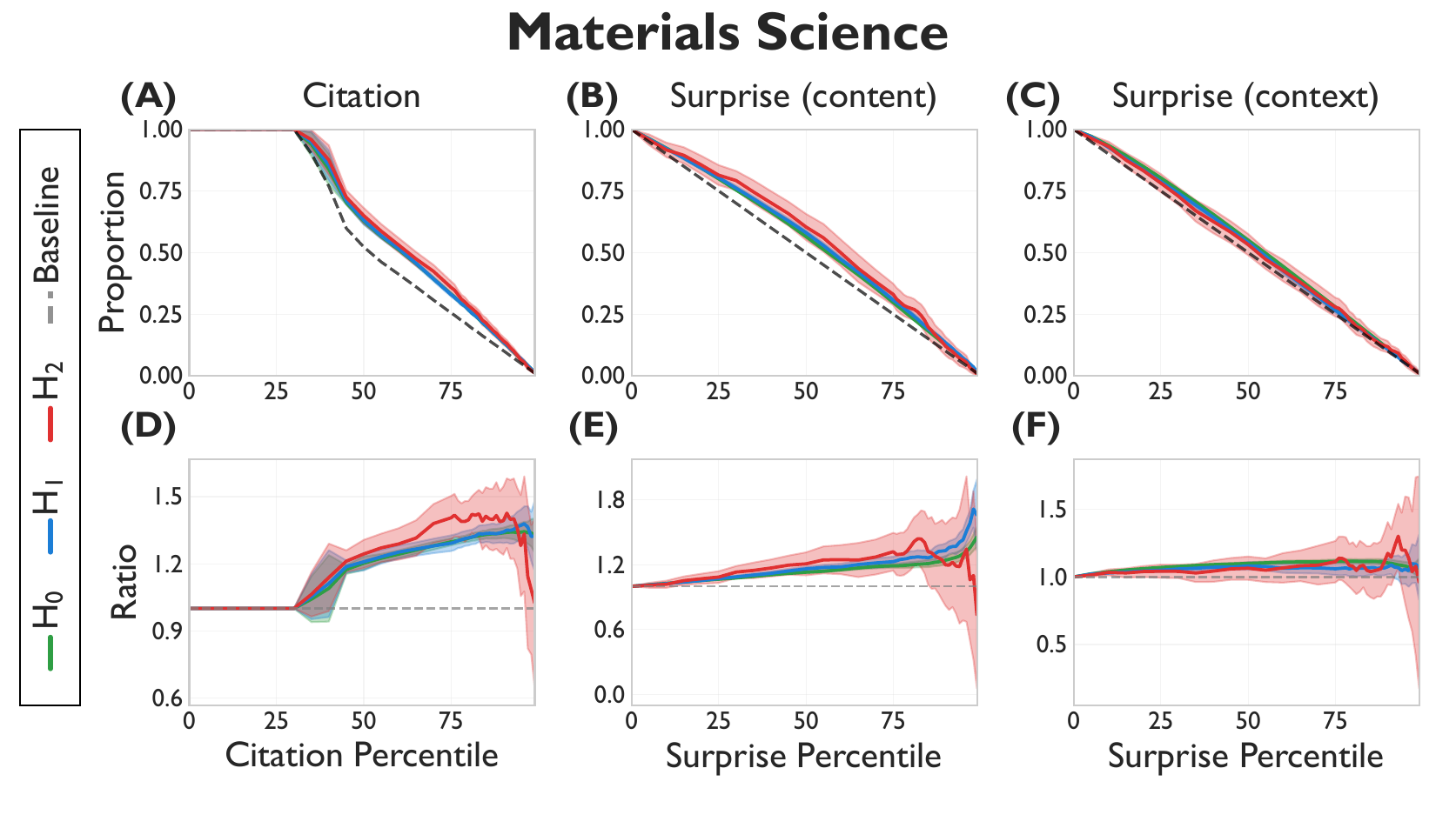}

	\continuedfigure
	\caption{\textbf{Properties of works that fill in holes, by discipline.}
    This page: materials science.
    }
	\label{fig:fill_papers_materials_science}
\end{figure}

\begin{figure}
	\centering
	\includegraphics[width=0.8\textwidth]{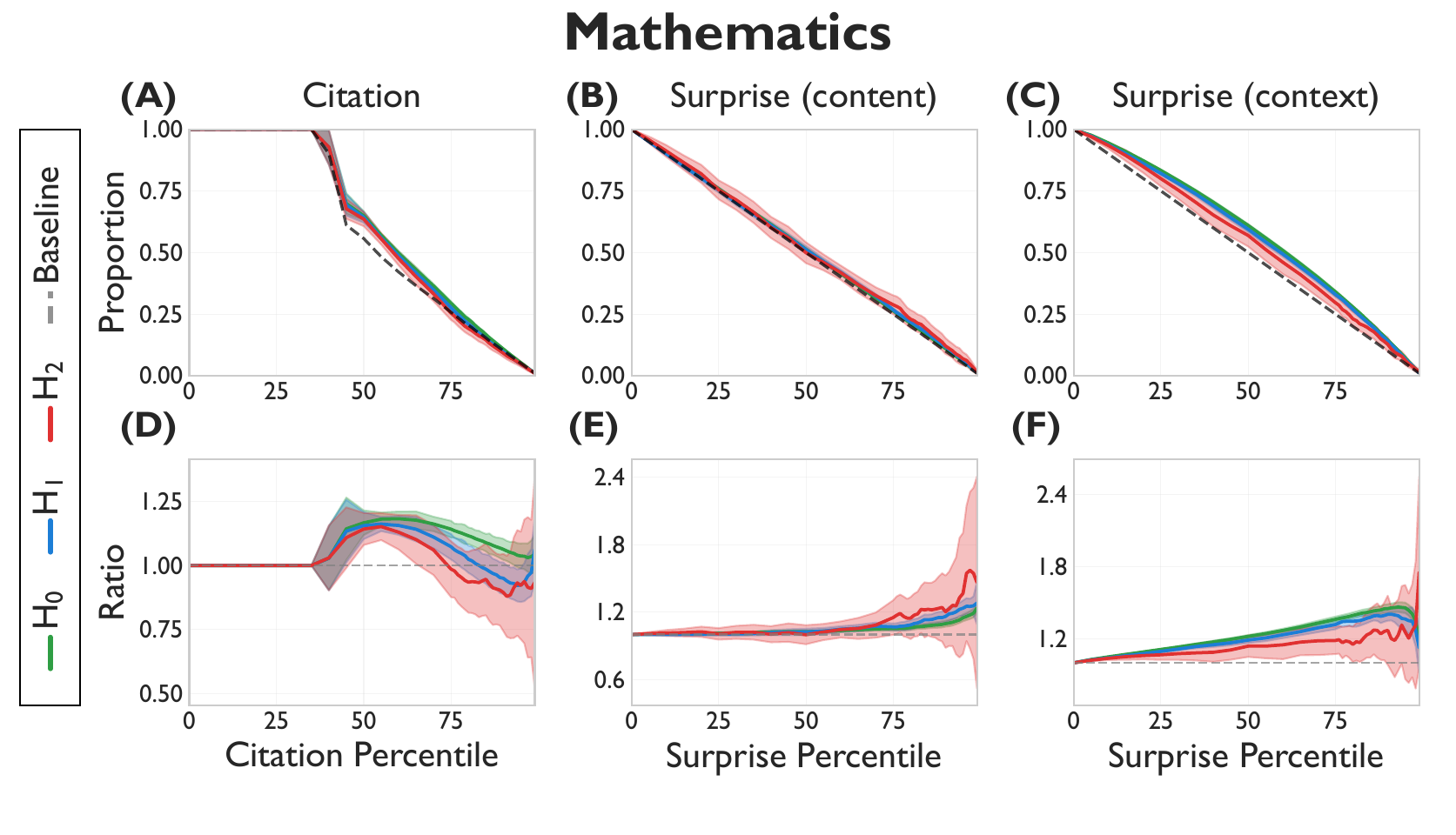}

	\continuedfigure
	\caption{\textbf{Properties of works that fill in holes, by discipline.}
    This page: mathematics.
    }
	\label{fig:fill_papers_mathematics}
\end{figure}

\begin{figure}
	\centering
	\includegraphics[width=0.8\textwidth]{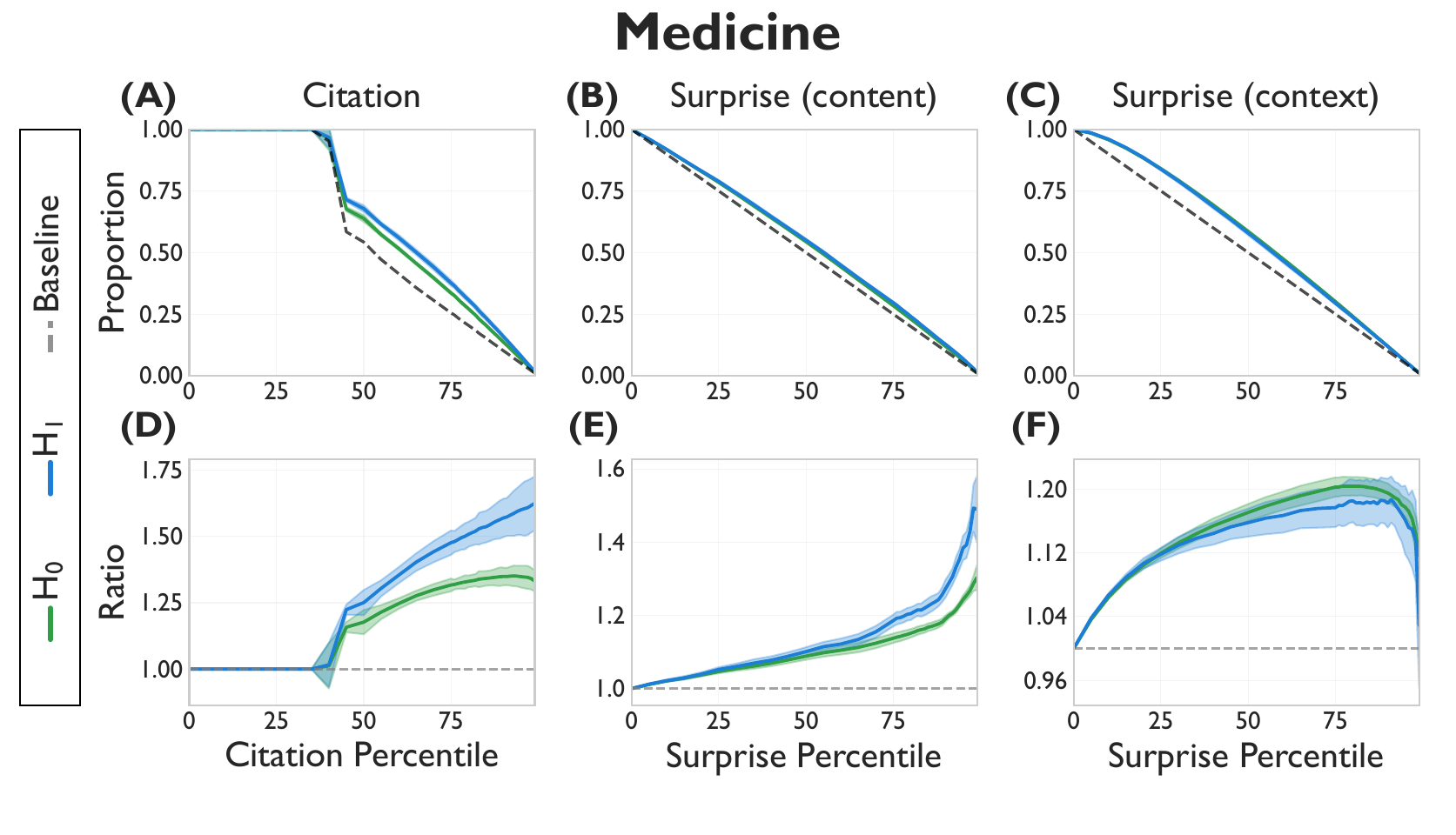}

	\continuedfigure
	\caption{\textbf{Properties of works that fill in holes, by discipline.}
    This page: medicine.
    }
	\label{fig:fill_papers_medicine}
\end{figure}

\begin{figure}
	\centering
	\includegraphics[width=0.8\textwidth]{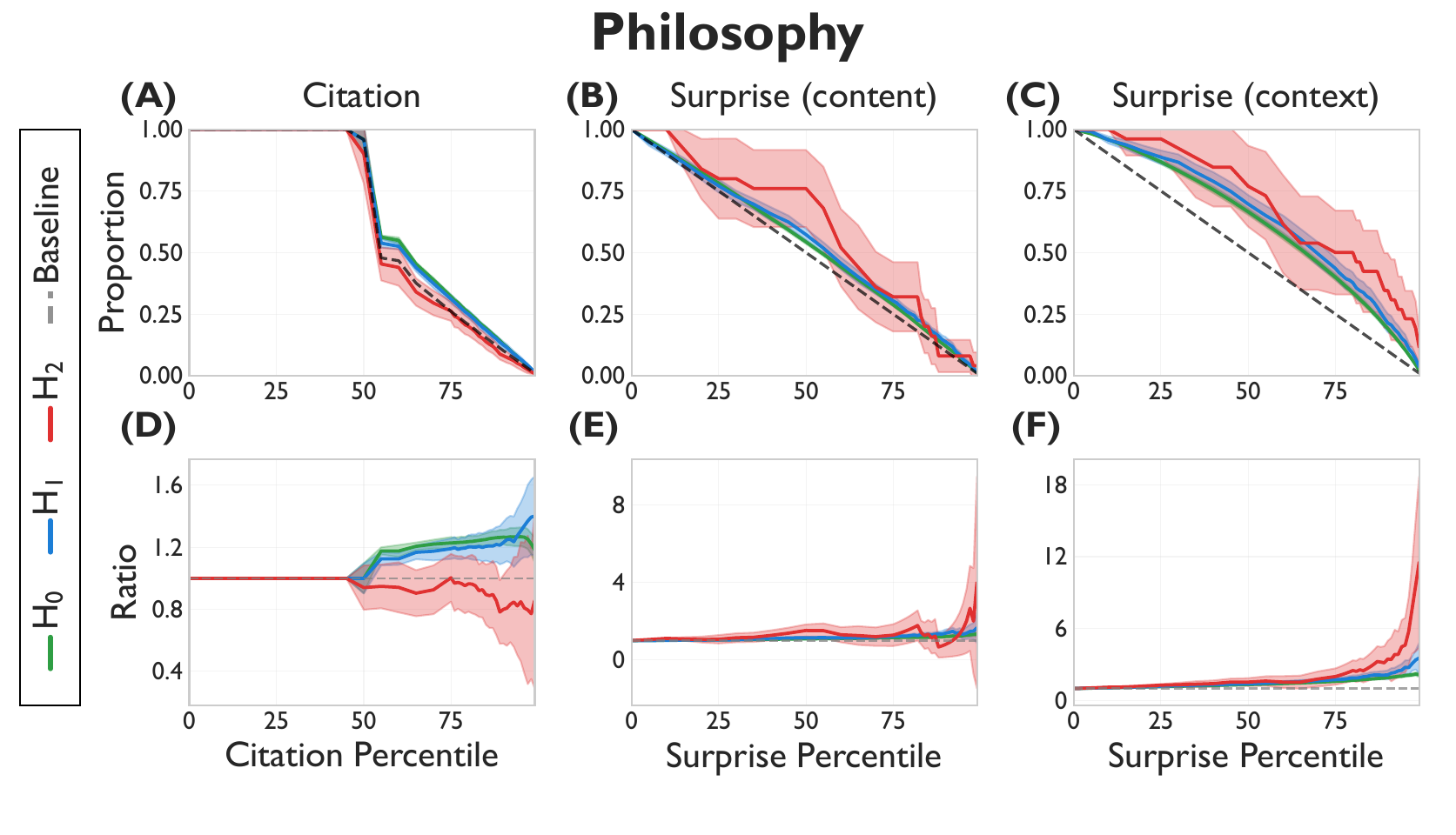}

	\continuedfigure
	\caption{\textbf{Properties of works that fill in holes, by discipline.}
    This page: philosophy.
    }
	\label{fig:fill_papers_philosophy}
\end{figure}

\begin{figure}
	\centering
	\includegraphics[width=0.8\textwidth]{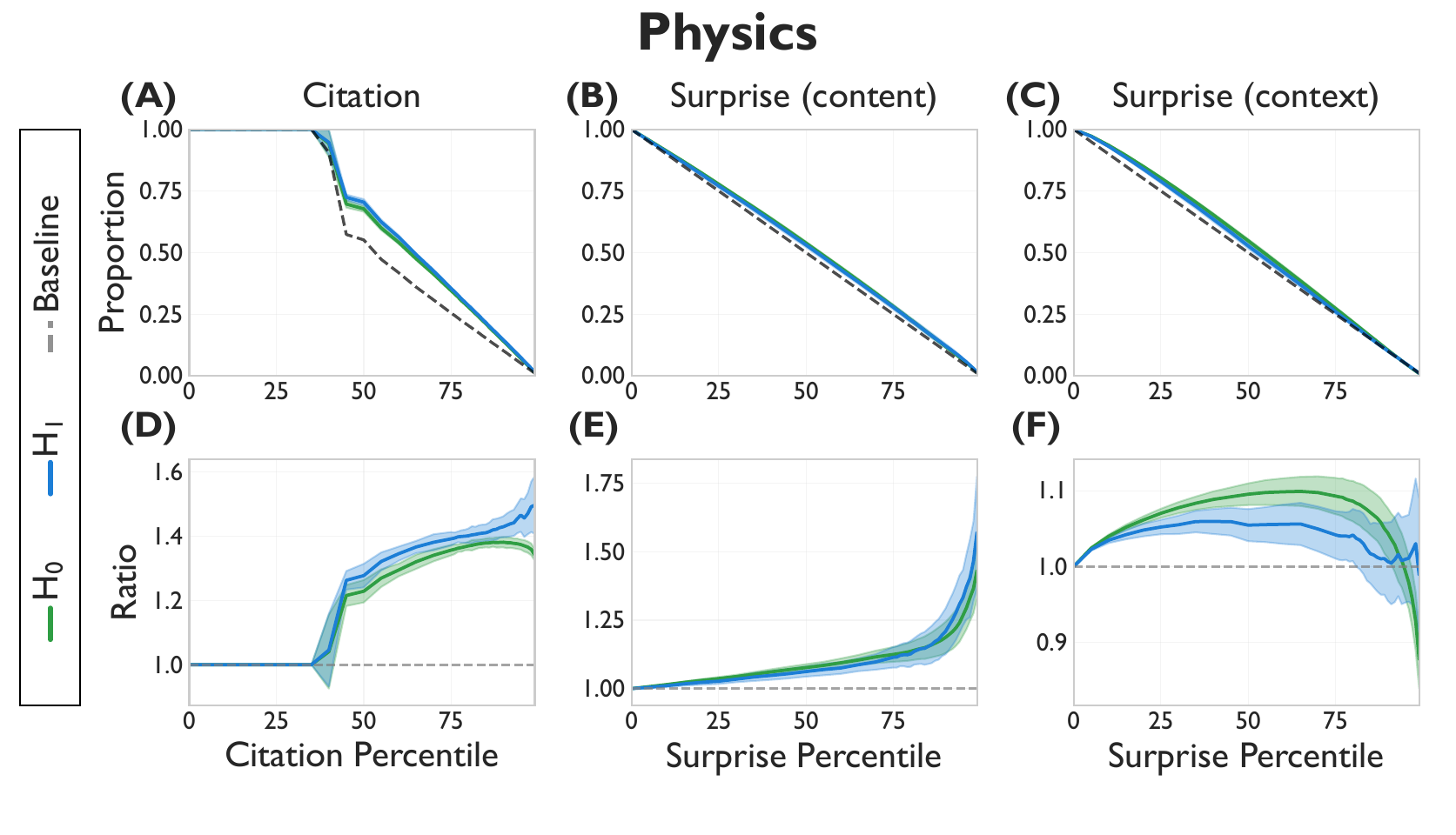}

	\continuedfigure
	\caption{\textbf{Properties of works that fill in holes, by discipline.}
    This page: physics.
    }
	\label{fig:fill_papers_physics}
\end{figure}

\begin{figure}
	\centering
	\includegraphics[width=0.8\textwidth]{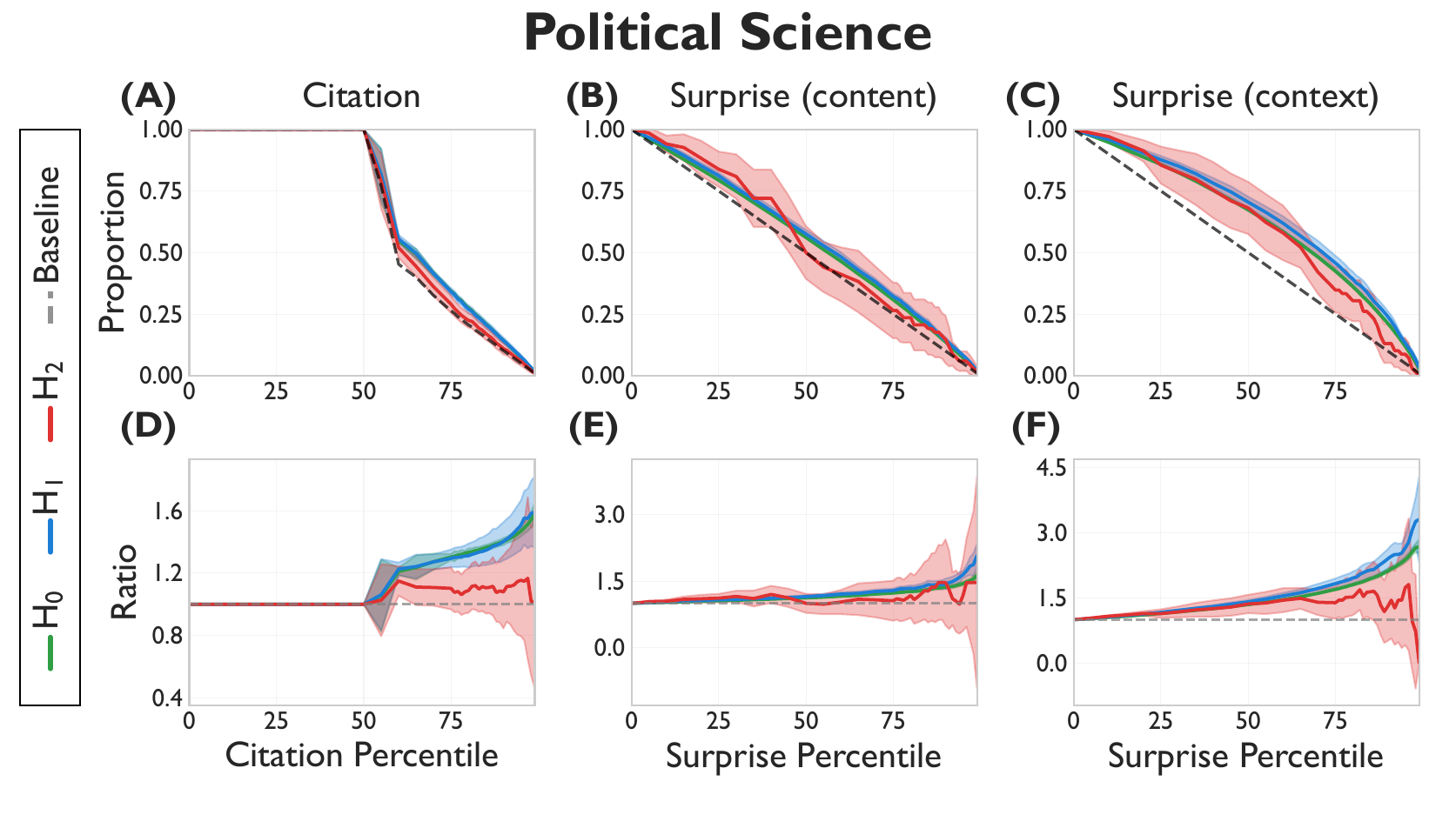}

	\continuedfigure
	\caption{\textbf{Properties of works that fill in holes, by discipline.}
    This page: political science.
    }
	\label{fig:fill_papers_political_science}
\end{figure}

\begin{figure}
	\centering
	\includegraphics[width=0.8\textwidth]{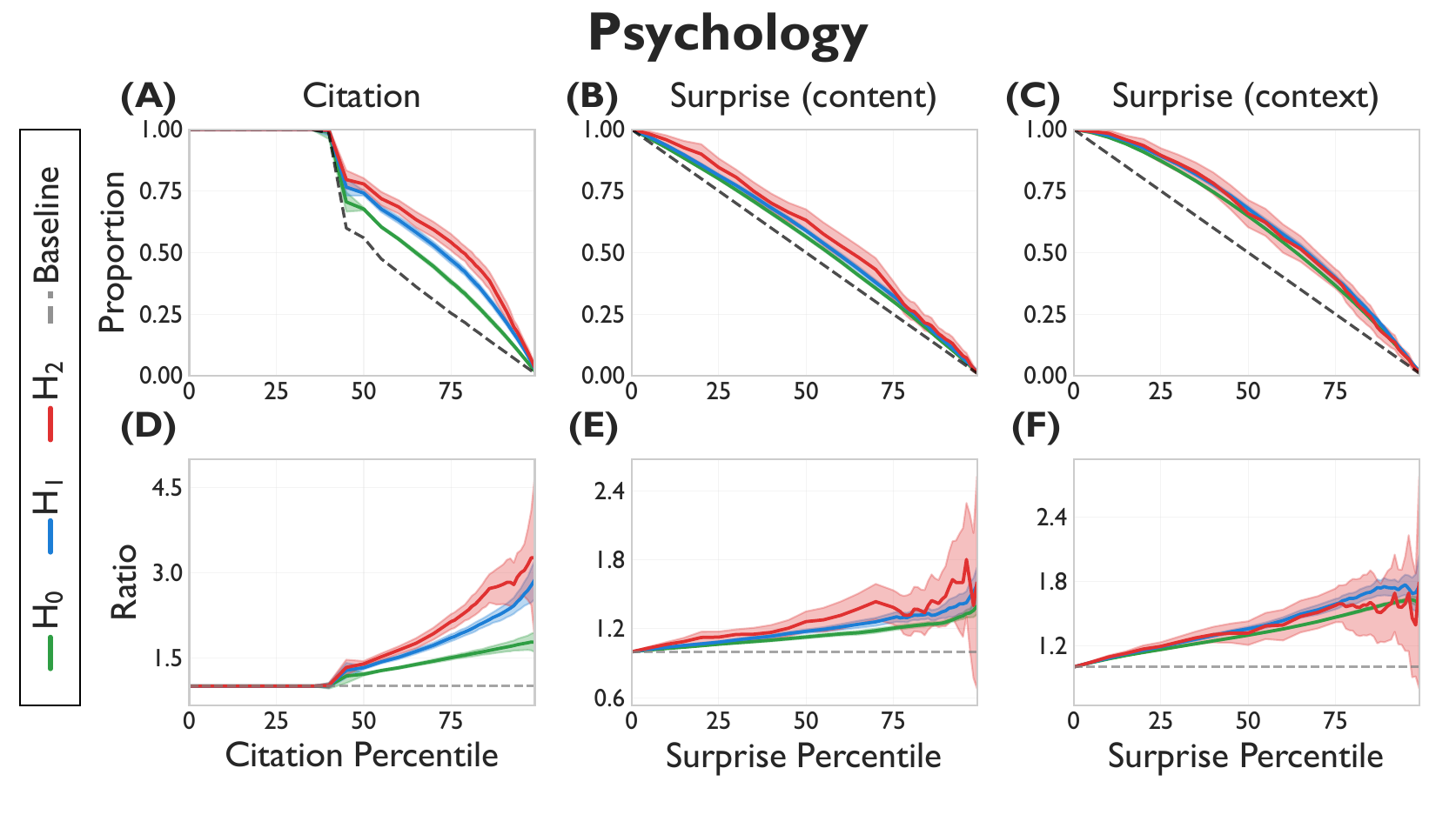}

	\continuedfigure
	\caption{\textbf{Properties of works that fill in holes, by discipline.}
    This page: psychology.
    }
	\label{fig:fill_papers_psychology}
\end{figure}

\begin{figure}
	\centering
	\includegraphics[width=0.8\textwidth]{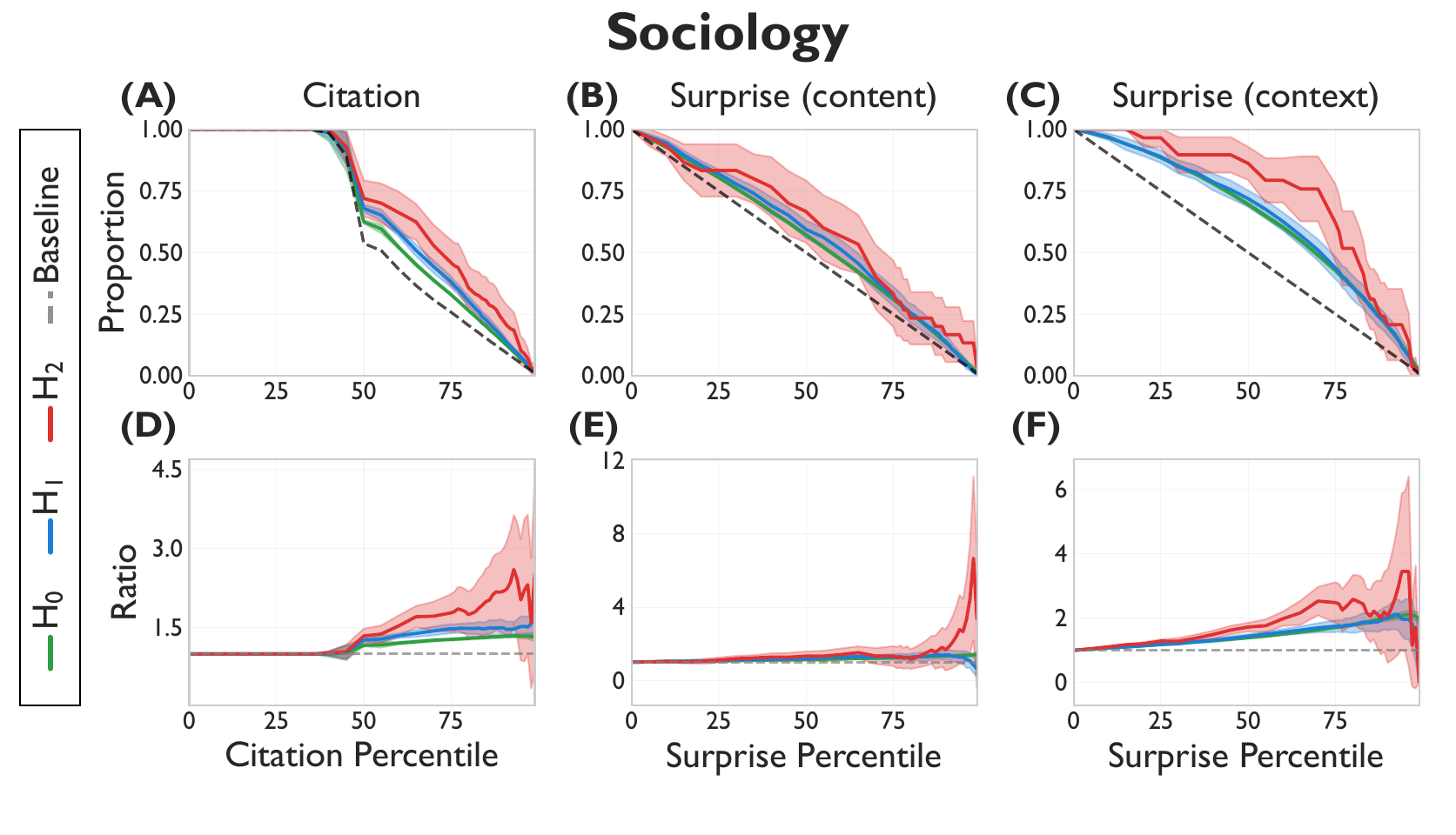}

	\continuedfigure
	\caption{\textbf{Properties of works that fill in holes, by discipline.}
    This page: sociology.
    }
	\label{fig:fill_papers_sociology}
\end{figure}

\newpage

\begin{figure}
	\centering
	\includegraphics[width=0.8\textwidth]{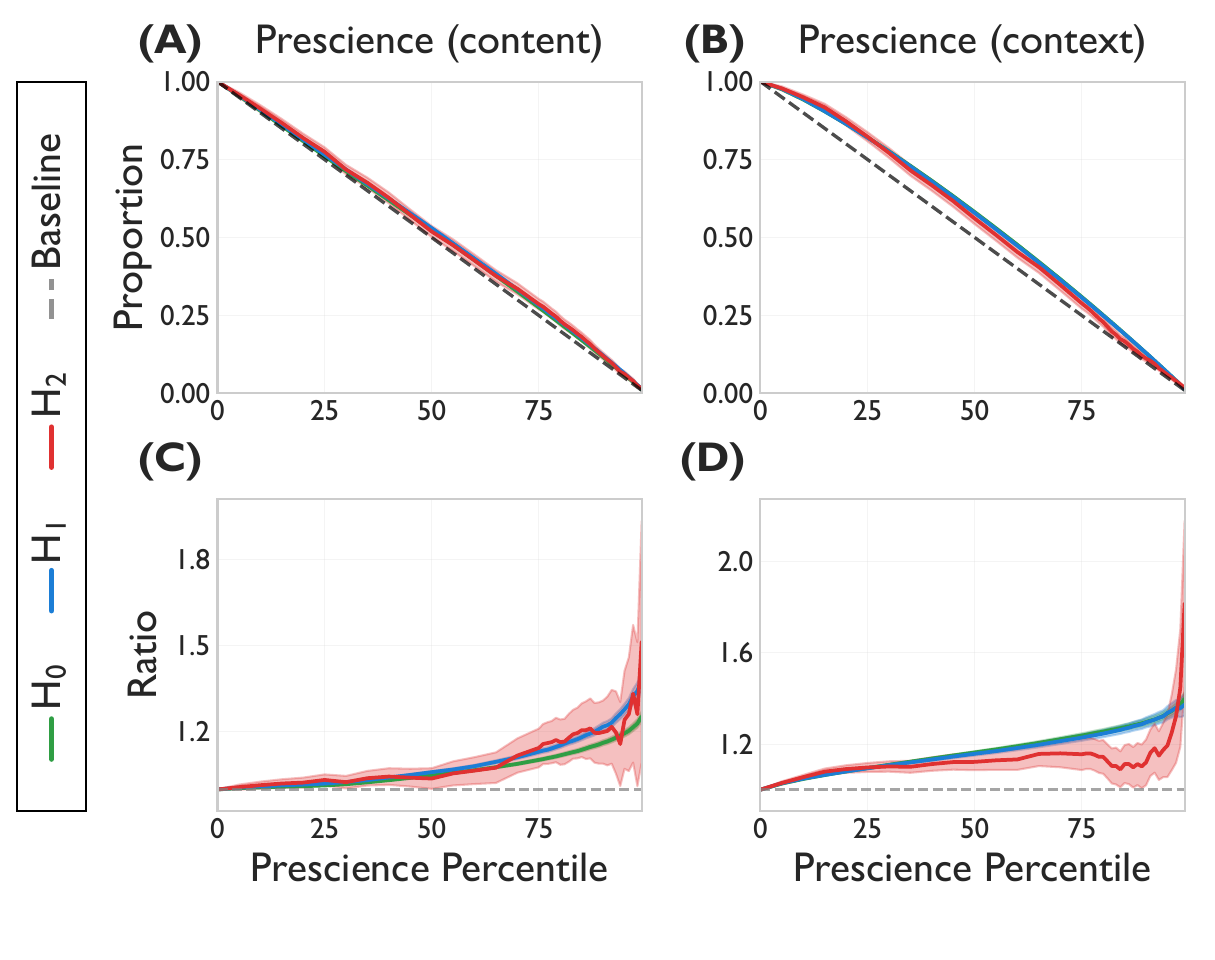}

	\caption{\textbf{Prescience of works that fill in holes.}
    In Panels $A$ and $B$, we plot, for dimensions $0$, $1$, and $2$, the proportion of hole-filling works at or above a given prescience percentile threshold for content and context prescience, respectively. For dimension $k$ and threshold $N$, this is the proportion of works filling $k$-dimensional holes with prescience percentile threshold $\geq N$. Panel $C$ and $D$ plots, for context and content prescience, respectively, the corresponding ratio: the proportion of hole-filling works at or above threshold $N$ divided by the proportion of all works at that threshold.
    }
	\label{fig:fill_papers_prescience}
\end{figure}

\begin{figure}
	\centering
	\includegraphics[width=0.8\textwidth]{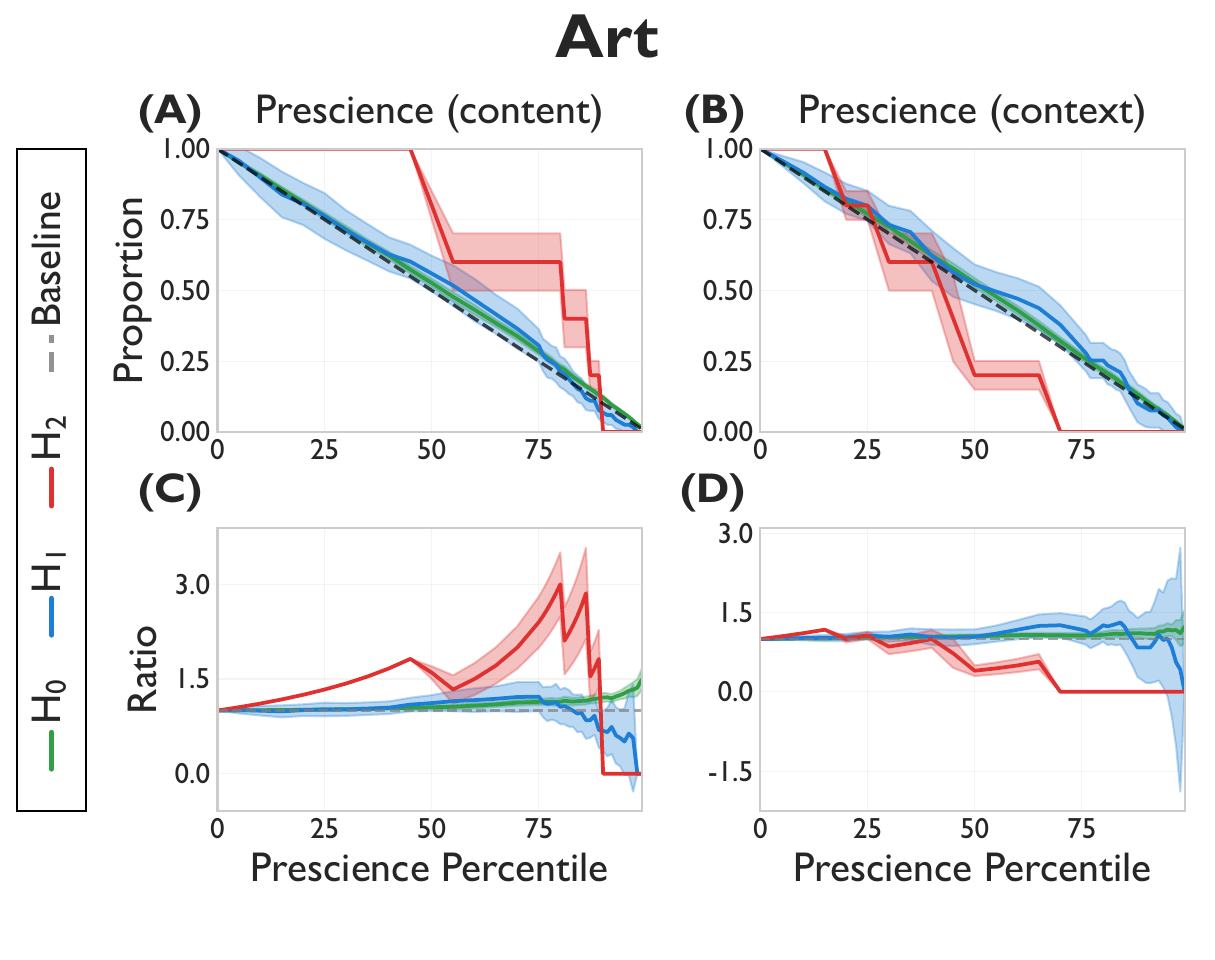}

	\caption{\textbf{Prescience of works that fill in holes, by discipline.}
    Same as Figure~\ref{fig:fill_papers_prescience}, but for individual disciplines, one discipline per page. This page: art.
    }
	\label{fig:fill_papers_art_prescience}
\end{figure}

\newpage

\begin{figure}
	\centering
	\includegraphics[width=0.8\textwidth]{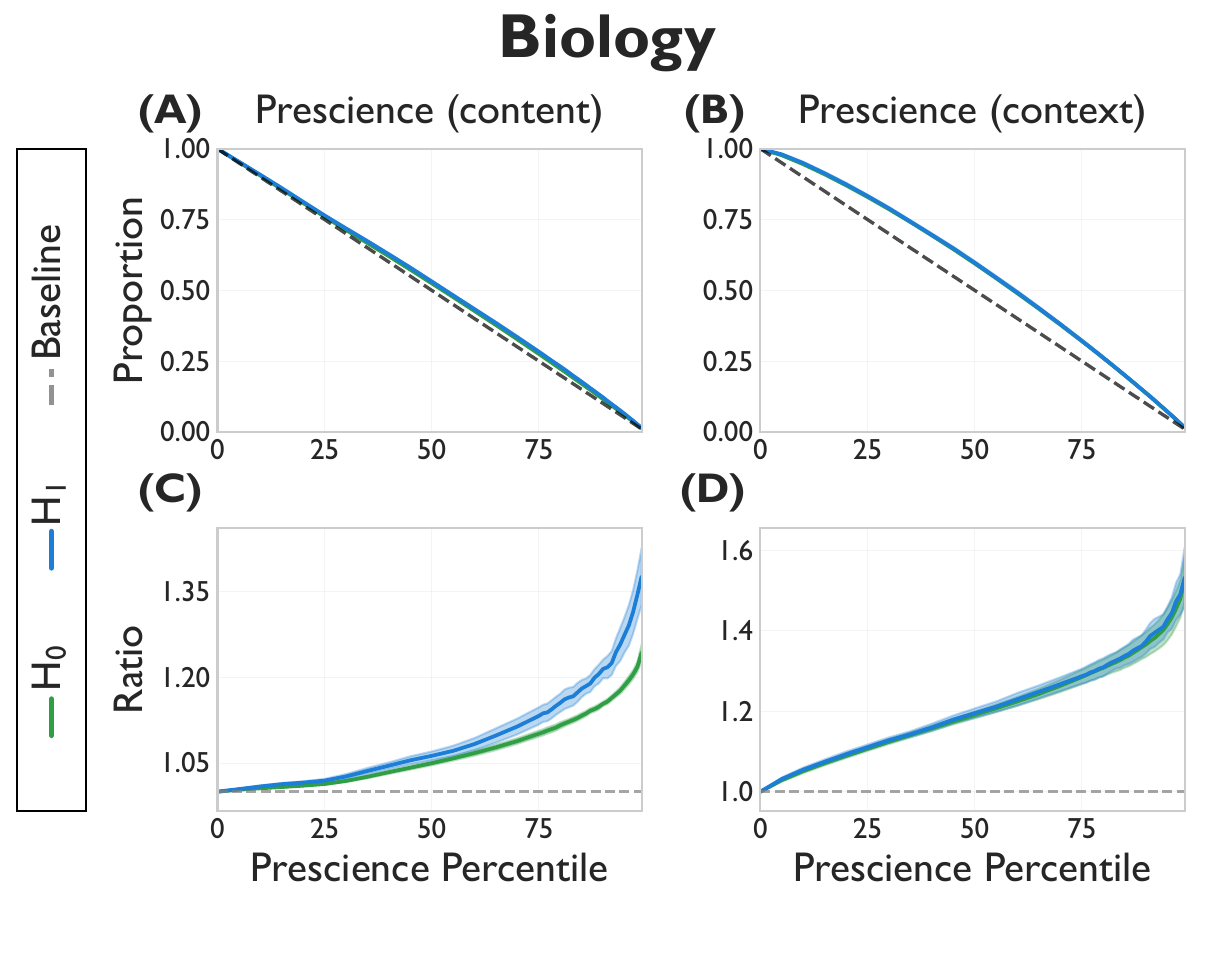}

	\continuedfigure
	\caption{\textbf{Prescience of works that fill in holes, by discipline.}
    This page: biology.
    }
	\label{fig:fill_papers_biology_prescience}
\end{figure}

\begin{figure}
	\centering
	\includegraphics[width=0.8\textwidth]{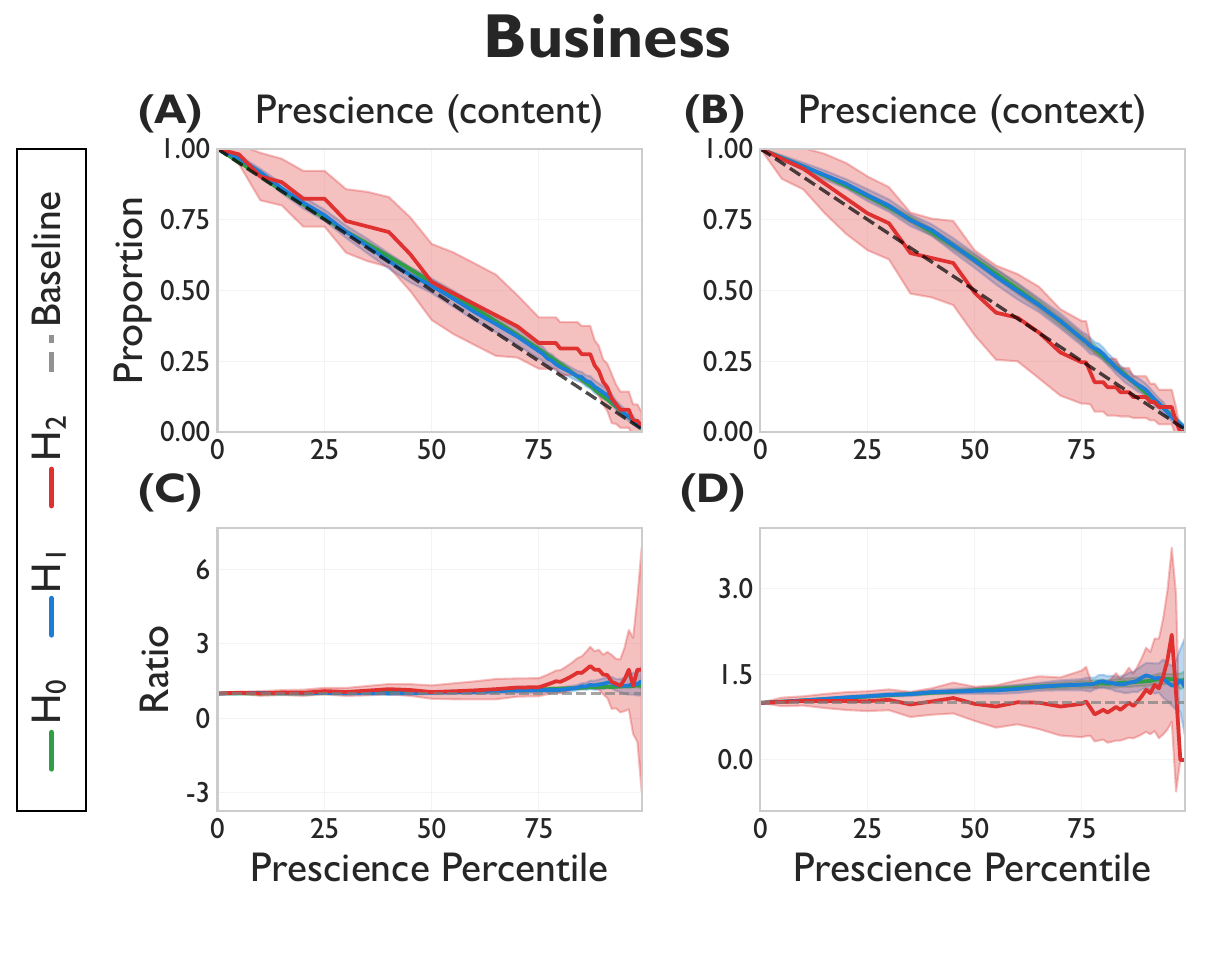}

	\continuedfigure
	\caption{\textbf{Prescience of works that fill in holes, by discipline.}
    This page: business.
    }
	\label{fig:fill_papers_business_prescience}
\end{figure}

\newpage

\begin{figure}
	\centering
	\includegraphics[width=0.8\textwidth]{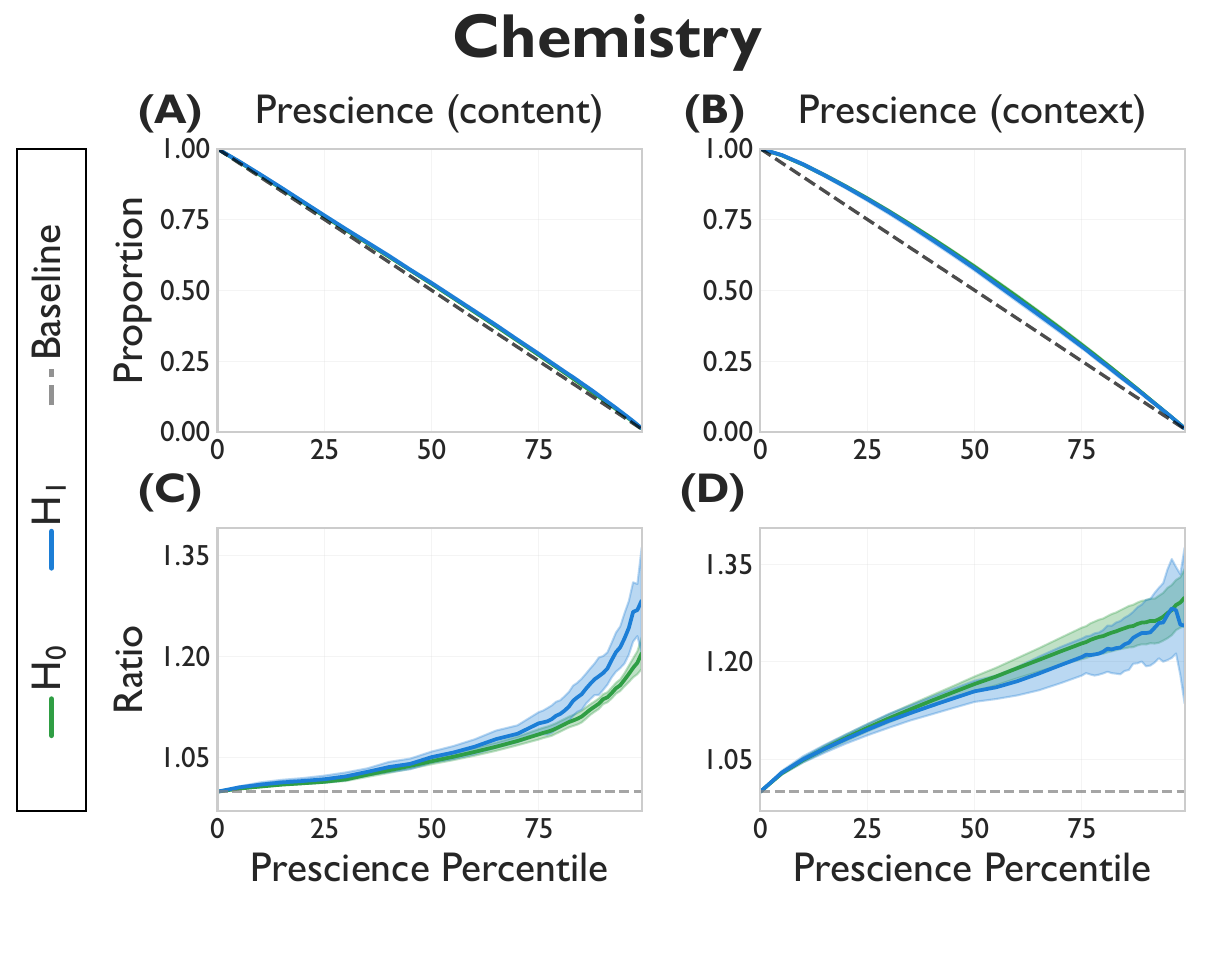}

	\continuedfigure
	\caption{\textbf{Prescience of works that fill in holes, by discipline.}
    This page: chemistry.
    }
	\label{fig:fill_papers_chemistry_prescience}
\end{figure}

\begin{figure}
	\centering
	\includegraphics[width=0.8\textwidth]{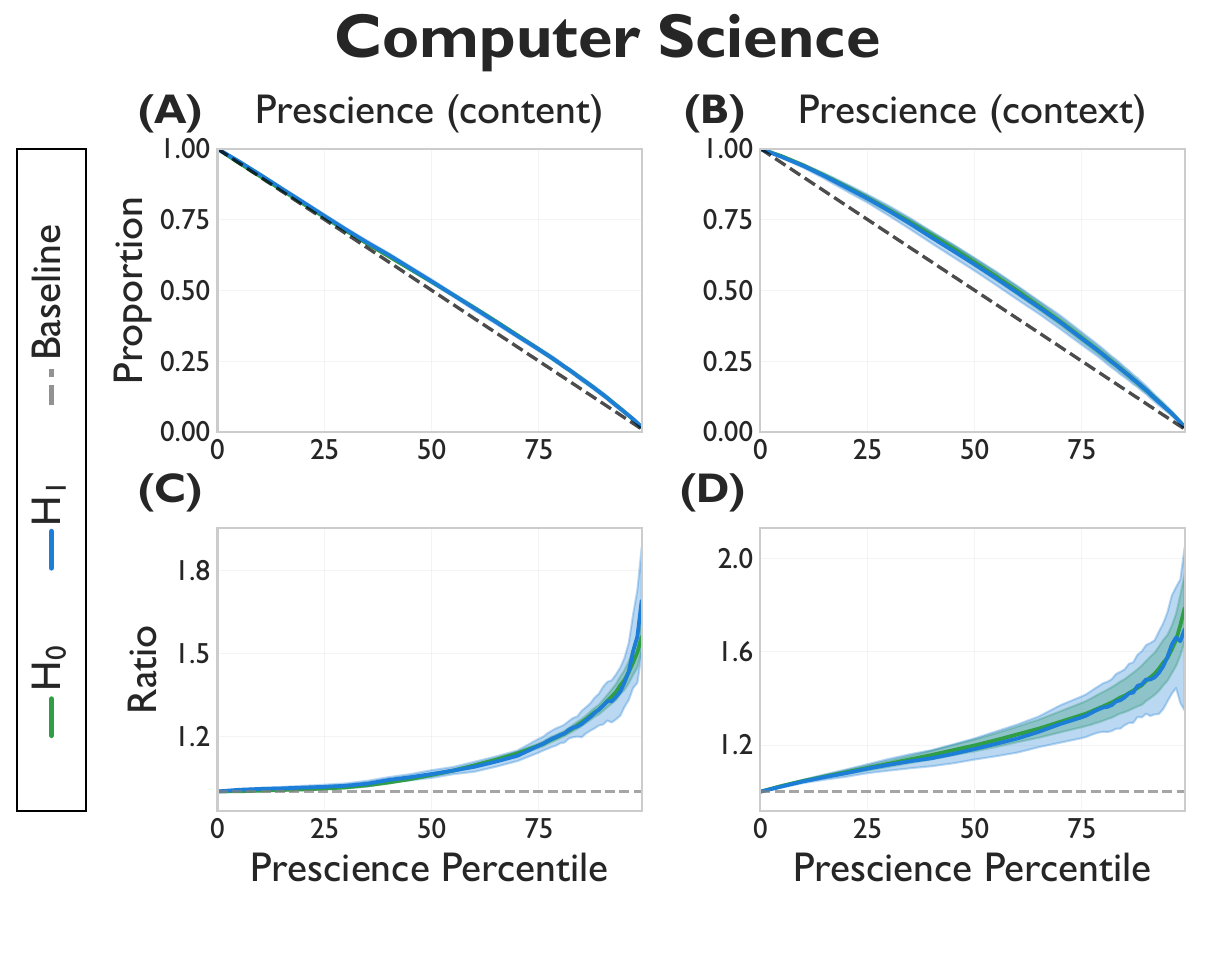}

	\continuedfigure
	\caption{\textbf{Prescience of works that fill in holes, by discipline.}
    This page: computer science.
    }
	\label{fig:fill_papers_computer_science_prescience}
\end{figure}

\newpage

\begin{figure}
	\centering
	\includegraphics[width=0.8\textwidth]{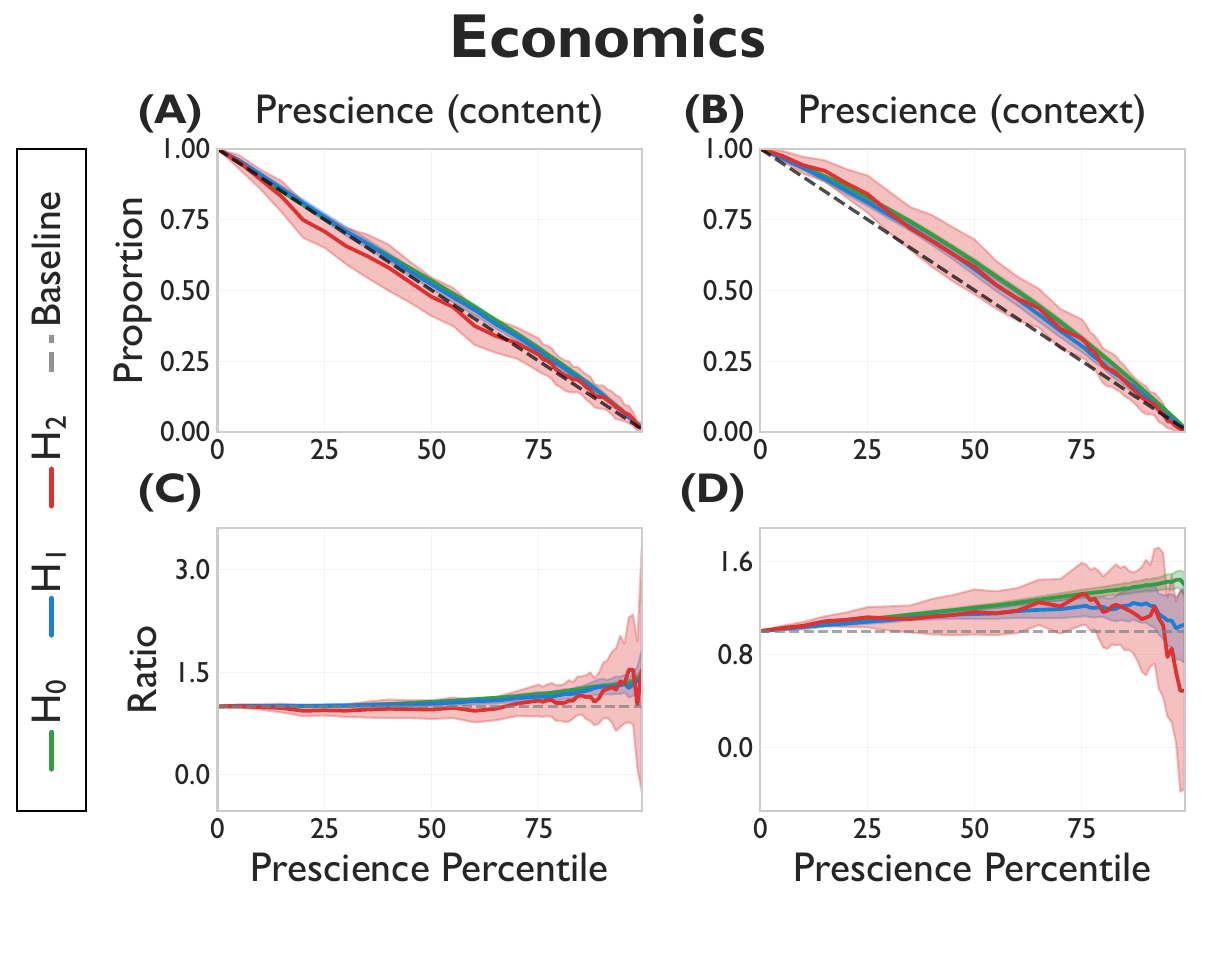}

	\continuedfigure
	\caption{\textbf{Prescience of works that fill in holes, by discipline.}
    This page: economics.
    }
	\label{fig:fill_papers_economics_prescience}
\end{figure}

\begin{figure}
	\centering
	\includegraphics[width=0.8\textwidth]{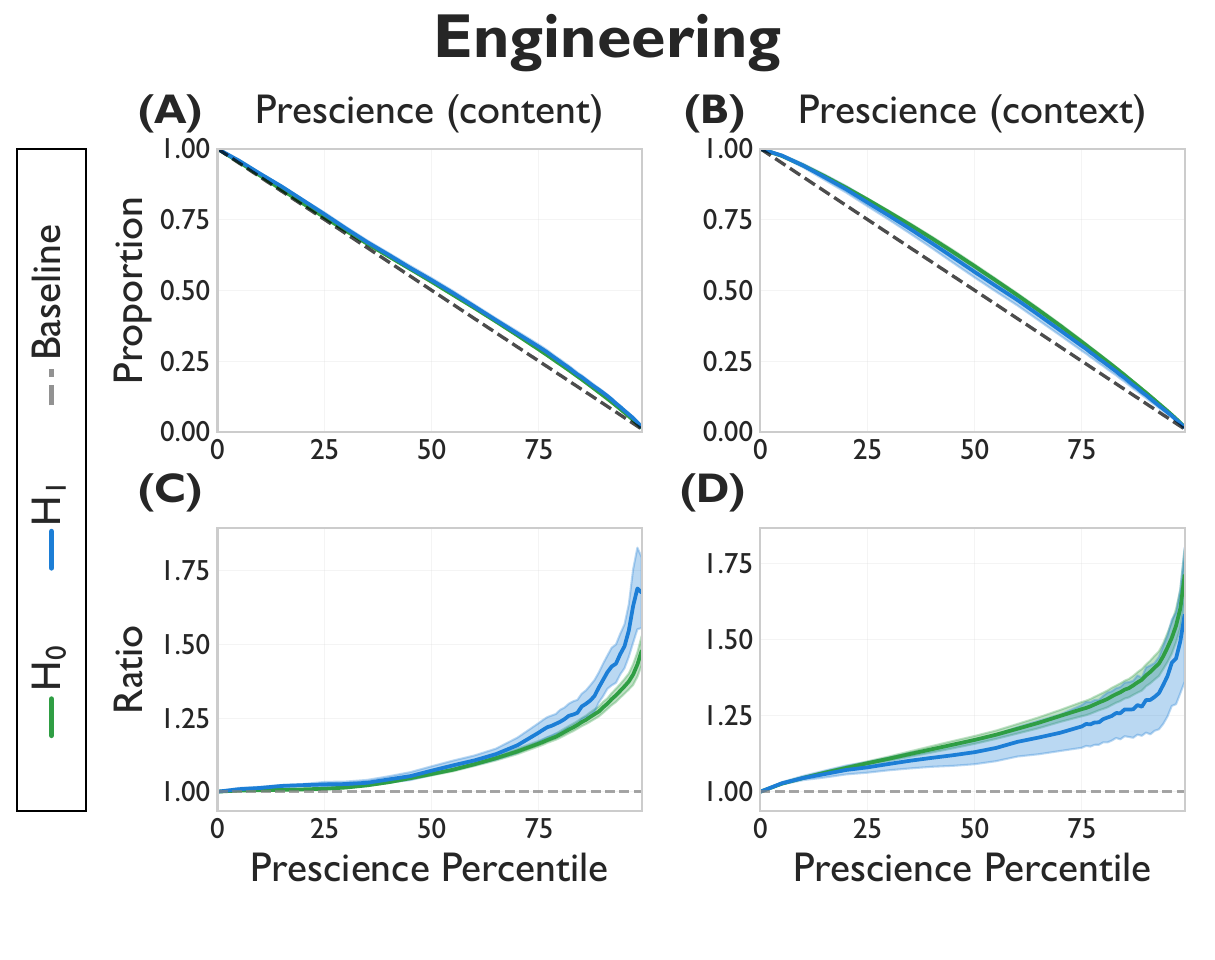}

	\continuedfigure
	\caption{\textbf{Prescience of works that fill in holes, by discipline.}
    This page: engineering.
    }
	\label{fig:fill_papers_engineering_prescience}
\end{figure}

\newpage

\clearpage

\begin{figure}
	\centering
	\includegraphics[width=0.8\textwidth]{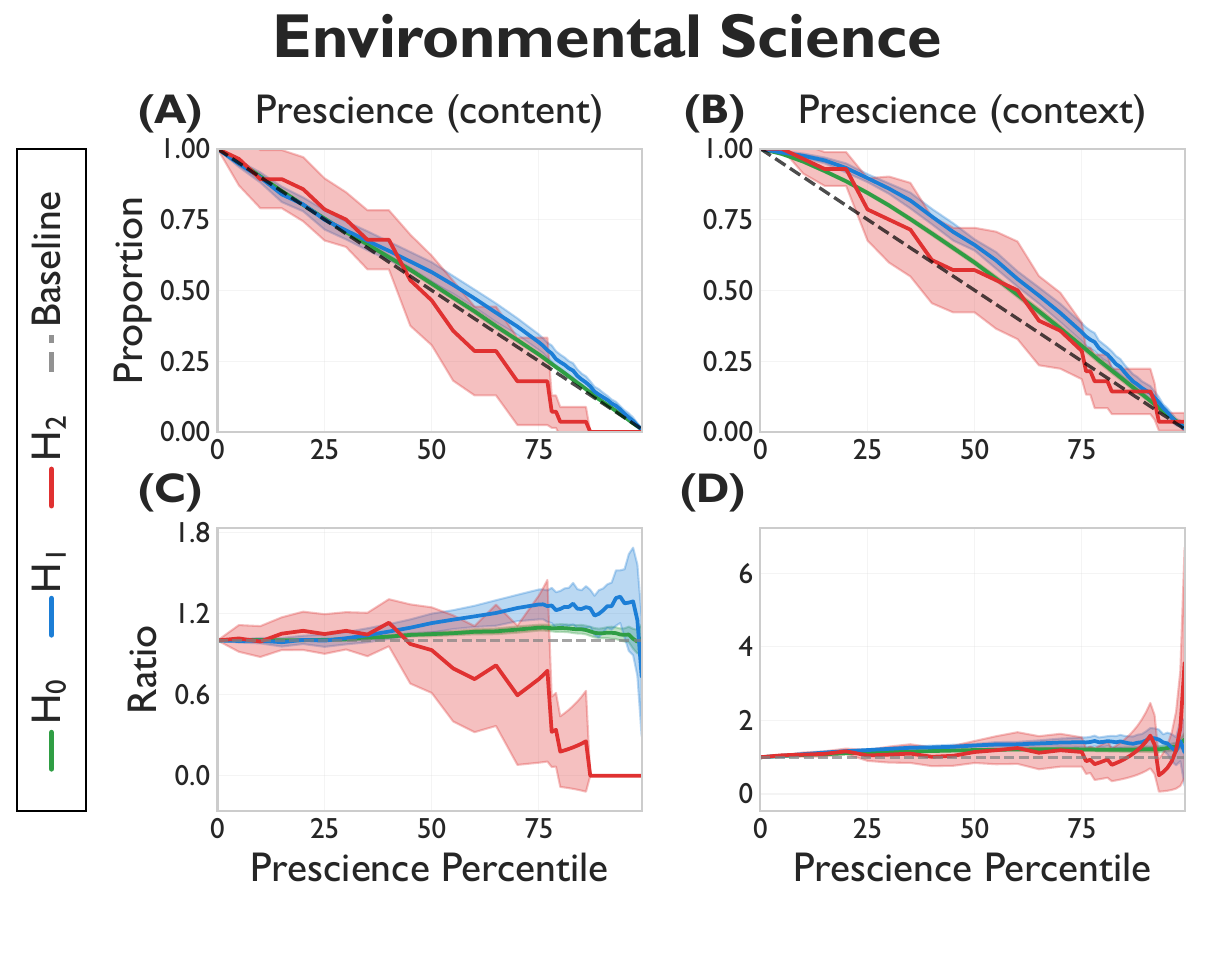}

	\continuedfigure
	\caption{\textbf{Prescience of works that fill in holes, by discipline.}
    This page: environmental science.
    }
	\label{fig:fill_papers_environmental_science_prescience}
\end{figure}

\begin{figure}
	\centering
	\includegraphics[width=0.8\textwidth]{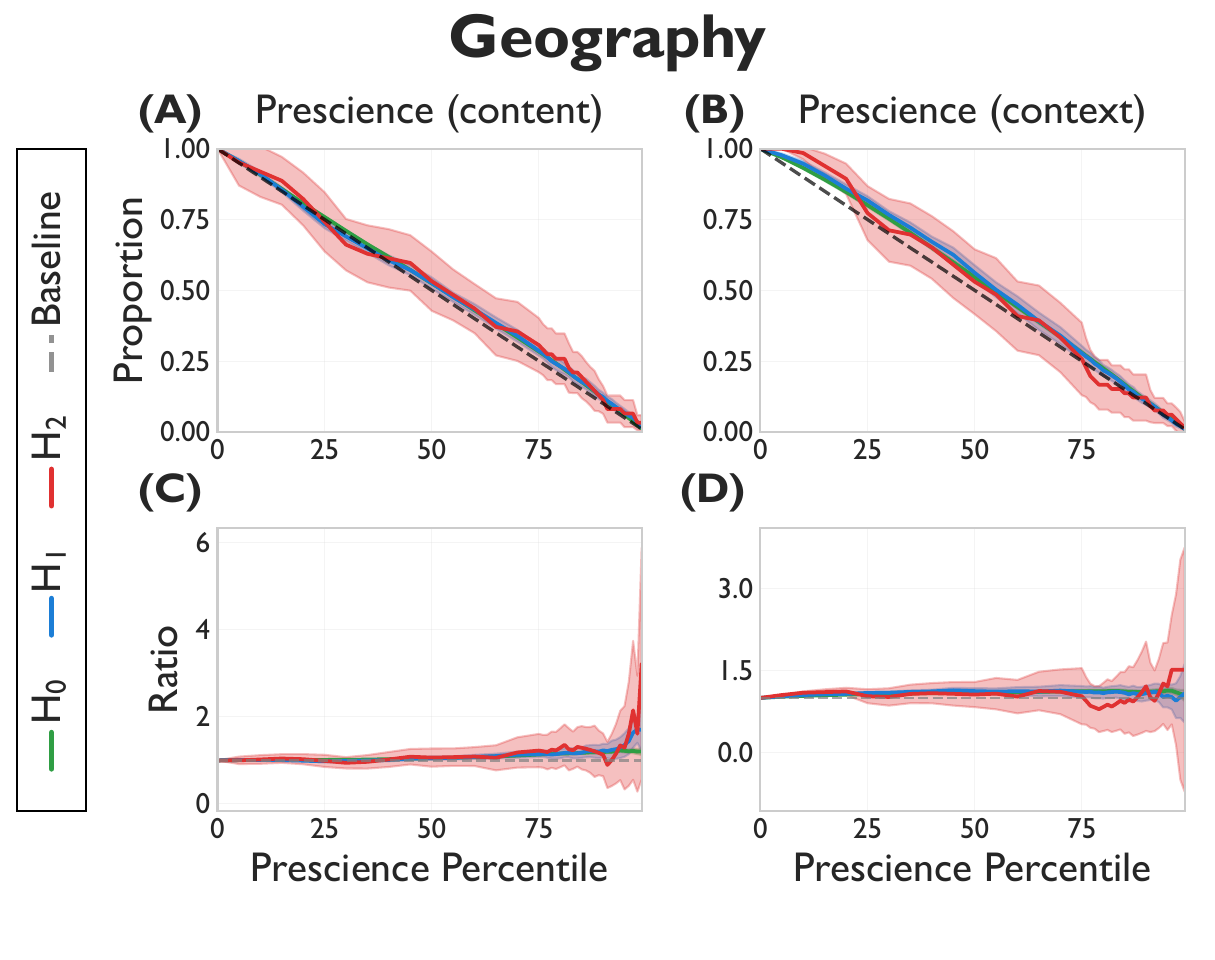}

	\continuedfigure
	\caption{\textbf{Prescience of works that fill in holes, by discipline.}
    This page: geography.
    }
	\label{fig:fill_papers_geography_prescience}
\end{figure}

\newpage
\clearpage

\begin{figure}
	\centering
	\includegraphics[width=0.8\textwidth]{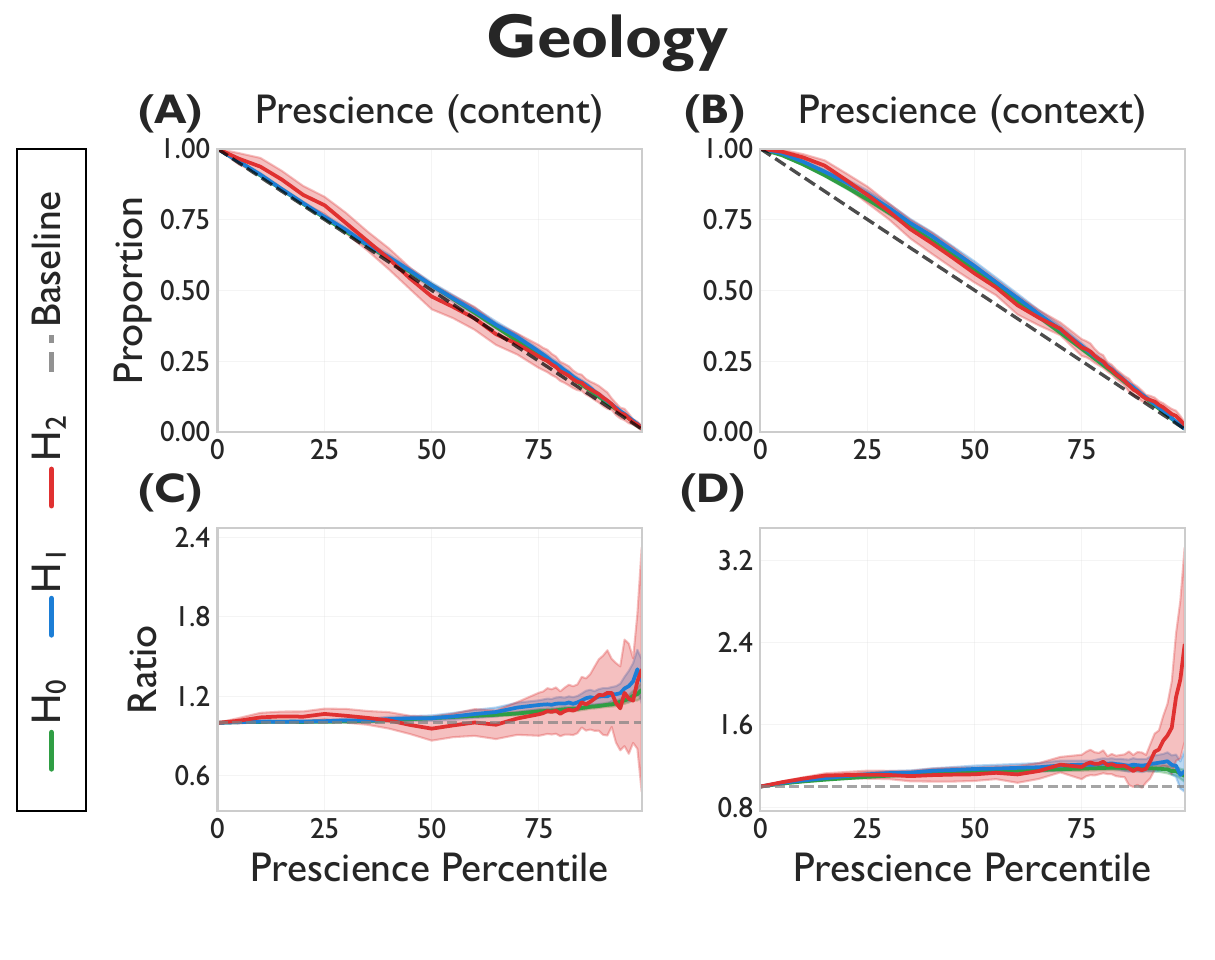}

	\continuedfigure
	\caption{\textbf{Prescience of works that fill in holes, by discipline.}
    This page: geology.
    }
	\label{fig:fill_papers_geology_prescience}
\end{figure}

\begin{figure}
	\centering
	\includegraphics[width=0.8\textwidth]{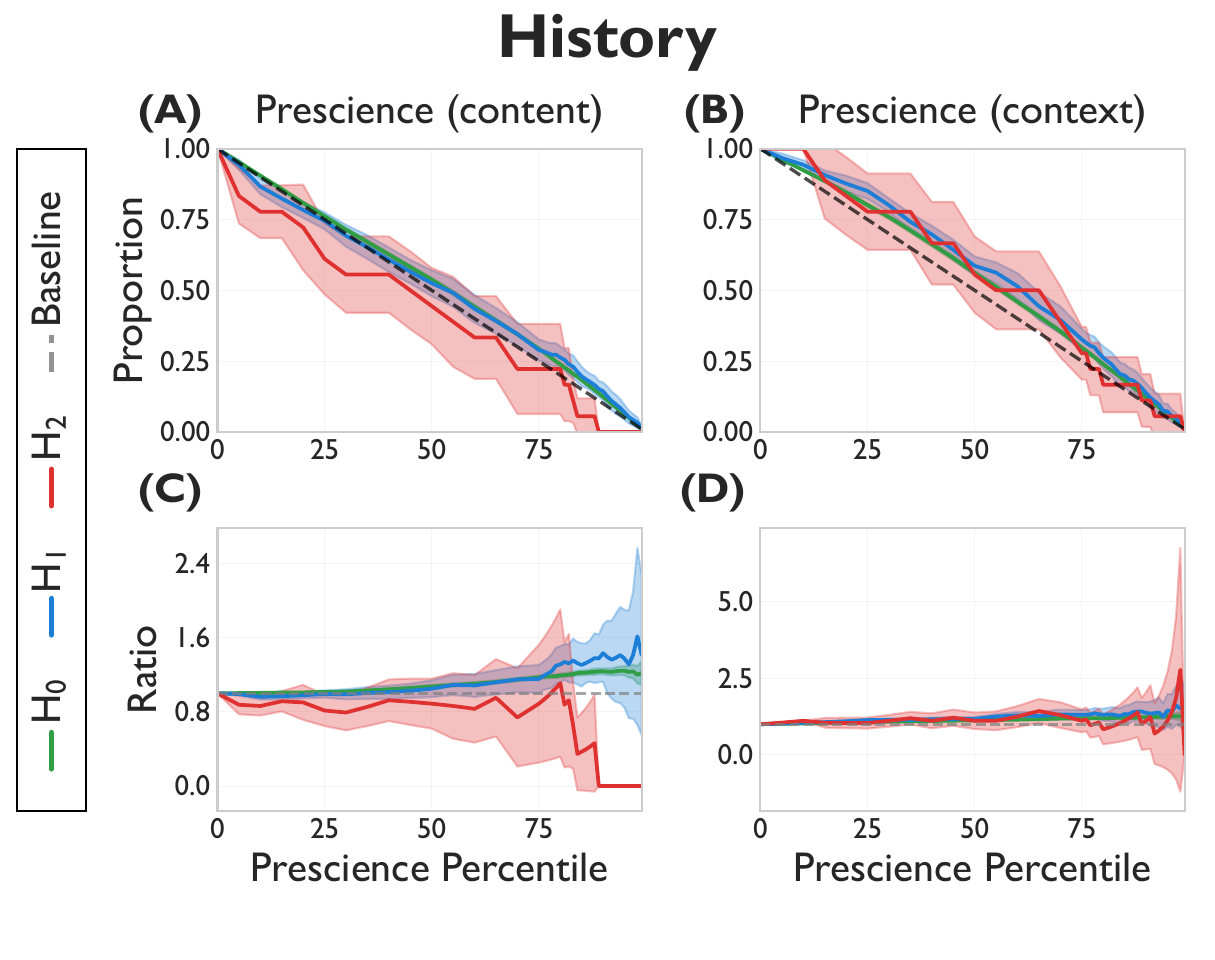}

	\continuedfigure
	\caption{\textbf{Prescience of works that fill in holes, by discipline.}
    This page: history.
    }
	\label{fig:fill_papers_history_prescience}
\end{figure}

\newpage
\clearpage

\begin{figure}
	\centering
	\includegraphics[width=0.8\textwidth]{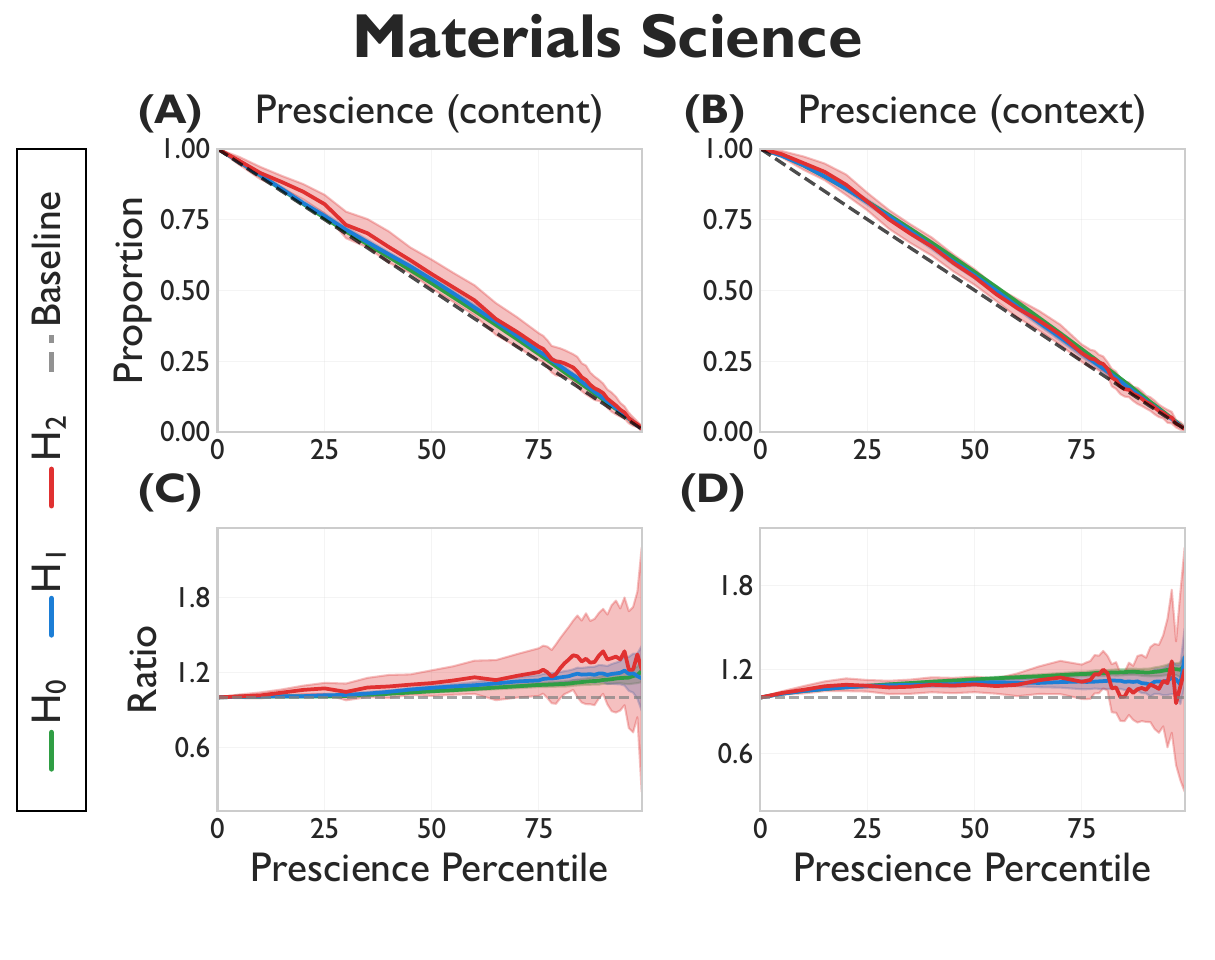}

	\continuedfigure
	\caption{\textbf{Prescience of works that fill in holes, by discipline.}
    This page: materials science.
    }
	\label{fig:fill_papers_materials_science_prescience}
\end{figure}

\begin{figure}
	\centering
	\includegraphics[width=0.8\textwidth]{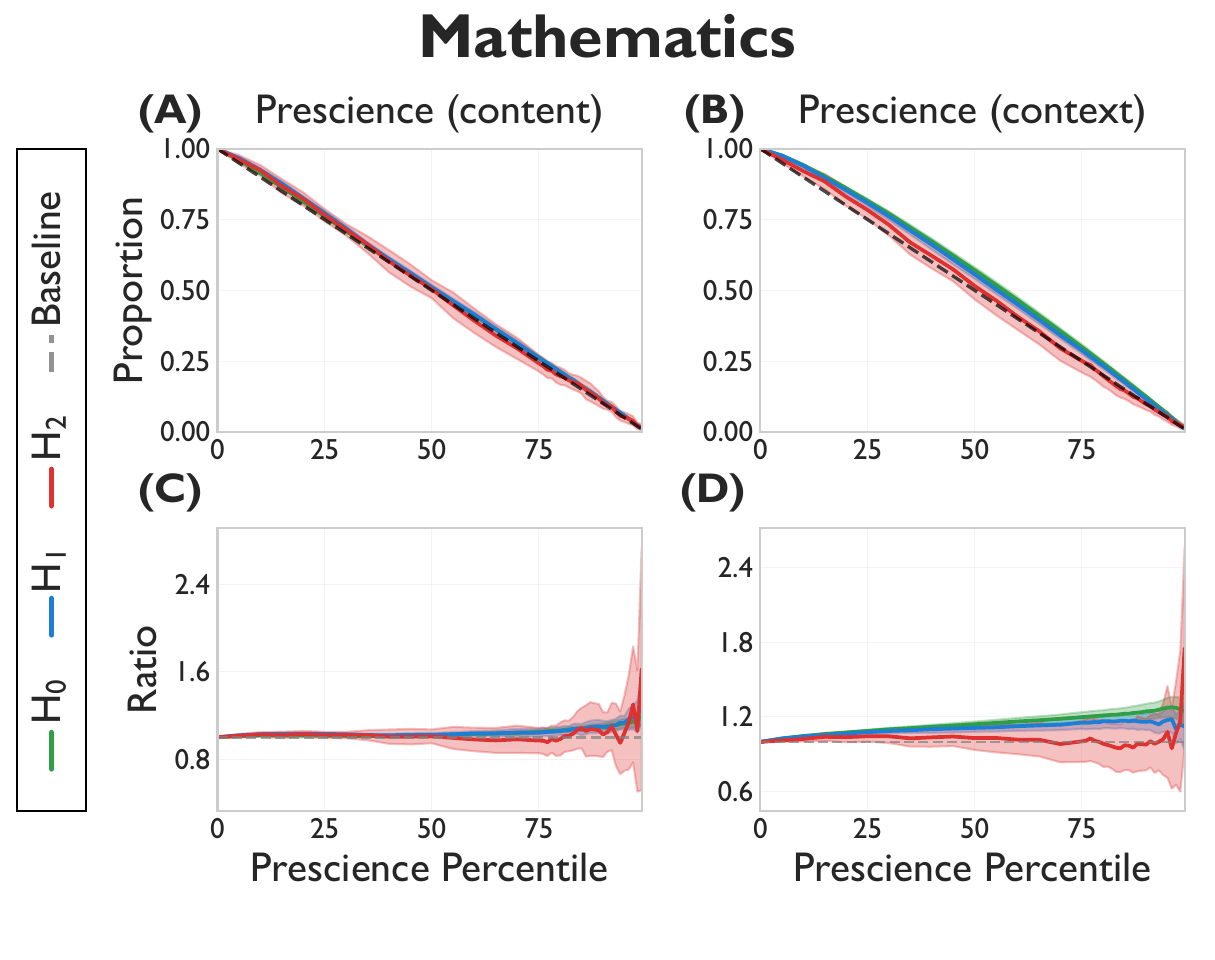}

	\continuedfigure
	\caption{\textbf{Prescience of works that fill in holes, by discipline.}
    This page: mathematics.
    }
	\label{fig:fill_papers_mathematics_prescience}
\end{figure}

\newpage
\clearpage

\begin{figure}
	\centering
	\includegraphics[width=0.8\textwidth]{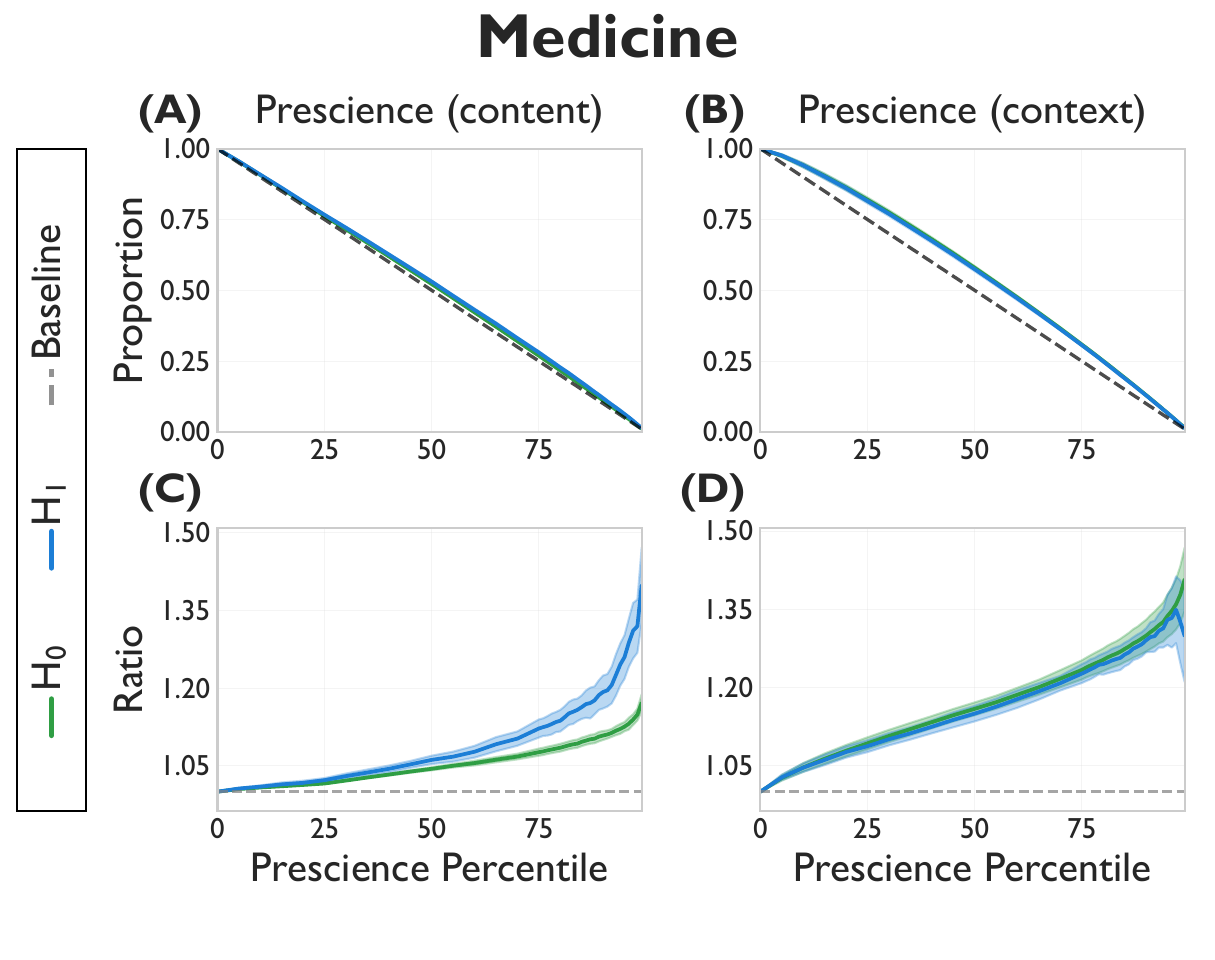}

	\continuedfigure
	\caption{\textbf{Prescience of works that fill in holes, by discipline.}
    This page: medicine.
    }
	\label{fig:fill_papers_medicine_prescience}
\end{figure}

\begin{figure}
	\centering
	\includegraphics[width=0.8\textwidth]{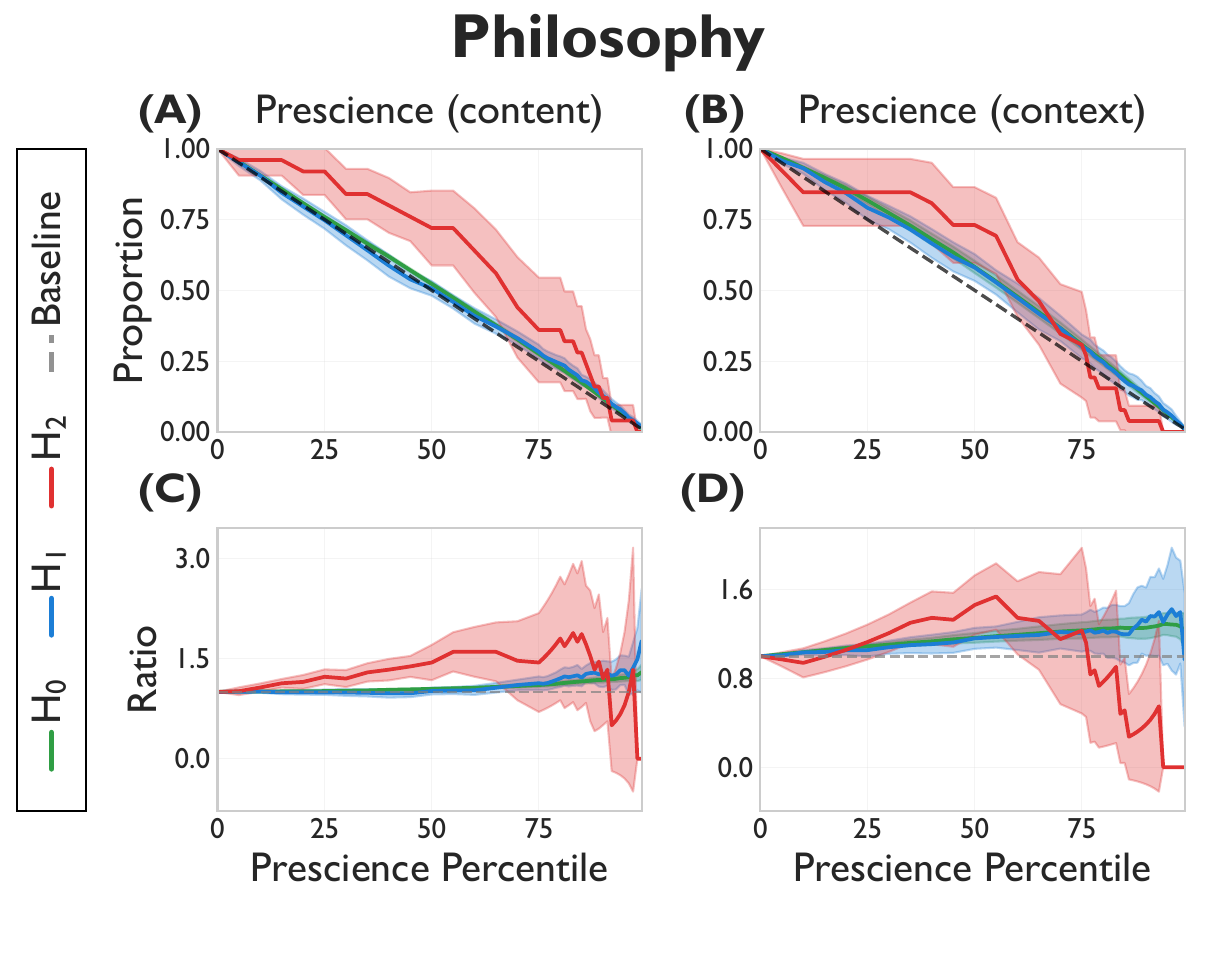}

	\continuedfigure
	\caption{\textbf{Prescience of works that fill in holes, by discipline.}
    This page: philosophy.
    }
	\label{fig:fill_papers_philosophy_prescience}
\end{figure}

\newpage

\begin{figure}
	\centering
	\includegraphics[width=0.8\textwidth]{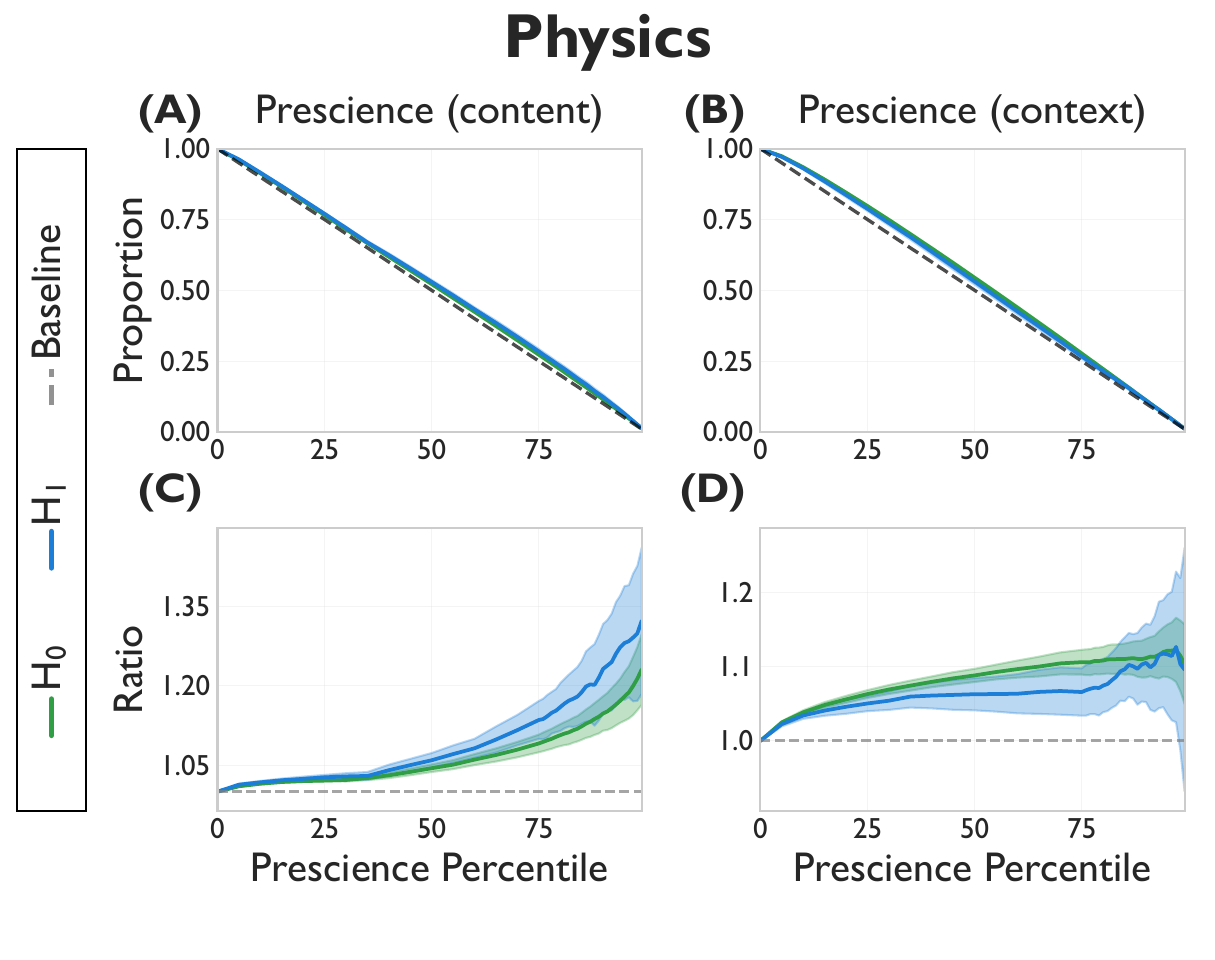}

	\continuedfigure
	\caption{\textbf{Prescience of works that fill in holes, by discipline.}
    This page: physics.
    }
	\label{fig:fill_papers_physics_prescience}
\end{figure}

\begin{figure}
	\centering
	\includegraphics[width=0.8\textwidth]{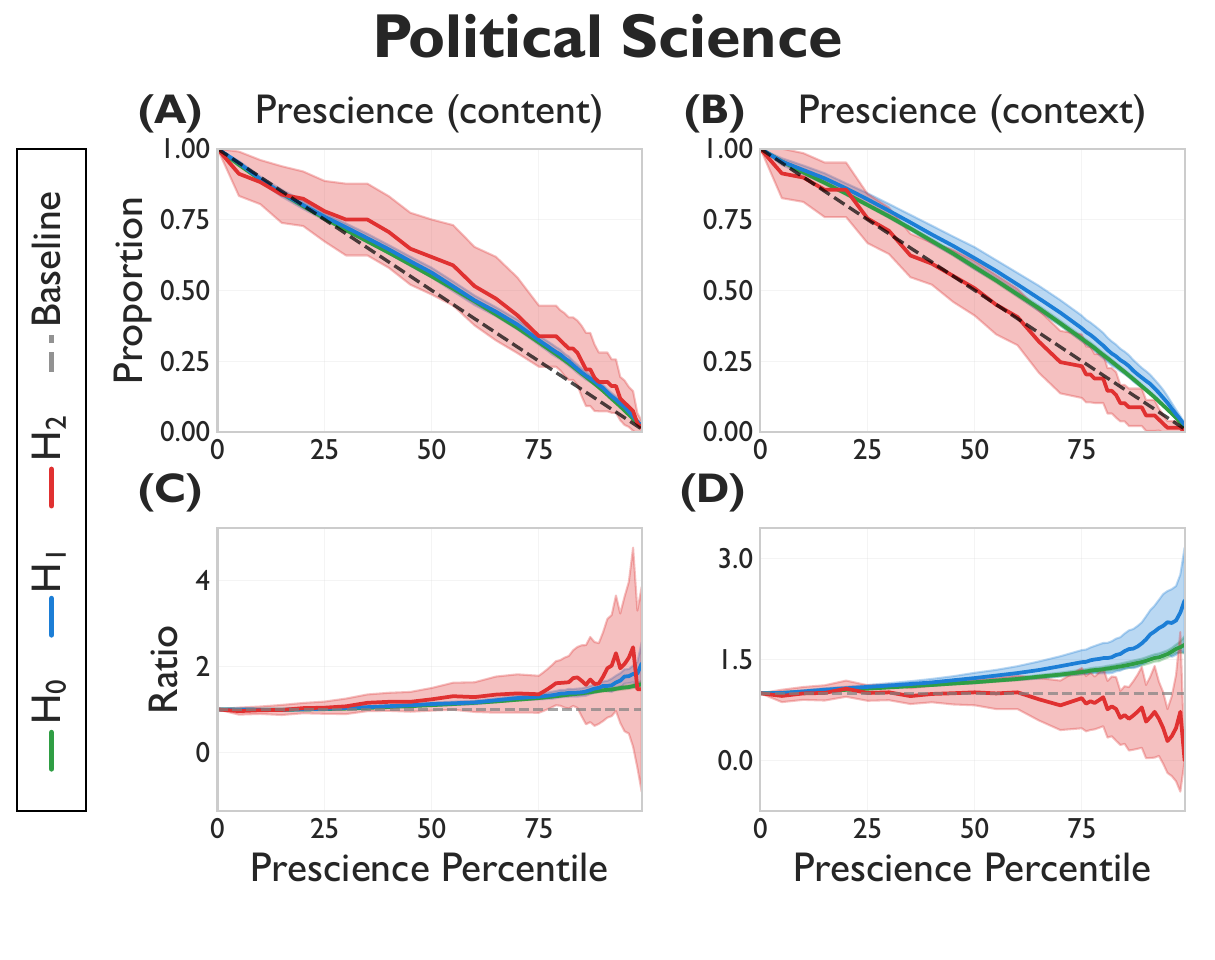}

	\continuedfigure
	\caption{\textbf{Prescience of works that fill in holes, by discipline.}
    This page: political science.
    }
	\label{fig:fill_papers_political_science_prescience}
\end{figure}

\newpage
\clearpage

\begin{figure}
	\centering
	\includegraphics[width=0.8\textwidth]{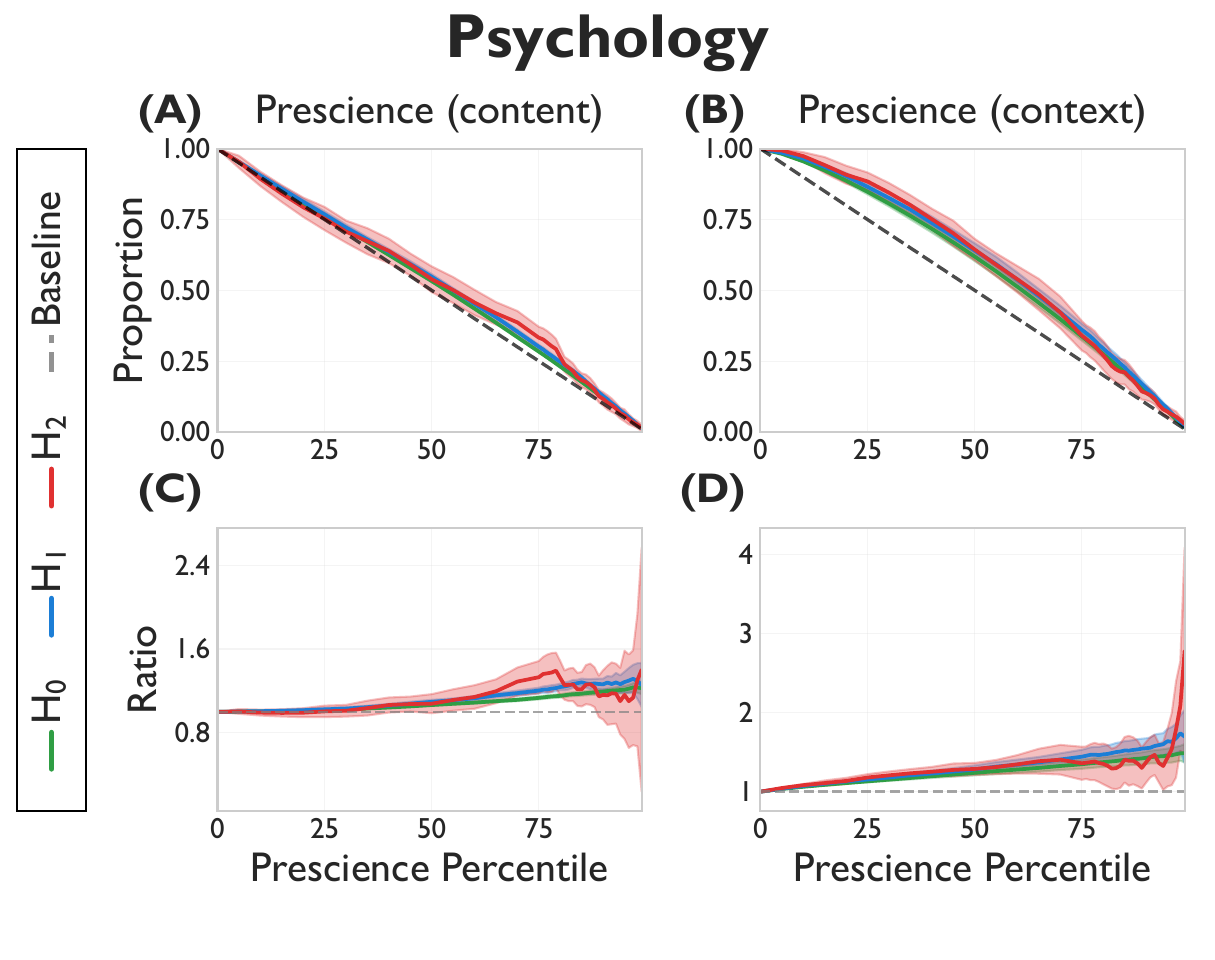}

	\continuedfigure
	\caption{\textbf{Prescience of works that fill in holes, by discipline.}
    This page: psychology.
    }
	\label{fig:fill_papers_psychology_prescience}
\end{figure}

\begin{figure}
	\centering
	\includegraphics[width=0.8\textwidth]{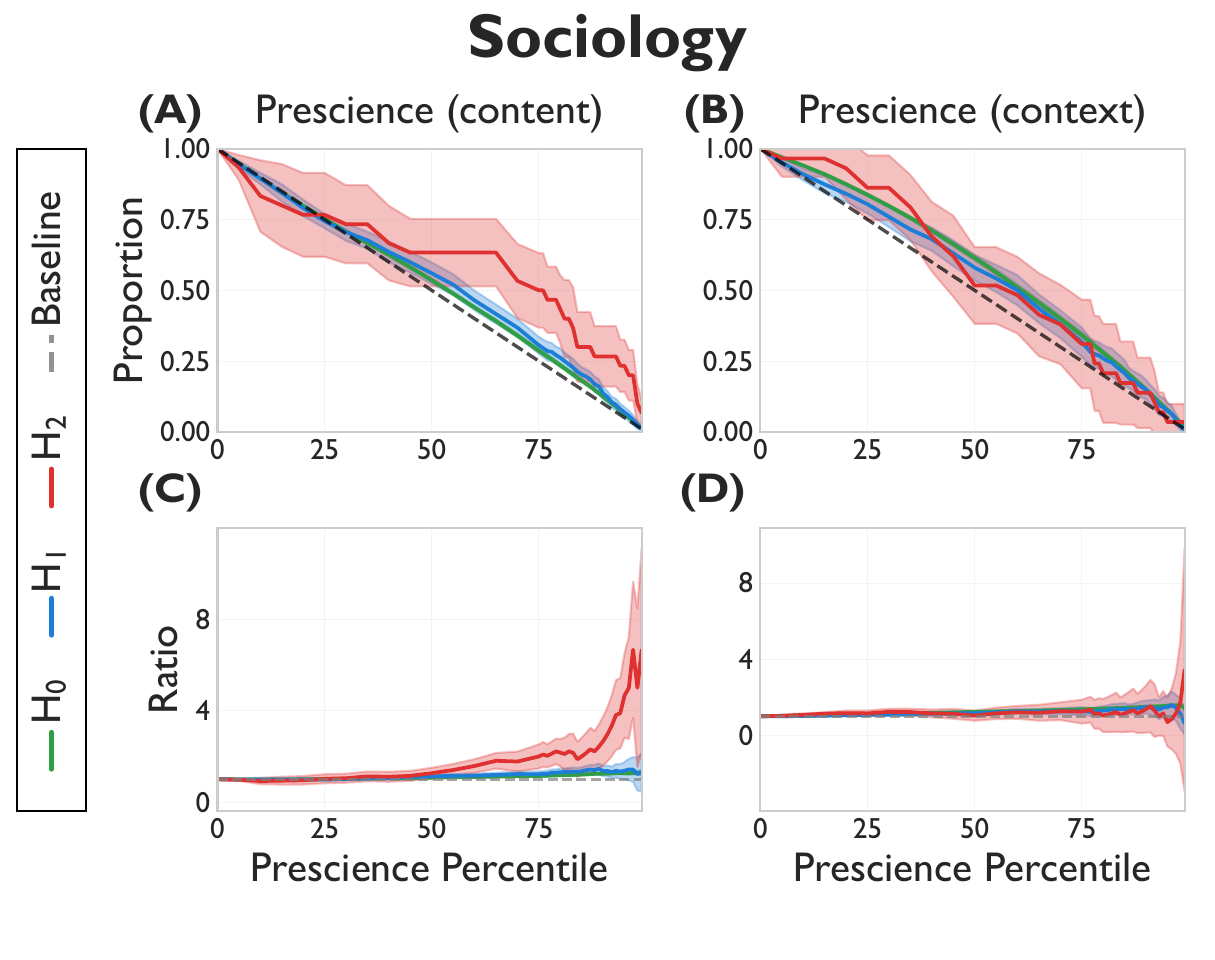}

	\continuedfigure
	\caption{\textbf{Prescience of works that fill in holes, by discipline.}
    This page: sociology.
    }
	\label{fig:fill_papers_sociology_prescience}
\end{figure}

\newpage

\begin{figure}[htbp]
\centering
\includegraphics[width=\textwidth]{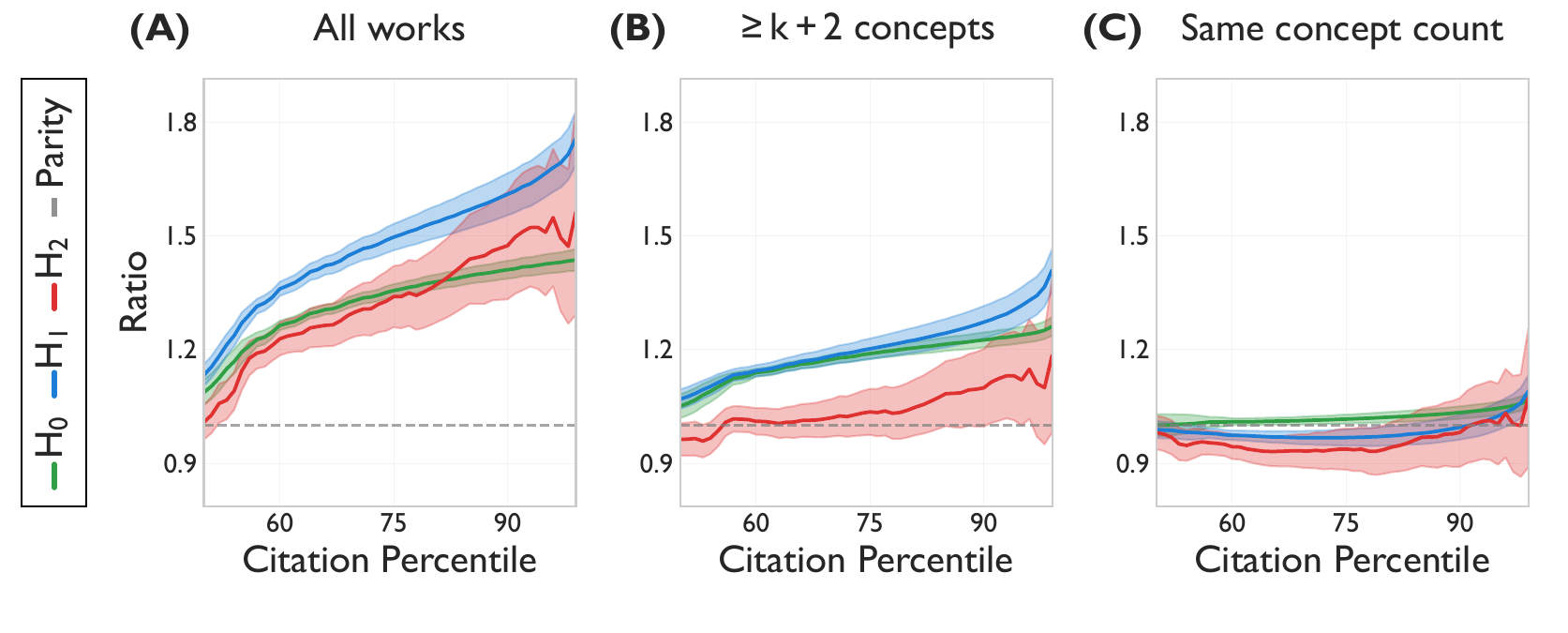}
\caption{\textbf{Restricting the comparison population to works that could have closed a hole of
the given order shrinks the citation premium, and at dimension 2 removes it.} Citation enrichment of hole-closing works
--- the proportion at or above a citation percentile, divided by the same proportion in the
comparison population --- against three progressively stricter comparison populations:
\textbf{(A)} all works, \textbf{(B)} works carrying at least $k+2$ of the discipline--year's
concepts, and \textbf{(C)} works carrying the same number of such concepts as the closing works
themselves. Panel (A) is the comparison already reported in the main text --- the same filler set
and the same all-works baseline as Figure \ref{fig:hole_fill_papers}D --- and is included so that
the effect of each restriction can be read against it. Bands are $95\%$ intervals from a standard error treating each discipline--year as one
observation. The premium survives the eligibility restriction at dimensions $0$ and $1$ but not at
dimension $2$, whose band includes parity at every threshold we tabulate and over almost the whole
sweep --- it clears parity only marginally, in a narrow window between the $91$st and $96$th
percentiles where the lower bound reaches $1.007$ to $1.018$. Under the strictest population
(C) almost nothing survives, but not quite nothing: dimension $0$ keeps a small premium that still
clears one at every threshold shown ($1.017\pm0.012$ at the top $25\%$ rising to $1.068\pm0.020$ at
the top $1\%$), dimension $1$ runs from \emph{below} parity at the top $25\%$ ($0.968\pm0.021$) to a
slight premium at the top $1\%$ ($1.091\pm0.042$), and dimension $2$ is indistinguishable from
parity except at the top $25\%$, where it falls below it ($0.938\pm0.054$). Concept count therefore
accounts for most of the premium but not all of it. All three
panels share one $y$ range, so the columns are directly comparable. Dimension $2$ pools the
thirteen disciplines for which it was computed, which is also why its comparison population differs
from the other two in panel (A).}
\label{fig:claude-eligible}
\end{figure}

\begin{figure}[htbp]
\centering
\includegraphics[width=\textwidth]{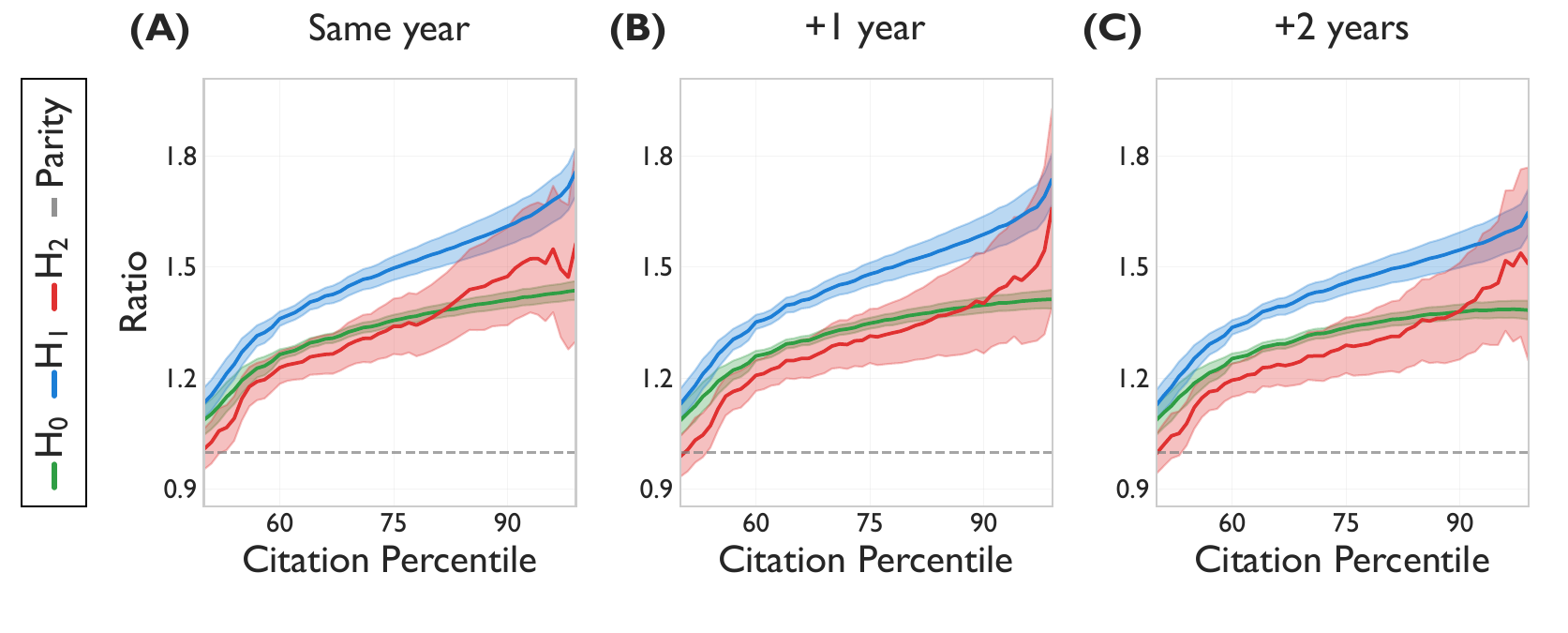}
\caption{\textbf{The citation premium does not depend on the closing work being inside the corpus
that trained the embedding.} Citation enrichment of hole-closing works, with the closing work
required to appear \textbf{(A)} in the same year as the hole, \textbf{(B)} one year later, or
\textbf{(C)} two years later. A work published after year $y$ cannot belong to the cumulative
corpus through $y$, so panels (B) and (C) cannot be produced by the closing work having shaped the
geometry it is then measured against. The three panels are indistinguishable, and share one $y$
range. The comparison population is \emph{all} works, so panel (A) is exactly the quantity plotted in
Figure \ref{fig:hole_fill_papers}D of the main text and reproduces its top-$1\%$ values of
$1.437$, $1.758$ and $1.564$; panels (B) and (C) are that same measurement with the filler cohort
moved forward one and two years. Reading the same measurement against the eligible baseline
instead lowers every panel uniformly, as Figure \ref{fig:claude-eligible}(B) shows for the
same-year case, and does not change the comparison across horizons.}
\label{fig:claude-horizon}
\end{figure}

\begin{figure}[htbp]
\centering
\includegraphics[width=\textwidth]{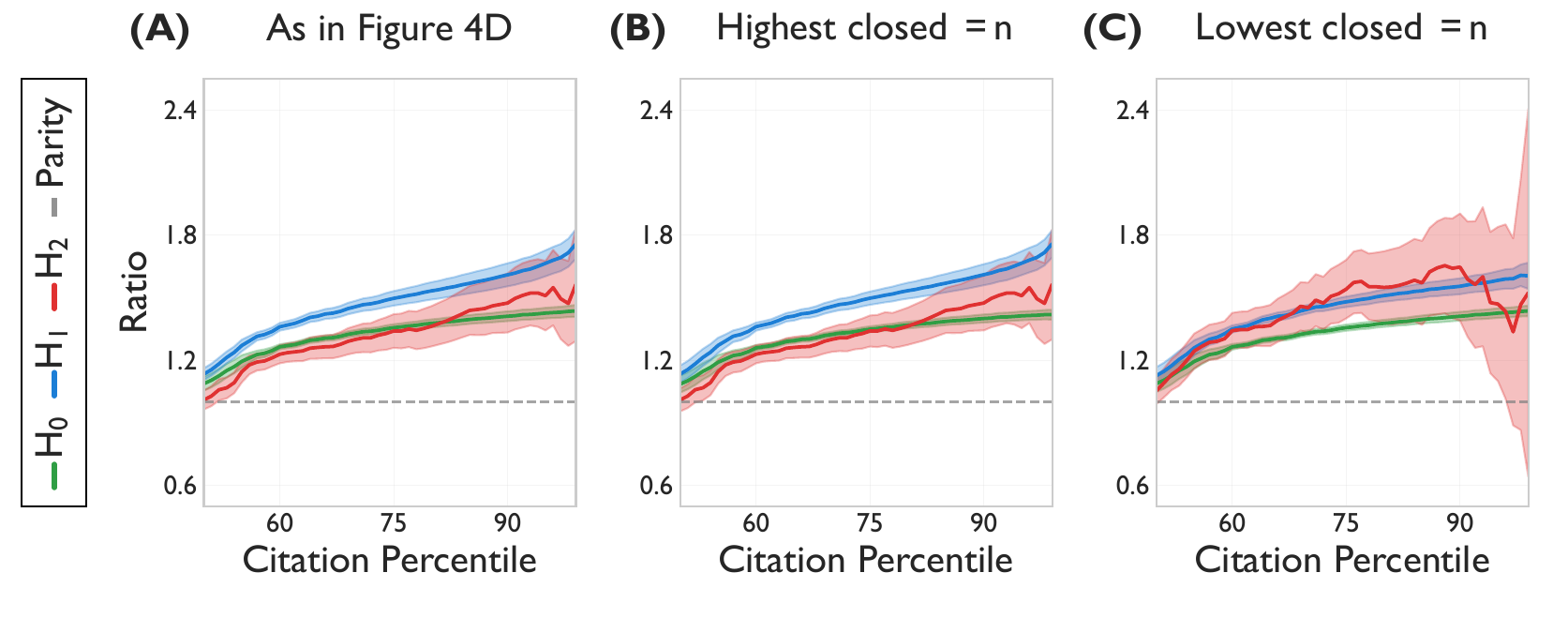}
\caption{\textbf{The step from gaps to loops survives restricting the filler sets so that they
share no works at all.} Citation enrichment against all works. \textbf{(A)} is the enrichment
already reported in the main text, Figure \ref{fig:hole_fill_papers}D, reproduced here so the two
restrictions can be read against a curve the reader has seen: a work enters the dimension-$n$ curve
if it closes a hole of order $n$, whatever else it closes. \textbf{(B)} and \textbf{(C)} reclassify
each work by the \emph{highest} and by the \emph{lowest} order hole it closes. Both are genuine
partitions --- every work has exactly one highest and one lowest --- and taking dimension 1 as the
example, (B) is built from works closing a loop but no void, gaps allowed, while (C) is built from
works closing a loop but no gap, voids allowed. Only (C) makes the orders disjoint: in (A) and (B)
the dimension-0 and dimension-1 curves still share $1{,}645{,}748$ works, in (C) they share none.
(A) and (B) are nearly identical, which is itself informative --- it says the dimension-1 premium is
not carried by the rare works that also close voids. In (C) the dimension-1 premium still exceeds
the dimension-0 premium ($1.606\pm0.063$ against $1.437\pm0.025$ in the top $1\%$), so that step is
not produced by the two filler sets overlapping; the dimension-2 curve rests on $914$ works and is
uninformative ($1.524\pm0.885$). All curves are weighted: each point is the filler-count-weighted
rate across discipline--years, and the bands are $95\%$ intervals from a standard error treating
each discipline--year as one observation. All three panels share one $y$ range. One caveat specific
to (C): a set defined as ``closes a loop but no gap'' is compared against a population $85\%$ of
which does close gaps, so the comparison mixes the order-1 effect with the unusual property of
closing nothing lower. Restricting the comparison population to works that likewise close nothing
below order $n$ raises the dimension-1 figure to $1.679\pm0.067$ and leaves the ordering unchanged.
Overlap percentages are in Table \ref{tab:claude-overlap}.}
\label{fig:claude-order}
\end{figure}

\begin{figure}[htbp]
\centering
\includegraphics[width=\textwidth]{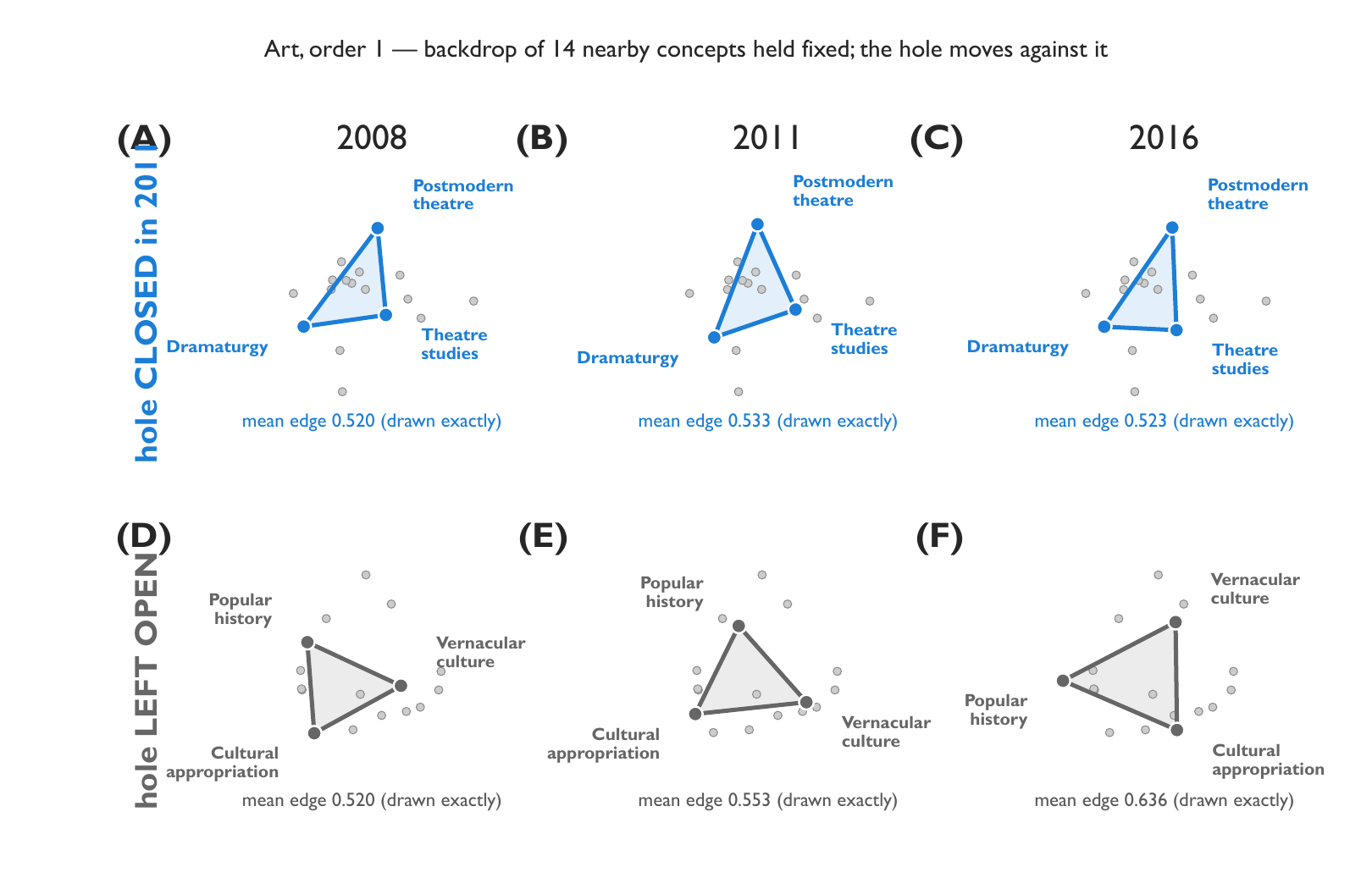}
\caption{\textbf{A hole that was closed and a comparable hole left open, against a fixed backdrop
of nearby concepts.} Art, order 1, three years before the hole was recorded, in the year it was
closed, and five years after. The coloured triangle joins the concepts of a hole closed in $2011$ by
a single work (\emph{Dramaturgy}, \emph{Theatre studies}, \emph{Postmodern theatre}); the grey
triangle joins those of a hole from the same discipline and year that nobody closed
(\emph{Popular history}, \emph{Cultural appropriation}, \emph{Vernacular culture}). Both had a
mean edge of $0.520$ in $2008$; by $2016$ the open hole's had grown to $0.636$, while the closed
hole's was back at $0.523$. Grey points are $14$ nearby concepts, drawn in the same position in
every panel; their mean pairwise distance changes by at most $0.03$ across the panels, and the
Supplementary Text shows that such neighbourhoods do not move on average. Each triangle is drawn
exactly, since three distances determine a triangle up to isometry, and the backdrop fixes only
where it sits and which way it faces: size and shape carry no projection error, position and
orientation are approximate. The pair illustrates the average pattern rather than evidencing it
--- it was chosen because it shows that pattern clearly, and one hole cannot speak to the size of
an average effect (Table \ref{tab:claude-drift}).}
\label{fig:claude-drift}
\end{figure}

\begin{figure}[htbp]
\centering
\includegraphics[width=\textwidth]{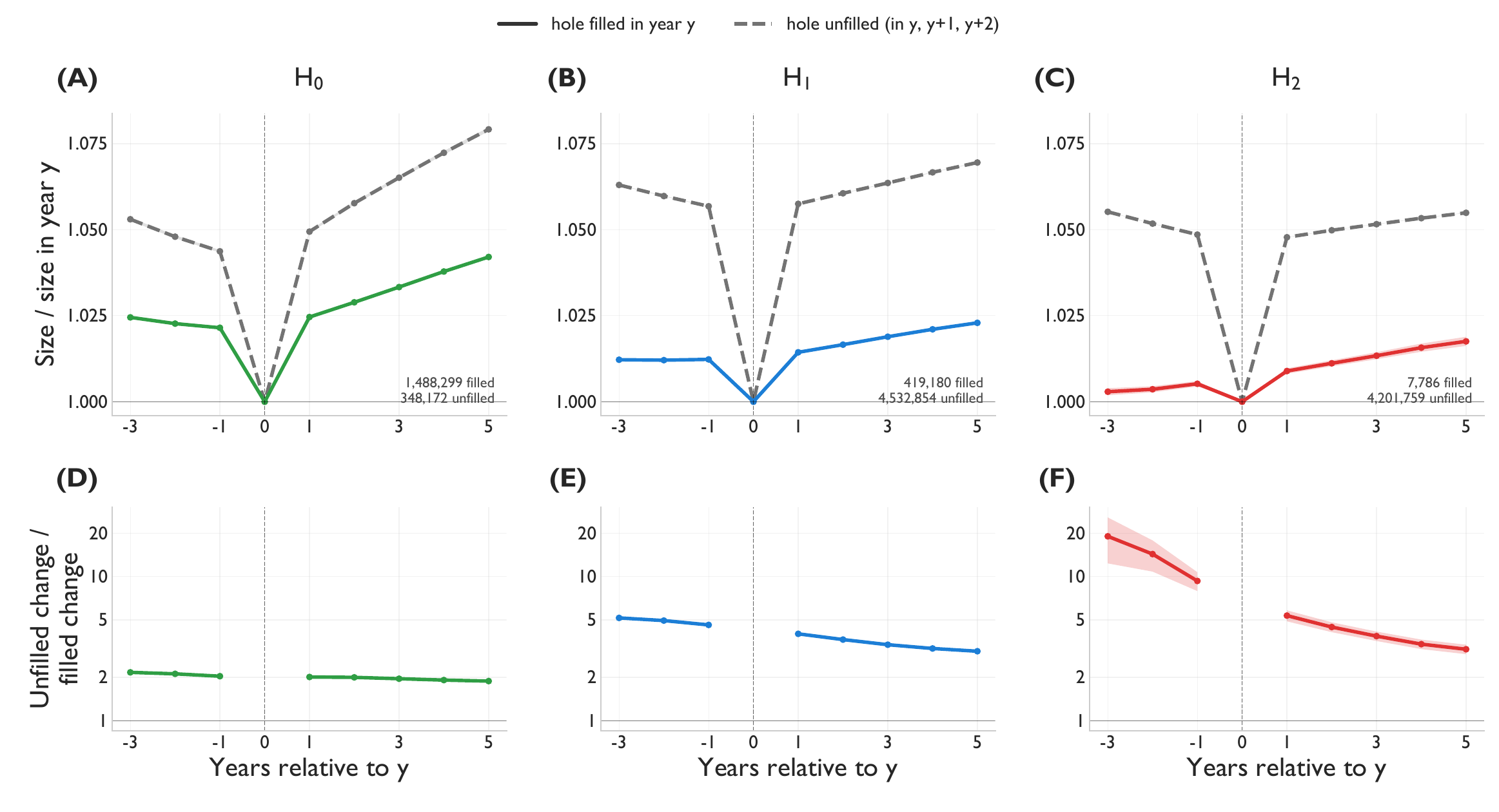}
\caption{\textbf{The concepts of holes that are filled stay closer to their configuration in the
year the hole is recorded, before filling as well as after.} Every hole recorded in a year $y$ of
2005--2019, across the nineteen disciplines (thirteen at dimension 2), followed from $y-3$ to
$y+5$. \textbf{(A--C)} Mean pairwise distance among the concepts of the hole's death simplex,
divided by its value in year $y$ and averaged over holes filled in $y$ (solid) and over holes
filled in none of $y$, $y+1$ and $y+2$ (dashed). \textbf{(D--F)} How many times further an
unfilled hole moves from its year-$y$ size than a filled one, $(\bar\rho_{\mathrm{unfilled}}-1) /
(\bar\rho_{\mathrm{filled}}-1)$, where $\bar\rho$ is the group mean plotted above; it is undefined
in $y$ itself, and is drawn on a logarithmic axis. Bands are $95\%$ intervals treating holes as
independent, and so are narrower than an interval treating each discipline--year as one
observation would be. No matching is applied; Table \ref{tab:claude-drift} gives the matched
comparison.}
\label{fig:claude-size}
\end{figure}

\begin{figure}
	\centering
	\includegraphics[width=0.8\textwidth]{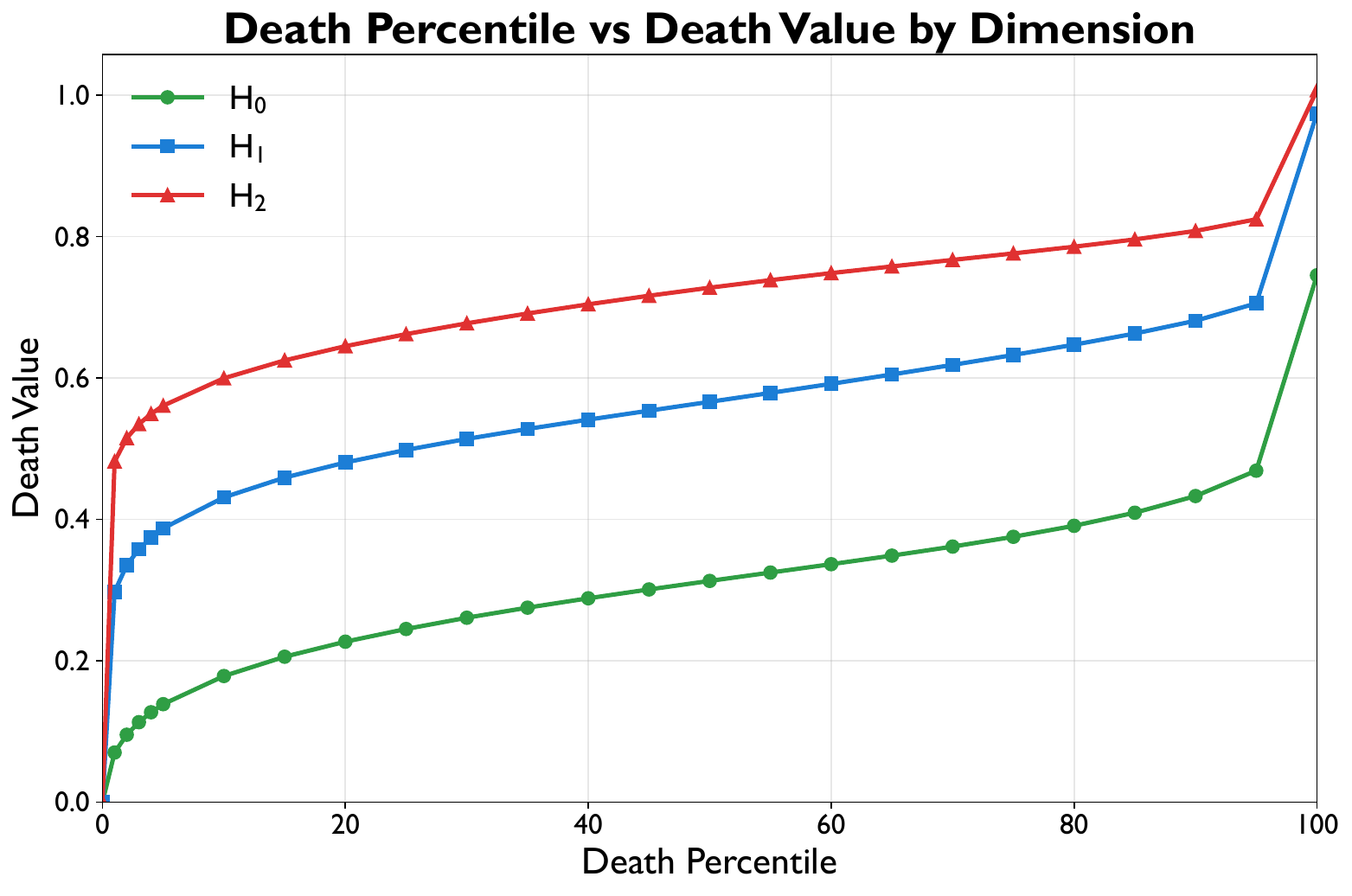}

	\caption{\textbf{Death percentile to value.}
    We show the correspondence between the death percentiles and their values, for dimension $0$, $1$, and $2$.
    }
	\label{fig:death_percentile_to_value}
\end{figure}

\newpage

\begin{figure}
	\centering
	\includegraphics[width=0.8\textwidth]{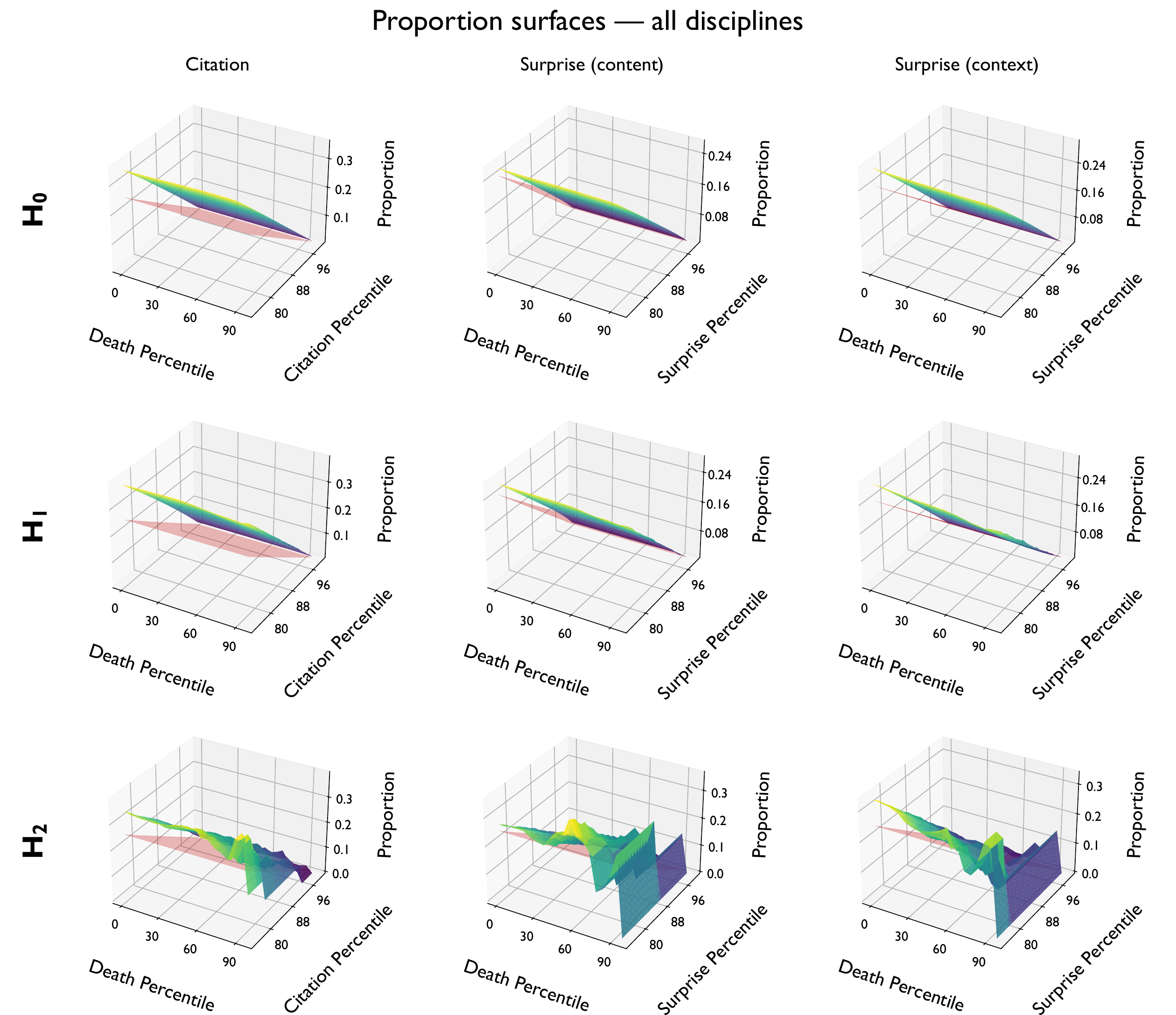}

	\caption{\textbf{3-dimensional surface plots of the proportion of hole-filling papers by metric percentile and death percentile, pooled across all disciplines.}
    We show a 3x3 grid of 3D surface plots. Rows correspond to homology dimension and columns to the paper-ranking metric. In each panel, the horizontal axes are the death percentile at which a hole is filled ($x$-axis) and the metric percentile ($y$-axis, spanning the 75th--99th percentiles), and the vertical axis is the proportion of hole-filling papers at that combination. The yellow-to-purple surface is the observed proportion, colored by height; the translucent red plane is the baseline proportion of papers at or above that metric. All surfaces pool over the 19 disciplines, weighted by paper count.
    }
	\label{fig:3d_proportion_plots}
\end{figure}

\begin{figure}
	\centering
	\includegraphics[width=0.8\textwidth]{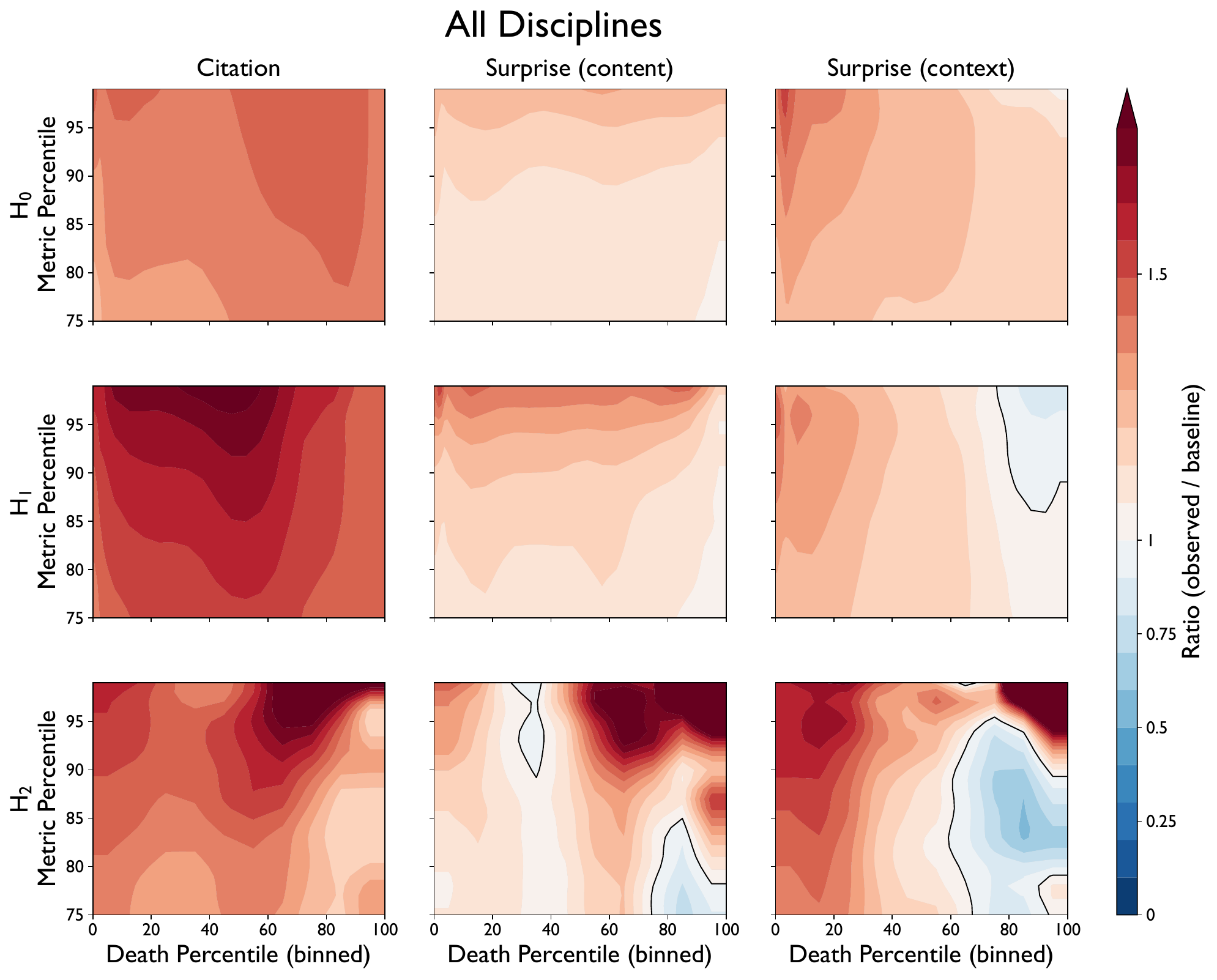}

	\caption{\textbf{Plots of the ratios between the proportion of hole-filling papers by metric percentile and death percentile and the baseline proportion, pooled across all disciplines.}
    Consider the 3D plots in Figure~\ref{fig:3d_proportion_plots}. For each metric and homology dimension, we take the ratio between the proportion of papers above a fixed percentile (yellow-to-purple surface) and the baseline proportion (red surface).
    The black contour line at $z=1$ separates the regions that are above $1$ and below $1$.
    }
	\label{fig:2D_ratio_plots_all_disciplines}
\end{figure}

\begin{figure}
	\centering
	\includegraphics[width=0.8\textwidth]{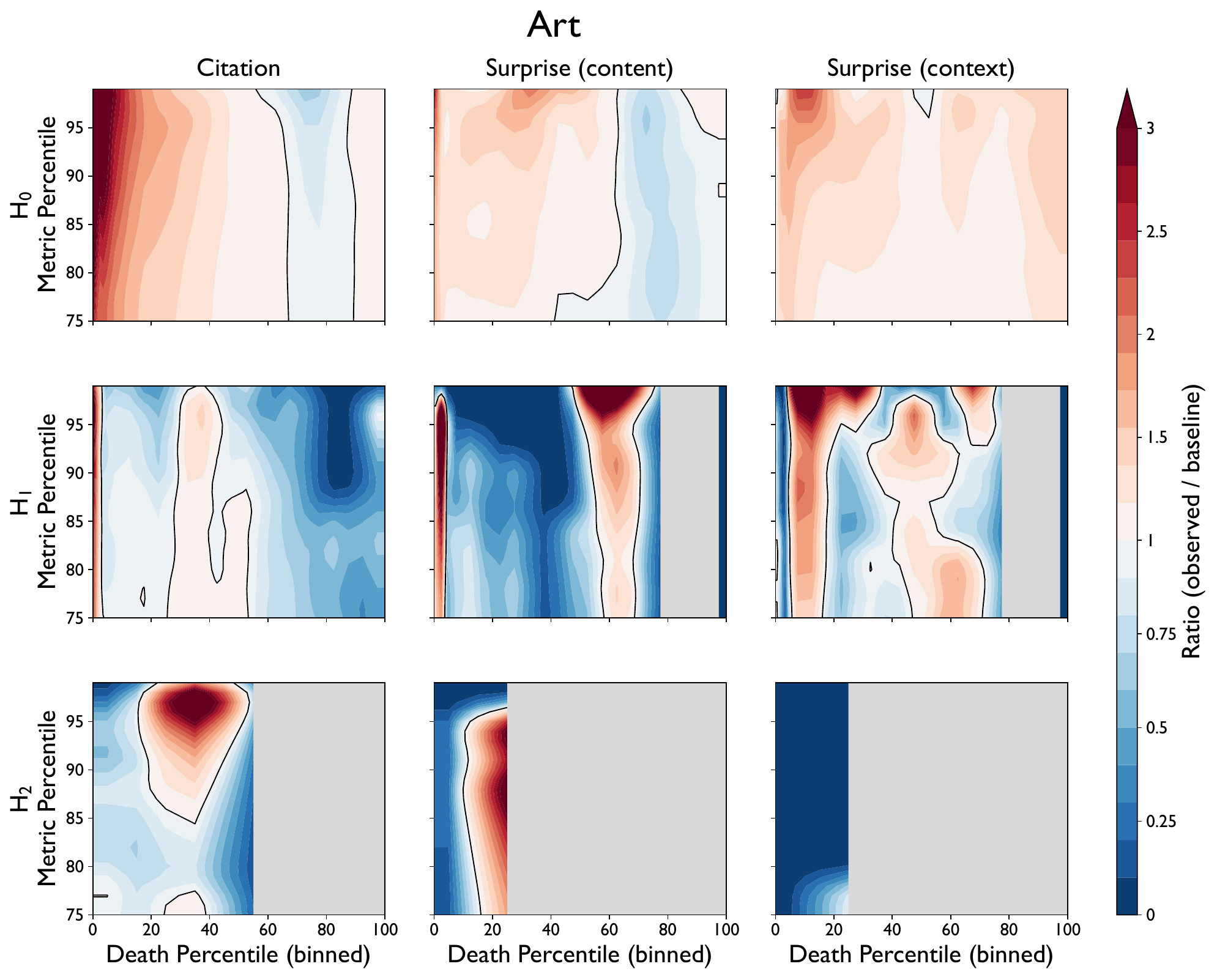}

	\caption{\textbf{Plots of the ratios between the proportion of hole-filling papers by metric percentile and death percentile and the baseline proportion, by discipline.}
    Same as Figure~\ref{fig:2D_ratio_plots_all_disciplines}, but for individual disciplines, one discipline per page. This page: art.
    }
	\label{fig:2D_ratio_plots_all_disciplines_art}
\end{figure}

\begin{figure}
	\centering
	\includegraphics[width=0.8\textwidth]{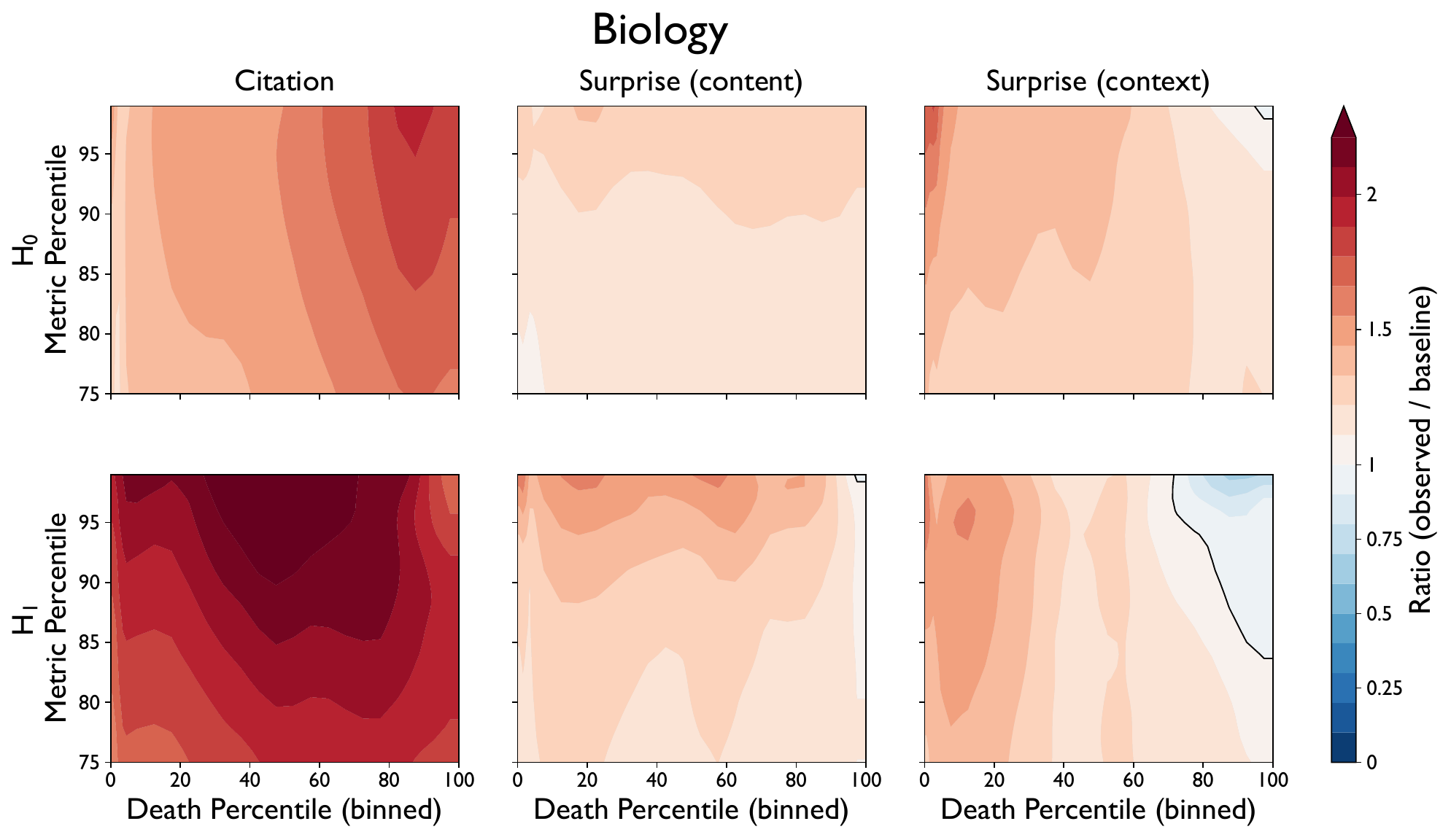}
	\continuedfigure
	\caption{\textbf{Plots of the ratios between the proportion of hole-filling papers by metric percentile and death percentile and the baseline proportion, by discipline.}
    This page: biology.
    }
	\label{fig:2D_ratio_plots_all_disciplines_biology}
\end{figure}

\begin{figure}
	\centering
	\includegraphics[width=0.8\textwidth]{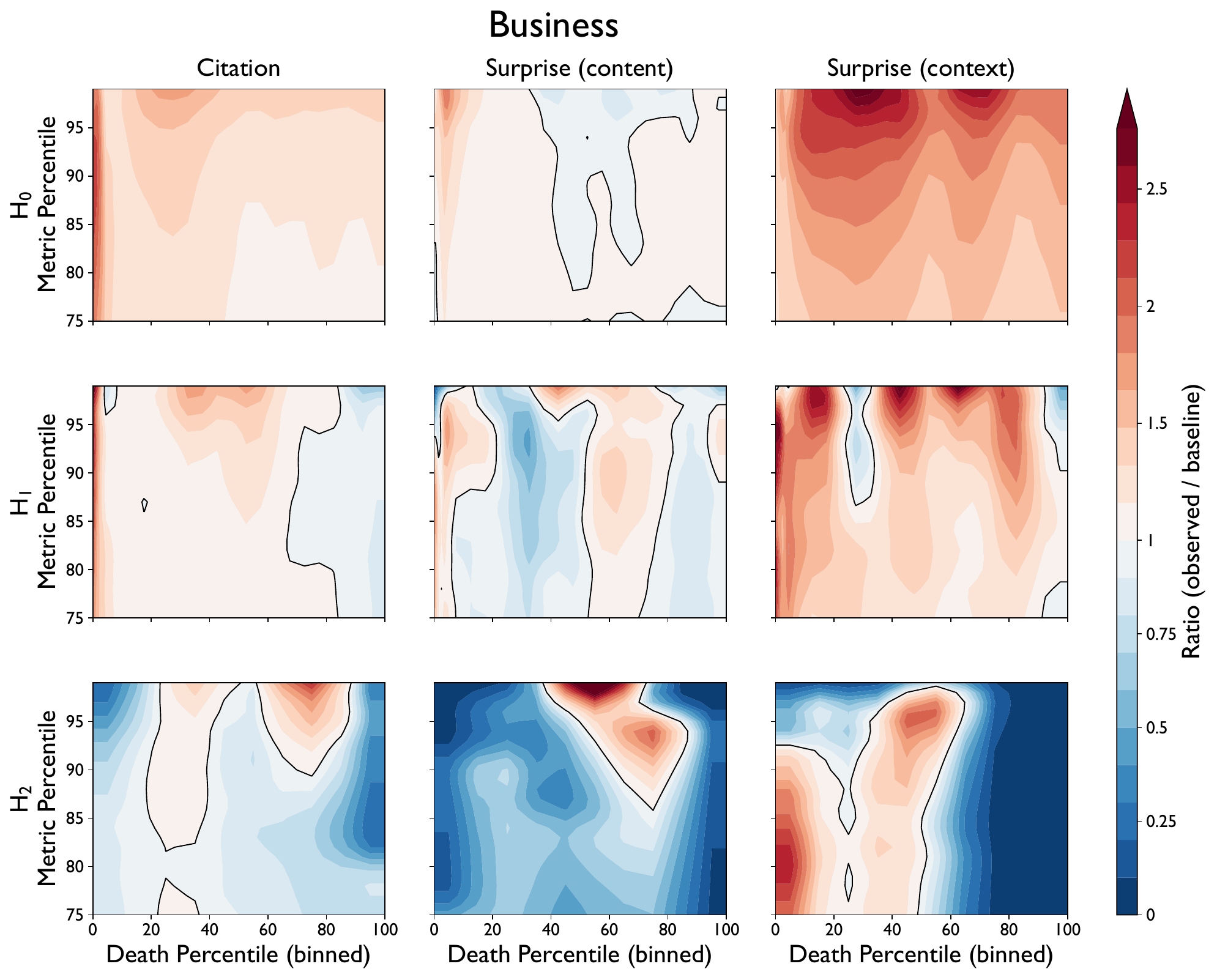}
	\continuedfigure
	\caption{\textbf{Plots of the ratios between the proportion of hole-filling papers by metric percentile and death percentile and the baseline proportion, by discipline.}
    This page: business.
    }
	\label{fig:2D_ratio_plots_all_disciplines_business}
\end{figure}

\begin{figure}
	\centering
	\includegraphics[width=0.8\textwidth]{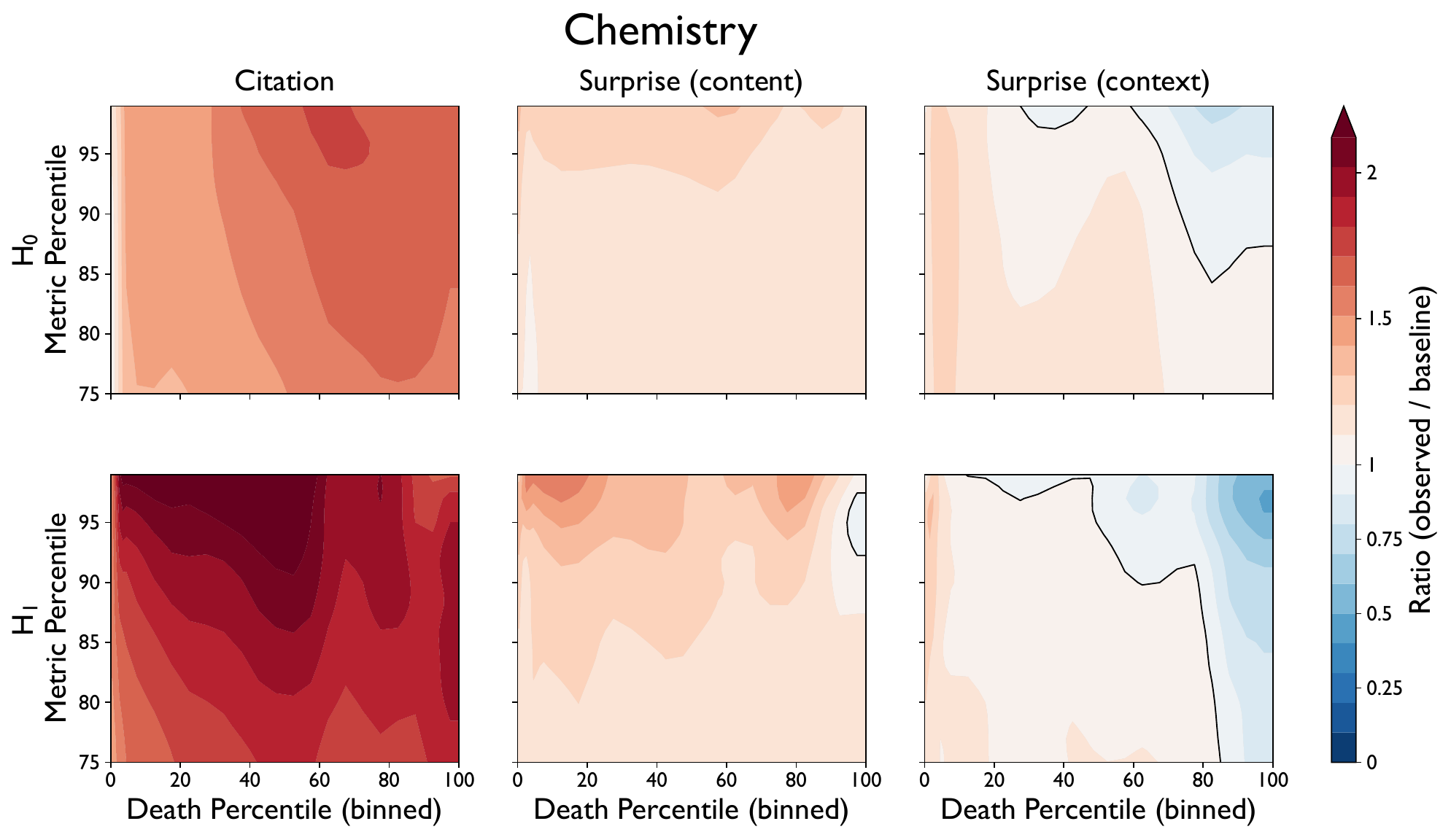}
	\continuedfigure
	\caption{\textbf{Plots of the ratios between the proportion of hole-filling papers by metric percentile and death percentile and the baseline proportion, by discipline.}
    This page: chemistry.
    }
	\label{fig:2D_ratio_plots_all_disciplines_chemistry}
\end{figure}

\begin{figure}
	\centering
	\includegraphics[width=0.8\textwidth]{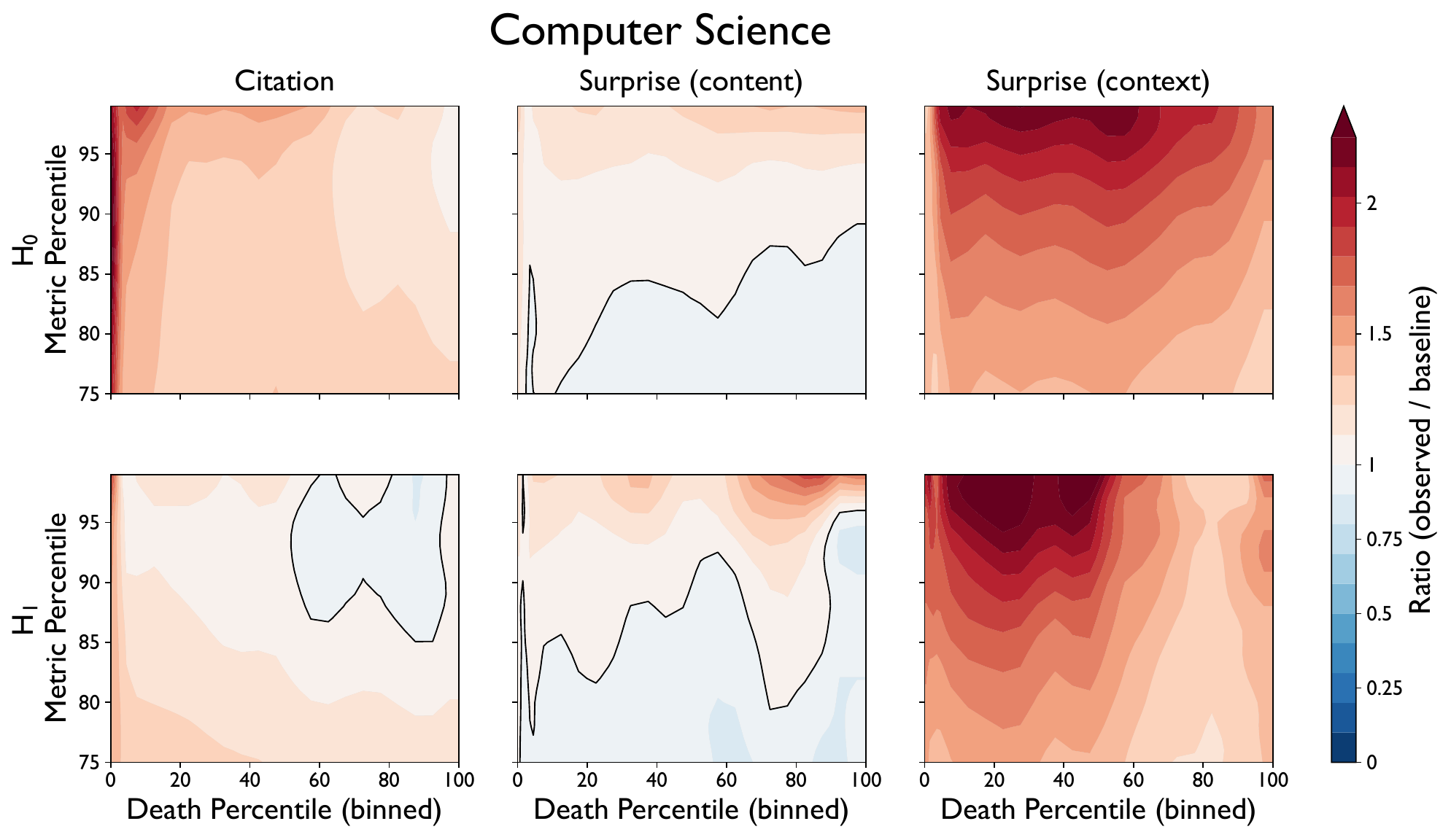}
	\continuedfigure
	\caption{\textbf{Plots of the ratios between the proportion of hole-filling papers by metric percentile and death percentile and the baseline proportion, by discipline.}
    This page: computer science.
    }
	\label{fig:2D_ratio_plots_all_disciplines_computer_science}
\end{figure}

\begin{figure}
	\centering
	\includegraphics[width=0.8\textwidth]{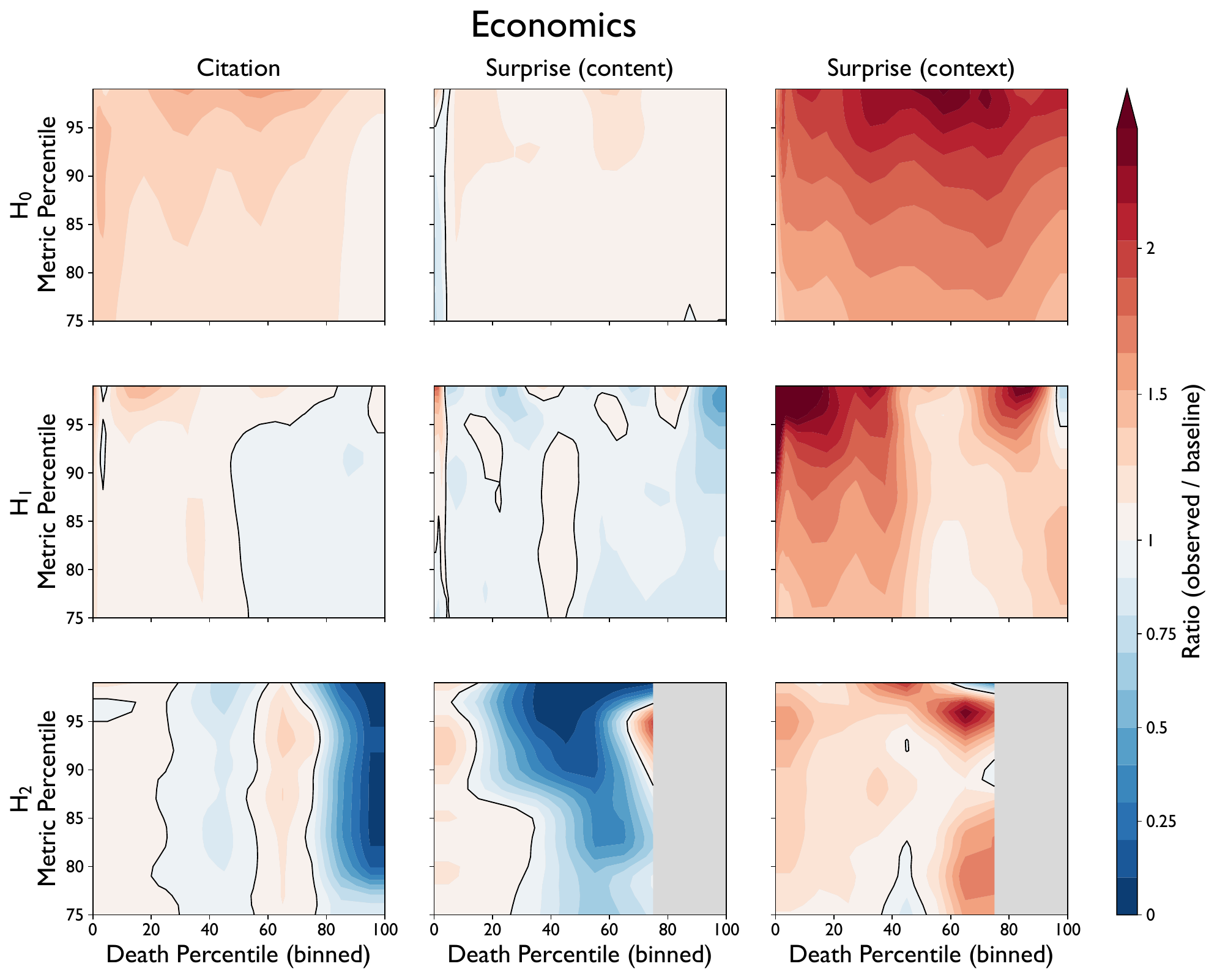}
	\continuedfigure
	\caption{\textbf{Plots of the ratios between the proportion of hole-filling papers by metric percentile and death percentile and the baseline proportion, by discipline.}
    This page: economics.
    }
	\label{fig:2D_ratio_plots_all_disciplines_economics}
\end{figure}

\clearpage

\begin{figure}
	\centering
	\includegraphics[width=0.8\textwidth]{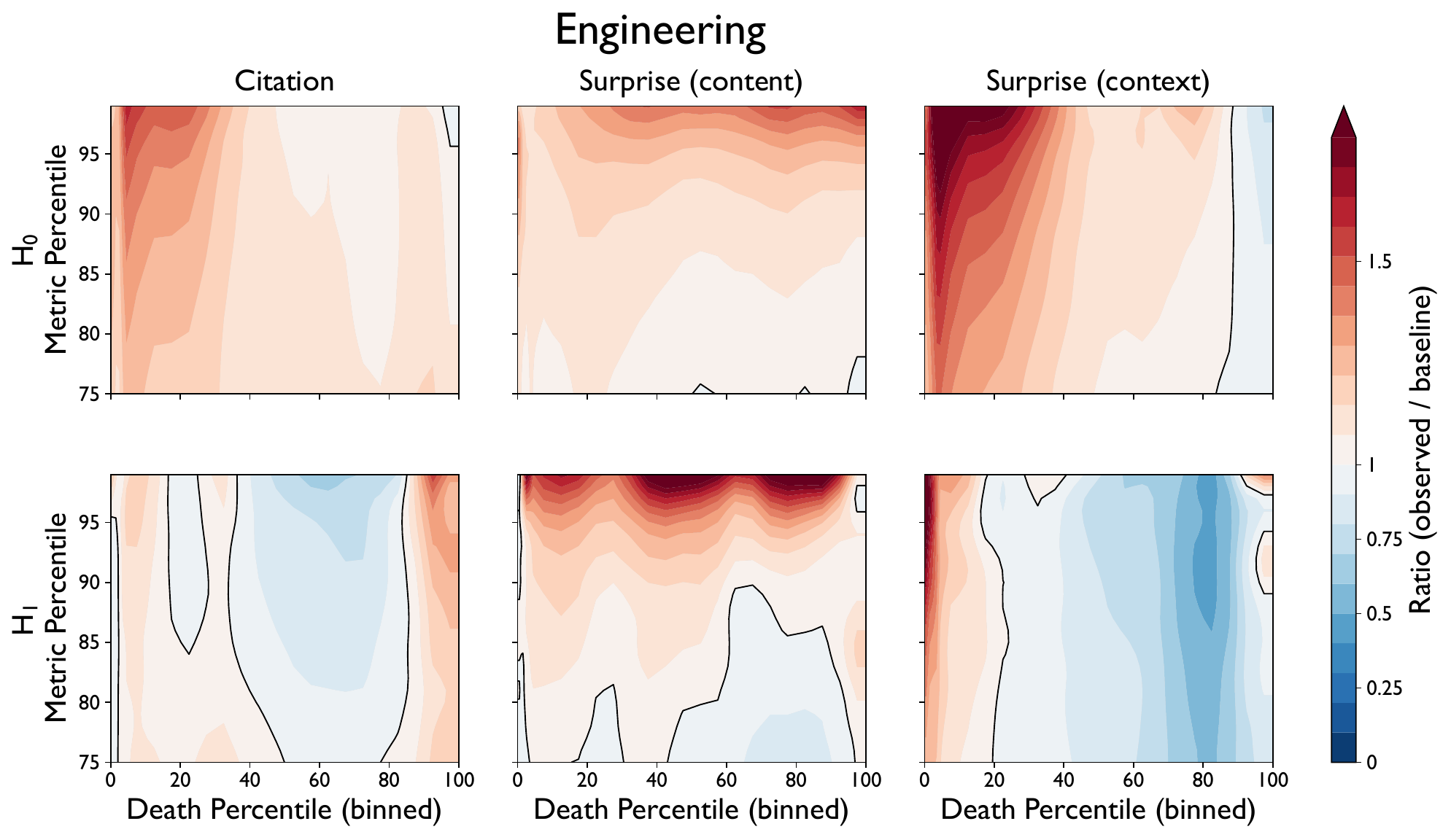}
	\continuedfigure
	\caption{\textbf{Plots of the ratios between the proportion of hole-filling papers by metric percentile and death percentile and the baseline proportion, by discipline.}
    This page: engineering.
    }
	\label{fig:2D_ratio_plots_all_disciplines_engineering}
\end{figure}

\begin{figure}
	\centering
	\includegraphics[width=0.8\textwidth]{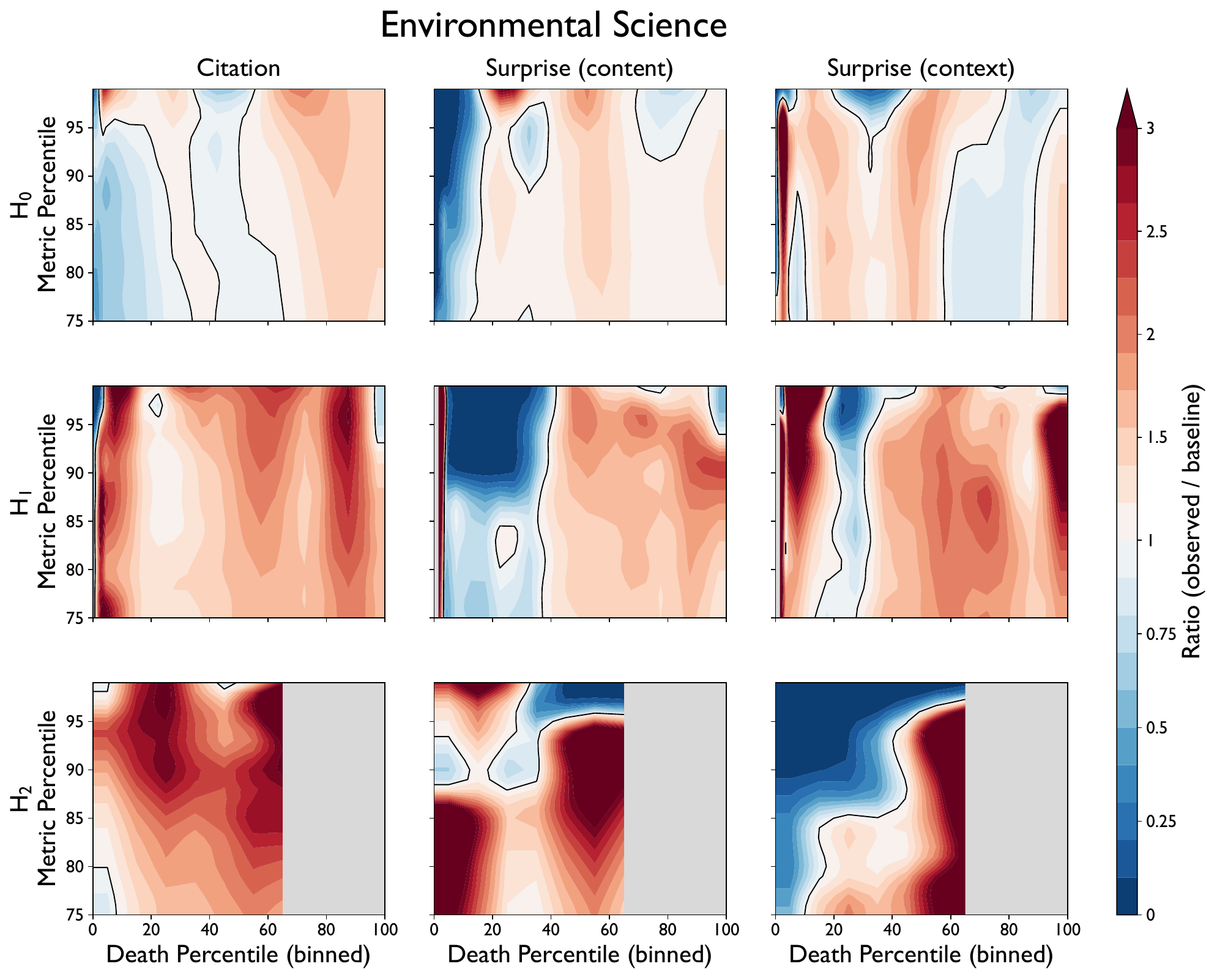}
	\continuedfigure
	\caption{\textbf{Plots of the ratios between the proportion of hole-filling papers by metric percentile and death percentile and the baseline proportion, by discipline.}
    This page: environmental science.
    }
	\label{fig:2D_ratio_plots_all_disciplines_environmental_science}
\end{figure}

\begin{figure}
	\centering
	\includegraphics[width=0.8\textwidth]{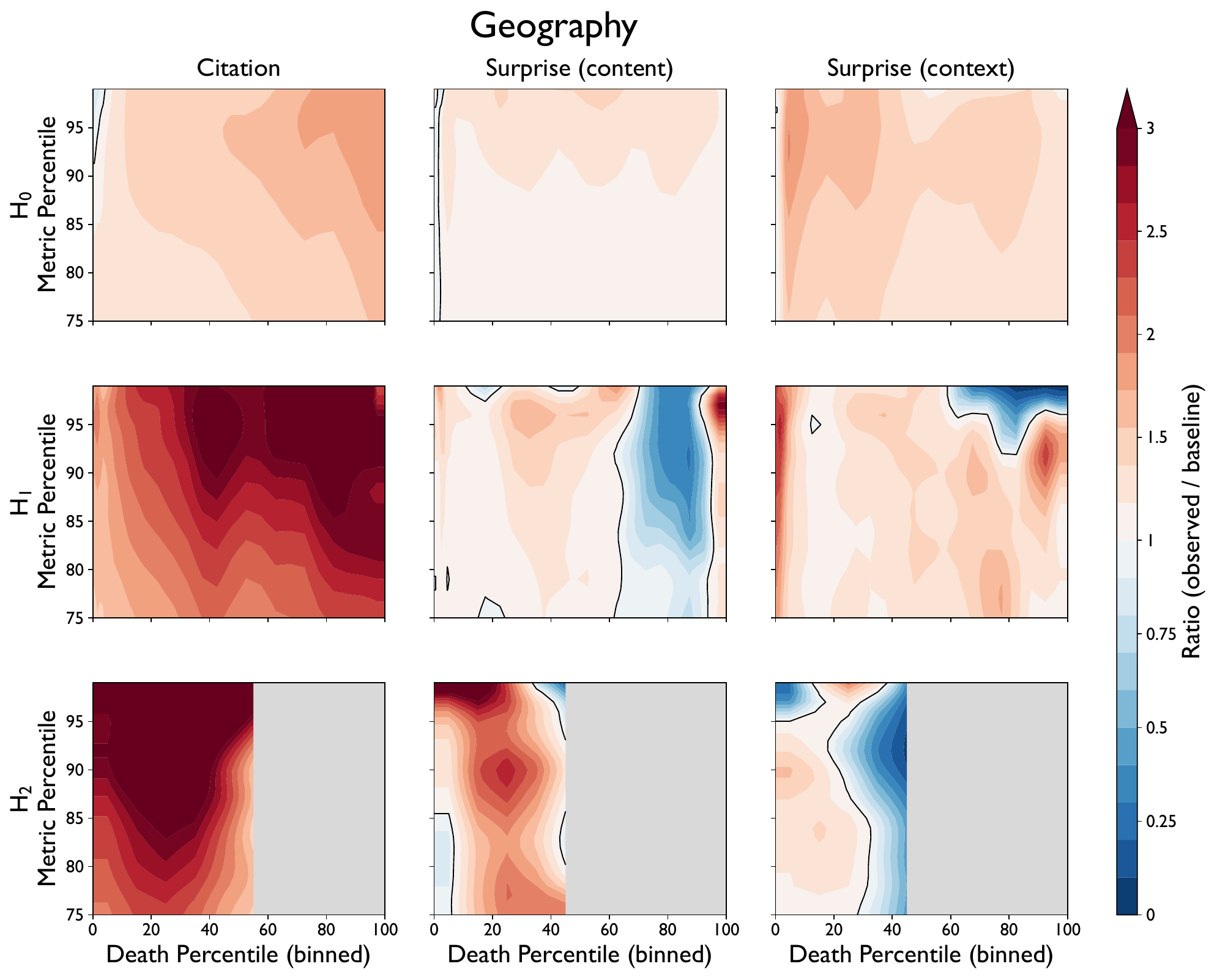}
	\continuedfigure
	\caption{\textbf{Plots of the ratios between the proportion of hole-filling papers by metric percentile and death percentile and the baseline proportion, by discipline.}
    This page: geography.
    }
	\label{fig:2D_ratio_plots_all_disciplines_geography}
\end{figure}

\begin{figure}
	\centering
	\includegraphics[width=0.8\textwidth]{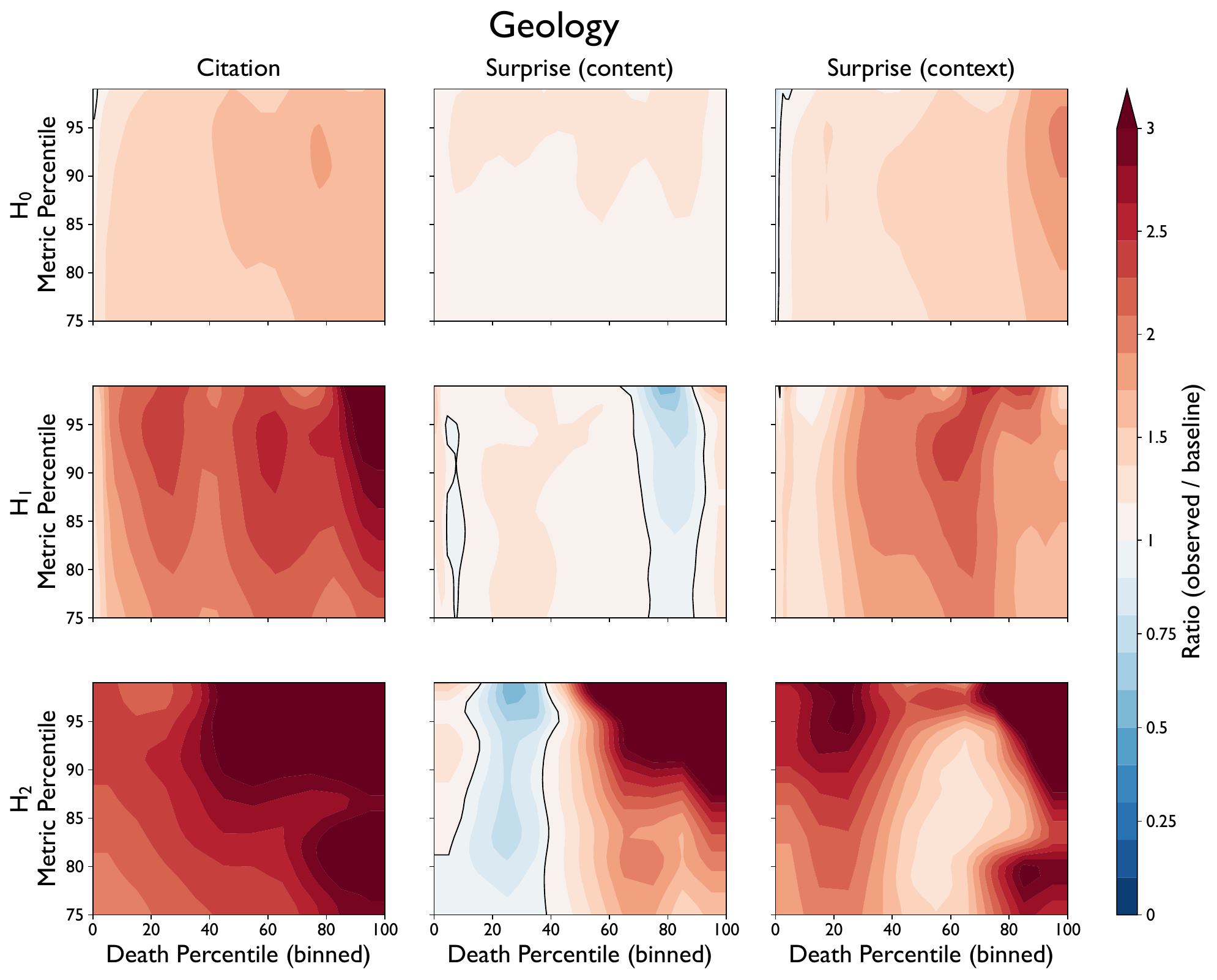}
	\continuedfigure
	\caption{\textbf{Plots of the ratios between the proportion of hole-filling papers by metric percentile and death percentile and the baseline proportion, by discipline.}
    This page: geology.
    }
	\label{fig:2D_ratio_plots_all_disciplines_geology}
\end{figure}

\begin{figure}
	\centering
	\includegraphics[width=0.8\textwidth]{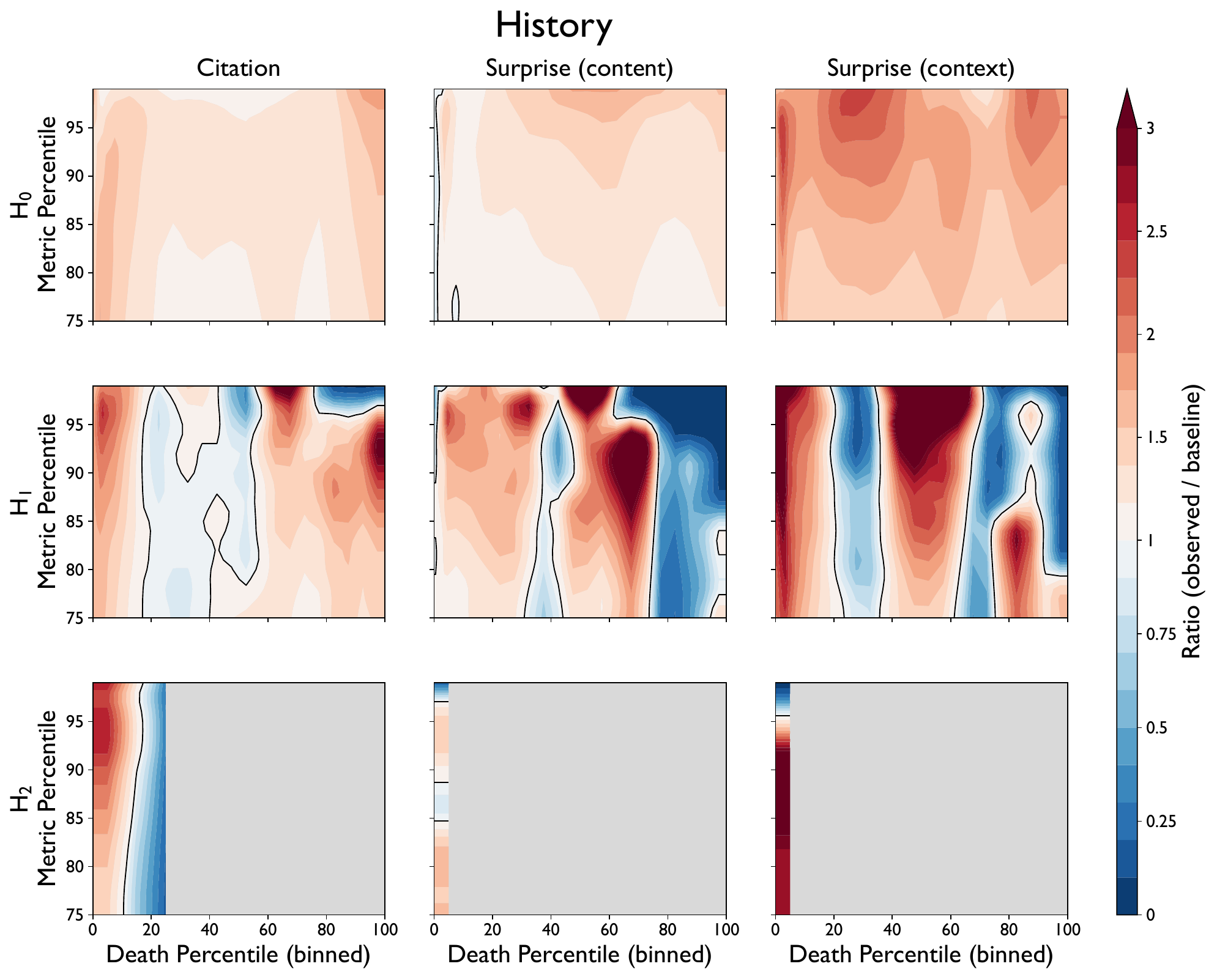}
	\continuedfigure
	\caption{\textbf{Plots of the ratios between the proportion of hole-filling papers by metric percentile and death percentile and the baseline proportion, by discipline.}
    This page: history.
    }
	\label{fig:2D_ratio_plots_all_disciplines_history}
\end{figure}

\clearpage

\begin{figure}
	\centering
	\includegraphics[width=0.8\textwidth]{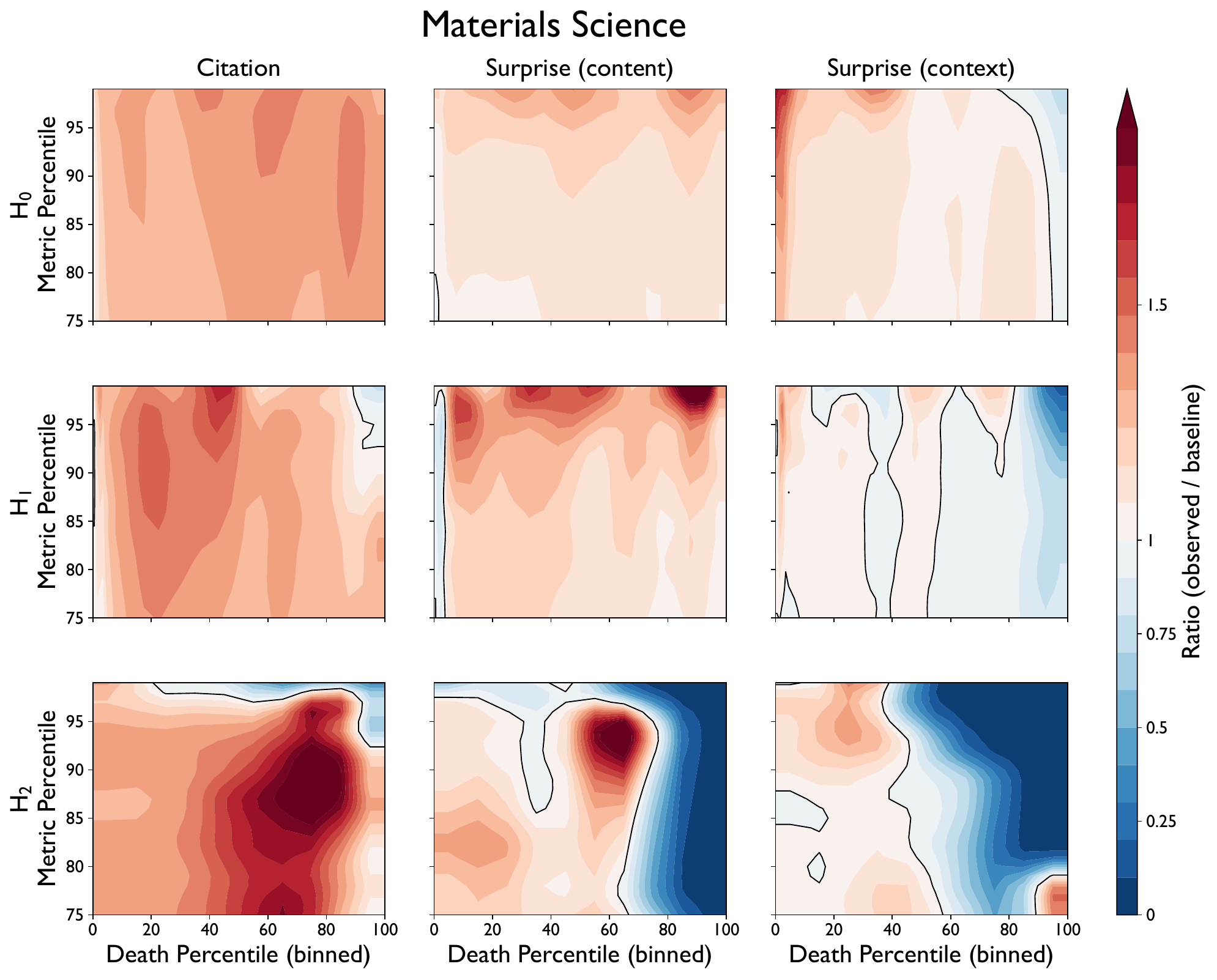}
	\continuedfigure
	\caption{\textbf{Plots of the ratios between the proportion of hole-filling papers by metric percentile and death percentile and the baseline proportion, by discipline.}
    This page: materials science.
    }
	\label{fig:2D_ratio_plots_all_disciplines_materials_science}
\end{figure}

\begin{figure}
	\centering
	\includegraphics[width=0.8\textwidth]{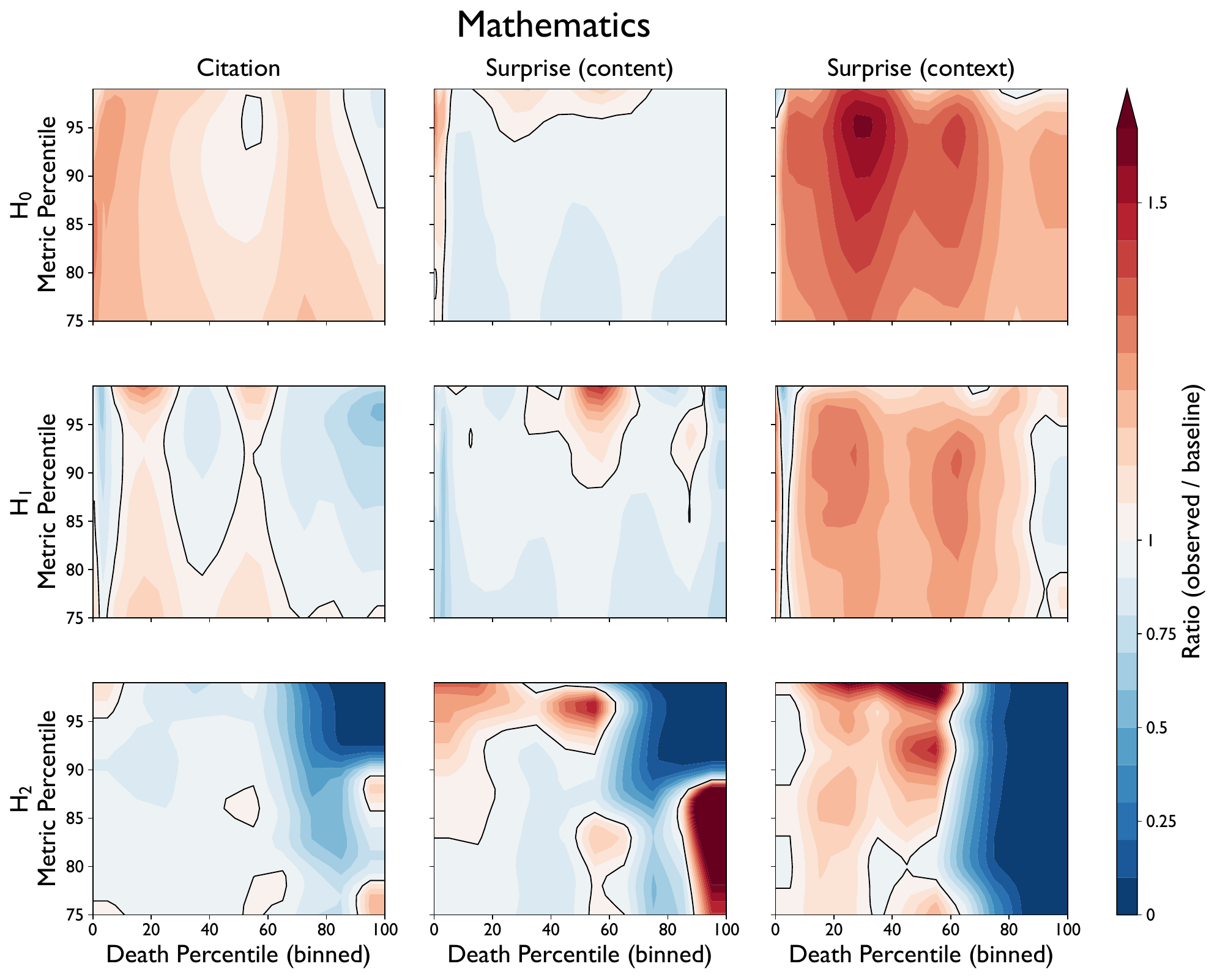}
	\continuedfigure
	\caption{\textbf{Plots of the ratios between the proportion of hole-filling papers by metric percentile and death percentile and the baseline proportion, by discipline.}
    This page: mathematics.
    }
	\label{fig:2D_ratio_plots_all_disciplines_mathematics}
\end{figure}

\begin{figure}
	\centering
	\includegraphics[width=0.8\textwidth]{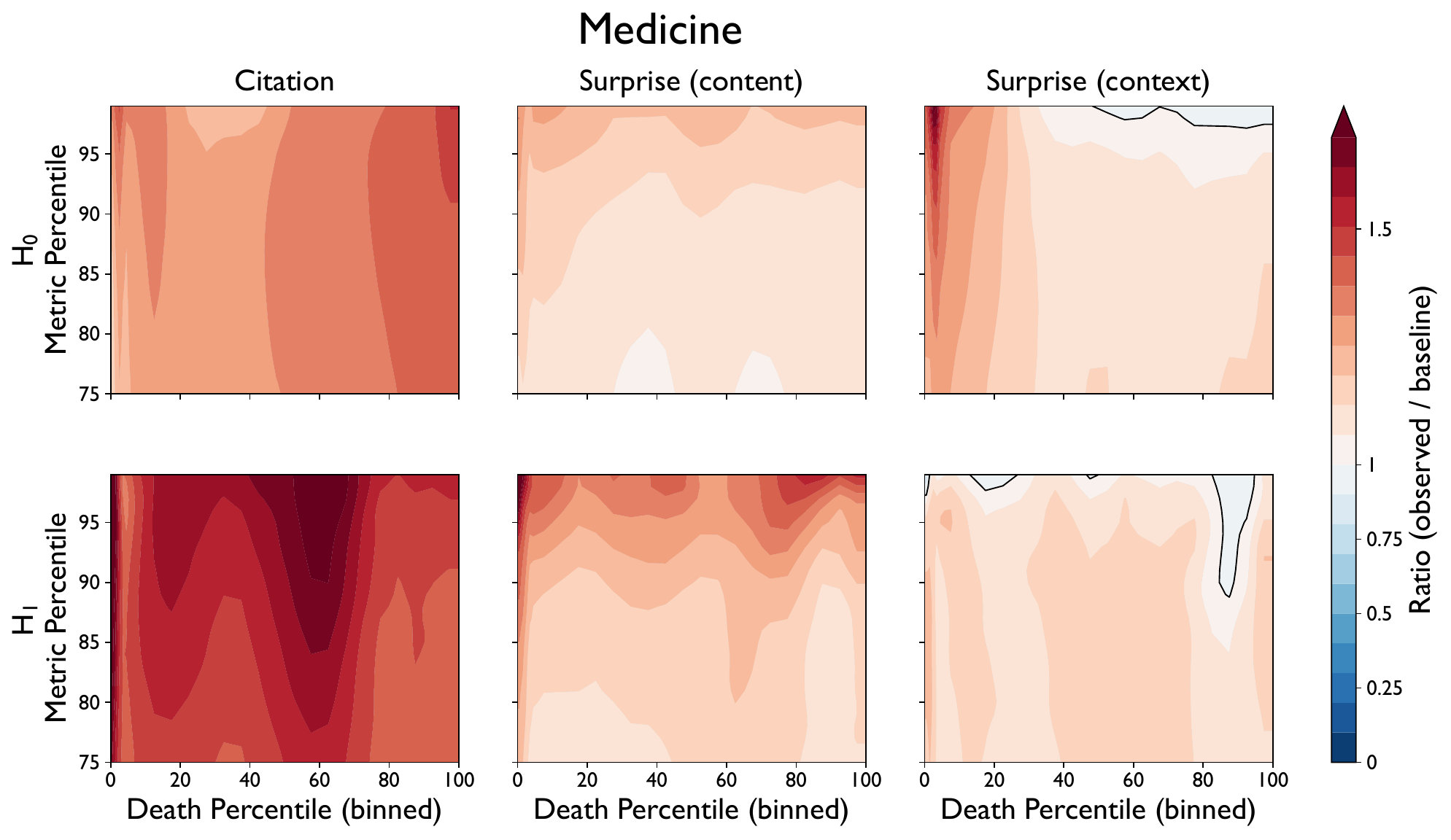}
	\continuedfigure
	\caption{\textbf{Plots of the ratios between the proportion of hole-filling papers by metric percentile and death percentile and the baseline proportion, by discipline.}
    This page: medicine.
    }
	\label{fig:2D_ratio_plots_all_disciplines_medicine}
\end{figure}

\begin{figure}
	\centering
	\includegraphics[width=0.8\textwidth]{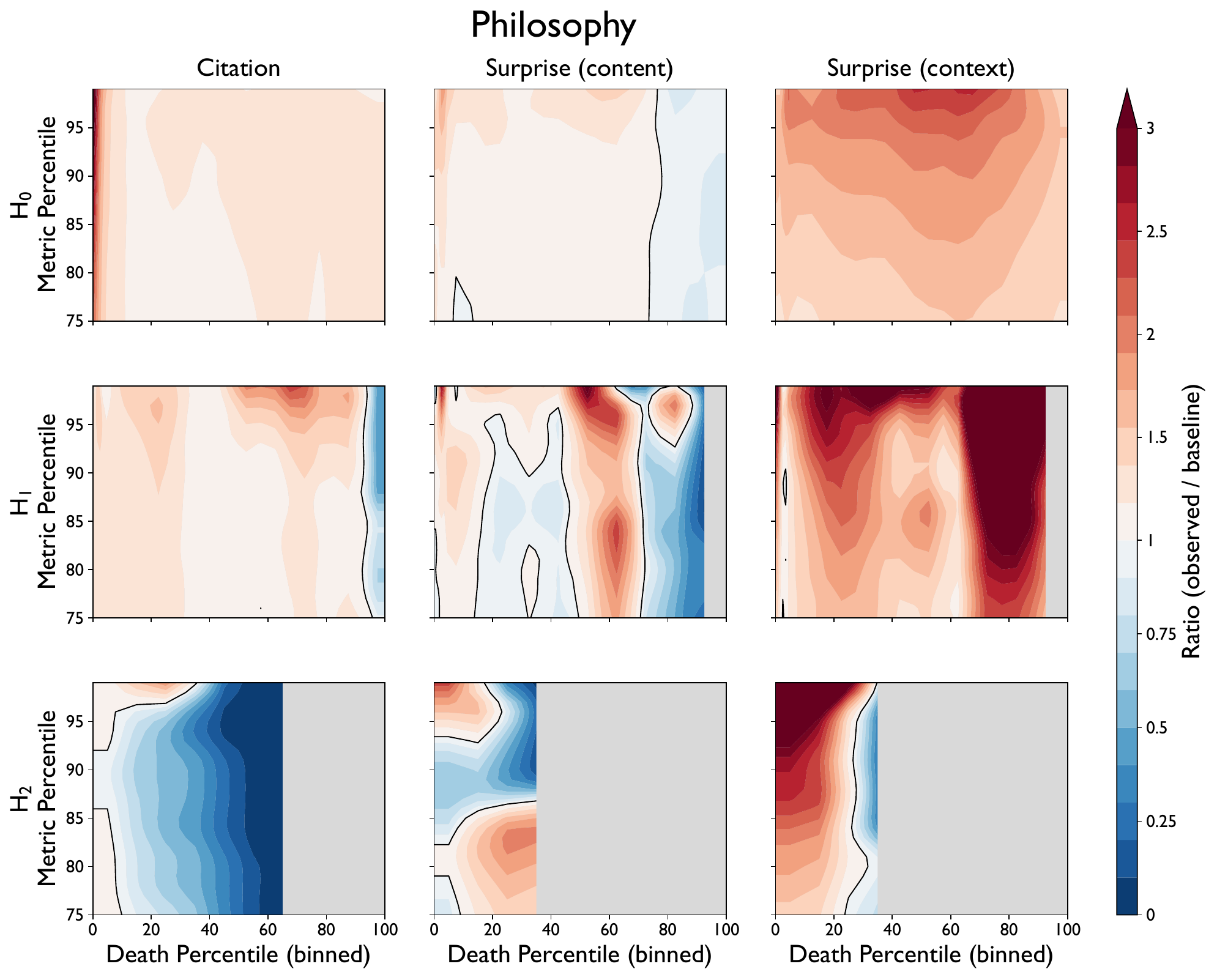}
	\continuedfigure
	\caption{\textbf{Plots of the ratios between the proportion of hole-filling papers by metric percentile and death percentile and the baseline proportion, by discipline.}
    This page: philosophy.
    }
	\label{fig:2D_ratio_plots_all_disciplines_philosophy}
\end{figure}

\begin{figure}
	\centering
	\includegraphics[width=0.8\textwidth]{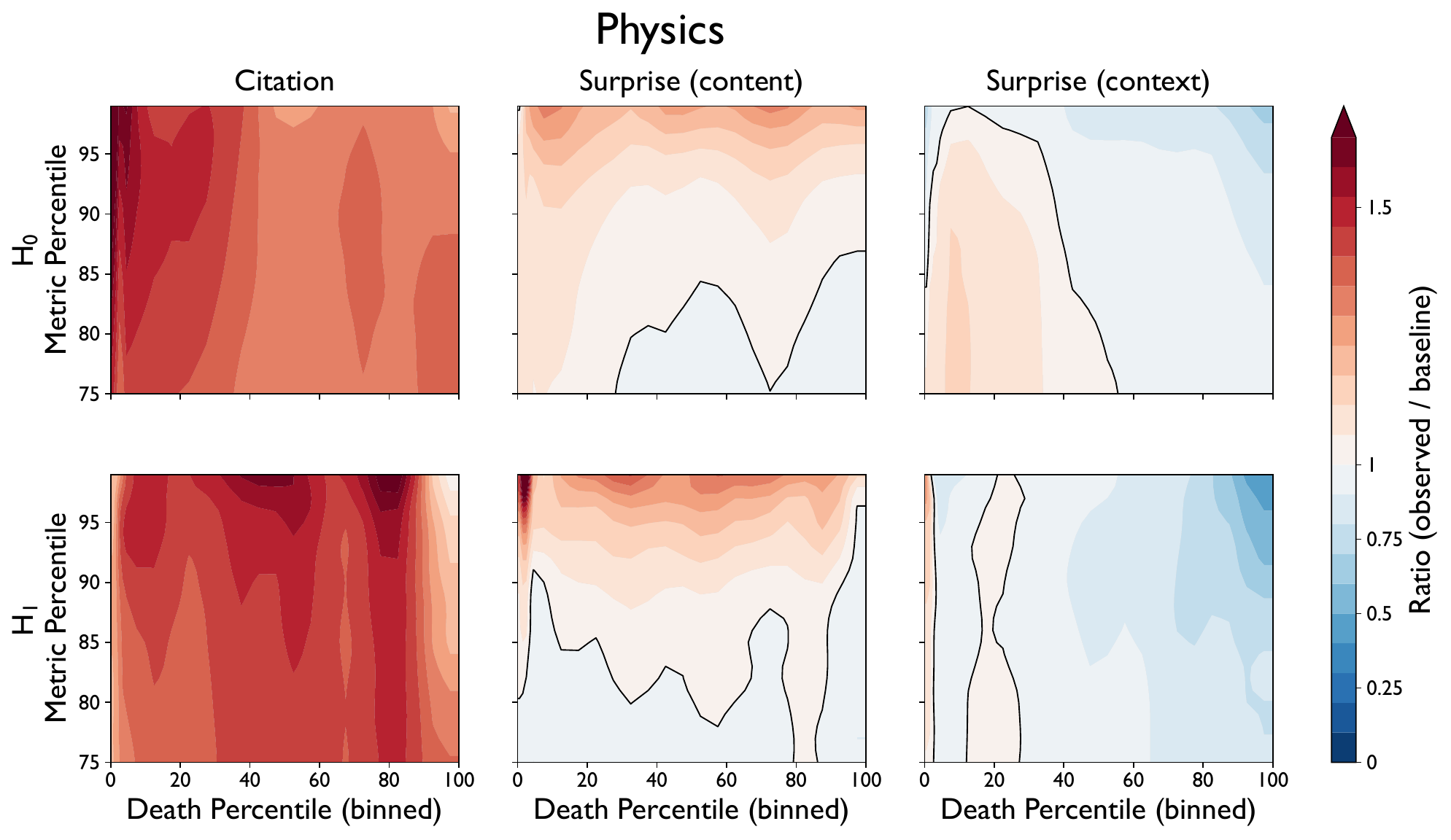}
	\continuedfigure
	\caption{\textbf{Plots of the ratios between the proportion of hole-filling papers by metric percentile and death percentile and the baseline proportion, by discipline.}
    This page: physics.
    }
	\label{fig:2D_ratio_plots_all_disciplines_physics}
\end{figure}

\clearpage

\begin{figure}
	\centering
	\includegraphics[width=0.8\textwidth]{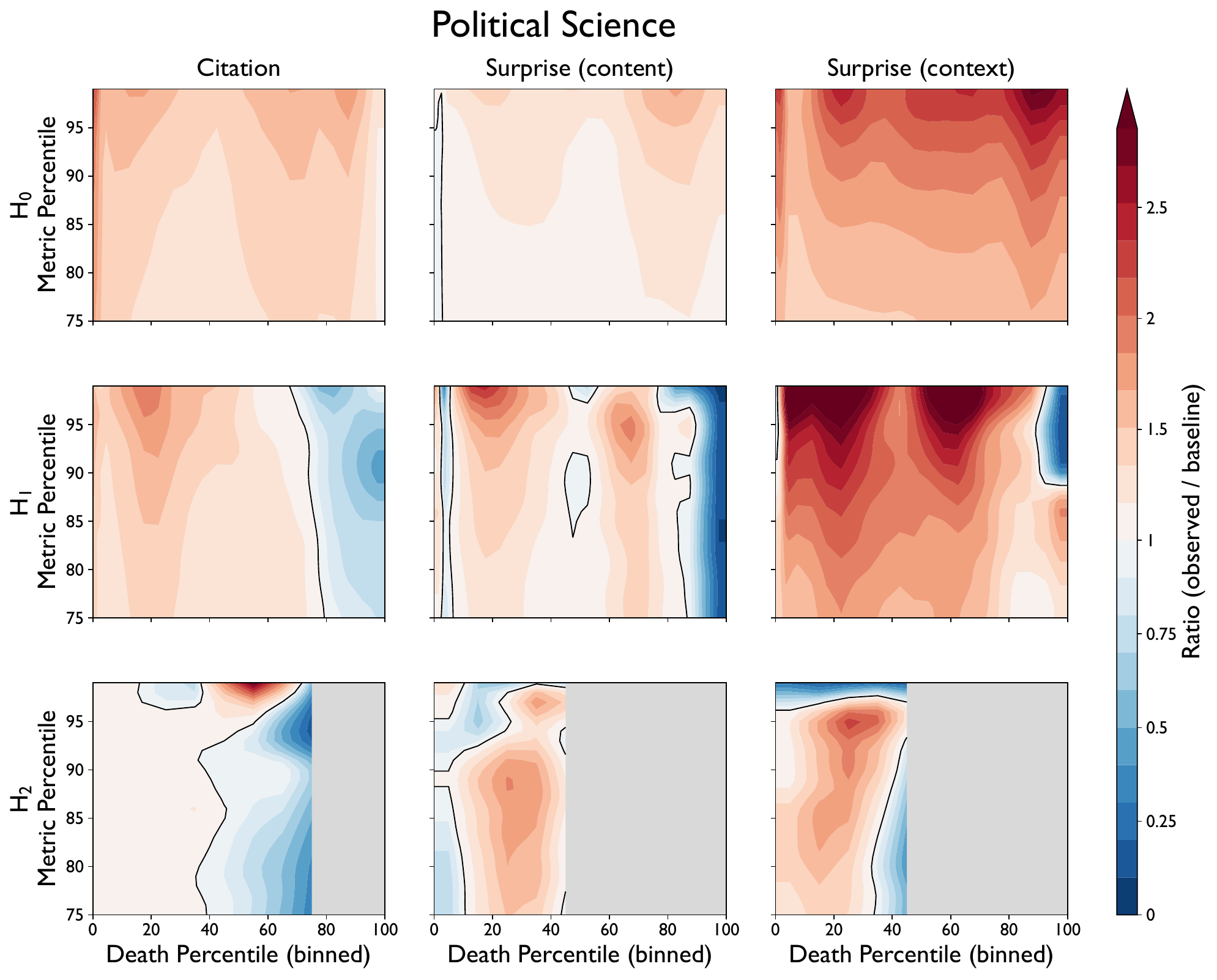}
	\continuedfigure
	\caption{\textbf{Plots of the ratios between the proportion of hole-filling papers by metric percentile and death percentile and the baseline proportion, by discipline.}
    This page: political science.
    }
	\label{fig:2D_ratio_plots_all_disciplines_political_science}
\end{figure}

\begin{figure}
	\centering
	\includegraphics[width=0.8\textwidth]{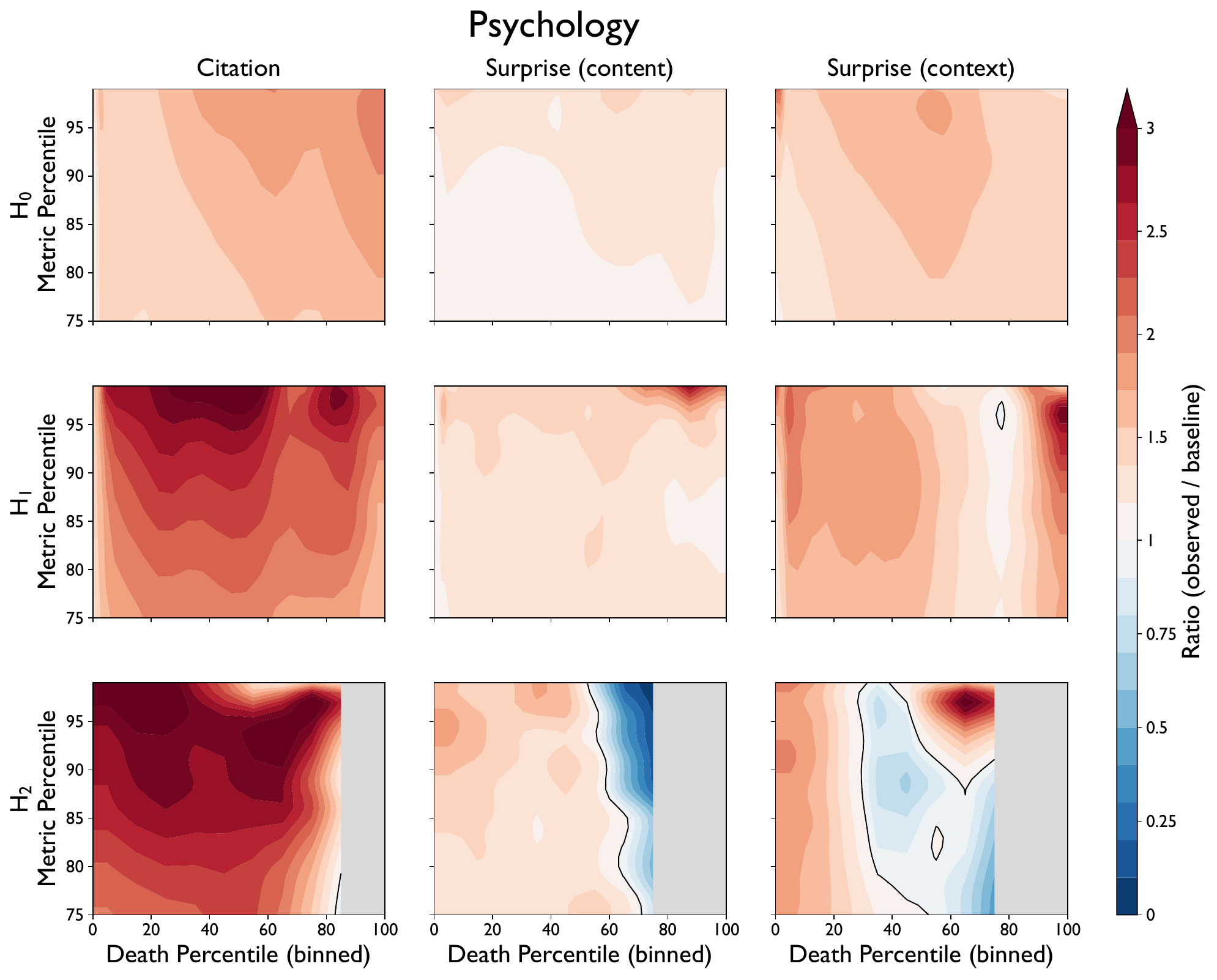}
	\continuedfigure
	\caption{\textbf{Plots of the ratios between the proportion of hole-filling papers by metric percentile and death percentile and the baseline proportion, by discipline.}
    This page: psychology.
    }
	\label{fig:2D_ratio_plots_all_disciplines_psychology}
\end{figure}

\begin{figure}
	\centering
	\includegraphics[width=0.8\textwidth]{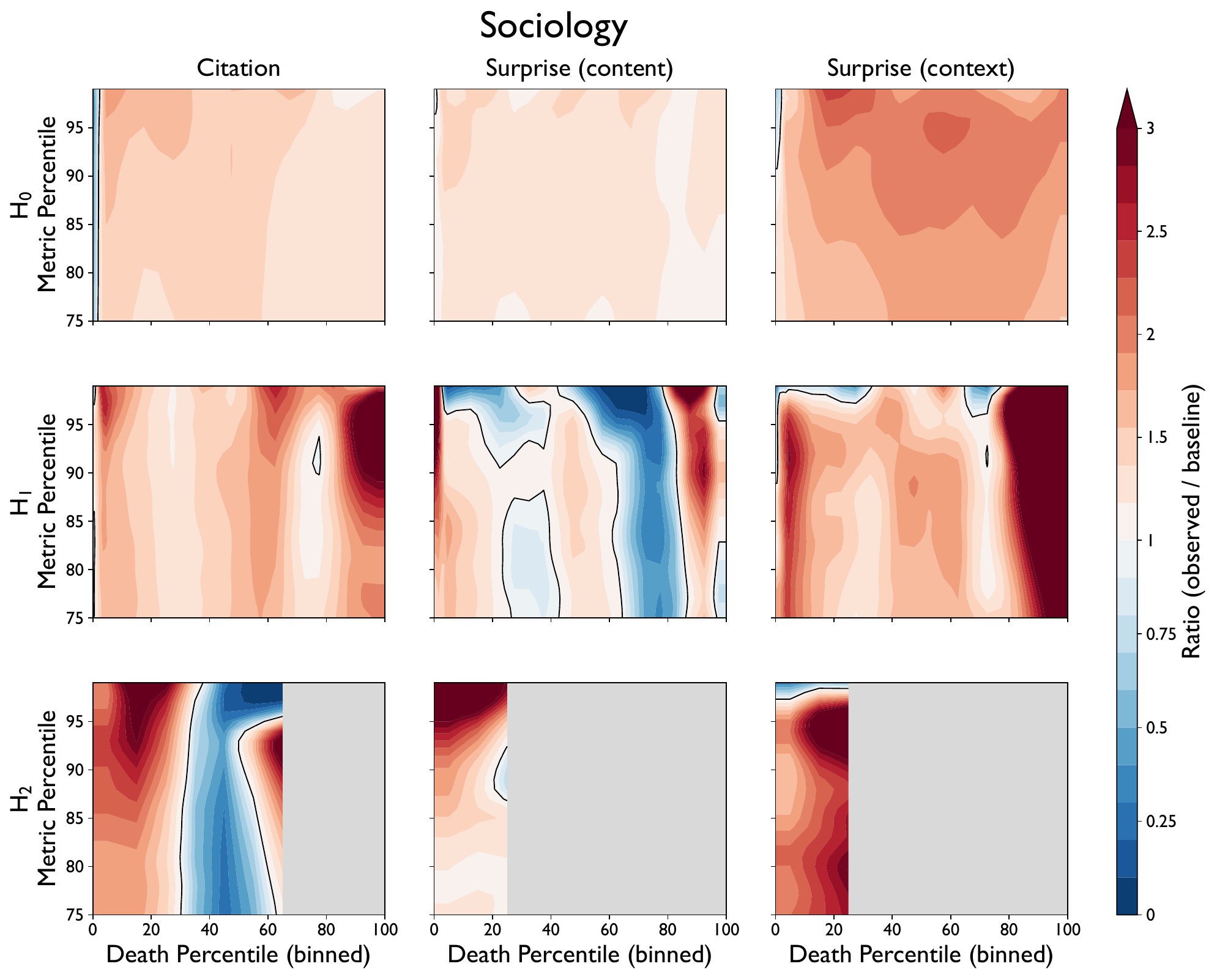}
	\continuedfigure
	\caption{\textbf{Plots of the ratios between the proportion of hole-filling papers by metric percentile and death percentile and the baseline proportion, by discipline.}
    This page: sociology.
    }
	\label{fig:2D_ratio_plots_all_disciplines_sociology}
\end{figure}

\begin{figure}
	\centering
	\includegraphics[width=0.8\textwidth]{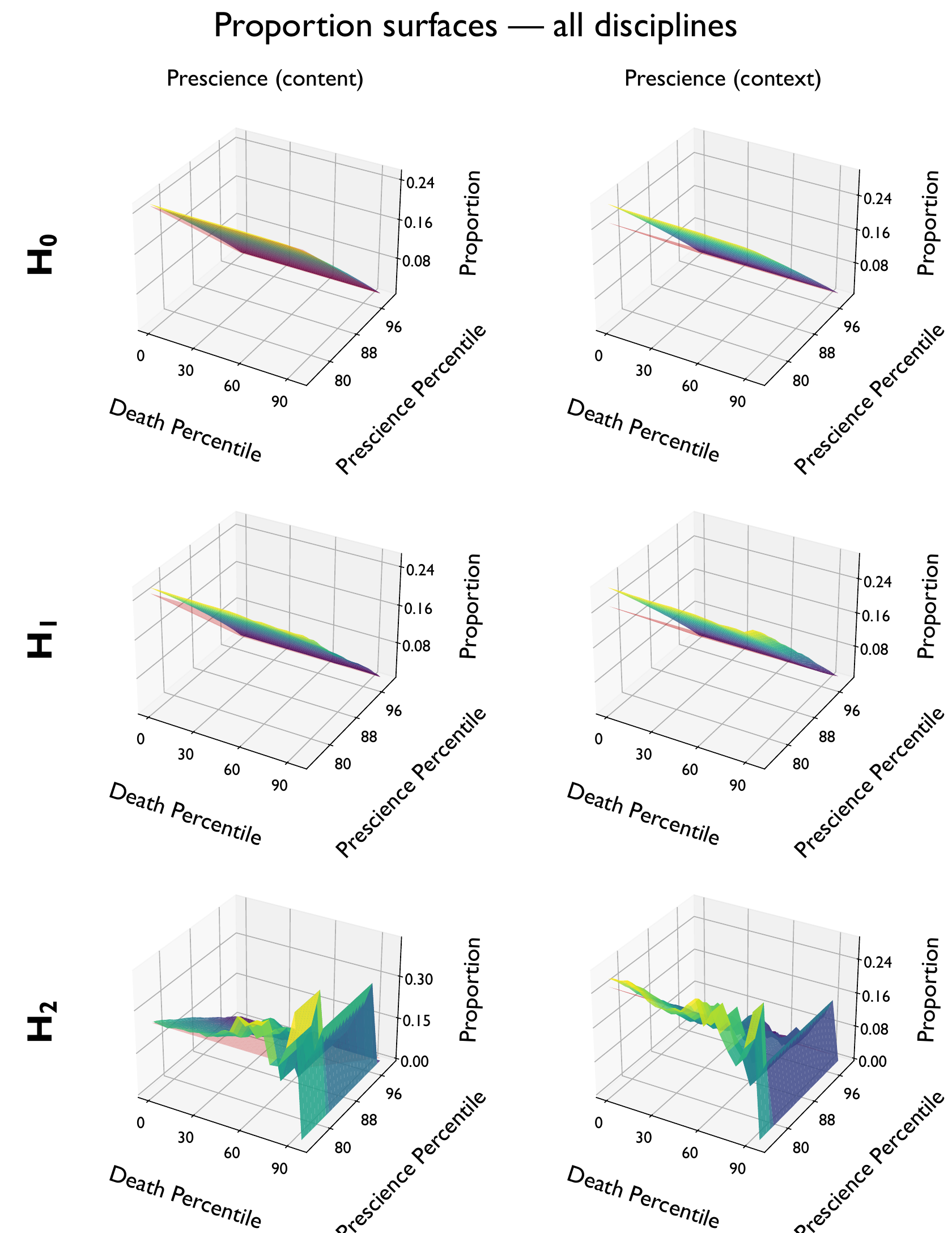}

	\caption{\textbf{3-dimensional surface plots of the proportion of hole-filling papers by prescience percentile and death percentile, pooled across all disciplines.}
    Same as Figure~\ref{fig:3d_proportion_plots}, but for content- and context-prescience.
    }
	\label{fig:3d_proportion_plots_prescience}
\end{figure}

\begin{figure}
	\centering
	\includegraphics[width=0.8\textwidth]{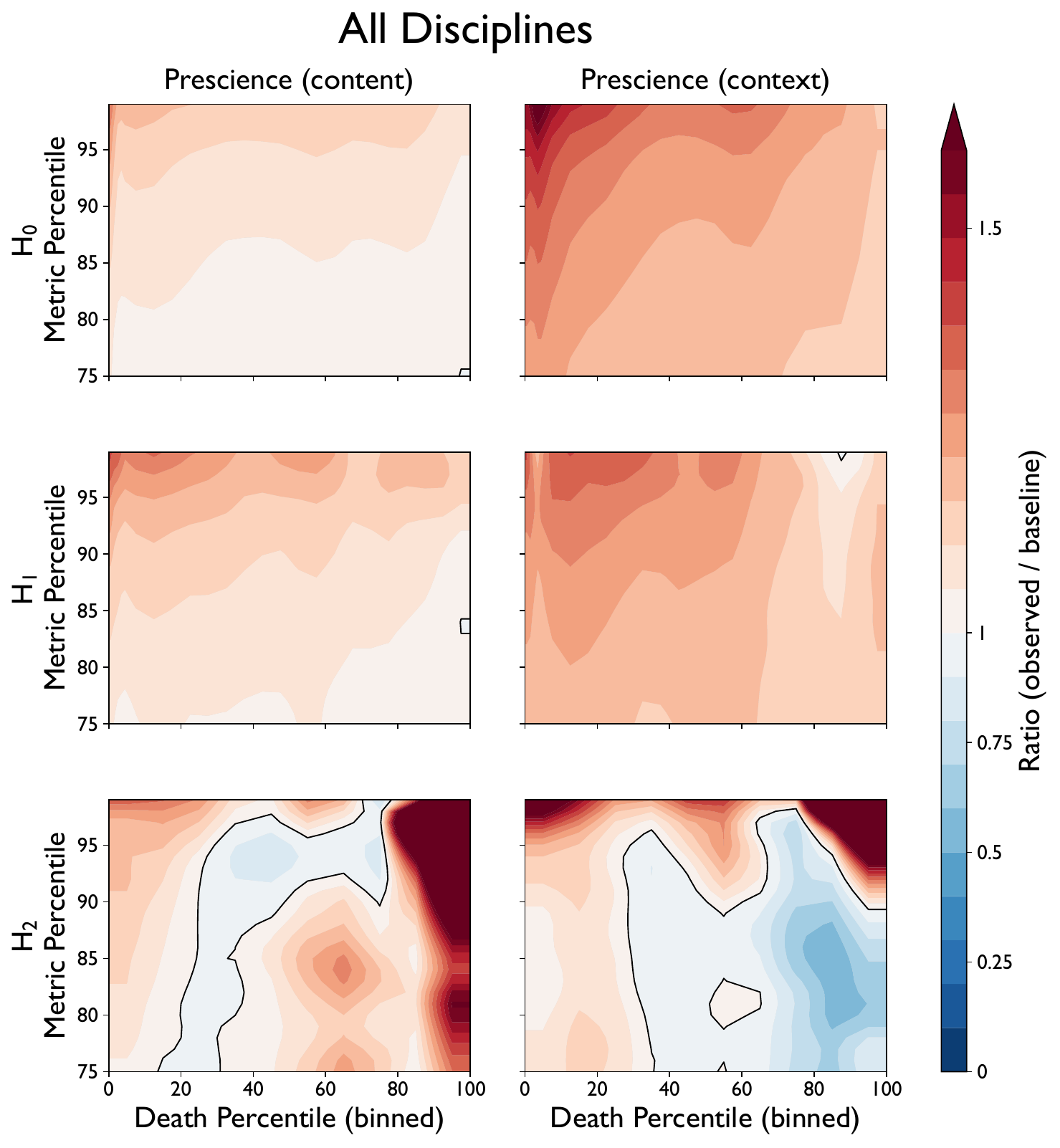}

	\caption{\textbf{Plots of the ratios between the proportion of hole-filling papers by metric percentile and death percentile and the baseline proportion, pooled across all disciplines.}
    Consider the 3D plots in Figure~\ref{fig:3d_proportion_plots_prescience}. For each metric and homology dimension, we take the ratio between the proportion of papers above a fixed percentile (yellow-to-purple surface) and the baseline proportion (red surface).
    The black contour line at $z=1$ separates the regions that are above $1$ and below $1$.
    }
	\label{fig:2D_ratio_plots_all_disciplines_prescience}
\end{figure}

\begin{figure}
	\centering
	\includegraphics[width=0.8\textwidth]{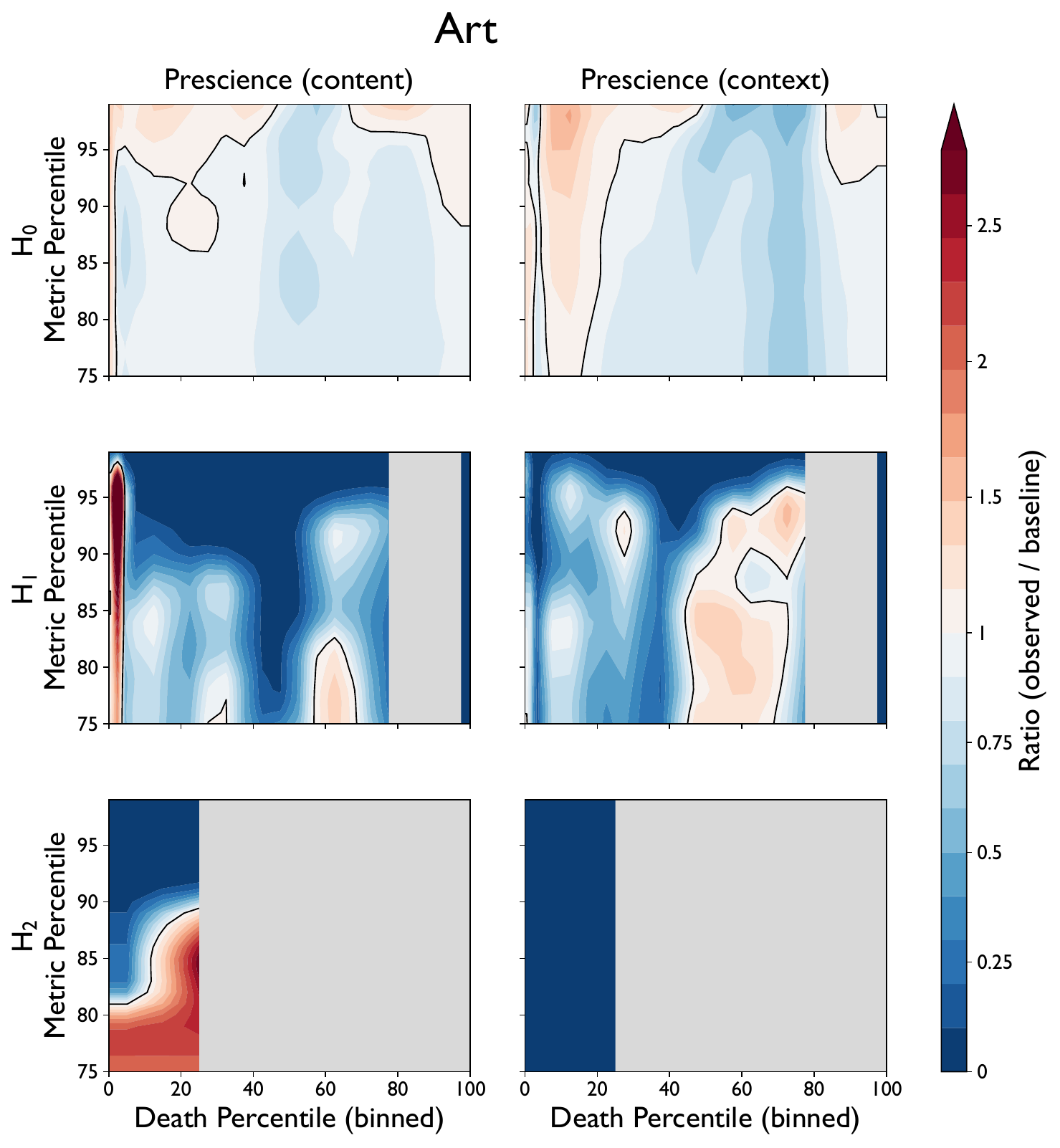}

	\caption{\textbf{Plots of the ratios between the proportion of hole-filling papers by metric percentile and death percentile and the baseline proportion, by discipline.}
    Same as Figure~\ref{fig:2D_ratio_plots_all_disciplines_prescience}, but for individual disciplines, one discipline per page. This page: art.
    }
	\label{fig:2D_ratio_plots_all_disciplines_art_prescience}
\end{figure}

\begin{figure}
	\centering
	\includegraphics[width=0.8\textwidth]{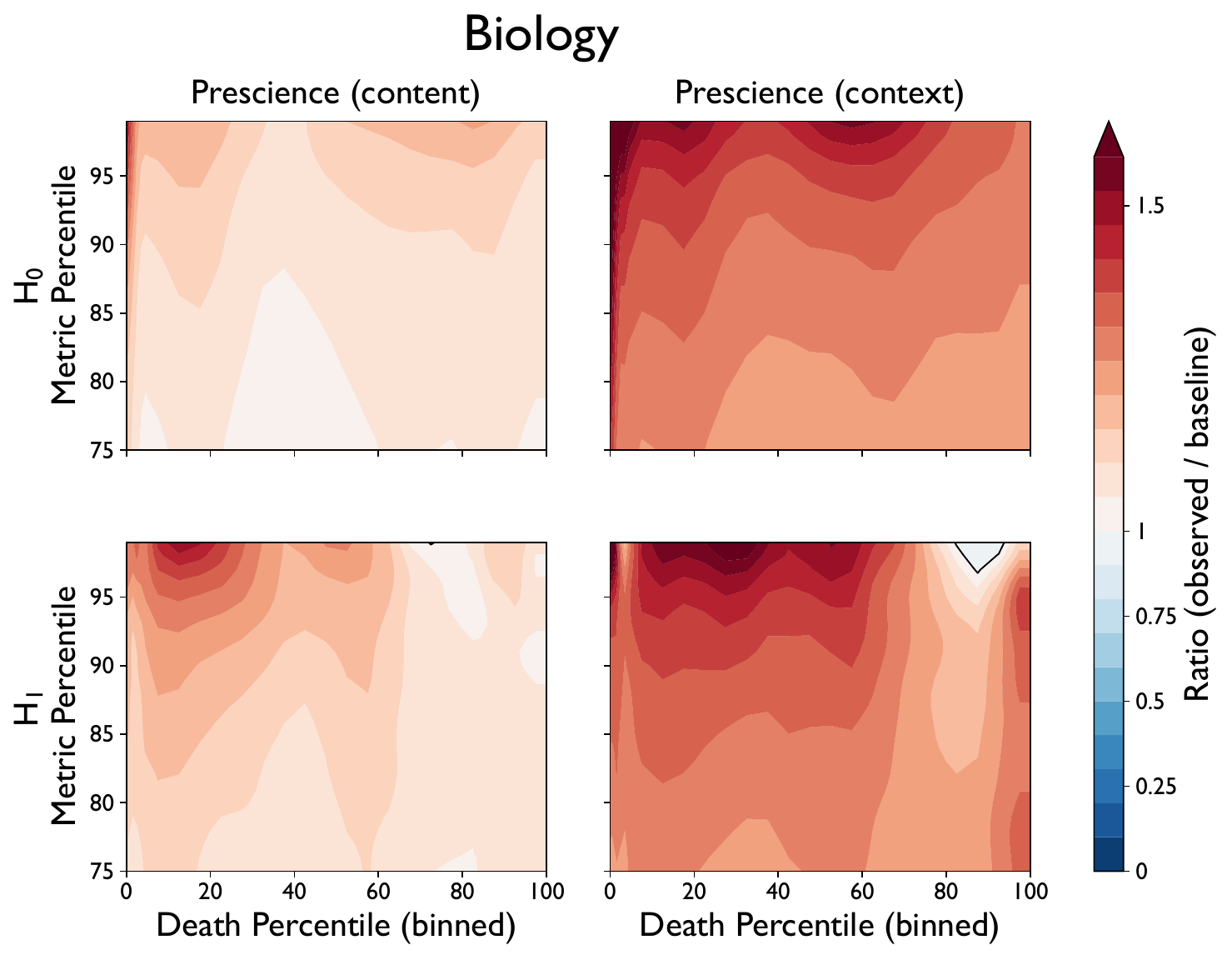}
	\continuedfigure
	\caption{\textbf{Plots of the ratios between the proportion of hole-filling papers by metric percentile and death percentile and the baseline proportion, by discipline.}
    This page: biology.
    }
	\label{fig:2D_ratio_plots_all_disciplines_biology_prescience}
\end{figure}

\begin{figure}
	\centering
	\includegraphics[width=0.8\textwidth]{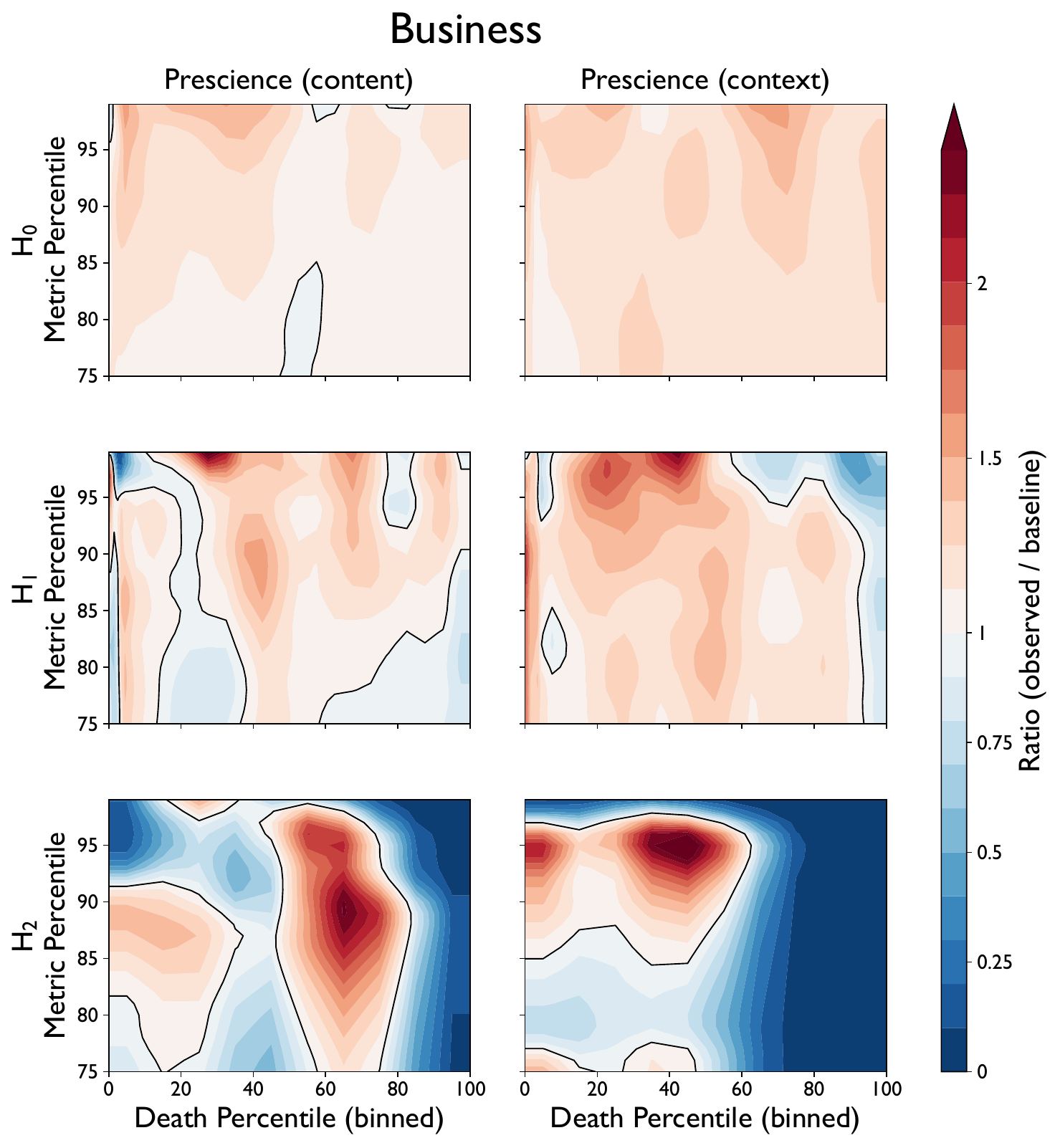}
	\continuedfigure
	\caption{\textbf{Plots of the ratios between the proportion of hole-filling papers by metric percentile and death percentile and the baseline proportion, by discipline.}
    This page: business.
    }
	\label{fig:2D_ratio_plots_all_disciplines_business_prescience}
\end{figure}

\begin{figure}
	\centering
	\includegraphics[width=0.8\textwidth]{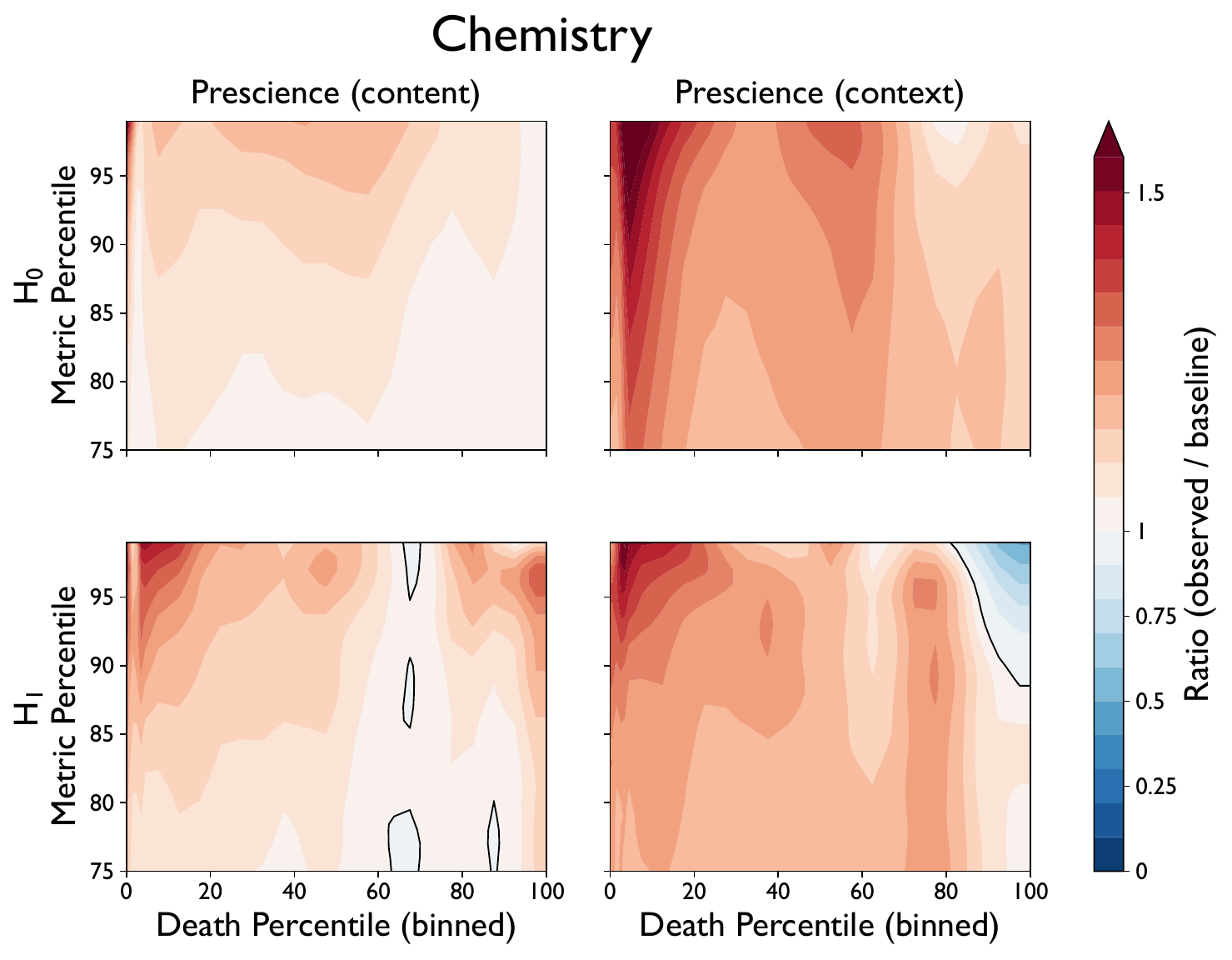}
	\continuedfigure
	\caption{\textbf{Plots of the ratios between the proportion of hole-filling papers by metric percentile and death percentile and the baseline proportion, by discipline.}
    This page: chemistry.
    }
	\label{fig:2D_ratio_plots_all_disciplines_chemistry_prescience}
\end{figure}

\begin{figure}
	\centering
	\includegraphics[width=0.8\textwidth]{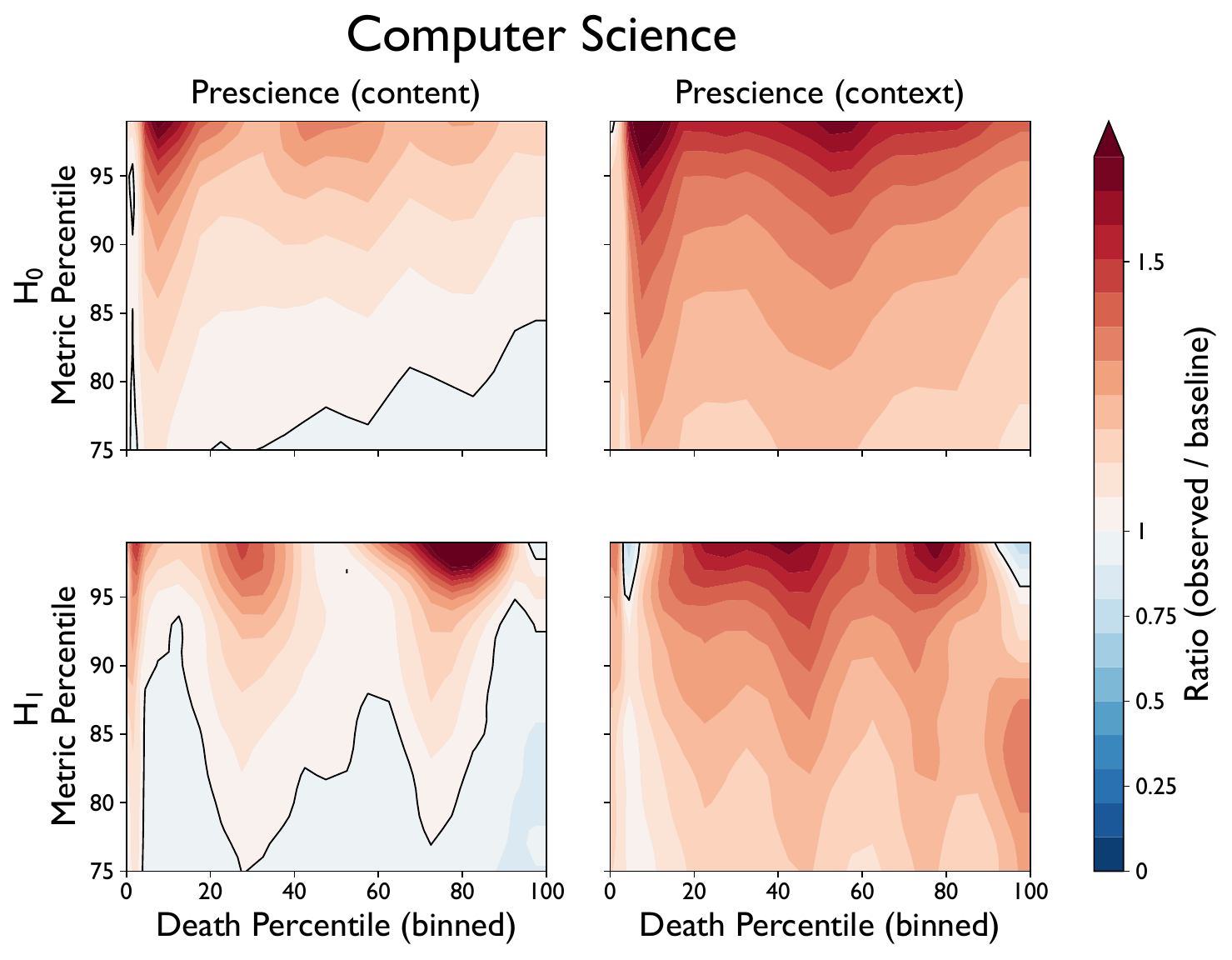}
	\continuedfigure
	\caption{\textbf{Plots of the ratios between the proportion of hole-filling papers by metric percentile and death percentile and the baseline proportion, by discipline.}
    This page: computer science.
    }
	\label{fig:2D_ratio_plots_all_disciplines_computer_science_prescience}
\end{figure}

\begin{figure}
	\centering
	\includegraphics[width=0.8\textwidth]{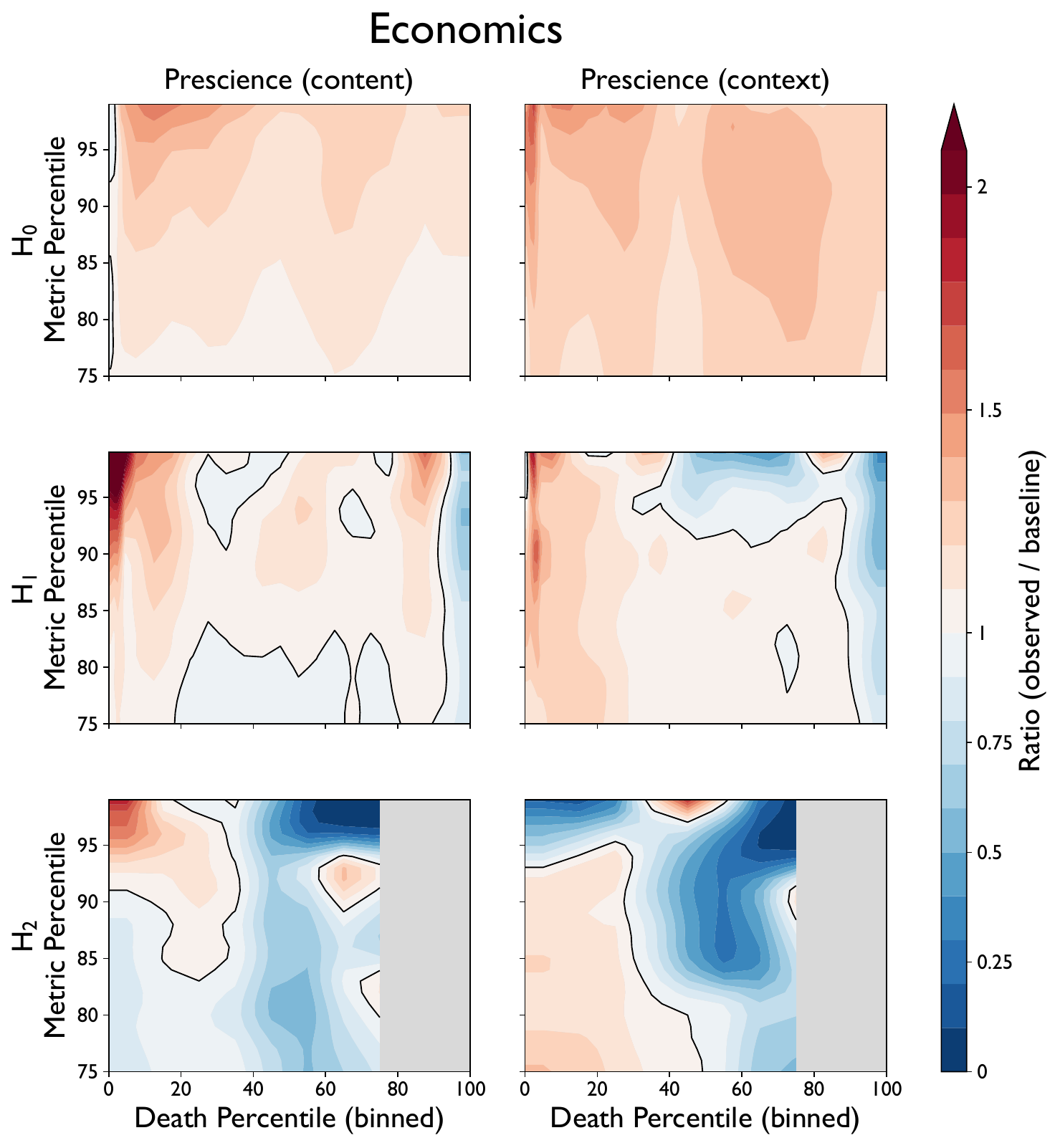}
	\continuedfigure
	\caption{\textbf{Plots of the ratios between the proportion of hole-filling papers by metric percentile and death percentile and the baseline proportion, by discipline.}
    This page: economics.
    }
	\label{fig:2D_ratio_plots_all_disciplines_economics_prescience}
\end{figure}

\clearpage

\begin{figure}
	\centering
	\includegraphics[width=0.8\textwidth]{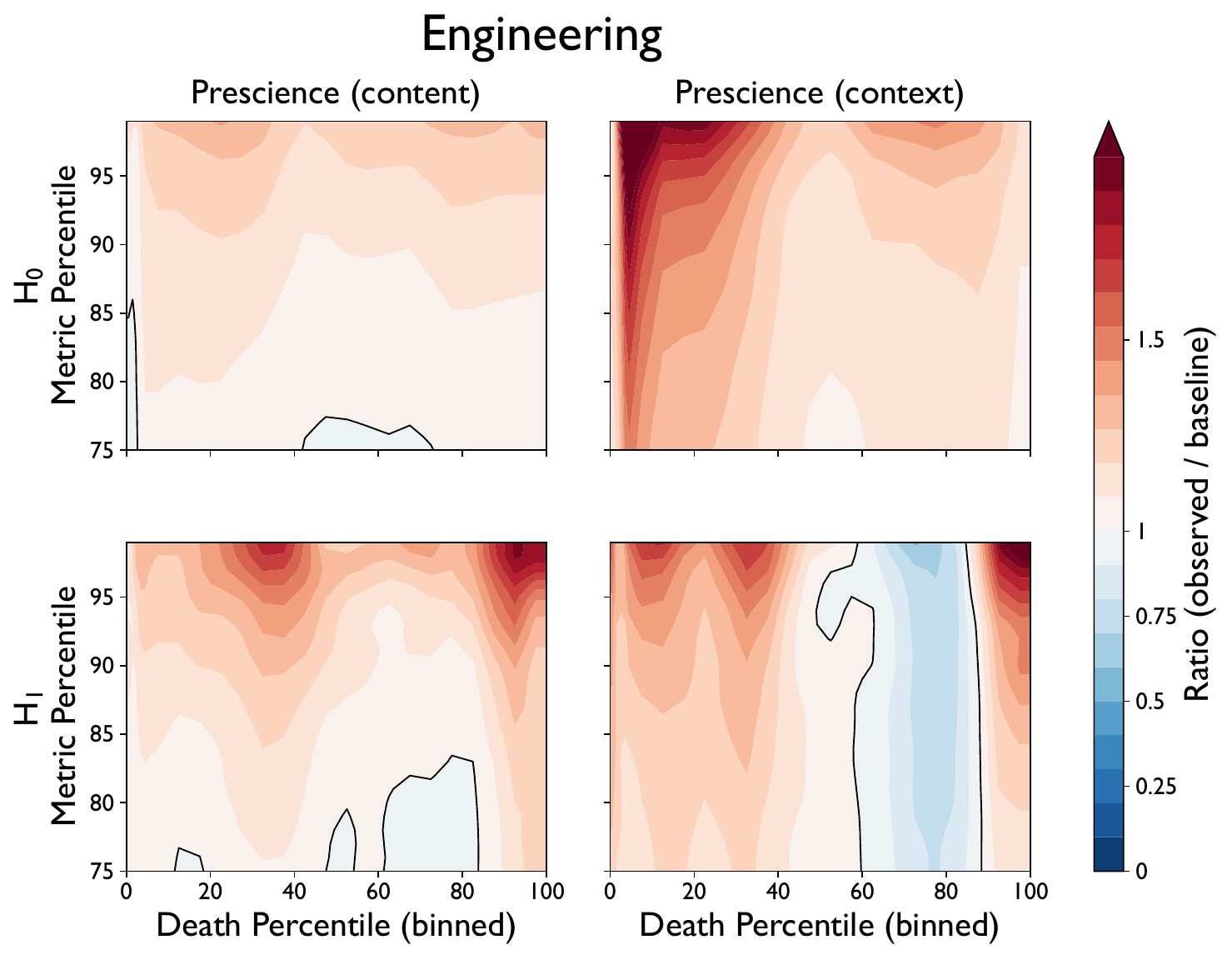}
	\continuedfigure
	\caption{\textbf{Plots of the ratios between the proportion of hole-filling papers by metric percentile and death percentile and the baseline proportion, by discipline.}
    This page: engineering.
    }
	\label{fig:2D_ratio_plots_all_disciplines_engineering_prescience}
\end{figure}

\begin{figure}
	\centering
	\includegraphics[width=0.8\textwidth]{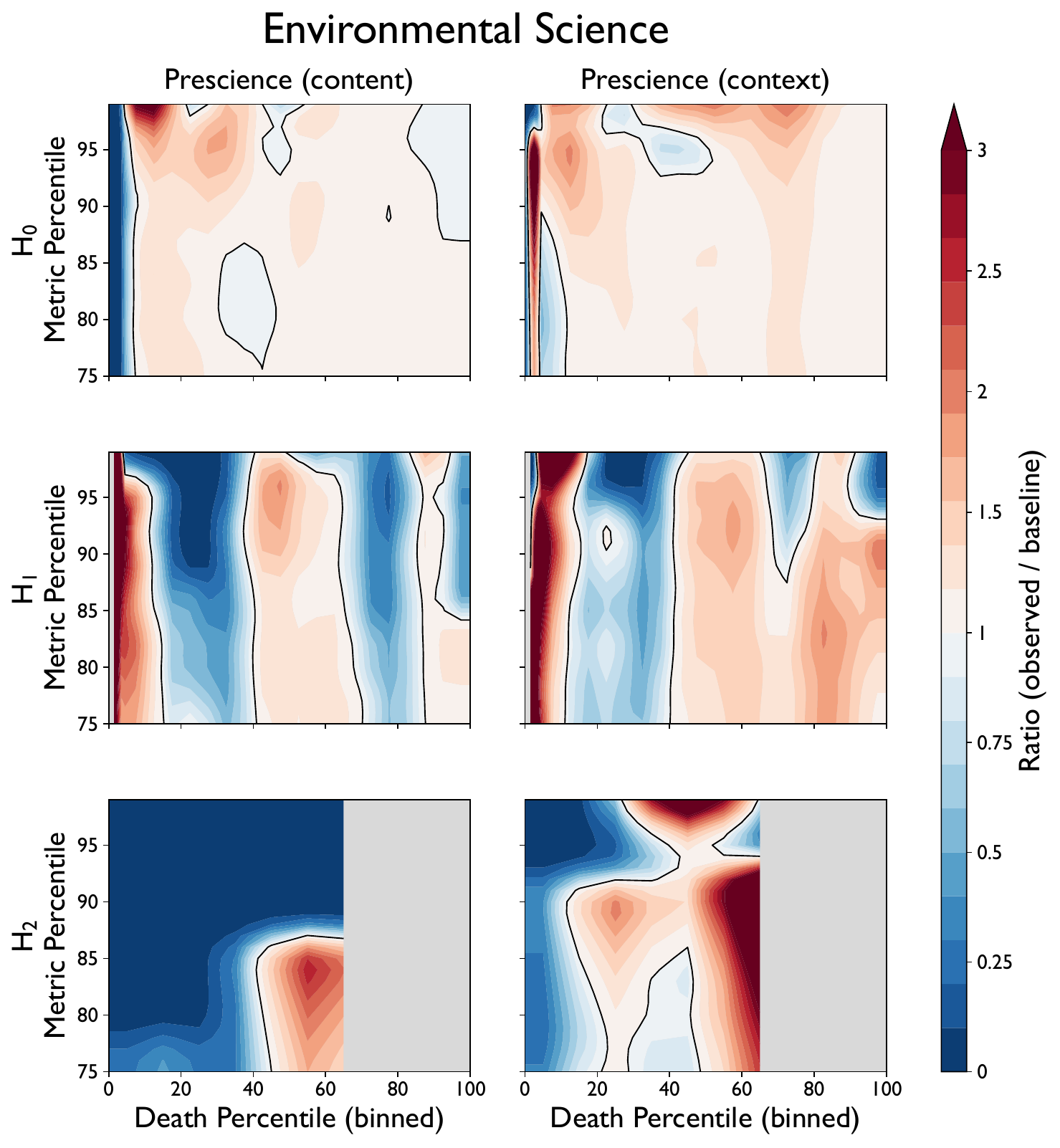}
	\continuedfigure
	\caption{\textbf{Plots of the ratios between the proportion of hole-filling papers by metric percentile and death percentile and the baseline proportion, by discipline.}
    This page: environmental science.
    }
	\label{fig:2D_ratio_plots_all_disciplines_environmental_science_prescience}
\end{figure}

\begin{figure}
	\centering
	\includegraphics[width=0.8\textwidth]{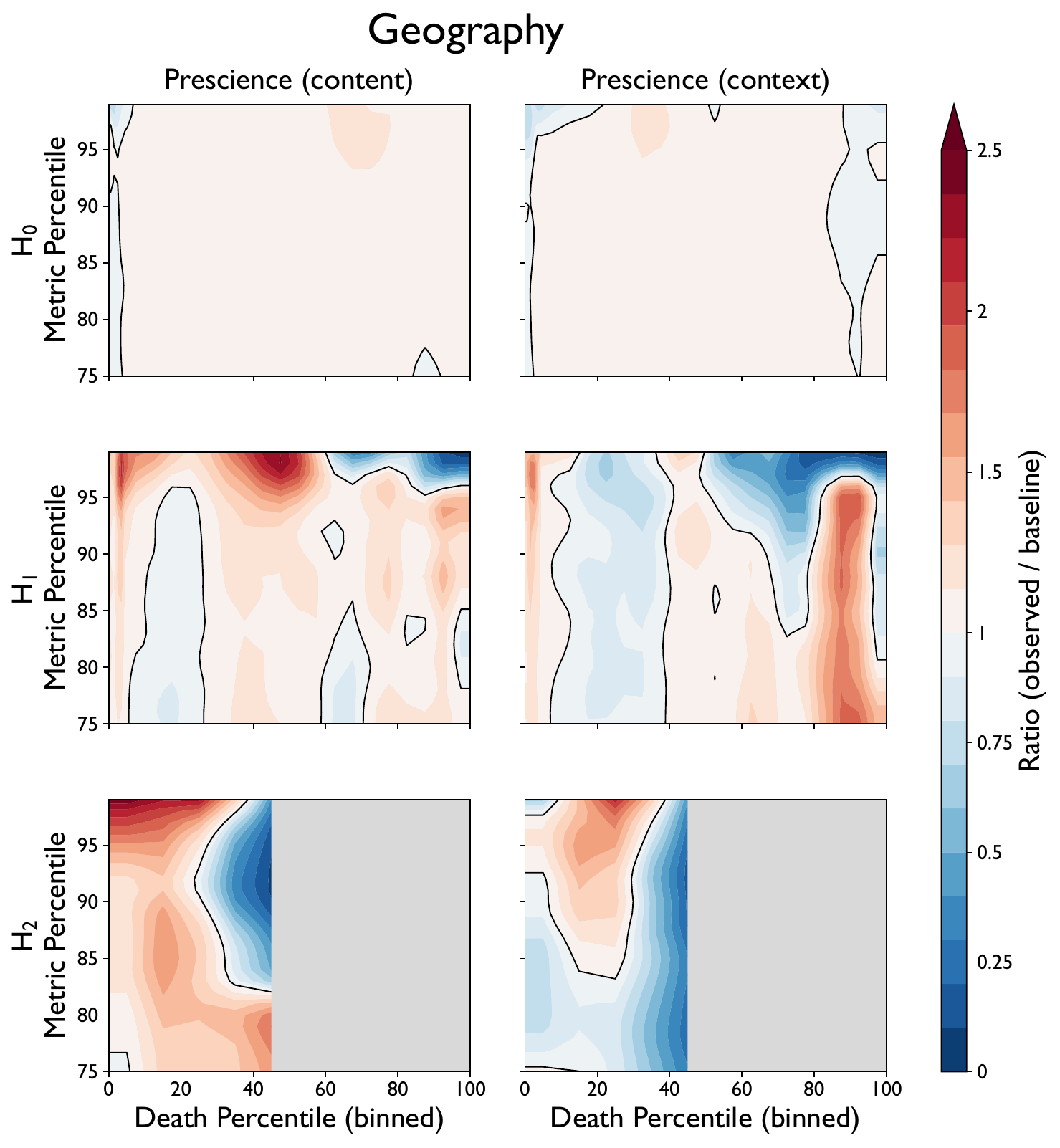}
	\continuedfigure
	\caption{\textbf{Plots of the ratios between the proportion of hole-filling papers by metric percentile and death percentile and the baseline proportion, by discipline.}
    This page: geography.
    }
	\label{fig:2D_ratio_plots_all_disciplines_geography_prescience}
\end{figure}

\begin{figure}
	\centering
	\includegraphics[width=0.8\textwidth]{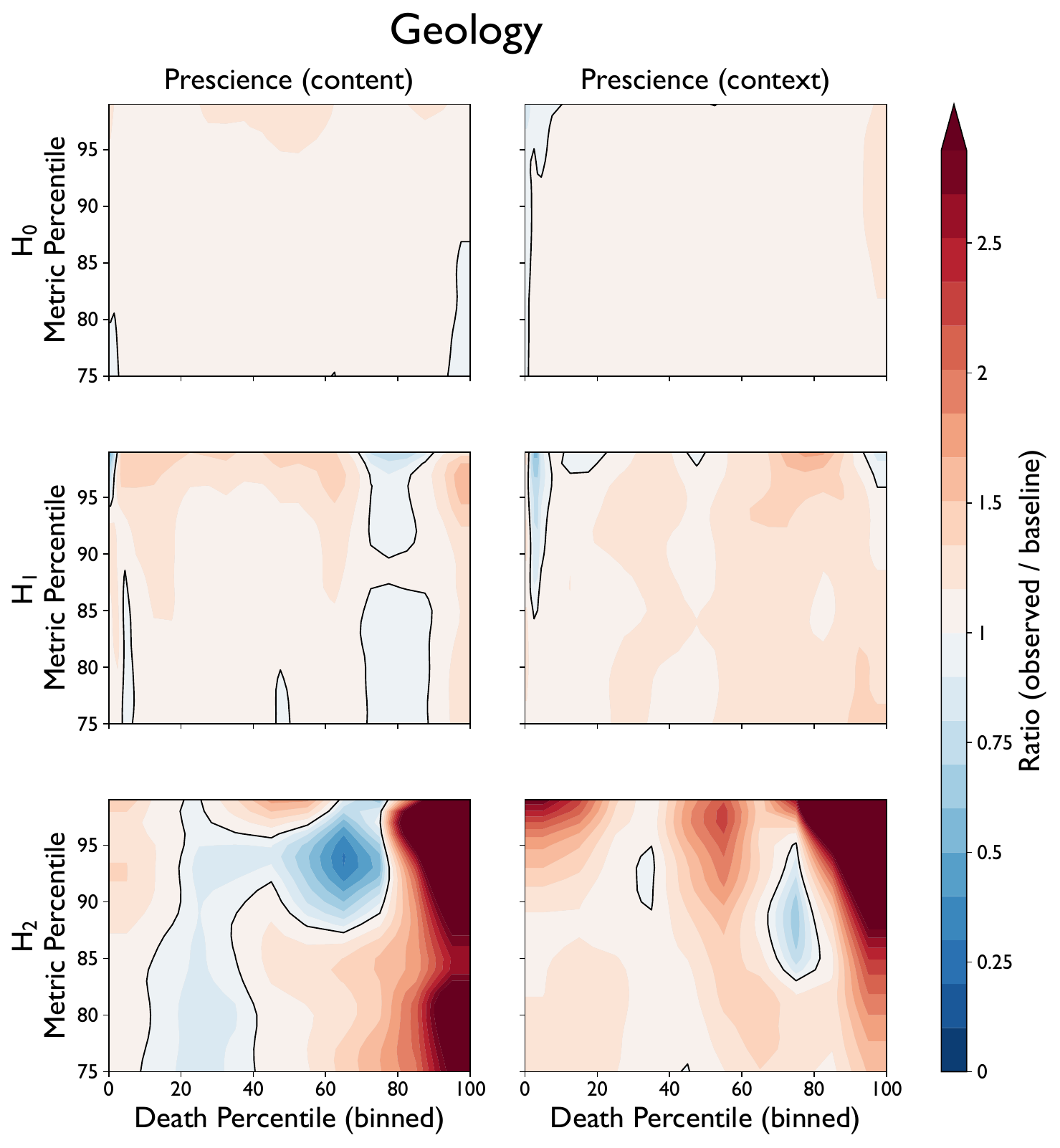}
	\continuedfigure
	\caption{\textbf{Plots of the ratios between the proportion of hole-filling papers by metric percentile and death percentile and the baseline proportion, by discipline.}
    This page: geology.
    }
	\label{fig:2D_ratio_plots_all_disciplines_geology_prescience}
\end{figure}

\begin{figure}
	\centering
	\includegraphics[width=0.8\textwidth]{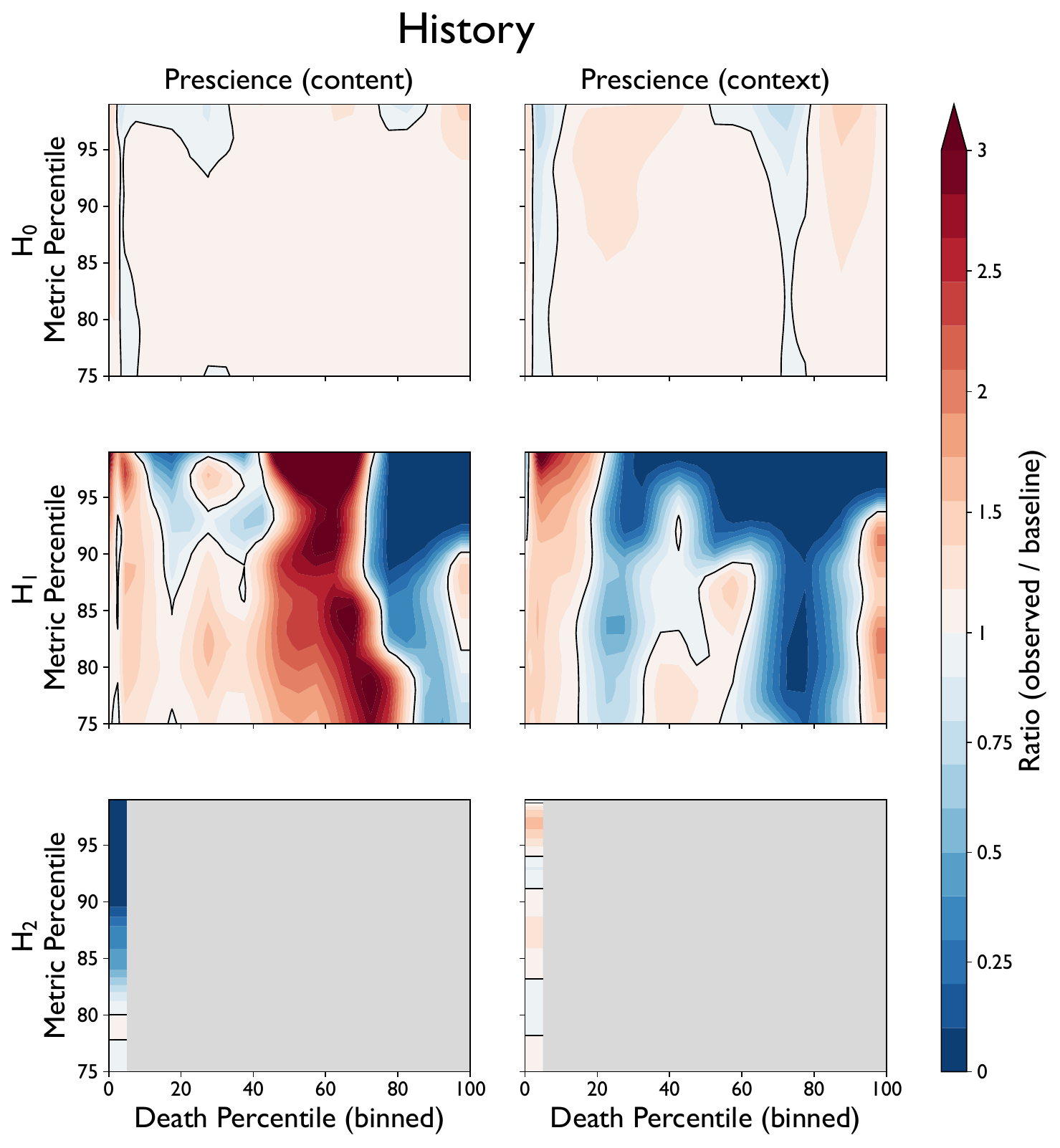}
	\continuedfigure
	\caption{\textbf{Plots of the ratios between the proportion of hole-filling papers by metric percentile and death percentile and the baseline proportion, by discipline.}
    This page: history.
    }
	\label{fig:2D_ratio_plots_all_disciplines_history_prescience}
\end{figure}

\clearpage

\begin{figure}
	\centering
	\includegraphics[width=0.8\textwidth]{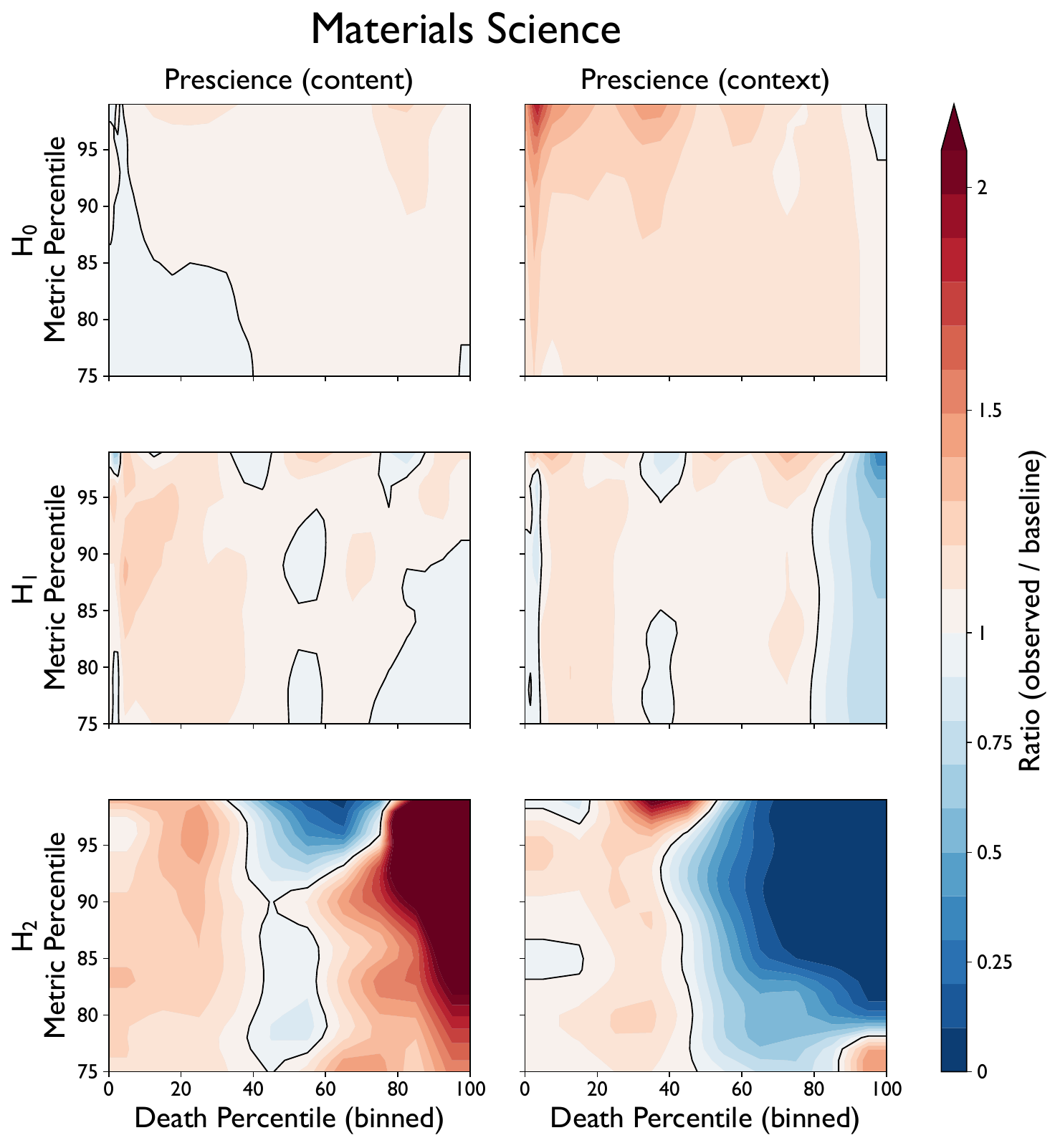}
	\continuedfigure
	\caption{\textbf{Plots of the ratios between the proportion of hole-filling papers by metric percentile and death percentile and the baseline proportion, by discipline.}
    This page: materials science.
    }
	\label{fig:2D_ratio_plots_all_disciplines_materials_science_prescience}
\end{figure}

\begin{figure}
	\centering
	\includegraphics[width=0.8\textwidth]{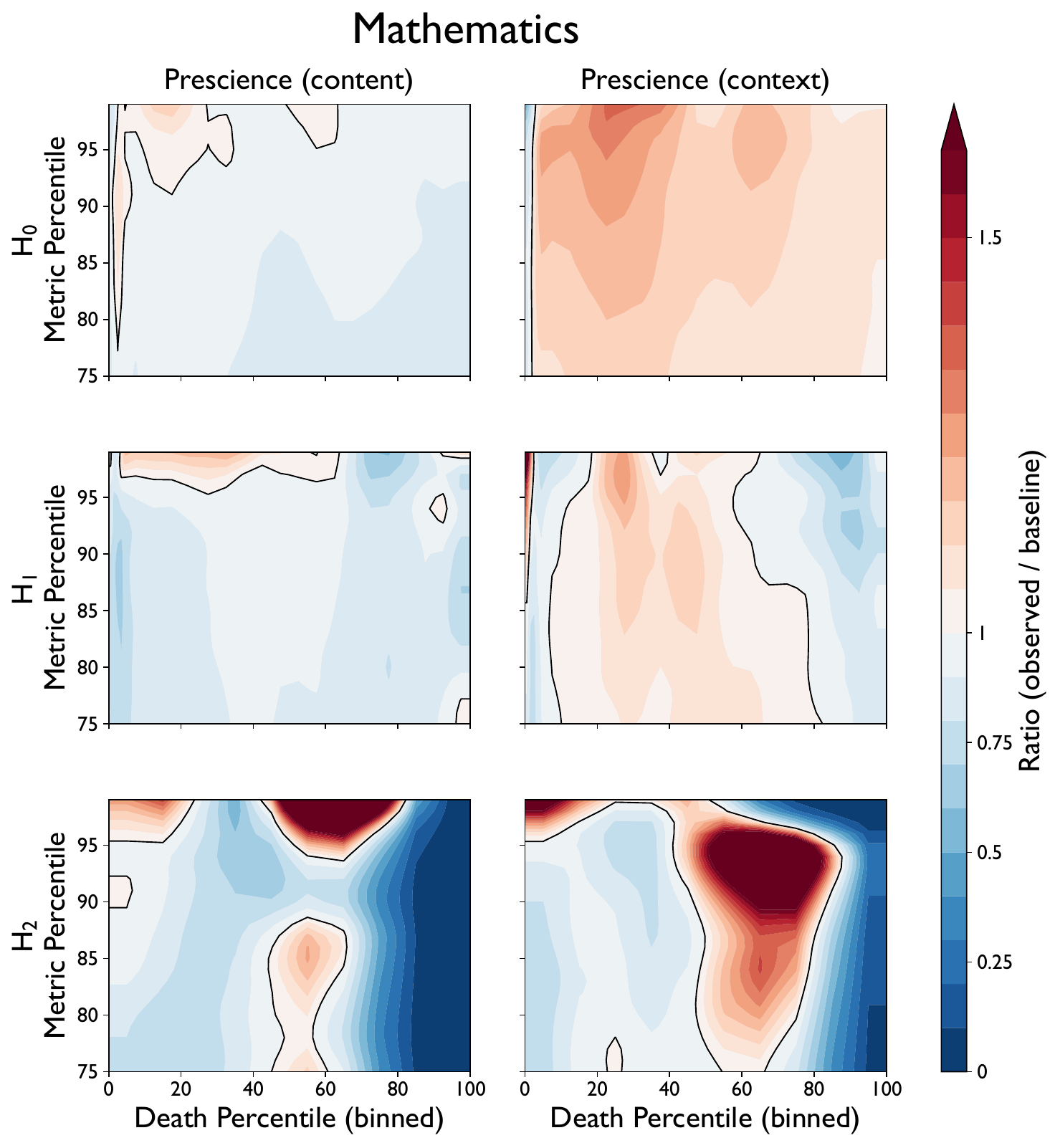}
	\continuedfigure
	\caption{\textbf{Plots of the ratios between the proportion of hole-filling papers by metric percentile and death percentile and the baseline proportion, by discipline.}
    This page: mathematics.
    }
	\label{fig:2D_ratio_plots_all_disciplines_mathematics_prescience}
\end{figure}

\begin{figure}
	\centering
	\includegraphics[width=0.8\textwidth]{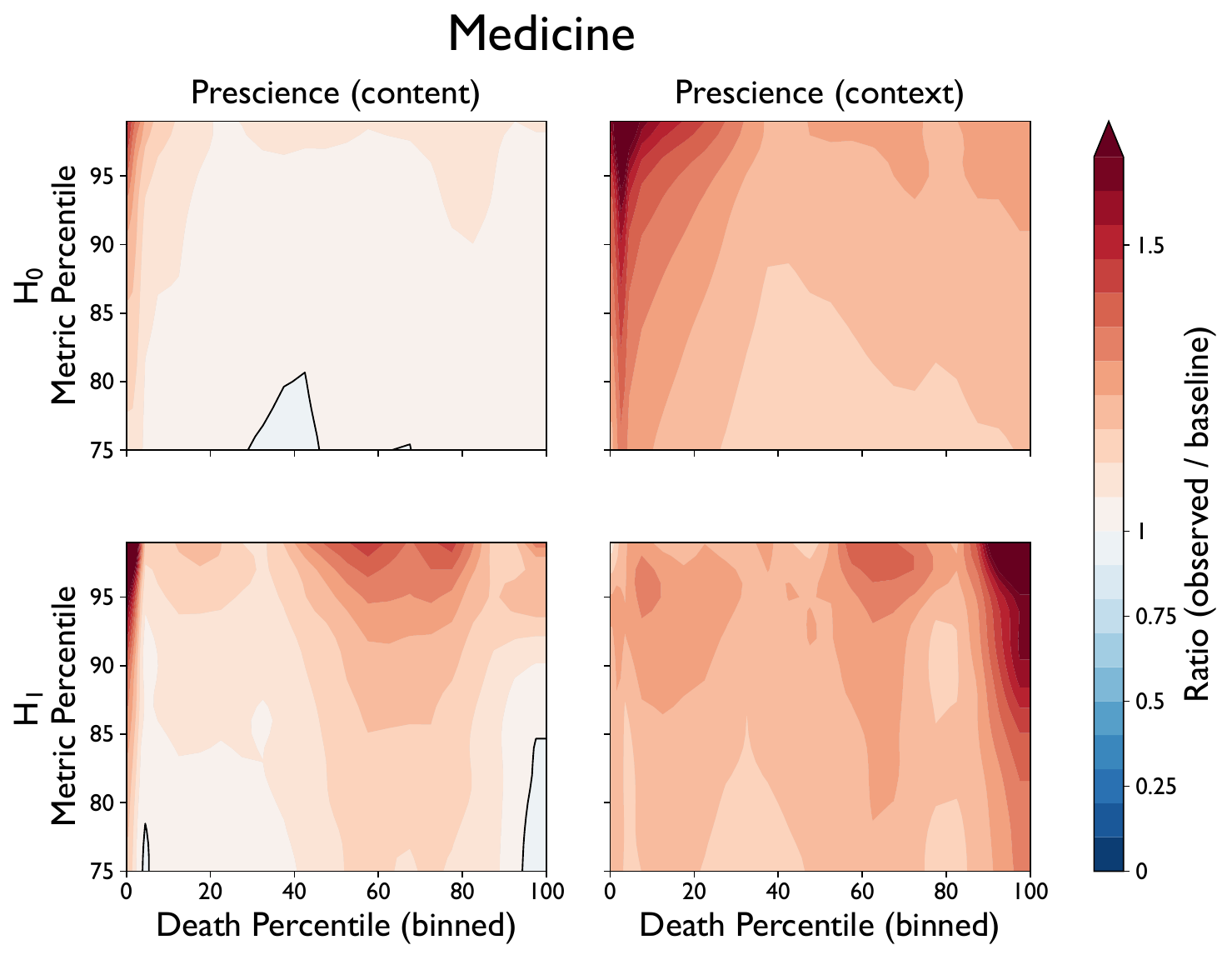}
	\continuedfigure
	\caption{\textbf{Plots of the ratios between the proportion of hole-filling papers by metric percentile and death percentile and the baseline proportion, by discipline.}
    This page: medicine.
    }
	\label{fig:2D_ratio_plots_all_disciplines_medicine_prescience}
\end{figure}

\begin{figure}
	\centering
	\includegraphics[width=0.8\textwidth]{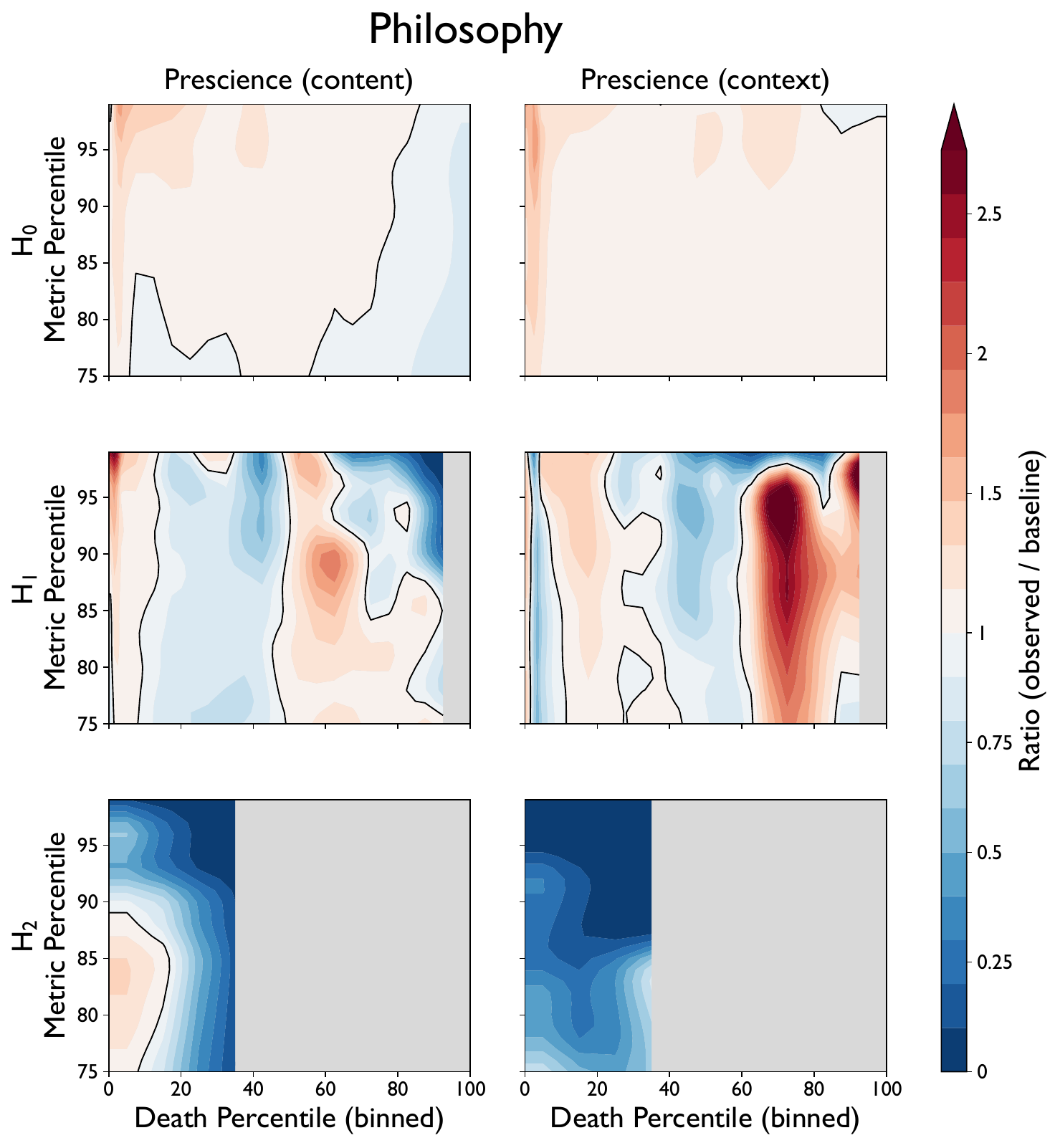}
	\continuedfigure
	\caption{\textbf{Plots of the ratios between the proportion of hole-filling papers by metric percentile and death percentile and the baseline proportion, by discipline.}
    This page: philosophy.
    }
	\label{fig:2D_ratio_plots_all_disciplines_philosophy_prescience}
\end{figure}

\begin{figure}
	\centering
	\includegraphics[width=0.8\textwidth]{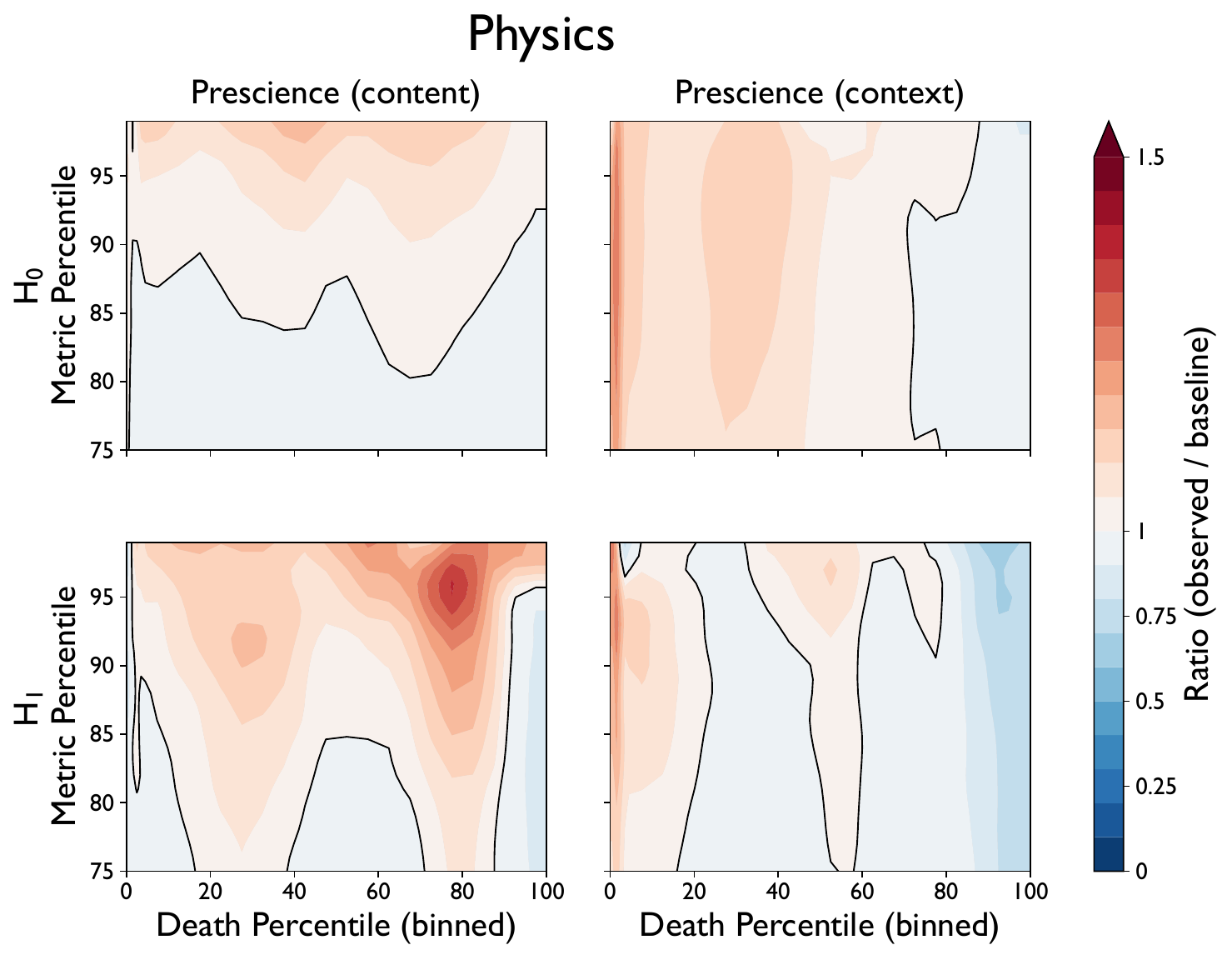}
	\continuedfigure
	\caption{\textbf{Plots of the ratios between the proportion of hole-filling papers by metric percentile and death percentile and the baseline proportion, by discipline.}
    This page: physics.
    }
	\label{fig:2D_ratio_plots_all_disciplines_physics_prescience}
\end{figure}

\clearpage

\begin{figure}
	\centering
	\includegraphics[width=0.8\textwidth]{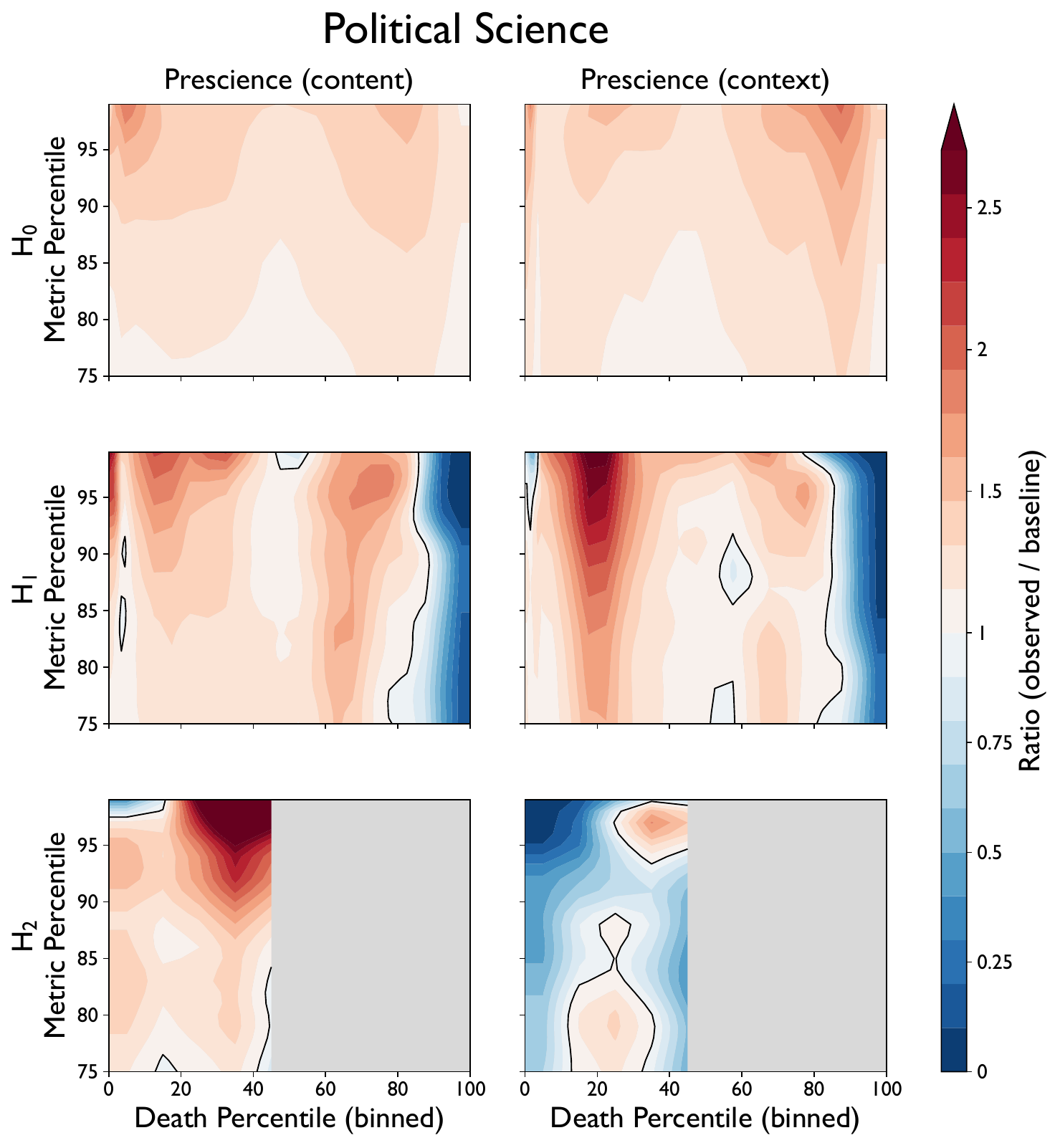}
	\continuedfigure
	\caption{\textbf{Plots of the ratios between the proportion of hole-filling papers by metric percentile and death percentile and the baseline proportion, by discipline.}
    This page: political science.
    }
	\label{fig:2D_ratio_plots_all_disciplines_political_science_prescience}
\end{figure}

\begin{figure}
	\centering
	\includegraphics[width=0.8\textwidth]{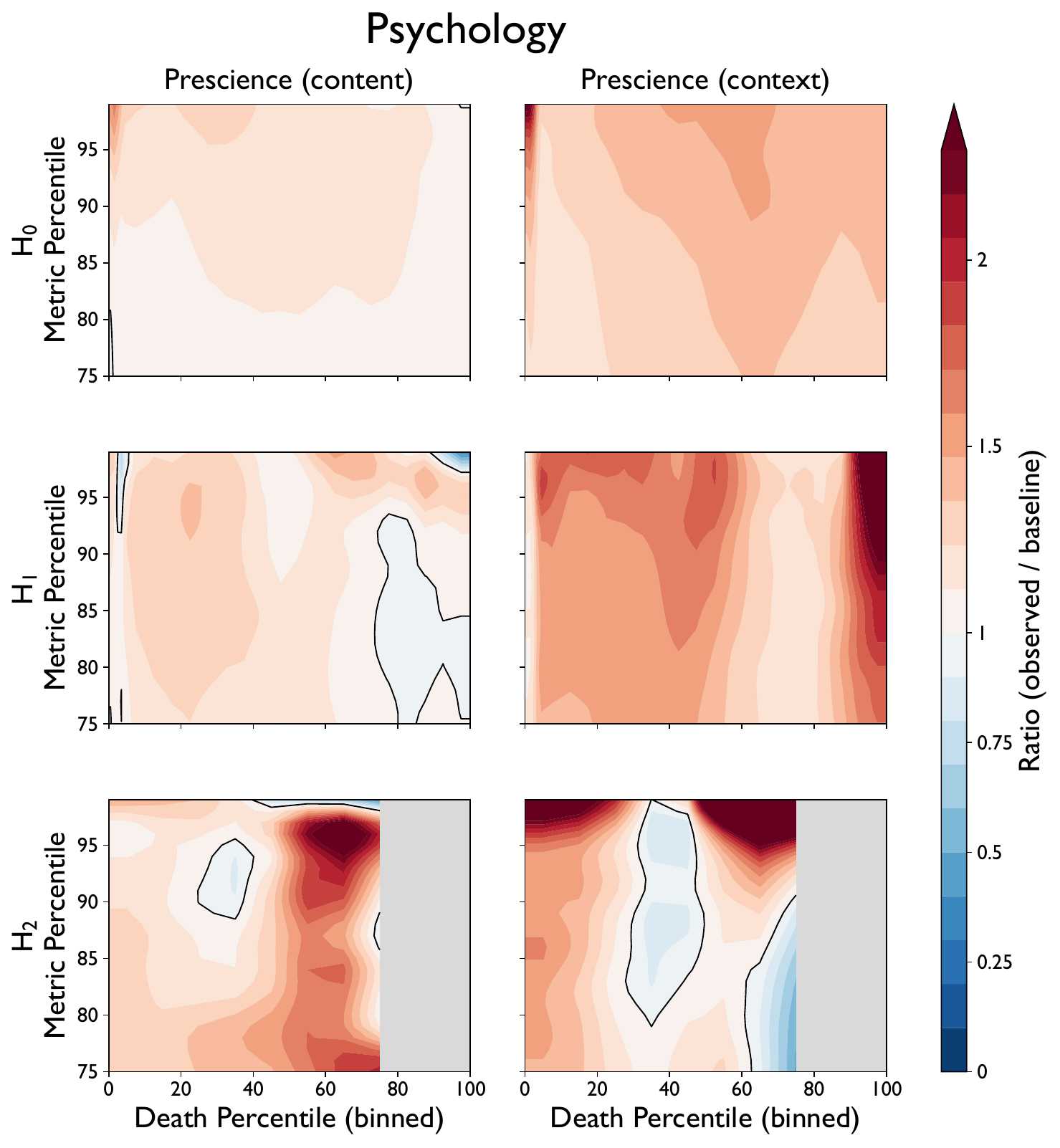}
	\continuedfigure
	\caption{\textbf{Plots of the ratios between the proportion of hole-filling papers by metric percentile and death percentile and the baseline proportion, by discipline.}
    This page: psychology.
    }
	\label{fig:2D_ratio_plots_all_disciplines_psychology_prescience}
\end{figure}

\begin{figure}
	\centering
	\includegraphics[width=0.8\textwidth]{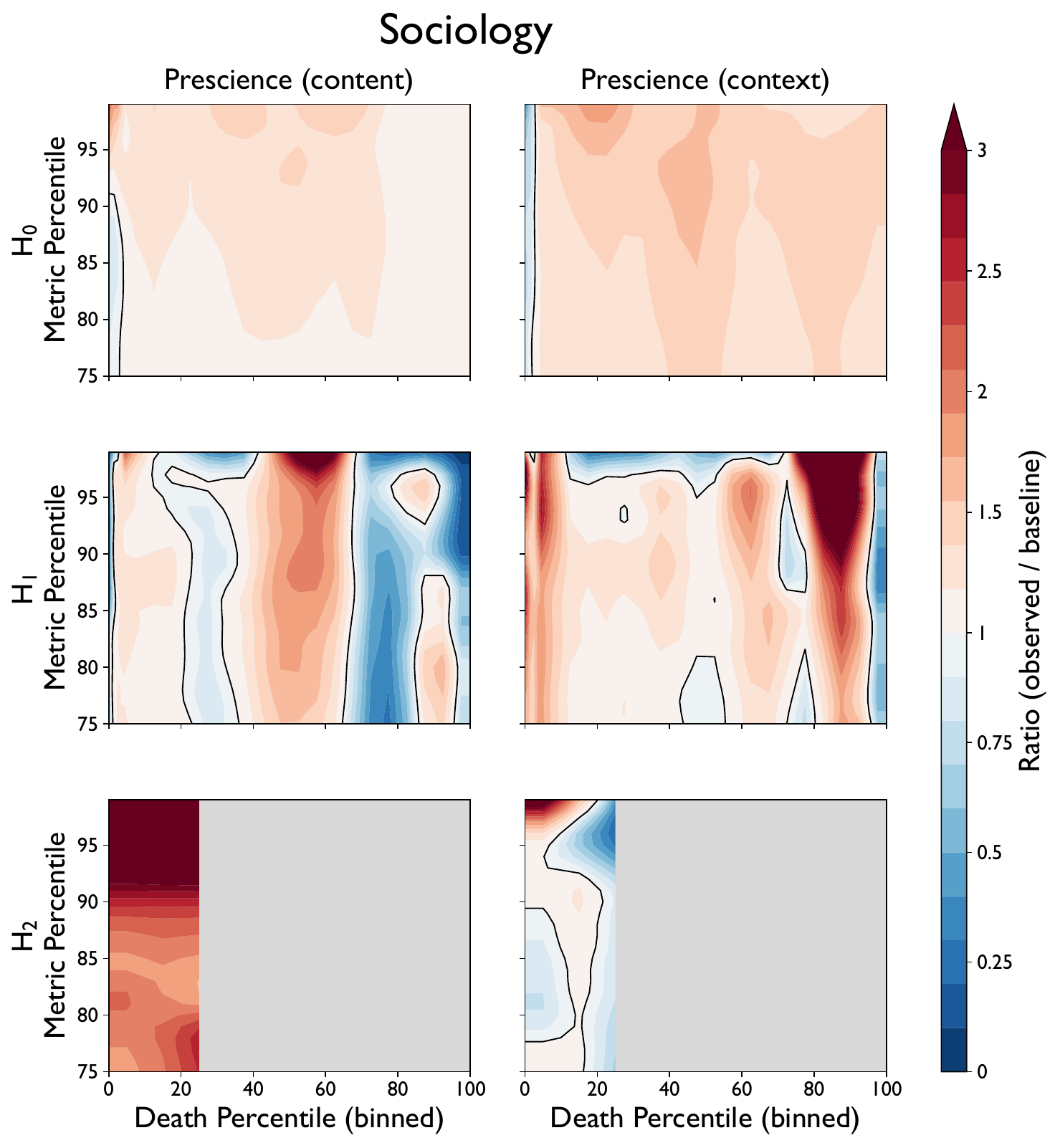}
	\continuedfigure
	\caption{\textbf{Plots of the ratios between the proportion of hole-filling papers by metric percentile and death percentile and the baseline proportion, by discipline.}
    This page: sociology.
    }
	\label{fig:2D_ratio_plots_all_disciplines_sociology_prescience}
\end{figure}

\end{document}